\documentclass[longauth]{aa} % for the long lists of affiliations 
\usepackage{amsmath}
\usepackage{graphicx}
\usepackage{txfonts}
\usepackage{natbib}
\usepackage{float}
\usepackage{lscape}
\usepackage{nicefrac}

\begin{document}
   \title{Water in star forming regions with \textit{Herschel} (WISH) \\
   III. Far-infrared cooling lines in low-mass young stellar objects}

   \author{A. Karska\inst{1,2}, G.J.~Herczeg\inst{1,3}, E.F.~van Dishoeck\inst{1,2},
   		   S.F.~Wampfler\inst{4,5}, L.E.~Kristensen\inst{2}, J.R.~Goicoechea\inst{6}, 
   		   R.~Visser\inst{7}, B.~Nisini\inst{8}, I.~San-Jose Garcia\inst{2}, S.~Bruderer\inst{1}, 
   		   P. \'Sniady\inst{11,12},	S.~Doty\inst{13}, D.~Fedele\inst{1},
   		   U.A.~Y{\i}ld{\i}z\inst{2}, A.O.~Benz\inst{4}, E.~Bergin\inst{7}, P.~Caselli\inst{9,10}, 
   		   F.~Herpin\inst{14,15}, M.R.~Hogerheijde\inst{2}, D.~Johnstone\inst{16,17}, J.K.~J{\o}rgensen\inst{5},
   		   R.~Liseau\inst{18}, M.~Tafalla\inst{19}, F.~van der Tak\inst{20,21}, and F.~Wyrowski\inst{22}}

   	\institute{
   			  $^{1}$ Max-Planck Institut f\"{u}r Extraterrestrische Physik (MPE),
   			  Giessenbachstr. 1, D-85748 Garching, Germany\\
   			  $^{2}$ Leiden Observatory, Leiden University, P.O. Box 9513,
          	  2300 RA Leiden, The Netherlands\\
          	  $^{3}$ Kavli Institut for Astronomy and Astrophysics, Yi He Yuan Lu 5, HaiDian Qu, Peking University, Beijing,
          	  100871, PR China\\
          	  $^{4}$ Institute of Astronomy, ETH Zurich, 8093 Zurich, Switzerland\\
          	  $^{5}$ Centre for Star and Planet Formation, Natural History Museum of Denmark, University of
				Copenhagen, {\O}ster Voldgade 5-7, DK-1350 Copenhagen K., Denmark\\
          	  $^{6}$ Centro de Astrobiolog\'{\i}a. Departamento de Astrof\'{\i}sica. CSIC-INTA. Carretera de Ajalvir,
				Km 4, Torrej\'{o}n de Ardoz. 28850, Madrid, Spain\\
              $^{7}$ Department of Astronomy, The University of Michigan, 500 Church Street, Ann Arbor, MI 48109-1042,
				USA \\
              $^{8}$ INAF - Osservatorio Astronomico di Roma, 00040 Monte Porzio catone, Italy \\
              $^{9}$ School of Physics and Astronomy, University of Leeds, Leeds LS2 9JT, UK \\
              $^{10}$ INAF - Osservatorio Astrofisico di Arcetri, Largo E. Fermi 5, 50125 Firenze, Italy\\
             $^{11}$  Institute of Mathematics, Polish Academy of Sciences, \mbox{ul.~\'Sniadec\-kich 8,} 
				00-956 Warszawa, Poland \\
 			$^{12}$ Institute of Mathematics, University of Wroclaw, \mbox{pl.\ Grunwaldzki~2/4,} 
			50-384 Wroclaw, Poland  \\
              $^{13}$ Department of Physics and Astronomy, Denison University, Granville, OH, 43023, USA\\
              $^{14}$ Universit\'{e} de Bordeaux, Observatoire Aquitain des Sciences de l'Univers, 2 rue de
l'Observatoire, BP 89, F-33271 Floirac Cedex, France \\
			  $^{15}$ CNRS, LAB, UMR 5804, F-33271 Floirac Cedex, France \\
              $^{16}$ National Research Council Canada, Herzberg Institute of Astrophysics, 5071 West Saanich Road,
Victoria, BC V9E 2E7, Canada \\
              $^{17}$ Department of Physics and Astronomy, University of Victoria, Victoria, BC V8P 1A1, Canada\\
              $^{18}$ Department of Radio and Space Science, Chalmers University of Technology, Onsala Space
Observatory, 439 92 Onsala, Sweden\\
              $^{19}$ Observatorio Astron\'{o}mico Nacional (IGN), Calle Alfonso XII,3. 28014, Madrid, Spain\\
               $^{20}$ SRON Netherlands Institute for Space Research, PO Box 800, 9700 AV, Groningen, The Netherlands\\
              $^{21}$ Kapteyn Astronomical Institute, University of Groningen, PO Box 800, 9700 AV, Groningen, The
Netherlands\\
              $^{22}$ Max-Planck-Institut f\"{u}r Radioastronomie, Auf dem H\"{u}gel 69, 53121 Bonn, Germany\\
             \email{karska@mpe.mpg.de}
             }
             
%             \thanks{The university of heaven temporarily does not
%                     accept e-mails}
   \date{Received July 16, 2012; accepted January 16, 2013}
	\titlerunning{Herschel PACS observations of low-mass YSOs}
	\authorrunning{A.~Karska et al. 2012}

% \abstract{}{}{}{}
% 5 {} token are mandatory
  \abstract   
  % context
  {}
  {Our aims are to quantify the far-infrared line emission from
    low-mass protostars and the contribution of different atomic and
    molecular species to the gas cooling budget, to determine the
    spatial extent of the emission and to investigate the underlying
    excitation conditions. Analysis of the line cooling will help us
     to characterize the evolution of the relevant physical processes as 
     the protostar ages.}
  % methods (mandatory)
  {Far-infrared Herschel-PACS spectra of 18 low-mass protostars of
    various luminosities and evolutionary stages are studied in
    the context of the WISH key program.}
   % For most targets, the spectra include many wavelength
   %  intervals selected to cover specific CO, H$_2$O, OH and atomic lines. 
   %  For four targets the spectra span the entire 55-200 $\mu$m region.
   %  The PACS field-of-view covers $\sim47''$ with the resolution of
   %  $9.4''$.}
   % results (mandatory)
  {Most of the protostars in our sample show strong atomic and
    molecular far-infrared emission. Water is detected in 17 out of 18
    objects (except TMC1A), including 5 Class I sources. 
  %  The high-excitation H$_2$O 8$_{18}$--7$_{07}$ 63.3 $\mu$m line 
  %  ($E_\mathrm{u}/k_\mathrm{B}=1071$ K) is detected in 7 sources. 
   CO
    transitions from $J=14-13$ up to $J=49-48$ are found and show two
    distinct temperature components on Boltzmann diagrams with 
    rotational temperatures of $\sim$350 K and $\sim$700 K. H$_2$O
    has typical excitation temperatures of $\sim$150 K. Emission 
    from both Class 0
    and I sources is usually spatially extended along the outflow
    direction but with a pattern depending on the species and the
    transition. In the \textit{extended} sources, emission is
    stronger off source and extended over $\geq$10,000 AU scales; in the 
    \textit{compact} sample, more than half of the flux originates within 1000
    AU of the protostar. The H$_2$O line fluxes correlate strongly
    with those of the high$-J$ CO lines, both for the full array and for the
    central position, as well as with the bolometric luminosity and envelope mass.
     They correlate less strongly with OH fluxes and not with
      [\ion{O}{i}] fluxes. 
     In contrast, [\ion{O}{i}] and OH often peak together at the central position. }
  {The PACS data probe at least two physical components. The
    H$_2$O and CO emission likely arises in non-dissociative (irradiated) shocks
    along the outflow walls with a range of pre-shock
    densities. Some OH is also associated with this component, likely 
    resulting from H$_2$O photodissociation. UV-heated gas contributes only a minor fraction to the
    CO emission observed by PACS, based on the strong correlation between 
    the shock-dominated CO 24-23 line and the CO 14-13 line. 
    [\ion{O}{i}] and some of the OH emission
    probe dissociative shocks in the inner envelope. The total
    far-infrared cooling is dominated by H$_2$O and CO, with the
    fraction contributed by [\ion{O}{i}] increasing for Class I
    sources. Consistent with previous studies, the ratio of total
    far-infrared line emission over bolometric luminosity decreases
    with evolutionary state.}
   

   \maketitle
%
%________________________________________________________________
 % {Understanding the physical phenomena involved in the earlierst
  %  stages of protostellar evolution requires knowledge of the
   % heating and cooling processes that occur in the surroundings of a
   % young stellar object. Spatially resolved information from
   % its constituent gas and dust provides the necessary constraints
   % to distinguish between different theories of accretion energy
   % dissipation into the envelope.}
%===========================
\section{Introduction}
%===========================
Stars form in collapsing dense molecular cores deep inside
interstellar clouds \citep[see reviews by][]{Fr7,BT07,L99}. Star formation is
associated with many physical phenomena that occur simultaneously:
infall from the envelope, action of jets and winds resulting in
shocks, outflows sweeping up surrounding material, and UV heating of
outflow cavity walls \citep{Sh87,Sp95,BT99}. In the
earliest phases of star formation \citep[Class 0 and I
objects;][]{An93,An00}, the interaction between the jet, wind and the
dense envelope is particularly strong and produces spectacular outflows
\citep{Ar07}.

Atomic and molecular tracers are needed 
to probe the physical conditions and to evaluate and disentangle
the energetic processes that occur in the Class 0/I young stellar
objects. Low$-J$ ($J\leq6$, $E_\mathrm{u}/k_\mathrm{B}\leq$116 K) rotational transitions of carbon monoxide
(CO) are among the most widely used tracers \citep{Bo96}, but are only
sensitive to the cold gas, \mbox{$T\leq100$} K, both from the envelope and
the entrained outflow material. Nevertheless, spectrally resolved
profiles of CO and $^{13}$CO 6--5 allowed \cite{vK09} and \cite{Yi12}
to attribute the narrow emission lines to the heating of the cavity
walls by UV photons \citep[see also][]{Sp95}. High-density tracers
 such as SiO ($n_\mathrm{H2}\geq10^5$~cm$^{-3}$)
have been used to study fast J-type shocks produced at bow shocks
where the jet plunges into the cloud \citep{Ba01}. At the same time,
theoretical studies of line cooling from dense cores predict that most
of the released energy is produced in between these two extreme
physical regimes and emitted mainly in atomic [\ion{O}{i}],
high$-J$ CO and H$_2$O rotational transitions in the far-infrared
spectral region in addition to H$_2$ mid-infrared emission
\citep{GL78,Ta83,NK93,CC96,DN97}. Therefore, in order to study the
energetics of young stellar objects (YSOs) and, in particular, the
relative importance of different energetic processes as a function of
the evolutionary state of a YSO, line observations in the $\sim 50-200$ $\mu$m
spectral region are necessary.

The Long-Wavelength Spectrometer on board the Infrared Space
Observatory offered for the first time spectral access to the complete far-IR
window \citep{ISO,LWS}. Many CO rotational transitions from $J=14-13$
to $J=29-28$ (for NGC1333-IRAS4) and several H$_2$O lines up to
$E_\mathrm{u}/k_\mathrm{B}\sim500$ K were detected in Class 0 sources
\citep{Gi01,Ma02}. On the other hand, H$_2$O remained undetected in
Class I sources, the exception being the outflow position of HH46. CO
emission was generally found to be weaker than H$_2$O, whereas the
fine structure lines of [\ion{O}{i}] and [\ion{C}{ii}] dominated the
ISO spectra \citep{Ni02}. The gas cooling budget calculations showed
similar contributions from lines of CO, H$_2$O, [\ion{O}{i}]
and to a smaller extent OH in Class 0 sources. Moreover, an
evolutionary trend of a gradual decrease in molecular luminosity and
total line luminosity was established as the objects evolve from the 
Class 0 to Class I phases. This trend was interpreted as the result of
weaker shocks and less shielded UV radiation in the later phase of protostellar
evolution \citep{Ni02}.

The Photodetector Array Camera and Spectrometer (PACS) \citep{Po10} on
board the \textit{Herschel} Space Observatory \citep{Pi10}\footnote{Herschel 
is an ESA space observatory with
  science instruments provided by European-led Principal Investigator
  consortia and with important participation from NASA.} with 25
$9\farcs4\times9\farcs4$ spatial pixels provides an $8\times$ improvement in
spatial resolution as compared to ISO/LWS. The PACS field of view of
$\sim47''$ is smaller than the $80''$ ISO beam but in many cases still
covers the full extent of the emission from nearby YSOs. For a typical
distance of 200 pc to our objects (Table \ref{catalog}), 
regions of $\sim9400$ AU are observed and resolved down to
$\sim1880$ AU. The higher sensitivity and better spectral 
resolution provides an important improvement in the quality of the spectra.
PACS is thus well suited for studies of atomic and molecular emission
in the Class 0/I objects, as demonstrated by PACS results on individual
 Class 0/I sources and their outflows \citep{vK10,vK10b,Ni10,He12,Be12,Go12}.
These results have already
indicated relative differences in the gas cooling budget from different sources and
 differences in spatial distributions of emission between different molecules.
\cite{Vi11} modelled these early data with a combination of shocks and
UV heating along the cavity wall. The strong [\ion{O}{i}] and OH
emission also suggests the presence of dissociative shocks in the
close vicinity of the protostar \citep{vK10}.

In our paper, we address the following questions: How does a YSO
affect its surrounding cloud and on what spatial scales? What are the
dominant gas cooling channels for deeply embedded YSOs? What do they
tell us about the physical components and conditions that cause
excitation of the observed lines?  How do all of these processes
change during the evolution from the Class 0 to the Class I stage?  To
this end, we present Herschel-PACS spectroscopy of 18 Class 0/I YSOs
targeting a number of CO, H$_2$O, OH and [\ion{O}{i}] lines obtained
as part of the `Water in star forming regions with Herschel' (WISH)
key program \citep{WISH}. WISH observes about 80 protostars at
different evolutionary stages (from prestellar cores to circumstellar
disks) and masses (low-, intermediate- and high-mass) with both the
Heterodyne Instrument for the Far-Infrared \citep[HIFI;][]{dG10} and
PACS.  Our paper focusses only on low-mass YSOs and is closely
associated with other WISH papers. Specifically, \citet{Kr12} studies
the spectrally resolved 557 GHz H$_2$O line observed towards all our
objects with HIFI. \citet{Wa12} analyzes the same sample of
sources but focuses on the excitation of OH in the Class 0/I sources,
whereas full PACS spectral scans of two sources are published by
Herczeg et al. (2012; NGC1333-IRAS4B) and Goicoechea et al. (2012; Ser
SMM1). A synthesis paper discussing the HIFI, PACS and SPIRE
  data being obtained in WISH and other programs is planned at the final stage
  of the program.

The paper is organized as follows: \S 2 introduces the source sample
and explains the observations and reduction methods; \S 3 presents
results that are derived directly from the observations; \S 4 focuses
on the analysis of the data; \S 5 provides the discussion of the
results in the context of the
available models and \S 6 summarizes the conclusions.\\

%__________________________________________________________________
\section{Observations}
%__________________________________________________________________
%===========================
\subsection{Sample selection}
%===========================

We used PACS to observe 18 out of 29 Class 0/I objects selected in the
low-mass part of the WISH key program. The WISH source list
  consists of nearby ($D\la450$ pc), well-known young stellar objects
  for which ample ground-based single dish and interferometer
  observations are available \citep[for details concerning the
    WISH program see][]{WISH}.  The remaining 11 sources, that were
  not targeted with PACS within WISH, were observed in the `Dust, Gas
  and Ice in Time' key program \citep[DIGIT, PI: N.\ Evans;][J{\o}rgensen et al., in prep., Lee et
  al., in prep.]{Gr12,Di12}.

Table \ref{catalog} presents our sample of objects together with their
basic properties. Bolometric luminosities and temperatures were
derived using our new PACS data supplemented with observations found
in the literature (see \S 2.4 for spectral energy distribution
discussion). Envelope masses are from \citet{Kr12}, which
  includes a discussion of the impact of new PACS measurements on the
  derived physical parameters.

\begin{table}
\begin{minipage}[t]{\columnwidth}
\caption{Catalog information and source properties.}
\label{catalog}
\centering
\renewcommand{\footnoterule}{}  % to avoid a line before footnotes
\begin{tabular}{lllrrrrrrrr}
\hline \hline
Nr & Object & $D$ & $L_\mathrm{bol}$ & $T_\mathrm{bol}$ &
$M_\mathrm{env}$\tablefootmark{a}\\

~ & ~ &  (pc) &  ($L_\mathrm{\odot}$) & (K) & ($M_\mathrm{\odot}$) & \\
\hline
1 & NGC1333-IRAS2A 	& 235 & 35.7 & 50 & 5.1 & \\
2 & NGC1333-IRAS4A 	& 235 & 9.1  & 33 & 5.2 & \\
3 & NGC1333-IRAS4B 	& 235 & 4.4  & 28 & 3.0 & \\
4 & L1527 			& 140 & 1.85 & 44 & 0.5 & \\
5 & Ced110-IRS4     & 125 & 0.8  & 56  & 0.2 & \\
6 & BHR71 			& 200 & 14.8 & 44 & 3.1 & \\
7 & IRAS15398\tablefootmark{b} & 130 & 1.6  & 52  & 0.5  & \\
8 & L483 			& 200 & 10.2 & 49 & 4.4 & \\
9 & Ser SMM1		& 230 & 30.4 & 39 & 16.1 & \\
10 & Ser SMM4 		& 230 & 1.9  & 26 & 1.9 & \\
11 & Ser SMM3 		& 230 & 5.1  & 38 & 3.2 & \\
12 & L723  			& 300 & 3.6  & 39 & 1.3 &\\
\hline
13 & L1489  			& 140 & 3.8  & 200 & 0.2 & \\
14 & TMR1   			& 140 & 3.8  & 133 & 0.2 & \\
15 & TMC1A 				& 140 & 2.7  & 118 & 0.3 & \\
16 & TMC1 				& 140 & 0.89 & 101 & 0.2  & \\
17 & HH46		 		& 450 & 27.9 & 104 & 4.4  & \\
18 & RNO91  			& 125 & 2.6  & 340 & 0.5 & \\
\hline
\end{tabular}
\end{minipage}
\tablefoot{
Sources above the horizontal line are Class 0, sources below are Class I. Source coordinates
 and references are listed in \cite{WISH}. Positional angles of CO 6-5 outflows 
 will be presented in Y{\i}ld{\i}z et al. (in prep.).  \\
\tablefoottext{a}{Envelope mass at 10 K from \citet{Kr12}.}
\tablefoottext{b}{The difference between the pointing 
coordinates and the coordinates derived from 2D Gaussian fits to PACS continuum observations 
in multiple wavelengths is ($9\farcs1\pm0.2$,$8\farcs0\pm0.3$).}}
\end{table}
%===========================
\subsection{Observing strategy}
%===========================
\begin{figure*}[tb]
\begin{center}
 \includegraphics[angle=90,height=13cm]{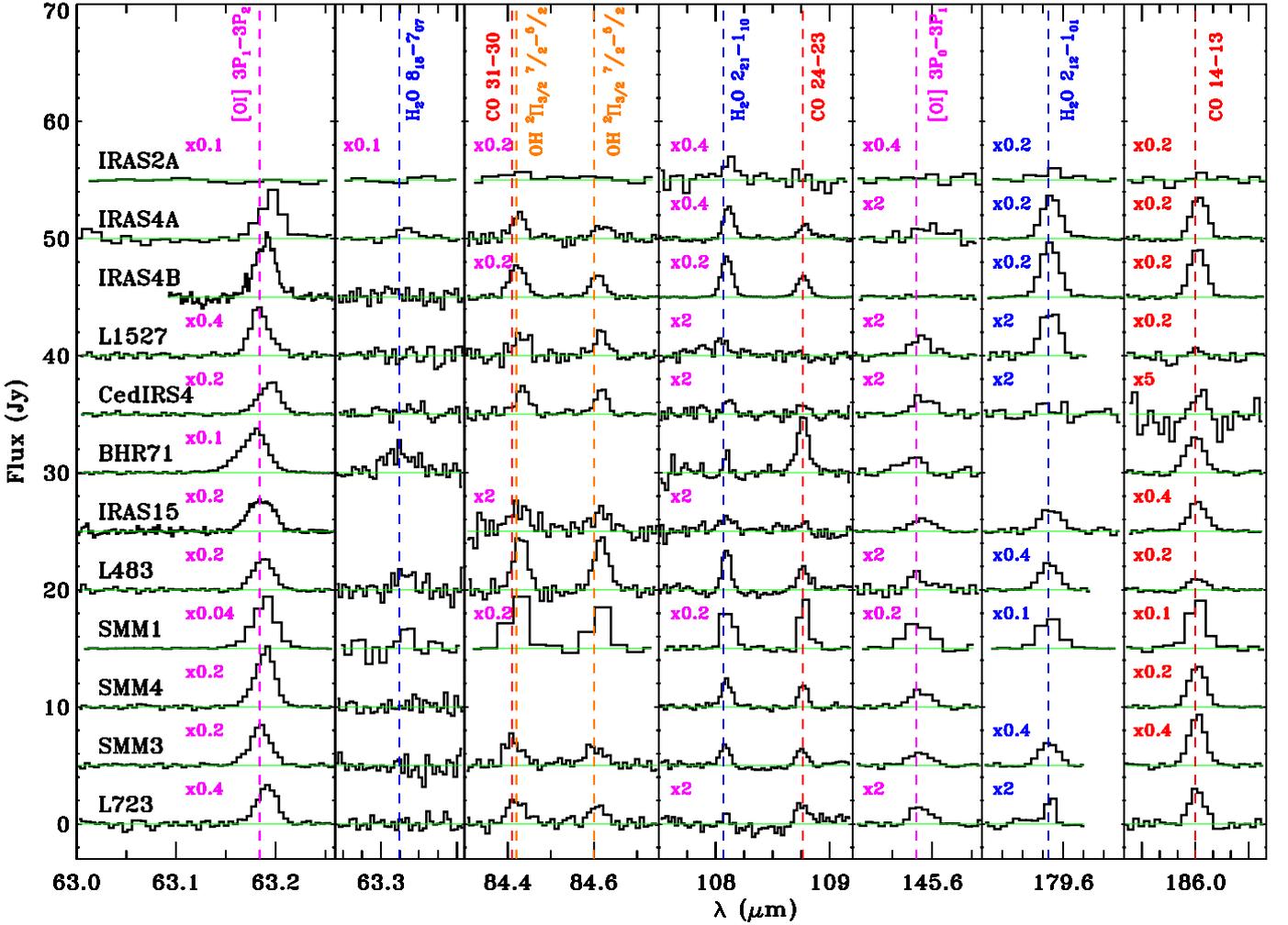}
\caption{\label{lines1}Line survey of Class 0 sources at the on-source position. Spectra are 
extracted from the central spaxel only for the well-pointed sources (for mispointed sources see text) 
and continuum subtracted. No correction for point spread function is made. Dashed lines show
 laboratory wavelengths of [\ion{O}{i}] (pink), OH (orange), CO (red) and H$_2$O (blue). 
 BHR71 and Ser SMM4 were not observed in all lines within our program (see text).}
\end{center}
\end{figure*}
\begin{figure*}[tb]
\begin{center}
\includegraphics[angle=90,height=11cm]{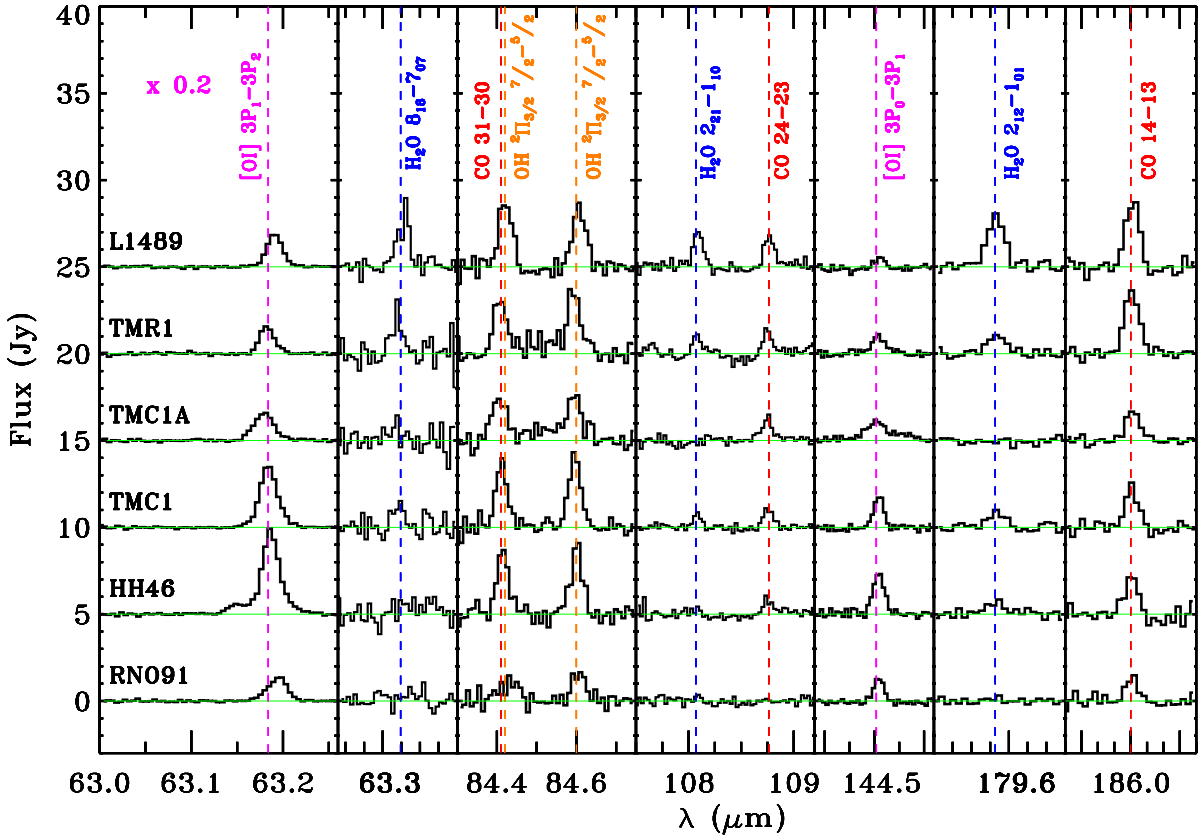} % AlllinesIn.eps
\caption{\label{lines2}Same as Figure \ref{lines1} but for Class I sources.}
\end{center}
\end{figure*}
%===========================

The far-IR spectra were obtained with PACS, an integral field unit
with a $5\times5$ array of spatial pixels (hereafter
\textit{spaxels}). Each spaxel covers $9\farcs4\times9\farcs4$,
providing a total field of view of $\sim47''\times47''$. The full
wavelength coverage consists of three grating orders (1st: 102-210
$\mu$m, 2nd: 71-105 $\mu$m or 3rd: 51-73 $\mu$m), two of which are
always observed simultanously (one in the blue, \mbox{$\lambda<105$
  $\mu$m}, and one in the red, \mbox{$\lambda>102$ $\mu$m}, parts of
the spectrum). The velocity resolution ranges from \mbox{$\sim$75 to
  300 km s$^{-1}$}, depending on the grating order and the
wavelength. The highest spectral resolution is obtained at the
  shortest wavelengths, below 65 $\mu$m. Two nod positions were used
for chopping 3$^\prime$ on each side of the source. Typical pointing
accuracy is better than 2$^{\prime\prime}$.

Two observing schemes were used in our program: line spectroscopy mode
to cover short spectral regions and range spectroscopy mode to cover
the full far-IR SED. Line spectroscopy mode uses small grating steps
to provide deep integrations and to fully sample the spectral resolution
over short (0.5-2 $\mu$m) wavelength intervals.
This mode was used to observe selected lines for 16 of 18 objects from
our sample (Ser SMM1 and NGC1333-IRAS2A are the exceptions). We
targeted 12 H$_{2}$O lines ($E_\mathrm{u}/k_\mathrm{B}\sim100-1320$
K), 12 CO lines ($E_\mathrm{u}/k_\mathrm{B}\sim580-3700$ K) and 4 OH
doublets ($E_\mathrm{u}/k_\mathrm{B}\sim120-291$ K) as well as the
[\ion{O}{i}] and [\ion{C}{ii}] lines (full list of available lines is
included in Table \ref{linestable}). BHR71 and Ser SMM4 were observed
only in a limited number of scans within the WISH program; range
spectroscopy observations of those sources are analyzed in DIGIT
(J{\o}rgensen et al. in prep. and Dionatos et
al. subm., respectively).

The range spectroscopy mode uses large
grating steps to quickly scan the full 50-210 $\mu$m wavelength range
with Nyquist sampling of the spectral resolution. This mode achieves a
spectral resolution of $R=\lambda/\Delta \lambda \approx$1000-1500 over
the full spectral range, which includes 37 high$-J$ CO transitions as
well as $~140$ H$_{2}$O transitions ($J<10$, $E_\mathrm{u}/k_\mathrm{B}<2031$ K)
and 11 OH doublets. NGC1333-IRAS2A, 4A, 4B, and Ser SMM1 were
observed with full range spectroscopy within WISH. The NGC1333-IRAS2A data 
were taken during the science demonstration phase, when the optimal 
PACS settings were not yet known. Hence the data are of poorer quality compared to the 
other full range spectroscopy observations.

%===========================
\subsection{Reduction methods}
%===========================

Both line spectroscopy and range spectroscopy basic data reduction was
performed with the Herschel Interactive Processing Environment v.8
\citep[HIPE,][]{Ot10}. The flux was normalized to the telescopic
background and calibrated using Neptune observations. Spectral flatfielding within HIPE was
used to increase the signal-to-noise (for details, see \citealt{He12} \citealt{Gr12}).
 The overall flux calibration is accurate to $\sim
30\%$, based on the flux repeatability for multiple observations of
the same target in different programs, cross-calibrations with HIFI
and ISO, and continuum photometry. The $5\times5$ datacubes were further
processed with IDL.

Since the spaxel size stays fixed whereas the Herschel beam size increases
with wavelength, the wavelength-dependent loss of radiation in a spaxel for a
well-centered point source is observed to be $\sim$30\% in the blue to
$\sim$60\% in red parts of the spectra (see PACS Observers Manual). Most of the radiation that
leaks outside a given spaxel is captured by the adjacent
ones. However, the far-IR emission from many Class 0/I objects is
spatially extended on scales of $>10^{\prime\prime}$, which are
resolvable by PACS. For these sources the central spaxel fluxes
corrected for the Point Spread Function (PSF) using the standard wavelength-dependent values
provided by the Herschel Science Center largely underestimate the
total emission from the source. Thus, in this paper, either a sum of 25 spaxels 
(for lines at $\lambda\geq100$ $\mu$m) or 
a sum of the spaxels with detected emission (for weak lines at $\lambda\leq100$ $\mu$m) are 
taken to calculate line fluxes used for most of the analysis. The only exceptions are
 in \S 3.1 and in \S 4.3, where central spaxel fluxes corrected for the PSF using the standard
factors are calculated in order to study the emission in the direct vicinity
of the YSOs. All line fluxes are listes in Tables \ref{obsn1} and \ref{obsn2}.

The approach to use the sum of the fluxes of all spaxels results in a
lower $S/N$ of the detected lines; some of the weak lines become
undetected. Therefore, we developed the `extended source
correction' method, which provides wavelength-dependent
correction factors for the brightest spaxel(s). This method is well suited for the
extended, Class 0/I sources. The details of the method are given in
 Appendix B; it is primarily applied to sources for which full line
scans are available.

The PACS maps show that Herschel was mispointed for some of our
objects. Continuum emission of BHR71, IRAS15398 and TMR1 peaks in
between a few spaxels. In the case of IRAS15398, our observations were
centered at the 2MASS position, which is offset by $\sim$10$''$ from
the far-infrared source position as determined from SCUBA maps by
\citet{Sh00}. Continuum emission from TMC1 and TMC1A peaks off-center
in the PACS array but is well confined to a single spaxel.
%===========================
\subsection{Spectral energy distributions}
%===========================

Radiation from the inner regions of a YSO is absorbed by dust in the
envelope and re-emitted in the far-IR. As the evolution proceeds, the
spectral energy distribution (SED) due to cold dust of a young Class 0
source evolves to a warmer Class I SED with emission observed also at
shorter wavelengths. Therefore, SEDs are a useful starting guide for
estimating the evolutionary stage of a YSO \citep{L99,An00}. PACS
continuum observations cover the SED peak of these embedded sources
and thus provide a more accurate determination of bolometric
luminosity ($L_\mathrm{bol}$) and bolometric temperature
($T_\mathrm{bol}$) than previously available.
  
Based on our PACS continuum measurements and the literature
measurements, new values of $L_\mathrm{bol}$ and
$T_\mathrm{bol}$ have been calculated. They are included in Table
\ref{catalog} and were also presented by \citet{Kr12}. The details of
the calculations, the continuum values at different PACS wavelengths
and the actual SEDs are presented in Appendix C.
\begin{figure*}[!tb]
  \begin{minipage}[t]{.33\textwidth}
  \begin{center}  
      \includegraphics[angle=90,height=7cm]{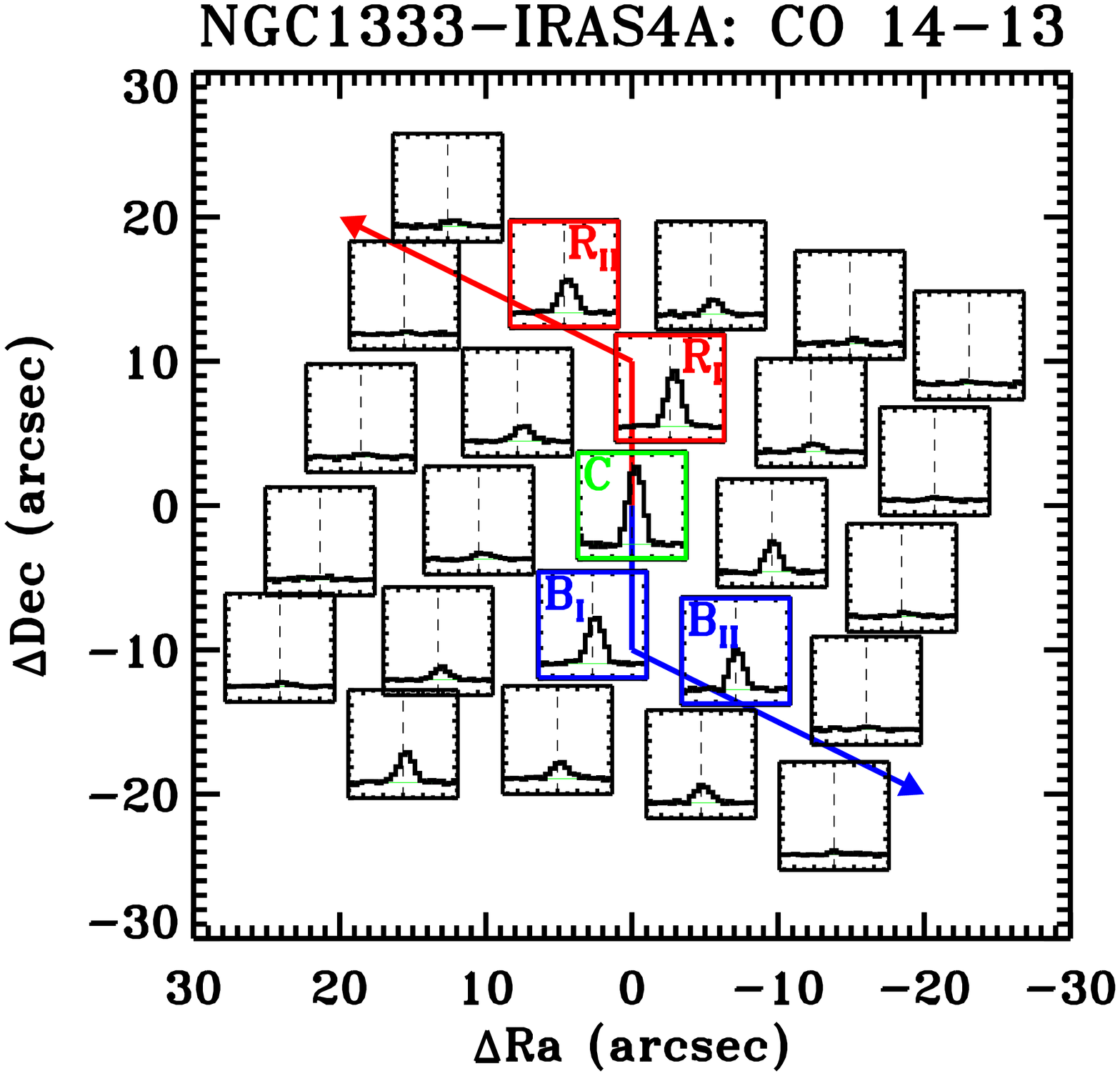}
     \vspace{+3ex}
       
       \includegraphics[angle=90,height=7cm]{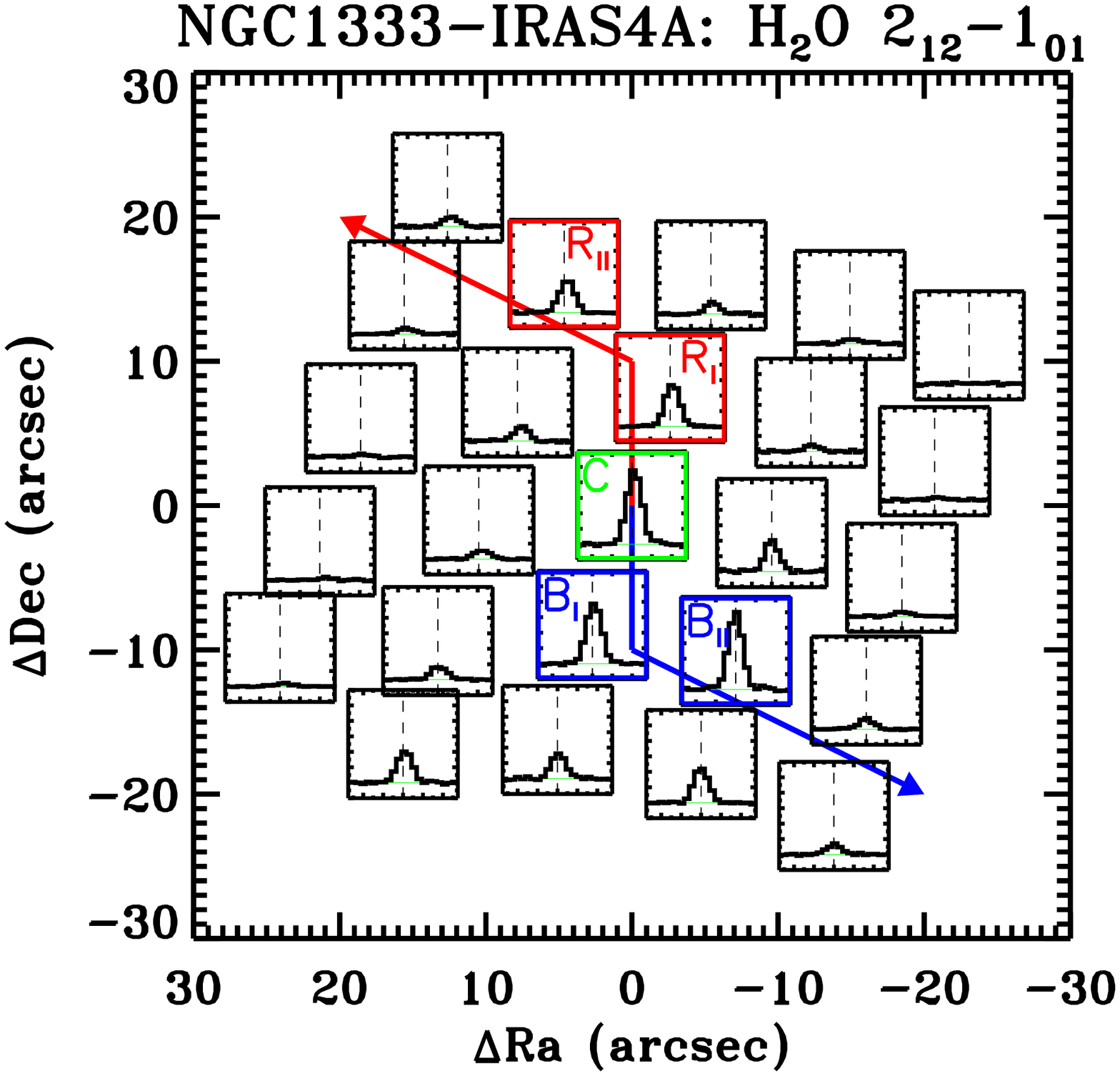}
    \end{center}
  \end{minipage}
  \hfill
  \begin{minipage}[t]{.33\textwidth}
  \begin{center}         
    \includegraphics[angle=90,height=7cm]{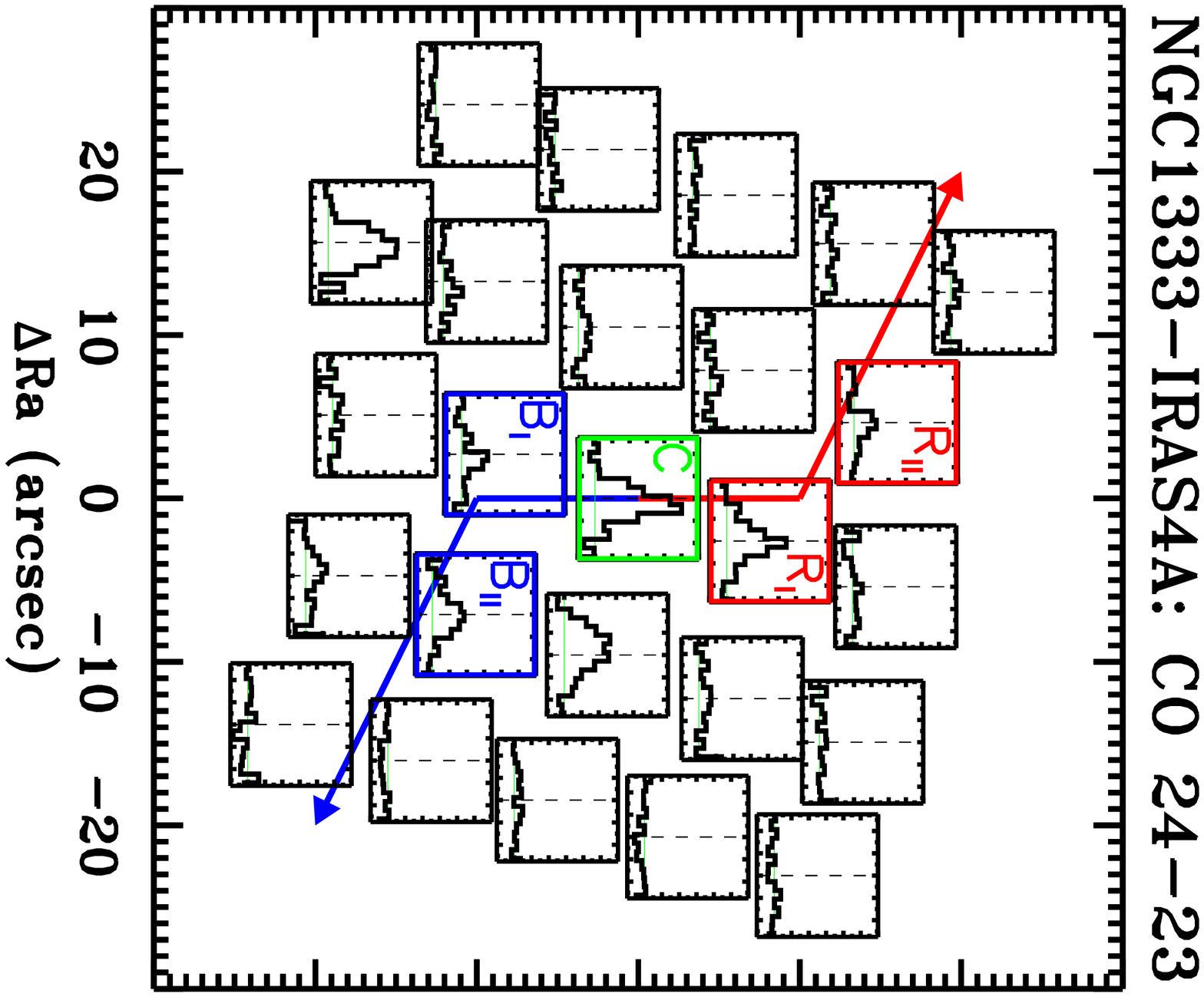} 
         \vspace{+3ex}
       
     \includegraphics[angle=90,height=7cm]{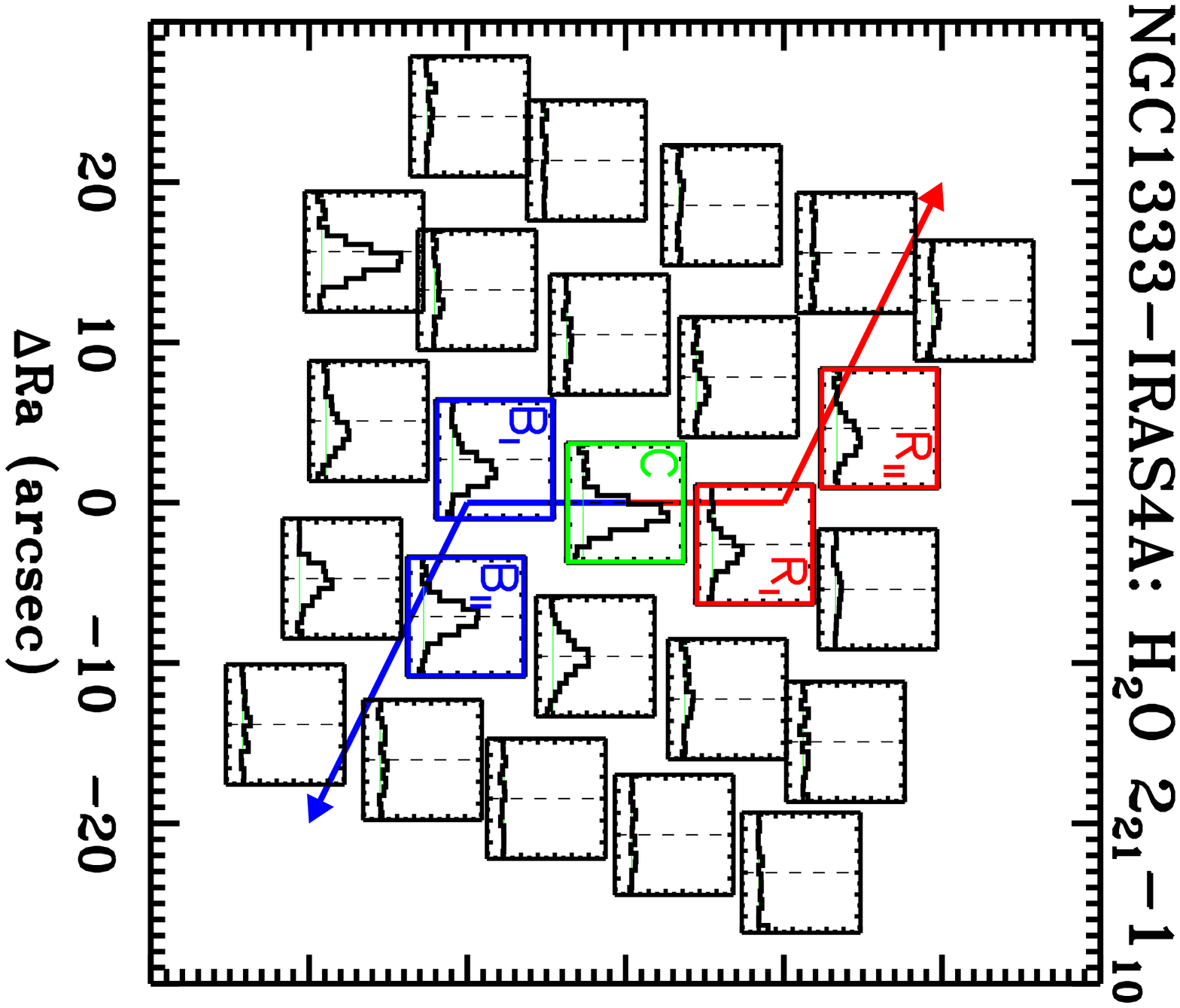}
    \end{center}
  \end{minipage}
    \hfill
  \begin{minipage}[t]{.33\textwidth}
  \begin{center}  
     \includegraphics[angle=90,height=7cm]{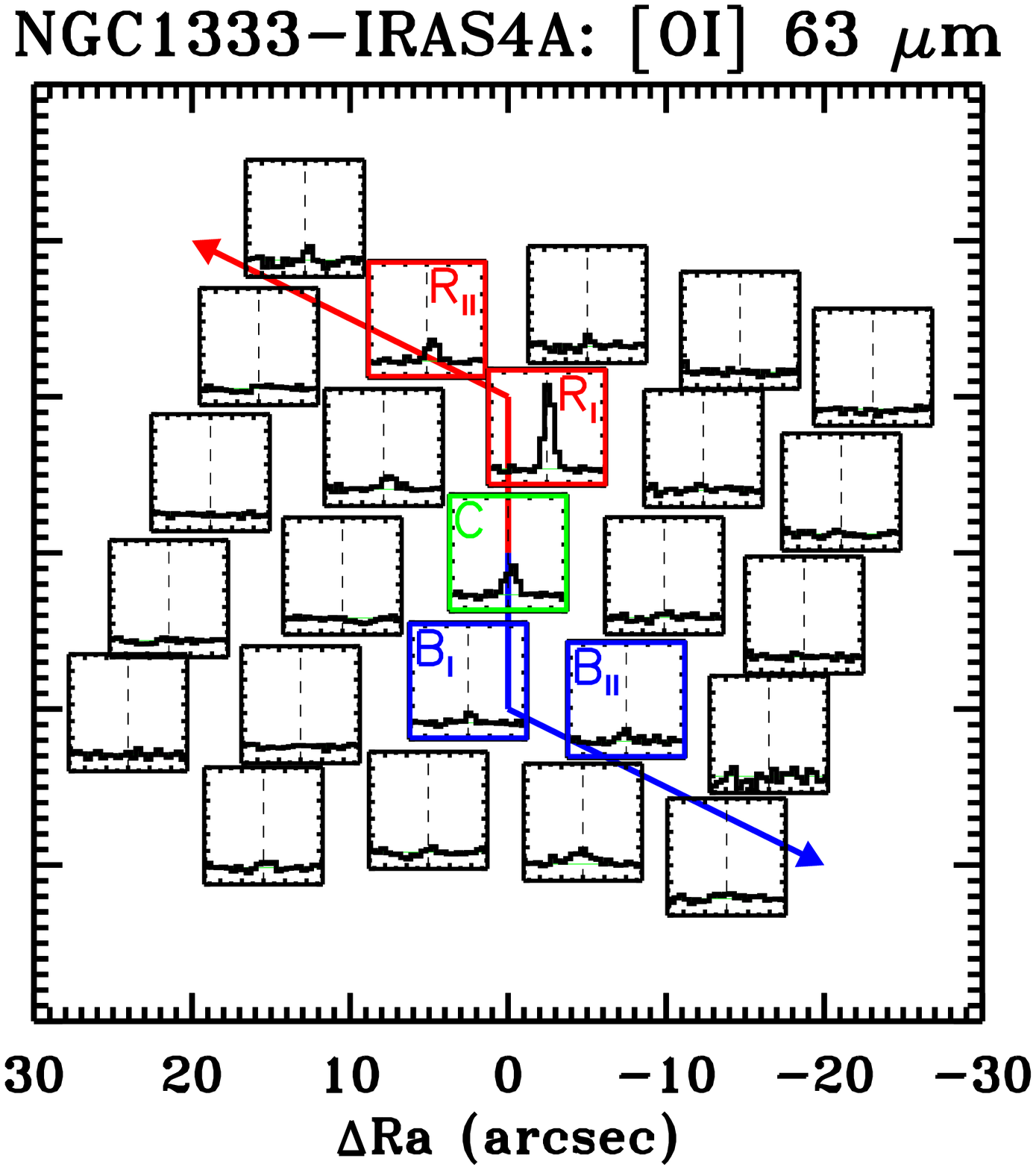}
          \vspace{+3ex}
       
     \includegraphics[angle=90,height=7cm]{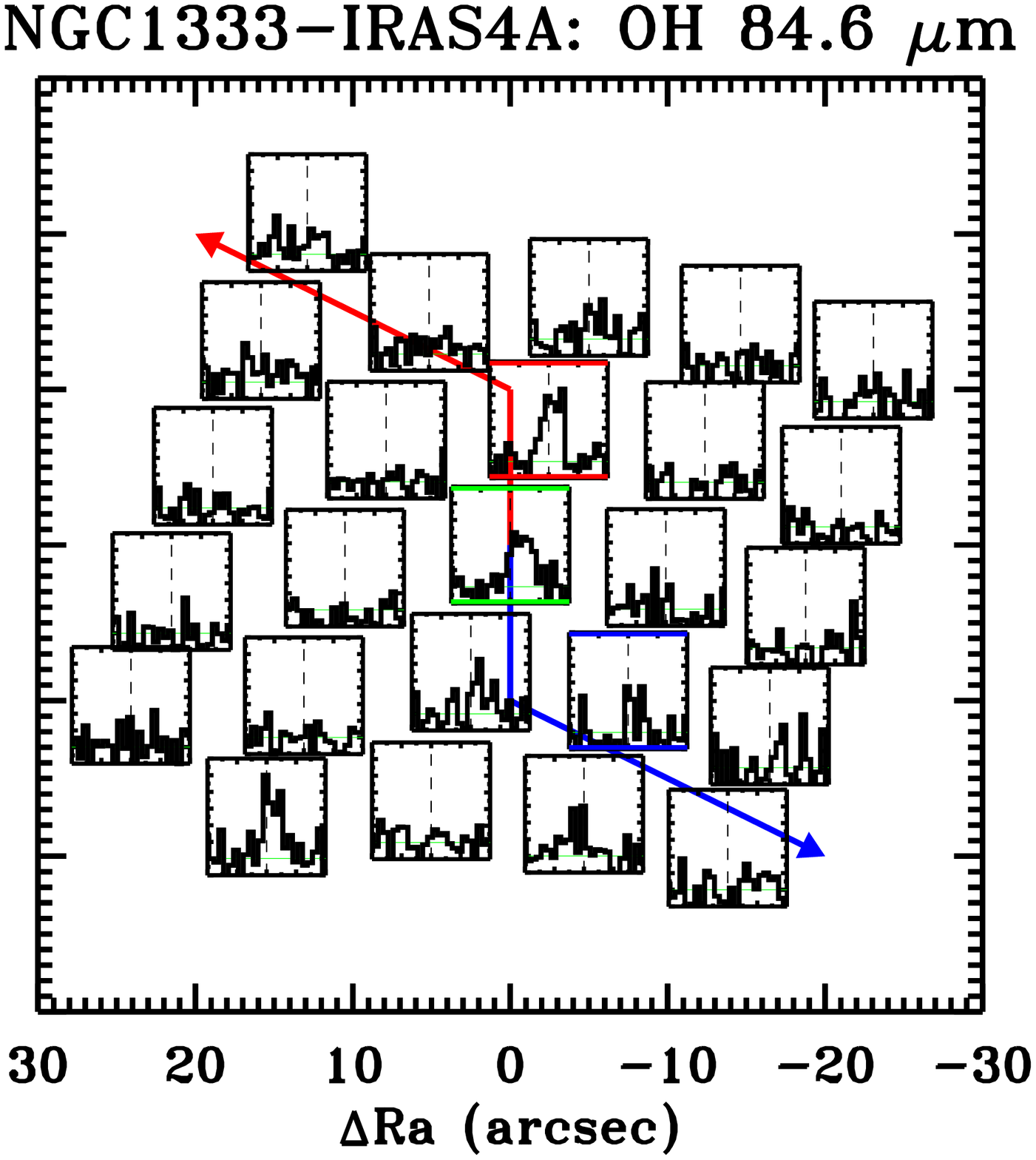}
    \end{center}
  \end{minipage}
 %  \vspace{+3ex}
        \caption{\label{specmap}PACS spectral maps of the Class 0 source NGC1333-IRAS4A 
        in the CO 14-13, CO 24-23, [\ion{O}{i}] 63.2 $\mu$m, H$_2$O 2$_{12}$-1$_{01}$, 
        H$_2$O 2$_{21}$-1$_{10}$ and OH 84.6  $\mu$m lines. The center of each spaxel box 
        corresponds to its position on the sky 
    with respect to the pointed source coordinates from \cite{WISH}; shown boxes are smaller than 
    the actual spaxel sizes. Wavelengths in 
    microns are translated to the velocity scale on the X-axis using laboratory wavelengths of the
species and cover the range from -550 to 550 km s$^{-1}$, except for the OH 84.6  $\mu$m lines
where -400 to 400 km s$^{-1}$ is shown. The Y-axis shows fluxes normalized to the 
    brightest spaxel on the map separately for each species in a range -0.2 to 1.2. Outflow directions
are drawn in blue and red lines based on CO 6-5 APEX CHAMP$^{+}$ sub-mm maps (Y{\i}ld{\i}z et al. 
2012 and in prep.) that traces the warm entrained gas ($T\sim100$ K). 
Two red outflow (R$_\mathrm{I}$, R$_\mathrm{II}$), on-source (C) and blue outflow (B$_\mathrm{I}$, B$_\mathrm{II}$) 
spaxels are marked with letters. IRAS4A spectra at those positions in different species are shown
     in Appendix D. The contribution from NGC1333-IRAS4B, located at (22.5$^{''}$,-22.8$^{''}$) with respect to IRAS4A, 
     is seen in  the S-E part of the map.}
\end{figure*}

\begin{figure*}[!tb]
  \begin{minipage}[t]{.33\textwidth}
  \begin{center}  
      \includegraphics[angle=90,height=7cm]{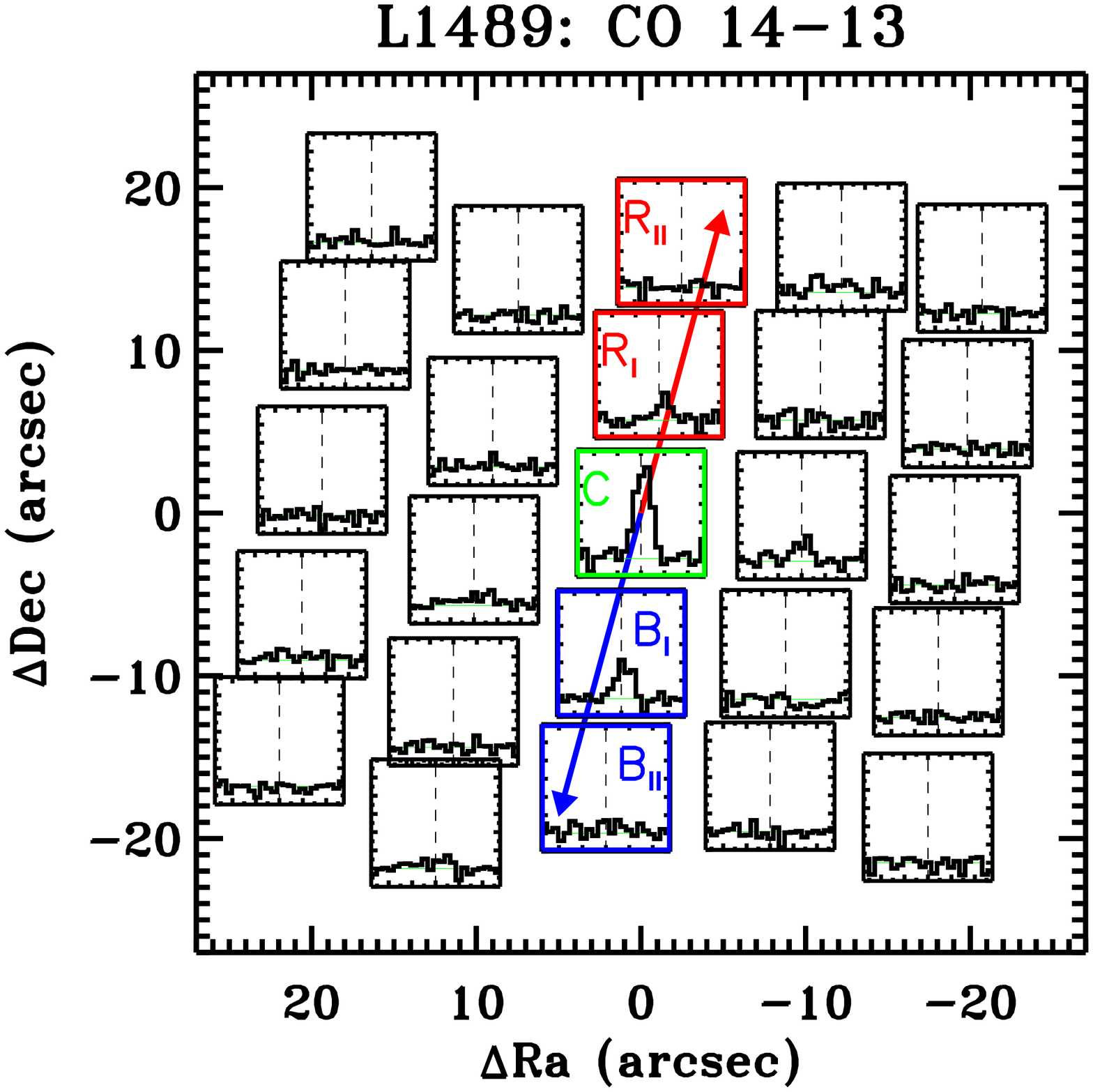}
     \vspace{+3ex}
       
       \includegraphics[angle=90,height=7cm]{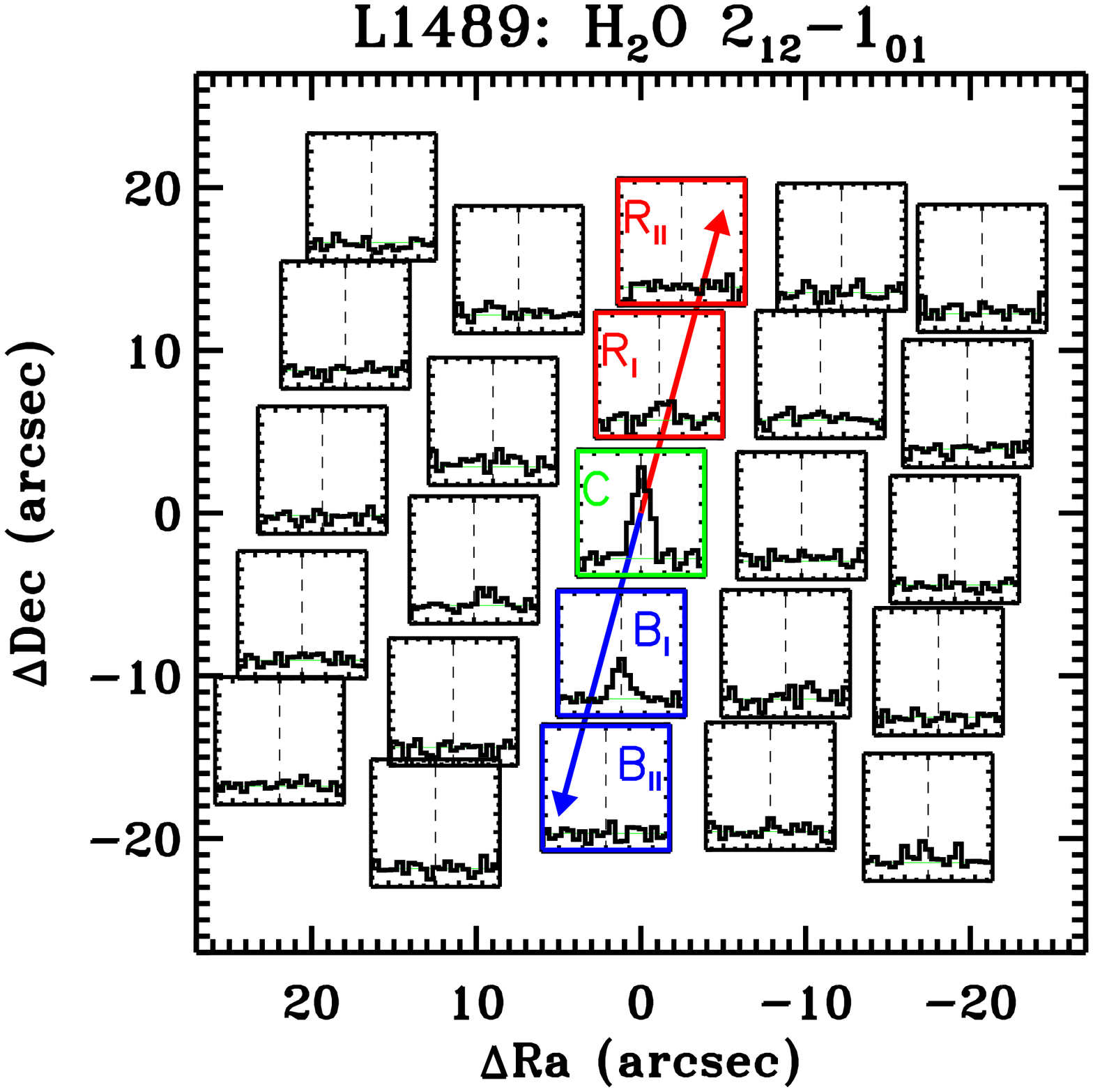}
    \end{center}
  \end{minipage}
  \hfill
  \begin{minipage}[t]{.33\textwidth}
  \begin{center}         
    \includegraphics[angle=90,height=7cm]{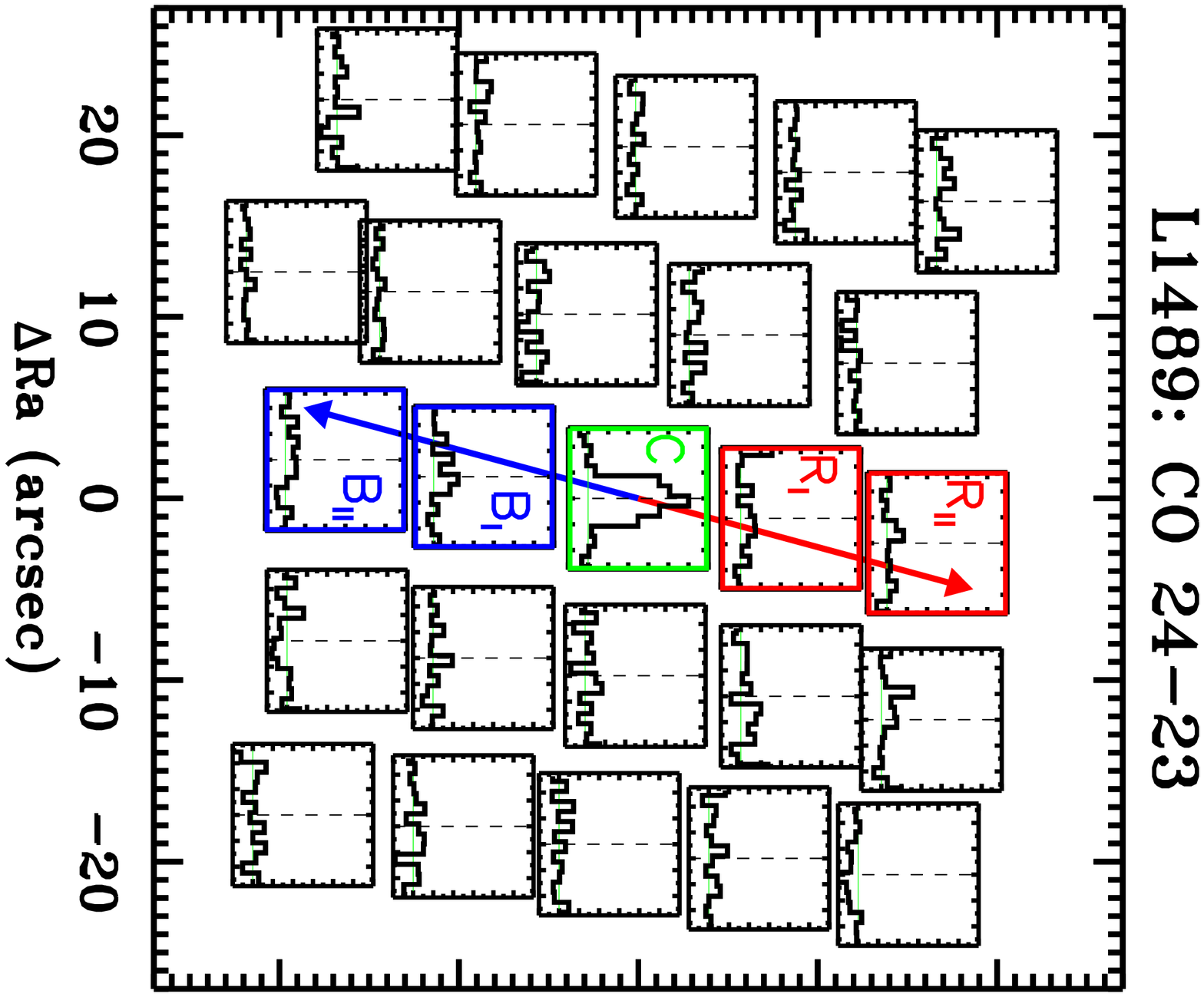} 
         \vspace{+3ex}
       
     \includegraphics[angle=90,height=7cm]{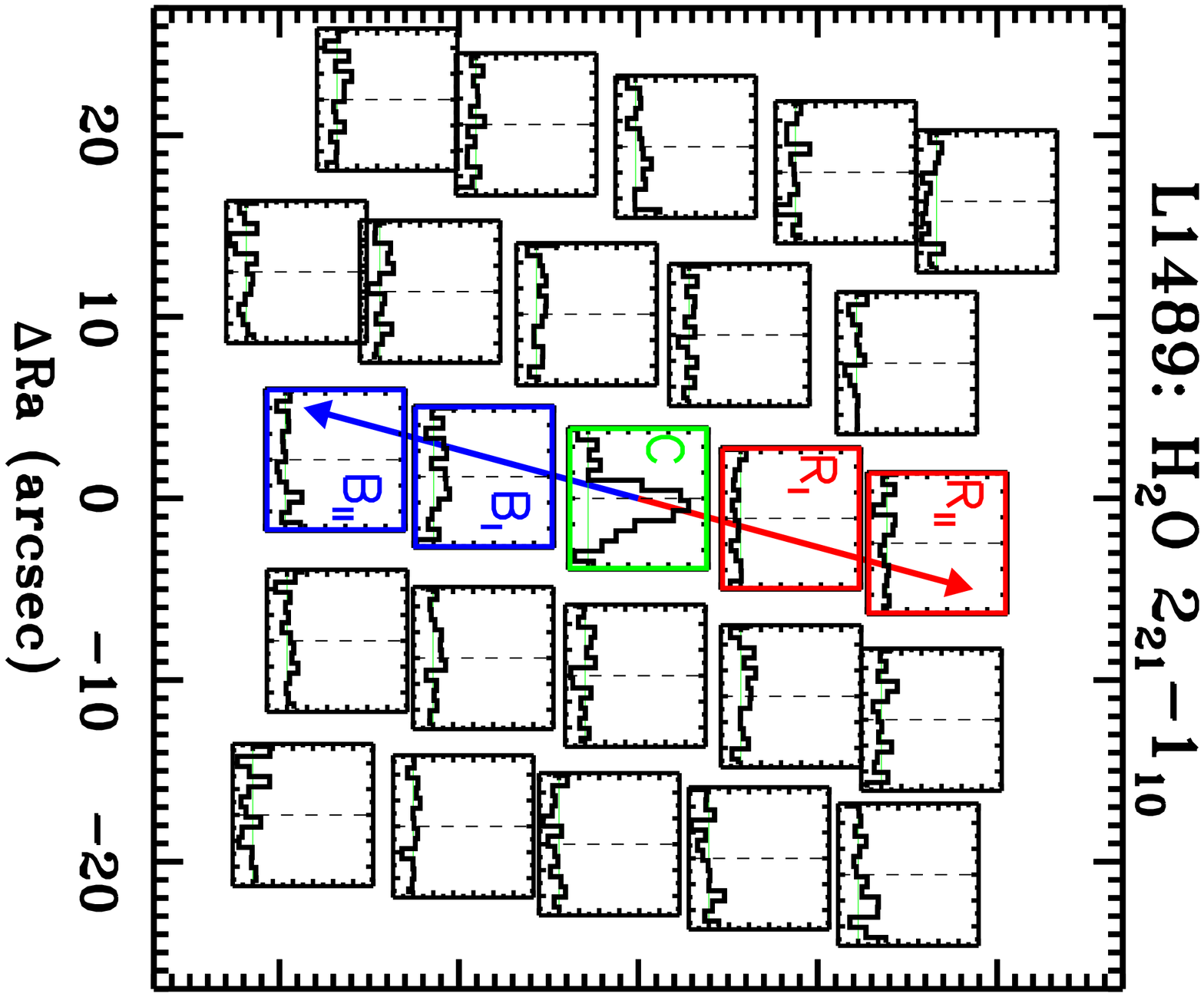}
    \end{center}
  \end{minipage}
    \hfill
  \begin{minipage}[t]{.33\textwidth}
  \begin{center}  
     \includegraphics[angle=90,height=7cm]{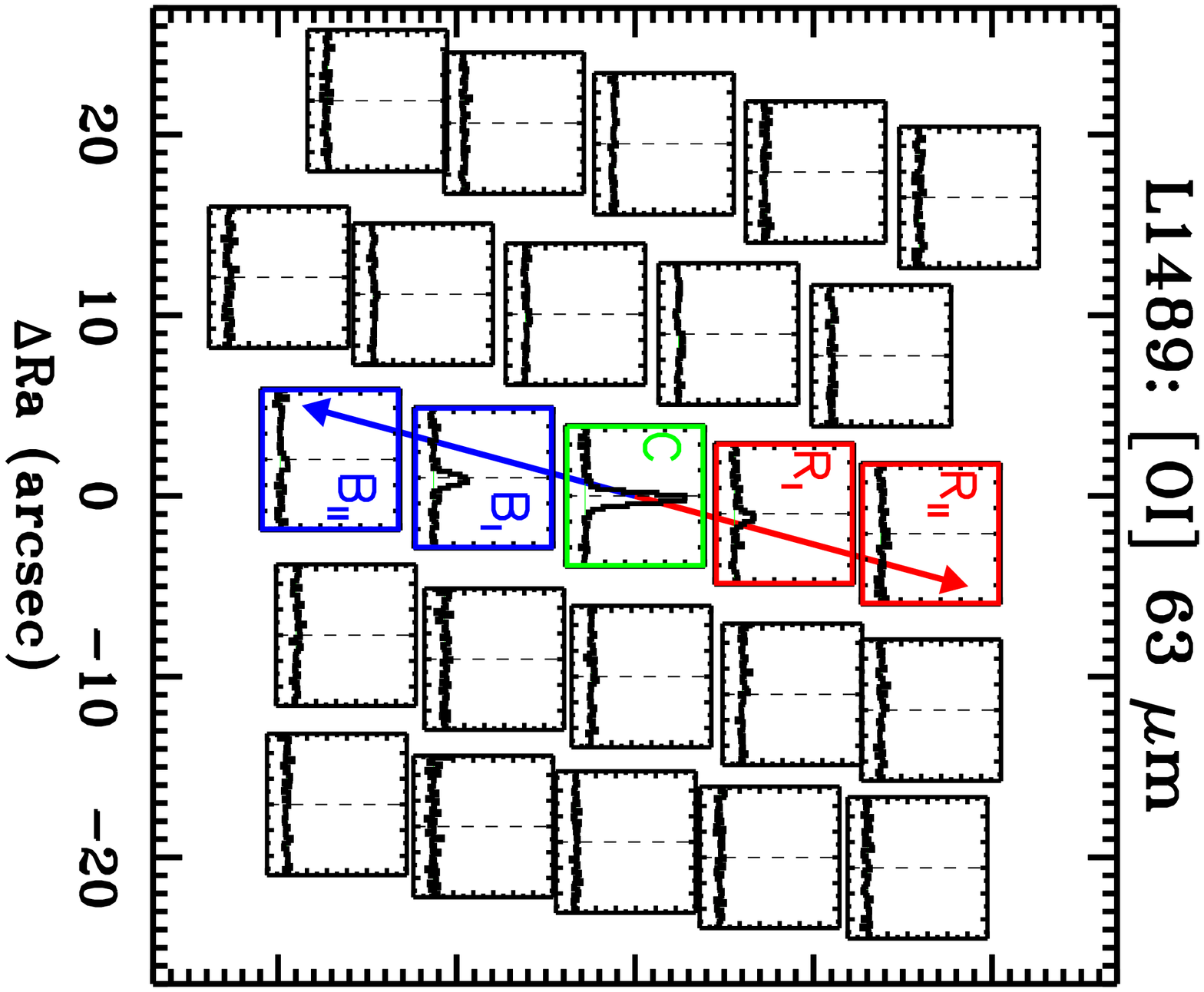}
          \vspace{+3ex}
       
     \includegraphics[angle=90,height=7cm]{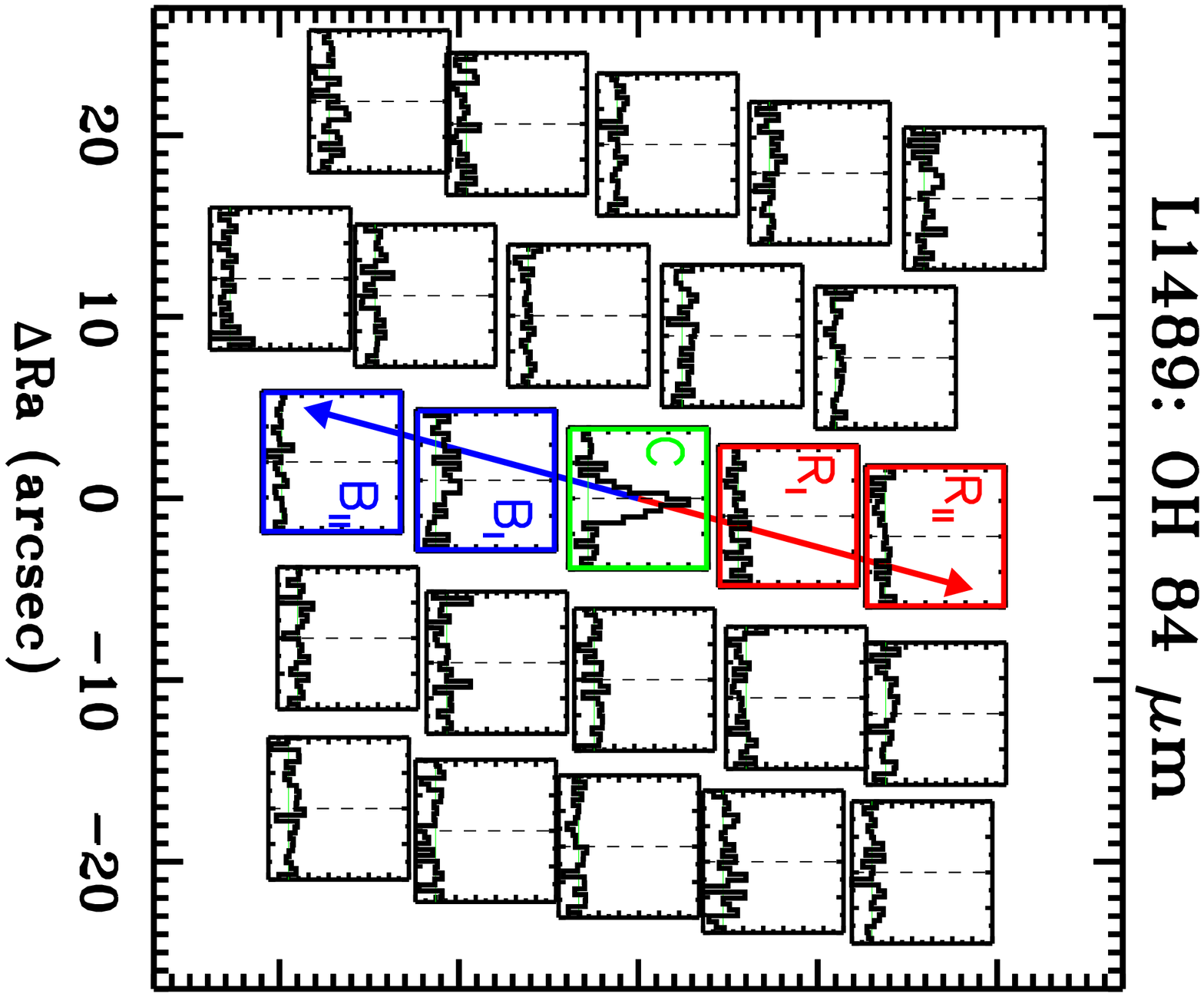}
    \end{center}
  \end{minipage}
 %  \vspace{+3ex}
        \caption{\label{specmap2} The same as Figure \ref{specmap} but for 
   the Class I source L1489.}
\end{figure*}

%===================
%__________________________________________________________________
\section{Results}
%_________________________________________________________________
%===========================
\subsection{Emission spectra}
%===========================
\begin{figure}[tb]
	\begin{center}
	\includegraphics[angle=90,width=\columnwidth]{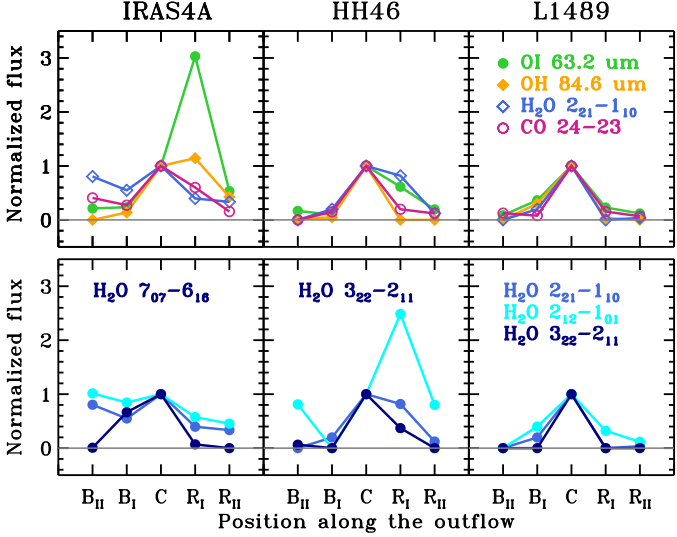}
	\end{center}	    
	\vspace{-6ex}
	\caption{\label{extent}Extent of line emission along the outflow direction for the 
	selected molecular and atomic lines. Top panel: The [\ion{O}{i}] 63.2 $\mu$m line 
	(green filled circles), the OH 84.6 $\mu$m line (orange filled diamonds),
	the H$_2$O 2$_{21}$-1$_{10}$ line at 108 $\mu$m (blue empty diamonds) and the CO \mbox{24-23} 
	line (violet empty circles) are shown for each object. Bottom panel: 
    The H$_2$O 2$_{21}$-1$_{10}$ line at 108 $\mu$m (blue), the H$_2$O 2$_{12}$-1$_{01}$
    line at 179 $\mu$m (light blue) are shown for all objects. Additionally, the H$_2$O 7$_{16}$-6$_{07}$ line
    at 71.9 $\mu$m is shown for IRAS4A and the H$_2$O 3$_{22}$-2$_{11}$ at 89.9 $\mu$m line is shown for 
    HH46 and L1489 (all in navy blue). The X-axis shows the selected spaxel names along the outflow direction 
(see Figure \ref{specmap}), whereas the Y-axis shows the flux normalized to the central spaxel
(C) value.}
\end{figure}

PACS spectroscopy of our sources reveals rich emission line
spectra superposed on the dust continuum emission. Several
transitions of the CO, H$_2$O and OH molecules as well as atomic emission
from [\ion{O}{i}] are detected. Emission in the [\ion{C}{ii}] line is 
only rarely detected and associated with the young stellar object.

Figure \ref{lines1} presents a line inventory at the on-source
position for Class 0 sources (central spaxel\footnote{For mispointed sources: TMR1, TMC1A and TMC1 
spaxel 32, corresponding to the continuum peak, is shown; for IRAS15398, where continuum emission falls 
into a few spaxels, only spaxel 23 is shown.}). The Class 0 spectra
show detections of at least one line of H$_2$O, CO, OH and
[\ion{O}{i}] each for every object (all except NGC1333-IRAS2A, which has a high upper limit). 
The H$_2$O 2$_{12}$-1$_{01}$ line
at 179.5 $\mu$m is the strongest observed water line and often the
strongest far-IR line in general, comparable only with CO 14-13 and
[\ion{O}{i}] 63.2 $\mu$m lines. CO transitions from $J=14-13$ to $J=48-47$ are
detected in the richest spectra; typically CO emission from
transitions higher than $J=31-30$ is either weak or undetected.  The
OH $^{2}\Pi_{\nicefrac{3}{2}}$ $J=\nicefrac{7}{2}-\nicefrac{5}{2}$
doublet at 84 $\mu$m is detected for all sources, except NGC1333-IRAS2A. 
The discussion of other OH transitions can be found in \citet{Wa12}. The
[\ion{O}{i}] \mbox{$^3P_{1}-^{3}P_{2}$} and \mbox{$^3P_{0}-^{3}P_{1}$} lines at 63.2
$\mu$m and 145.5 $\mu$m are detected for all sources
except NGC1333-IRAS2A (both lines undetected) and NGC1333-IRAS4B (the 145.5 $\mu$m line undetected).

For Class I objects, on-source spectra are presented in Figure
\ref{lines2}. At least one water line is detected in all Class I
sources except TMC1A; H$_2$O in RNO91 is detected when a few lines are
co-added. Unlike the case of the Class 0 sources in our sample, the
H$_2$O 2$_{12}$-1$_{01}$ ($E_\mathrm{u}/k_\mathrm{B}=114$ K) line at
179.5 $\mu$m is no longer the strongest water or molecular
line. For all sources except HH46, the H$_2$O
  2$_{21}$-1$_{10}$ line ($E_\mathrm{u}/k_\mathrm{B}=194$ K) at 108.07
  $\mu$m or the H$_2$O 3$_{03}$-2$_{12}$ line
  ($E_\mathrm{u}/k_\mathrm{B}=196$ K) at 174.63 $\mu$m is the
  strongest water line, whereas CO 16-15 or CO 18-17 and OH 84.6
  $\mu$m lines are the strongest molecular lines.  The CO lines are
  typically weaker from Class I than from Class 0 objects, up to a
  factor of 10 compared with the brightest Class 0 sources; the CO 24-23
  is even undetected for one Class I object, RNO91.  On the other
hand, the OH 84.6 $\mu$m line and both fine-structure [\ion{O}{i}]
lines are seen in all sources. The [\ion{O}{i}] line at 63.2
$\mu$m is always the strongest emission line in the far-IR spectrum of
the Class I sources. The profiles of the [\ion{O}{i}] line at
  63.2 $\mu$m are discussed in \S 3.3.

%===========================
\subsection{Spatial extent of line emission}
%===========================

PACS maps of the line emission in the detected species show a variety
of patterns and thus allow us to spatially resolve the emission from
different components of a young stellar object.

The Class 0 source NGC1333-IRAS4A and the Class I source L1489 are
used here to demonstrate the differences in spatial distributions of
the emission from the objects in our sample. Figures \ref{specmap} 
and \ref{specmap2} show PACS $5\times5$
maps for the two sources in the [\ion{O}{i}] 63.2 $\mu$m,
H$_2$O 2$_{12}$-1$_{01}$, H$_2$O 2$_{21}$-1$_{10}$, \mbox{CO 14-13}, \mbox{CO 24-23} and 
OH $^{2}\Pi_{\nicefrac{3}{2}}$ $J=\nicefrac{7}{2}-\nicefrac{5}{2}$ lines.
 In each map the CO 6-5
blue and red outflow directions are overplotted for comparison
(Y{\i}ld{\i}z et al.\ in prep.). The same figures for the rest of our
objects are included in the Online Material.

The NGC1333-IRAS4A emission in [\ion{O}{i}], CO and H$_2$O cover the
outflow direction over the entire map, corresponding to a radius
of $25\arcsec$ or 5900 AU from the protostar. The [\ion{O}{i}]
emission peaks at the red outflow position. The CO and H$_2$O maps also
show a pattern of extended emission but are less concentrated than 
[\ion{O}{i}], although some of this apparent extent can be attributed to the larger PSF at longer wavelengths.
The CO emission, however, is rather symmetric and peaks in the center/red
outflow position, whereas H$_2$O, contrary to [\ion{O}{i}], is more
pronounced in the blue outflow lobe, including the peak of the
emission.  
The OH 84.6 $\mu$m line is detected both on-source and
off-source, but with a pattern that is difficult to compare with other 
lines because of low signal-to-noise (as well as at the neigbouring IRAS4B
position in the S-E corner of the map).  
OH follows the [\ion{O}{i}] emission by peaking at the center/red outflow position. On the other
hand, the maps of L1489 show that the emission from all species peaks
strongly on-source, i.e. within a $5\arcsec$ radius corresponding to
$700$ AU distance from the protostar. Weaker molecular and atomic emission is detected
along the outflow direction and is more pronounced in the blue outflow
position.

These differences are further shown in Figure \ref{extent}, which
illustrates the extent of line emission from various species and
transitions in NGC1333-IRAS4A, HH46 and L1489, including higher
excited H$_2$O lines. The distributions are normalized to the emission
in the central spaxel.
%The peak of [\ion{O}{i}] and OH emission peaks
%in the red-outflow position (R$_\mathrm{I}$), whereas H$_2$O lines,
%including highly-excited 7$_{16}$-6$_{07}$, are almost equally strong
%on-source (S) and in the blue-outflow position (B$_\mathrm{I}$). 
For HH46, [\ion{O}{i}] and H$_2$O are strong in the red-outflow
position R$_\mathrm{I}$, whereas OH and CO 24-23 are observed only
on-source. L1489 shows some extended emission in the blue outflow (in
particular in [\ion{O}{i}] and OH), but clearly most of the
emission originates in the central spaxel. Since L1489 is much closer to us 
than the other two sources (see Table 1), the extended emission in 
NGC1333-IRAS4A and HH46 indeed covers a much larger area on the sky.

NGC1333-IRAS4A and L1489 are thus the prototypes for the two
morphologically different groups of objects: sources with 
\textit{extended} emission and sources with 
\textit{compact} emission. Figures
\ref{liness2} and \ref{liness1} show the spectra in the four discussed
species in the blue outflow, on-source and red outflow positions for
those two groups. The adopted selection rule is based on
the ratio of the on-source and the outflow [\ion{O}{i}]: the sources
where the outflow [\ion{O}{i}] emission (in a selected position)
accounts for more than the half of the on-source emission form the
\textit{extended} group, whereas the sources for which the
off-source emission is $\leq50$\% compared with the on-source emission
form the \textit{compact} group.

Table \ref{pattern} summarizes the results of using the same criterion
for the [\ion{O}{i}] 63.2 $\mu$m, CO 14-13, H$_2$O
2$_{12}$-1$_{01}$ and OH $^{2}\Pi_{\nicefrac{3}{2}}$
$J=\nicefrac{7}{2}-\nicefrac{5}{2}$ lines in all objects\footnote{No correction 
for PSF is performed. As a result, the calculated ratio of the on-source and off-source 
emission is lowered. The effect is the strongest for the CO 14-13 and H$_2$O
2$_{12}$-1$_{01}$ lines. Also, no correction for different distances is made, but 
since our sources are located at a similar (mean) distance of 190$\pm$50 pc (excluding HH46), 
this does not change our conclusions.}.
The general trends are: (1) in the \textit{compact}
 group, [\ion{O}{i}] and OH emission dominate the central spaxel, whereas CO and
H$_2$O either follow the same pattern or are off-source--dominated; (2) in the \textit{extended} group
OH is often strong off-source (except L1527 and HH46 where it
dominates on-source), similar to CO and H$_2$O; (3) Most objects in 
the \textit{extended} group are Class 0 objects, with the exception of TMC1A and HH46.; 
(4) Class 0 and I sources are almost equally represented in the \textit{compact} group.

In a few cases both H$_2$O and CO are extended but in a different manner. For
example, L1527 and NGC1333-IRAS4A show a brighter CO line and a weaker
H$_2$O line in the red outflow position and the opposite in the blue
outflow position (Figure \ref{extent}). L483 shows similar
differences, but with the brighter CO and weaker H$_2$O line in the
blue outflow position. In those three cases the [\ion{O}{i}] line is
stronger at the position of weak H$_2$O; the same holds for the OH in
case of NGC1333-IRAS4A (OH is not detected off-source in L483 and
L1527). These differences are further discussed in \S 5.2.

For all objects, the [\ion{O}{i}] emission is seen from the young
stellar object and associated outflows rather than extended cloud
emission.  In the NGC1333-IRAS4A, 4B, Ser SMM3 and SMM4 regions,
spaxels where the emission originates from the nearby sources are
omitted. When detected, the [\ion{C}{ii}] emission is usually spread
across the entire detector and seen in different strengths in the two
nods, which both indicate that the emission is primarily produced by
the parent cloud. TMC1 is the only source with [\ion{C}{ii}] detected
from the central source (maps in both nods are shown in the Online Material). 
In Ser SMM1, [\ion{C}{ii}] emission follows the pattern of other species
along the outflow direction \citep{Go12}. The [\ion{C}{ii}] emission
is not discussed further in this paper.
%----------------------------
\begin{table}
\begin{minipage}[t]{\columnwidth}
\caption{\label{pattern}Patterns of emission in atomic and molecular species.}
\centering
\renewcommand{\footnoterule}{}  % to avoid a line before footnotes
\begin{tabular}{lcccccc}
\hline \hline
Source &  [\ion{O}{i}] & CO 14-13 & H$_2$O 2$_{12}$-1$_{01}$ & OH 84.6 \\
\hline
\multicolumn{5}{c}{\textit{Compact} emission} \\
\hline
NGC1333-IRAS2A	& \ldots & x & x & \ldots\\
Ced110-IRS4 	& x & b & x & x \\
BHR71 			& x & r & \ldots & \ldots \\
L483 			& x	& x & x & x \\
L723			& x & b & b & x \\
L1489 			& x & x & x & x \\
TMR1			& x & x & r & x \\
TMC1 			& x & b & b & x \\
RNO91		 	& x	& x & r & x \\
\hline
\multicolumn{5}{c}{\textit{Extended} emission} \\
\hline
NGC1333-IRAS4A & r  & rb & rb & r\\
NGC1333-IRAS4B & b  & b  & b  & b\\
L1527 		 & r  & b  & x  & x\\
Ser SMM1 	 & rb & x  & b  & b\\
Ser SMM4		 & b  & b  & \ldots  & \ldots\\
Ser SMM3		 & rb & rb & rb  & b \\
TMC1A		 & b  & b  & rb  & b \\
HH46 		 & r  & x  & rb  & x \\
IRAS15398 	 & rb & rb & rb  & rb\\
\hline
\end{tabular}
\end{minipage}
\tablefoot{Compact emission (see text) is denoted with `x'. Red and blue 
extended (outflow) emission that accounts for $\geq50$\% of the on-source flux is denoted with `r' and `b'.
BHR71 and Ser SMM4 were not observed in the H$_2$O 2$_{12}$-1$_{01}$ and OH 84.6 $\mu$m lines. 
NGC1333-IRAS2A shows non-detections of the above lines; the CO 15-14 line is used instead of CO 14-13 
and the H$_2$O 3$_{03}$-2$_{12}$ line instead of H$_2$O 2$_{12}$-1$_{01}$ line.}
\end{table}

%===========================
\subsection{Velocity shifts in [\ion{O}{i}] and OH lines}
%===========================

\begin{figure}[tb]
	\begin{center}
\includegraphics[angle=90,height=7.5cm]{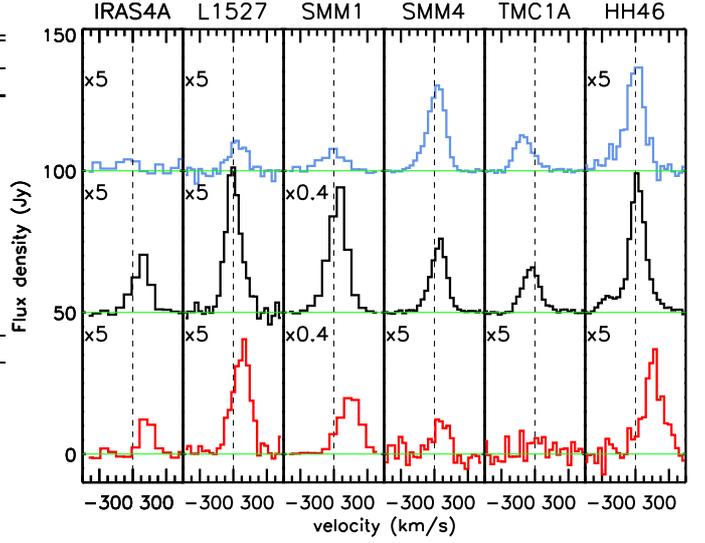}
	\end{center}
	\caption{\label{oshift}Velocity shifts and high-velocity line wings in the [\ion{O}{i}]
	 line at 63.2 $\mu$m for NGC1333-IRAS4A, L1527, Ser SMM1, SMM4, TMC1A and HH46. 
	Selected blue outflow, on-source and red outflow positions are shown for each object 
	from top to bottom in velocity range from -300 to 300 km s$^{-1}$. 
	The black dashed line shows the laboratory wavelength of [\ion{O}{i}].}
\end{figure}

Figure \ref{oshift} shows the [\ion{O}{i}] line at 63.2 $\mu$m
towards the Class 0 sources NGC1333-IRAS4A,
L1527, Ser SMM1 and SMM4 as well as the Class I sources TMC1A and HH46 
(for comparison between PACS and ISO fluxes of the [\ion{O}{i}] lines 
see Appendix E). The blue and red outflow profiles show significant line velocity
shifts and, in particular in the case of HH46 and Ser SMM4, high-velocity
line wings.\footnote{The velocity resolution of PACS is
 $\sim 90$ km s$^{-1}$ at 63 $\mu$m (or 0.02 $\mu$m).  In principle,
apparent velocity shifts can result from the location of the emission
in the dispersion direction within each spaxel.  This type of spatial
offset is ruled out for the velocity shifts presented here because the
velocity shifts are large and because we
 would expect to see stronger emission in neighboring spaxels than is
observed.}

Early results by \cite{vK10} showed that the bulk of the
[\ion{O}{i}] emission in HH46 comes from low-velocity gas. On top of
this `quiescent' profile, high-velocity gas was detected in the
blueshifted jet with a centroid velocity of about --170 km s$^{-1}$ and
in the redshifted jet with a centroid velocity of +100 km
s$^{-1}$.
Such velocity shifts, indicative of an optically invisible `hidden' atomic jet, 
are seen towards at least a third of our objects
(shown in Figure
\ref{oshift}).   High-velocity tails are detectable in a few
[\ion{O}{i}]-bright sources, but have a minor contribution to the
total line emission.  %Within our sample, HH46 remains the
%best example of a source with a velocity-resolved [\ion{O}{i}]
%profile.

The [\ion{O}{i}] 63.2 $\mu$m velocity shifts and profile wings may be
associated with similar features of the OH line at 84.6 $\mu$m.  For
NGC1333-IRAS4A, the OH 84.6 $\mu$m line from the source spaxel is
redshifted by 90 km s$^{-1}$, compared with 50 km s$^{-1}$ for the
[\ion{O}{i}] profile shift (see also Figure \ref{5spectra}).
HH46 also shows a tentative detection of blue-shifted
high-velocity OH material that resembles the [\ion{O}{i}]
pattern.  Within our sample, no other molecular lines have significant
centroid velocity shifts, with typical limits of $\sim 40$ km s$^{-1}$
at $<100$ $\mu$m and $\sim 100$ km s$^{-1}$ at $>100$ $\mu$m.
Such velocity shifts are at the velocity calibration limit and may be introduced
by emission that is spatially offset within the slit(s). 
Inclination affects the projected velocity
of the jet but is unlikely the explanation for why a majority of
sources do not show a velocity shift.
%Inclination can also play a role but is unlikely the explanation in the majority
%of sources that do not show a velocity shift.
%==========================
%===========================
\section{Analysis}
%===========================
\subsection{Rotational diagrams}
%===========================
\begin{figure}[tb]
\begin{center}
 \includegraphics[angle=90,height=6cm]{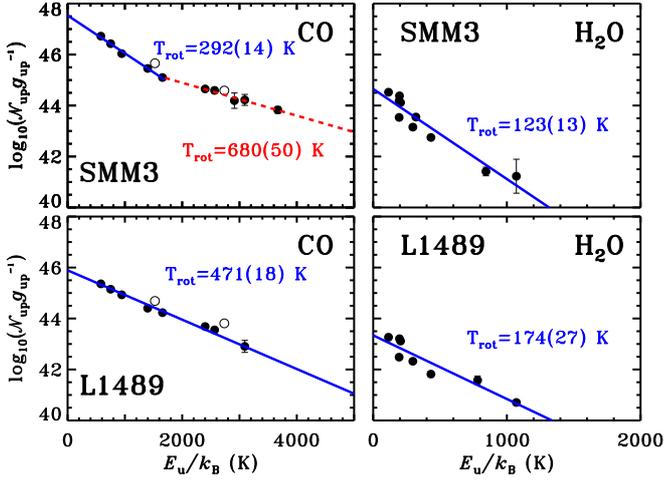}
  \vspace{+6ex}
\caption{\label{alldiag}CO (left panel) and H$_2$O (right panel) rotational diagrams for 
Ser SMM3 (Class 0) and L1489 (Class I) calculated using 
the total flux in lines measured in the PACS field-of-view. The base 10 logarithm  
of the number of emitting molecules from a level $u$, 
$\mathcal{N}_\mathrm{u}$, divided by 
the degeneracy of the level, 
$g_\mathrm{u}$, are written on the Y-axis.
Two-component fits in the CO diagrams cover the transitions below and above 
$E_\mathrm{u}/k_\mathrm{B}\sim1700$ K (CO 24-23) for
the `warm' and `hot component' (see text). Each data point corresponds to one 
observed transition of a molecule. The limited number of lines observed
in the line spectroscopy mode is responsible for the gaps in the otherwise linearly spaced CO
diagrams. The error bars reflect the uncertainties in the fit}.
%Dashed lines in Ser SMM3 diagrams, and the corresponding values of rotational 
%temperatures in brackets, show the fit results using the same lines as detected in L1489.
\end{center}
\end{figure}
Boltzmann (or rotational) diagrams are used to determine the rotational
temperatures $T_\mathrm{rot}$ from level populations for the Class 0/I
objects from our sample \citep[see][for Boltzmann diagrams]{GL78}. 
For optically thin thermalized lines, the natural
logarithm of the column density of the upper level $N_\mathrm{u}$ of a
given transition over its degeneracy $g_\mathrm{u}$ is related
linearly to the energy $E_\mathrm{u}$ of that level:
%\citep[Equation \ref{rot};after][]{Bl87}.
\begin{equation}
\label{rot}
\mathrm{ln}\frac{N_\mathrm{u}}{g_\mathrm{u}}=
\mathrm{ln}\frac{N_\mathrm{T}}{Q(T_{\mathrm{rot}})}-\frac{E_\mathrm{u}}{k_\mathrm{B}T_\mathrm{rot}}\\
\end{equation}
where $Q(T_\mathrm{rot})$ denotes the rotational partition function at a temperature
$T_\mathrm{rot}$ for a given molecule, $N_\mathrm{T}$ is the total column density 
and $k_\mathrm{B}$ is the Boltzmann constant. 

The emitting region is unresolved in the PACS data due to the low spatial resolution, thus
 the number of emitting molecules, $\mathcal{N}_\mathrm{u}$, is calculated for each transition, defined as:
\begin{equation}
\label{rot2}
\mathcal{N}_\mathrm{u}=\frac{4\pi d^{2}F_{\lambda}\lambda}{hcA}
\end{equation}
$F_{\lambda}$ denotes the flux of the line at wavelength $\lambda$, $d$ is the distance to the source,
$A$ is the Einstein coefficient, $c$ is the speed of light and $h$ is Planck's constant.

Figure \ref{alldiag} shows CO and H$_2$O rotational diagrams
calculated using the fluxes measured over the entire $5\times5$ PACS
array for the Class 0 source Ser SMM3 and the Class I
source L1489. Diagrams for all objects are included in the Appendix 
(Figures \ref{codiag} and \ref{wdiag}).

Full range scan observations cover many more
CO transitions than our targeted line scans \citep{vK10b,He12,Go12,Ma12,Gr12,Di12}. In those observations, two excitation
temperature components are clearly present. The lower-$T_\mathrm{rot}$
component of $\sim 250-300$ K dominates mid-$J$ transitions with 
$E_\mathrm{u}/k_\mathrm{B}$ below
$\sim1000-2000$ K. A higher-$T_\mathrm{rot}$ component of $\sim500-1000$ K 
dominates high-$J$ transitions with $E_\mathrm{u}/k_\mathrm{B}$ above $\sim 2000$ K. We
call these components \textit{warm} and \textit{hot}, respectively, in
order to distinguish them from the \textit{cool} component,
$T_\mathrm{rot}\sim100$ K, observed in the $J<14$ lines
\citep{vK09,Yi12,Go12}. Motivated by these
observations of complete CO ladders, we fit two linear components to
our more limited set of CO data. The exceptions are L723, L1489, TMR1 and TMC1A,
where there is no indication of the hot component in our dataset. The physical interpretation
 of these two components is discussed in \S 5.

As an example, the CO diagram of the Class 0 object Ser SMM3
  in Figure \ref{alldiag} show a break around
  $E_\mathrm{u}/k_\mathrm{B}\sim1200-2000$ K, with a rotational
  temperature for the warm component, $T_\mathrm{rot}$(warm), of
  $292\pm14$ K and a rotational temperature for the hot component,
  $T_\mathrm{rot}$(hot), of $670\pm50$ K. The error bars reflect the
  uncertainties given by the fit and include the uncertainties in
  individual line fluxes as given in Table A.2. The
temperature fits include only relative flux uncertainties between
lines and not the absolute flux uncertainty, which would shift the
total luminosity up and down but would not change the temperature.
   The corresponding values for the
  Class I object L1489 are $405\pm20$ K and $480\pm55$ K for the warm
  and hot components, respectively. However, since a single
  temperature fit to these latter data is valid within the errors, one
  component fit is used to derive the temperature of $471\pm20$ K. As
  a result, the ratio of the number of emitting molecules of the hot CO over the
  warm CO is $\sim0.1$ for Ser SMM3, whereas no value for hot CO can
  be given for L1489.

The rotational temperatures of warm and hot CO and H$_2$O, the numbers
of emitting CO molecules and the ratio of hot over warm CO for all
sources are tabulated in Table \ref{tab:exc}.  The uncertainties due to the
limited number of observed lines, associated with the fits and the
selection of the break point are discussed in Appendix \ref{sec:unc}.

Median rotational temperatures of the CO warm component are 325 and
420 K for Class 0 and Class I sources, respectively (calculated using
the unbracketed values from Table \ref{tab:exc}). The hot CO
average temperature is $\sim$700 K for the Class 0 sources;
for the Class I sources the temperatures seem to be 100--200 K lower,
but in general they are poorly constrained due to the limited
detections. The median logarithm of the number of emitting CO molecules for
the warm and hot components of Class 0 sources is 49.4 and 48.7,
respectively. Therefore, about 16\% of the CO molecules observed by
PACS in Class 0 sources are hot.

%====================================
\begin{table*}
%\begin{minipage}[t]{\columnwidth}
\caption{\label{tab:exc}CO and H$_2$O rotational excitation and number of emitting molecules 
$\mathcal{N}_\mathrm{u}$ based on the full array data.}
\centering
\renewcommand{\footnoterule}{}  % to avoid a line before footnotes
\begin{tabular}{llrrrrrrrrrrrrr}
\hline \hline
Source & \multicolumn{2}{c}{Warm CO} & \multicolumn{2}{c}{Hot CO} & H$_2$O & 
$\mathcal{N^\mathrm{hot}_\mathrm{CO}}$/$\mathcal{N^\mathrm{warm}_\mathrm{CO}}$ \\
~ & $T_\mathrm{rot}$(K) & $\mathrm{log}_\mathrm{10}\mathcal{N}$ & $T_\mathrm{rot}$(K) & 
$\mathrm{log}_\mathrm{10}\mathcal{N}$ & $T_\mathrm{rot}$(K)  &  \\
 \hline
NGC1333-IRAS2A & 310 & 49.1 & \ldots  & \ldots  & (210) & \ldots \\
NGC1333-IRAS4A & 300 & 49.7 & (390)  & (49.4) & 90  & (0.5) \\
NGC1333-IRAS4B & 340 & 49.6 & 820  & 48.7 &  200  & 0.1 \\
L1527		   & (297) & (48.0) & (600) &  (47.2) &  70 & (0.2) \\
Ced110-IRS4    & (490) & (47.2) & (800) &  (46.9)  &  90	 & (0.4) \\
BHR71 		   & (370) & (49.4) & (550) &	(49.0) &  (140)	 & (0.4) \\
IRAS15398 	   & 280 & 48.9 & (530) &	(47.8)   &  50  & (0.1) \\
L483  		   & 360 & 48.6 & 620 &	48.1   &  150  & 0.3  \\
Serpens SMM1   & 350 & 49.9 & 690 &	49.0   &  150  & 0.1  \\
Serpens SMM4   & (260) & (49.5) & (690) & (48.3)   &  \ldots  & (0.1)  \\
Serpens SMM3   & 290 & 49.6 & 660 &	48.6   &  130  & 0.1 \\
L723		   & 431 & 48.5 & \ldots &	\ldots   &  130  & (0.4) \\
\hline
L1489      &  471 & 48.1 & \ldots & \ldots  &  170	 & (0.8)  \\
TMR1	   &  470 & 48.3 & \ldots & \ldots &  170	 & (0.4) \\
TMC1A	   &  420 & 47.9 & \ldots	& \ldots &  \ldots & \ldots \\
TMC1 	   &  350 & 48.1 & 510 & 47.7  &  160	& 0.4  \\
HH46       &  310 & 49.2 & (630) & (48.3)  &  50	 & (0.1)  \\
RNO91	   &  (250) & (47.8) & \ldots &	\ldots   &  (100)	& \ldots \\
\hline
\end{tabular}

\tablefoot{Rotational temperatures and corresponding numbers of 
emitting molecules measured using 
less than 5 points are written in brackets (see Figures \ref{codiag} and \ref{wdiag}). 
Non-detections are marked with ellipsis dots (\ldots). 
In case of L723, L1489, TMR1 and TMC1A one component fit is used.}
\end{table*}

The H$_2$O rotational diagrams of Class 0/I sources in Figure
\ref{alldiag} show scatter in the single-temperature fits that
significantly exceeds the measurement errors and is due to subthermal
excitation and optical depth effects \citep[][see also \citealt{Jo03}
  for the case of CH$_\mathrm{3}$OH]{He12}. 
We refrain from
calculating of the number of emitting H$_2$O molecules because the
high optical depths require orders of magnitude
correction factors. 
Highly excited H$_2$O
emission from at least some Class 0 and Class I sources is seen in the
H$_2$O 8$_{18}$-7$_{07}$ line at 63.3 $\mu$m
($E_\mathrm{u}/k_\mathrm{B}$=1071 K). The single rotational
temperatures obtained from the fit to the H$_2$O diagrams in Figure
\ref{alldiag} are $\sim$120 K for the Class 0 source and $\sim$170 K
for the Class I source. Similarly low values of H$_2$O
  rotational temperatures are also obtained from full spectroscopy
  observations \citep{He12,Go12}. The fact that the H$_2$O rotational
temperature is 100~K or higher already indicates that H$_2$O cannot be
in the entrained outflow gas seen in $^{12}$CO low$-J$ lines.
%====================================
\begin{figure}
 \includegraphics[angle=90,height=6cm]{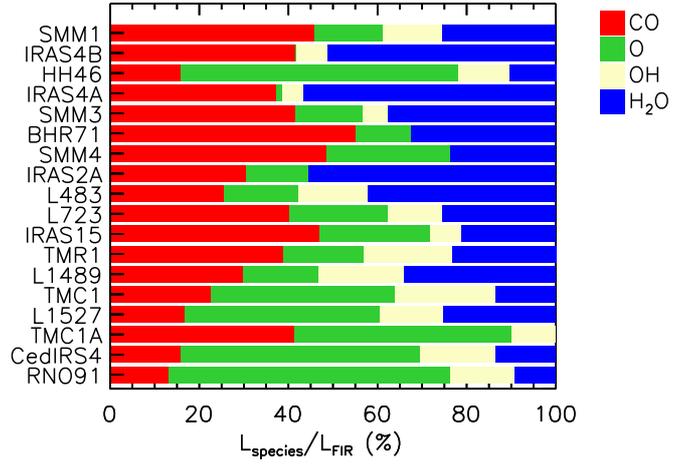}
\caption{\label{coolingfig}Relative contributions of CO (red), [\ion{O}{i}] (green), OH
(yellow) and H$_2$O (blue) to the total far-IR gas cooling integrated over the
 entire PACS array are shown from left to right horizontally for each source. The objects are 
 in the sequence of decreasing total gas far-IR cooling, 
 L$_\mathrm{FIR}$. Note that OH cooling is not available for BHR71 and Ser SMM4 and not 
 measured in NGC1333-I2A. Water cooling is not calculated for TMC1A.}
\end{figure}
%====================================

 %===========================
\subsection{Far-infrared line cooling}
%===========================
The CO rotational temperatures of Class 0/I sources presented in \S
4.1 are used to estimate the flux in non-observed lines and to
calculate the total far-infrared CO cooling. The extrapolation of the 
fluxes is limited to the transitions in the PACS range, from $J$=14-13 to $J$=49-48.
This accounts for $\sim80$ \% of the CO cooling calculated for the first 60 rotational
 transitions of CO ($J\leq60$, $E_\mathrm{u}/k_\mathrm{B}\leq10006$ K), used in CO cooling 
 calculations by \cite{Ni02} (see 
 Appendix I). Additional line emission arising from the $T_\mathrm{rot}\sim100$ K
component seen in the CO $J\leq13$ with SPIRE \citep{Go12} and ground-based data \citep{Yi12}
 is not included in this estimation, because
these lines probe a different physical component, the entrained outflow gas.

The far-IR cooling in H$_2$O lines is calculated by scaling the total
H$_2$O flux observed over the full PACS range towards NGC1333-IRAS4B
and Serpens SMM1 (Herczeg et al. 2012, Goicoechea et al. 2012) to the
limited number of lines observed here in the line spectroscopy
mode. These two sources, even though both classified as Class
  0, have very different water spectra, with IRAS4B showing numerous
  high-excitation water lines that are absent in Ser SMM1.

In the water rich source NGC1333-IRAS4B, the total luminosity of the
water lines equals 2.6 $10^{-4}$ L$_\mathrm{\odot}$, whereas the
luminosity calculated from the selected lines equals 1.0 $10^{-4}$
L$_\mathrm{\odot}$. For Serpens SMM1, where H$_2$O lines are much
weaker as compared to CO, the total water luminosity observed in the
PACS range equals 2.4 $10^{-4}$ L$_\mathrm{\odot}$, of which
1.1 $10^{-4}$ L$_\mathrm{\odot}$ is detected in the small set of lines 
observed in the line scan observations. Thus, the scaling factor from
the line scan observations to the total far-IR water cooling, based on
these two sources, is $\sim2.4$\footnote{For Ser SMM4 and BHR71, a 
scaling factor of $\sim10$ is used. These two objects were observed 
in a limited number of settings and therefore the correction for the 
missing lines, and hence uncertainty, is larger with respect to other sources, with many more 
lines observed. The two sources are excluded from the analysis in \S 5.4.}.

Despite the obvious limitations of the method, which assumes similar
gas properties for all the sources, it provides more reasonable values
of the cooling than the extrapolation using the H$_2$O rotational
temperature (see Appendix \ref{sec:unc} and \ref{sec:cool}).
Indeed, as argued in Appendix I, the adopted scaling of the
  H$_2$O luminosity should be robust for a broad range of objects
  within the quoted uncertainties of $\sim$30 \%. As a further
  validation of our approach, the values derived for Class I sources
  in Taurus agree within 30\% with the full range spectroscopy
  observations obtained in the DIGIT program (Lee et al., in prep.).
H$_2$O cooling in the PACS range accounts for $\sim86$ \% of the total
cooling in this molecule (from non-LTE large velocity gradient model
of Serpens SMM1, Goicoechea et al.\ 2012).

For OH, a scaling factor of $\sim1.5$ is derived based on the full
scan observations of Ser SMM1 and IRAS4B, calculated in the same way
as for H$_2$O.  Fluxes of OH for all our sources are from \citep{Wa12}.

The cooling in the [\ion{O}{i}] lines is calculated as the sum of the
observed fluxes in the 63 $\mu$m and 145 $\mu$m lines. Cooling in
[\ion{C}{ii}] is omitted because the line emission is a minor
contributor to the cooling budget \citep{Go12} and
usually originates from extended cloud emission.

We use the definition of the total far-IR line cooling,
$L_\mathrm{FIR}$, adopted by \citet{Ni02}:
$L_\mathrm{FIR}=L_{\ion{O}{i}}+L_\mathrm{CO}+L_\mathrm{H_{2}O}+L_\mathrm{OH}$.
The results of the calculations are summarized in Table \ref{cool} and 
illustrated in Figure \ref{coolingfig}, which shows the relative contributions
 to the total far-IR gas cooling, $L_\mathrm{FIR}$, by different atomic and 
 molecular species, individually for each object from our sample. 
 Note that this definition does not include the contribution of H$_2$
  near- and mid-infrared lines to the cooling. At off-source shock
  positions, H$_2$ can be a significant, and even dominant, gas
  coolant \citep{Ne09,Ni02b}. However, at the much more
  obscured central source position, data on H$_2$ are generally not
  available. In case where they are, the H$_2$ mid-infrared emission
  (after extinction correction) is found to have a negligible
  contribution to the total cooling for Class 0 sources
  \citep{He12,Go12}.

%====================================
\begin{table}
%\begin{minipage}[t]{\columnwidth}
\caption{\label{cool} Atomic and molecular far-IR line luminosities based on the full array data.}
\centering
\renewcommand{\footnoterule}{}  % to avoid a line before footnotes
\begin{tabular}{lrrrrrrrrrrrrrr}
\hline \hline
Source & $L_\mathrm{CO}$ &  $L_\mathrm{H2O}$  & $L_\mathrm{OH}$ & $L_\mathrm{OI}$  & $L_\mathrm{FIR}$  \\
% ~  & (K)  & (1) & (K) & (1) & (K) & (1) & ($10^{-3}$ L$_{\odot}$) & ($10^{-3}$ L$_{\odot}$)  & 
%  ($10^{-3}$ L$_{\odot}$) & ($10^{-3}$ L$_{\odot}$) & ($10^{-3}$ L$_{\odot}$)\\
\hline
NGC1333-I2A & 3.9  & 7.1 	&	\ldots   &  1.6	   &  12.6  \\
NGC1333-I4A & 13.2 & 20.0 	&	1.7   &  0.5	   &  35.4  \\
NGC1333-I4B & 22.5 & 27.6    &	5     &  0.3       &  54.1 \\
L1527		    & 0.3 & 0.5 	&	0.3   &  0.8     &  1.9\\
Ced110-IRS4    & 0.1 & 0.1 	&	0.2   &  0.5     &  0.9\\
BHR71 		   & 12.1 & 7.1 	&	\ldots   &  2.8     &  22.0 \\
IRAS15398 	   & 2.0 & 0.9 	&	0.3   &  1.1	   &  4.3  \\
L483  		   & 1.7 & 2.8 	&	1.0   &  1.1   &  6.7 \\
Ser SMM1    & 42.8 & 23.6  &	11.5  &  14.3   &  93.1  \\
Ser SMM4   & 7.4 & 3.6	&	\ldots   &  4.2	   &  15.2 \\
Ser SMM3   & 10.5 & 9.5 &	1.4   &  3.9	   &  25.3  \\
L723		   & 1.9 & 1.2 	&	0.6   &  1.0	   &  4.7  \\
\hline
L1489      & 0.7 & 0.8 	& 0.5  &  0.4   &  2.4\\
TMR1	   & 1.0 & 0.6 	& 0.5  &  0.5   &  2.6\\
TMC1A	   & 0.5 & \ldots 	& 0.1  &  0.6  &  1.2\\
TMC1 	  & 0.5 & 0.3	& 0.5   &  0.9   &  2.2\\
HH46       & 5.8 & 3.8 	& 4.1   &  22.8   &  36.5\\
RNO91	  & 0.1 & 0.1 	& 0.1   &  0.5  &  0.8 \\
\hline 
\end{tabular}
\tablefoot{All luminosities are in $10^{-3}$~L$_{\odot}$. Typical uncertainties
 are better than 30\% of the quoted values. BHR71 and Ser SMM4 were not 
observed in the OH lines in our program, thus their OH luminosities are not available. 
OH is not detected in NGC1333-I2A and H$_2$O is not detected in TMC1A.}
\end{table}

H$_2$O, CO and [\ion{O}{i}] are the most important gas cooling channels among the 
considered species. For the sources with the largest $L_\mathrm{FIR}$ in our sample 
(most of them are young Class 0 sources, except HH46), H$_2$O and CO clearly dominate the far-IR 
emission. [\ion{O}{i}] is the most important coolant for those objects with relatively
 small $L_\mathrm{FIR}$ (and more evolved, except Ced110-IRS4). The relative contribution 
 of H$_2$O is the highest for NGC1333-IRAS4A and IRAS4B, where it accounts for 
 $\sim50$ \% of $L_\mathrm{FIR}$. The relative contribution of CO is the largest for 
 Ser SMM1 and SMM3, where $\sim40$ \% of $L_\mathrm{FIR}$ is emitted in this channel. 
 [\ion{O}{i}] dominates the far-IR emission in HH46 and RNO91 (both are Class I sources) 
 with a relative contribution of $\sim60$ \%. The OH contribution to the total cooling is the
  highest for the Class I Taurus sources, 
in particular L1489, TMR1 and TMC1, where $\sim20$ \% of the total cooling is done by OH. 
For the rest of the sources the typical OH contribution is $\sim5-10$ \%.

The Class I source HH46 was studied by \citet{vK10}, who used the
central spaxel emission to determine relative contributions of 28,
22, 40 and 10\% for CO, H$_2$O, [\ion{O}{i}] and OH. 
Our results for this object, but using the full PACS
array and correcting for the missing lines, yield: 16, 10, 63
and 11\% for those species. This example illustrates that depending on
the extent of emission in different molecular and atomic species, the
derived values of the relative cooling can vary significantly, both
between species and across the protostellar envelope. In HH46, a
significant amount of the [\ion{O}{i}] emission is found
off-source and therefore in our calculations it is an even more significant 
cooling channel as compared to van Kempen et al. results (see
Table \ref{pattern} and Figure \ref{liness2} in Appendix).

%===========================
\subsection{Flux correlations}

To gain further insight into the origin of the emission of the various
species, the correlation of various line fluxes is investigated. In
addition, correlations between the far-infrared line cooling and other
source parameters are studied. The strength of the correlations is
quantified using the Pearson coefficient, $r$. The $3\sigma$
correlation corresponds to the Pearson coefficient of
$r=3/\sqrt{N-1}$, where $N$ is the number of sources used for the
calculation. For our sample of $N$=18 sources, $|r|<0.7$ denotes a
lack of correlation, $0.7<|r|<0.8$ a weak ($3 \sigma$) correlation and
$|r|>0.8$ a strong correlation.
%===========================
\begin{figure}[tb]
\begin{center}
\vspace{+3ex}

 \includegraphics[angle=0,height=12cm]{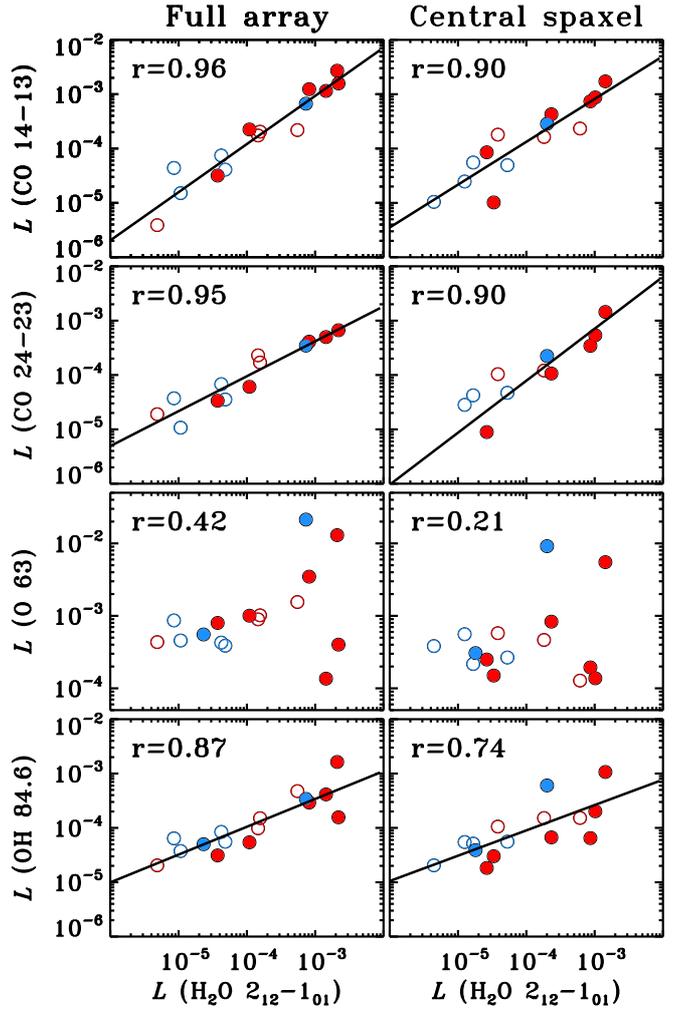}
 \vspace{+9ex}
 
 \caption{\label{new} Luminosity correlations between the H$_2$O
   2$_{12}$-1$_{01}$ line at 179.527 $\mu$m and (from top to bottom) 
   CO 14-13, CO 24-23, [\ion{O}{i}] at 63.18 $\mu$m and OH at 84.6
   $\mu$m, for 16 out of 18 objects (all luminosities in units of $L_\mathrm{\odot}$). 
   Full array and central spaxel luminosities are shown in the left and right columns,
   respectively. Red circles correspond to Class 0 sources and blue
   circles to Class I sources. Empty symbols show the \textit{compact}
   sources, whereas filled symbols denote the
   \textit{extended} sources. The Pearson coefficient of the correlation
   ($r$) is shown.}
\end{center}
\end{figure}

%===========================
\begin{figure}[!tb]
\begin{center}
\vspace{+3ex}

 \includegraphics[angle=0,height=12cm]{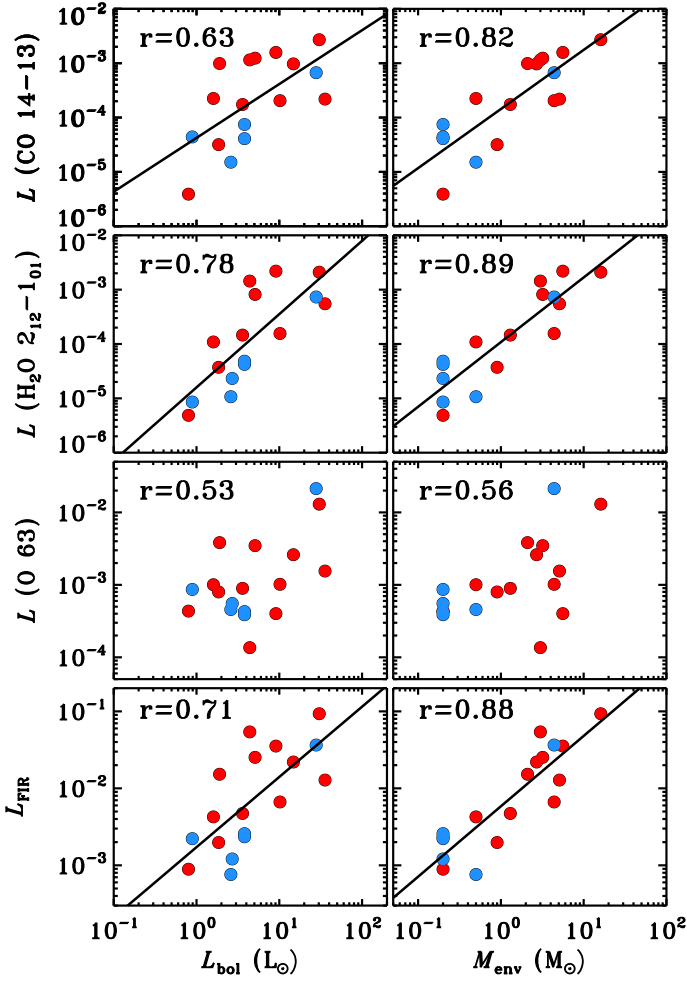}
 \vspace{+9ex}
 
 \caption{\label{flx2} Correlations of full array line emission with bolometric luminosity
   (left column) and envelope mass (right column) from top to
   bottom: CO 14-13, H$_2$O 2$_{12}$-1$_{01}$, [\ion{O}{i}] at 63.2
   $\mu$m line luminosities and total far-IR gas cooling $L_\mathrm{FIR}$.}
\end{center}
\end{figure}
%===========================
\begin{figure}[tb]
\begin{center}
\vspace{+3ex}

 \includegraphics[angle=90,height=5cm]{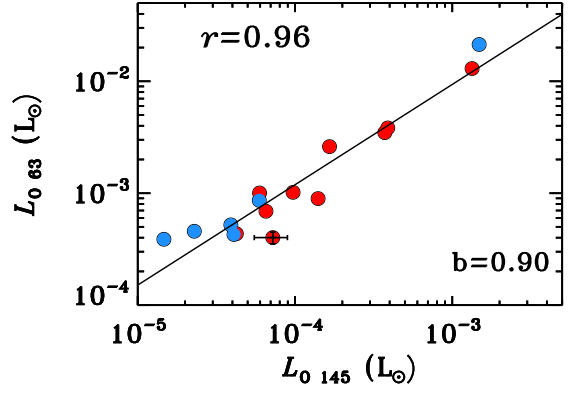}
 \caption{\label{oxyg} Full array correlation between the
   [\ion{O}{i}] line luminosities at 145 and 63 $\mu$m for 16 out of 18 
   objects (the [\ion{O}{i}] 145 line is not detected in NGC1333-IRAS2A and 4B). 
   Class 0 sources are shown in red and Class I sources in blue. A typical 
   error is shown on the NGC1333-IRAS4A observation.}
\end{center}
\end{figure}
%===========================

Figure \ref{new} compares the luminosities in the H$_2$O 2$_{12}$-1$_{01}$
line as a function of the CO 14-13, CO 24-23, [\ion{O}{i}] 63.2
$\mu$m and OH 84.6 $\mu$m line fluxes. Both the integrated luminosities
over the 5$\times 5$ array and the central spaxels corrected for the
missing flux according to standard PSF corrections are presented. Data
for BHR71 and Ser SMM4 are not included since some lines are
missing for these two sources.

A very tight correlation is found between the H$_2$O 2$_{12}$-1$_{01}$
and the CO lines, both on large and on small scales ($r\sim0.95$ for
5$\times$5 fluxes and $r=0.90$ for central spaxel luminosities). Thus,
H$_2$O and CO -- both in CO 14-13 and CO 24-23 -- likely arise in the
same gas. OH also strongly correlates with H$_2$O, in particular on
the larger scales ($r=0.87$), where it likely probes the same gas as
CO and H$_2$O. However, on the central spaxel, the correlation between
H$_2$O and OH weakens. The same results are found when the H$_2$O
3$_{03}$-2$_{12}$ line is used ($E_\mathrm{u}/k_\mathrm{B}\sim200$ K).

On the other hand, the [\ion{O}{i}] 63 $\mu$m line luminosities do not
correlate with the H$_2$O line luminosity, either on large or on small scales
($r=0.42$ and $r=0.21$, respectively). Significant variations in the
[\ion{O}{i}] luminosities are found in particular for the H$_2$O-bright Class
0 sources. Similar results are obtained for the [\ion{O}{i}] 145 $\mu$m line.

A strong correlation is also found between the two [\ion{O}{i}]
lines, at 63.2 and 145.5 $\mu$m ($r=0.96$, see Figure \ref{oxyg}). The
observed ratios of the [\ion{O}{i}] 63 $\mu$m / 145 $\mu$m lines vary
from $\sim$6 to $\sim$26 with a median value of $\sim$13 (see Table
\ref{oxytab}). Such a median line ratio corresponds to
$n_\mathrm{H}>10^{3}$ cm$^{-3}$ for optically thin lines excited
by collisions with H$_2$ \citep[see Figure 4 in][]{Li06}. Line ratios
of $\sim$20, as observed for example in IRAS15398, L1489 and RNO91,
are indicative of J-type shocks \citep{HM89}.

The median [\ion{O}{i}] 63.2 to 145.5 flux ratio of $\sim$10 for the
Class 0 sources and $\sim$16 for the Class I sources could imply
stronger shocks for the more evolved sources. However, the
[\ion{O}{i}] 63.2 flux may be suppressed for Class 0 sources because
they have higher envelope masses and therefore higher extinctions than
Class I sources (see also Herczeg et al. 2012, Goicoechea et
al. 2012).

Mass loss rates can be derived directly from the luminosity of
[\ion{O}{i}] at 63 $\mu$m using the \citet{Ho85} equation,
d$M$/dt=10~$L$(\ion{O}{i}~63), in units of $10^{-5}$ M$_{\odot}$
yr$^{-1}$. This formula assumes that the [\ion{O}{i}] line traces the
primary wind shock and that the shock is a major coolant, which is
valid for pre-shock densities of $\leq 10^{5}$ cm$^{-3}$. Mass loss
rates calculated from this simple formula for our sources vary from
$10^{-8}$ to $10^{-6}$ M$_{\odot}$ yr$^{-1}$ and are the highest for
Serpens SMM1 and HH46. Median values of the mass loss rate are 1
$10^{-7}$ M$_{\odot}$ yr$^{-1}$ and 0.5 $10^{-7}$ M$_{\odot}$ 
yr$^{-1}$ for Class 0 and
Class I sources, respectively, which reflects the lower [\ion{O}{i}]
63 $\mu$m line luminosities for the more evolved sources in Figure
\ref{oxyg}. In general, these values are in the same range as those of
the entrained outflow material derived from low-$J$ CO observations
(Y{\i}ld{\i}z et al. subm.). However, for some deeply-embedded and
heavily extincted sources, e.g. NGC1333-IRAS4A, the low-$J$ CO
mass-loss rates are up to 2 orders of magnitude higher than those
determined from the [\ion{O}{i}] 63.2 line (see Y{\i}ld{\i}z et
al. 2012 for low-$J$ CO spectroscopy of NGC1333-IRAS4A).
\begin{table}
\begin{minipage}[t]{\columnwidth}
\caption{Flux ratios of oxygen lines and inferred mass loss rates.}
\label{oxytab}
\centering
\renewcommand{\footnoterule}{}  % to avoid a line before footnotes
\begin{tabular}{lrrrrr}
\hline \hline
Object & $F$(O 63)/$F$(O 145) & d$M$/dt  \\
 ~ & ~ &  ($10^{-7}$~M$_{\odot}$~yr$^{-1}$)  \\
\hline
NGC1333-I2A &   $>45.0$ &  1.6  \\
NGC1333-I4A &   5.5 &  0.5  \\
NGC1333-I4B &    $>$15.8 &  0.1   \\
L1527  &   10.6 &  0.8  \\
Ced110-IRS4 &  10.3 &  0.5 \\
BHR71 &    15.7 &  2.8 \\
I15398	&  16.9 &  1.1\\
L483 &     10.5 &  1.1 \\
Ser SMM1 &     9.7  & 14.3 \\
Ser SMM4 &     9.8  & 4.2 \\
Ser SMM3 &     9.3  & 3.9 \\
L723 &     6.4  & 1.0 \\ 
\hline
L1489 &  26.4 	& 0.4  \\
TMR1 &  10.4  	& 0.5  \\
TMC1A &  13.3  	& 0.6  \\
TMC1 &  14.6  	& 0.9  \\
HH46 &  14.4  	& 22.8  \\
RNO91 &  20.0  	& 0.5  \\
\hline
\end{tabular}
\end{minipage}
\end{table}
%===========================
%\subsubsection{Correlations between line emission and source properties}
%===========================

Figure \ref{flx2} shows the line luminosities of the CO 14-13, H$_2$O
2$_{12}$-1$_{01}$ and [\ion{O}{i}] lines and the total far-IR line cooling
$L_\mathrm{FIR}$ as functions of physical parameters: bolometric luminosity,
$L_\mathrm{bol}$, and envelope mass, $M_\mathrm{env}$ (for correlations with 
bolometric temperature and density at 1000 AU see Appendix G). Values for these
parameters are taken from the physical models of \citet{Kr12} and the
results of this paper. The values of $L_\mathrm{bol}$ and $M_\mathrm{env}$
are not fully independent in our sample (a weak correlation exists, see 
Kristensen et al. 2012).

The CO and H$_2$O luminosities, as well as $L_\mathrm{FIR}$,
correlate with both $L_\mathrm{bol}$ and $M_\mathrm{env}$. The
correlations are particularly strong when the envelope mass is used.
However, no correlation is found for the [\ion{O}{i}] fluxes and the 
[\ion{O}{i}] line ratios with physical parameters.

%==========================
\section{Origin of far-IR line emission}
%==========================

The protostellar environment contains many physical components that
can give rise to far-infrared line emission. As illustrated in Figure
5 of \cite{WISH}, these include (i) the warm quiescent inner part of
the envelope passively heated by the luminosity of the source (the
`hot core'); (ii) the entrained outflow gas; (iii) UV-heated gas along
the cavity walls; (iv) shocks along the outflow cavity walls where the
wind from the young star directly hits the envelope; (v) bow shocks at
the tip of the jet where it impacts the surrounding cloud; (vi)
internal working surfaces within the jet; and (vii) a disk embedded in
the envelope.  In the case of shocks, both C- and J-type shocks are
possible. Spatially disentangling all of these components is not
possible with the resolution of Herschel, but our data combined with
velocity information from HIFI and physical-chemical models of the
molecular excitation provide some insight into which components most
likely dominate the emission (Visser et al.\ 2012, Herczeg et
al.\ 2012).

Velocity-resolved spectra of $^{12}$CO up to $J_\mathrm{u}$=10
\citep{Yi10,Yi12,Bj11,Be12} and of several H$_2$O
\citep{Kr10,Co10,Kr11,Kr12,Sa12,Va12} and OH \citep{Wa11} lines
indicate that the bulk of the far-infrared emission has broad line
wings with \mbox{$\Delta \varv > 15$ km s$^{-1}$}.  The quiescent warm
inner envelope is primarily seen in the narrow (\mbox{$\Delta\varv$
  $\lesssim$ 5 km s$^{-1}$}) $^{13}$CO and C$^{18}$O isotopolog
mid-$J$ spectra centered at the source position but this component
does not contribute to H$_2$O emission. The broad line wings combined
with the spatial extent of both CO and H$_{2}$O beyond the central
spaxel (i.e., beyond 1000~AU) argue against the hot core (option (i))
and the disk (option (vii)) being the main contributors, at least for
the {\it extended} sources. Internal jet shocks (option (vi)) have
large velocity shifts of 50 km s$^{-1}$ or more. While such
fast-moving gas is seen in some [\ion{O}{i}] \citep[][and \S
  3.3]{vK10} and HIFI H$_{2}$O spectra \citep{Kr11}, it is only a
minor fraction of the total emission for our sources (based on a lack
of extremely-high velocity (EHV)/ bullet emission in HIFI spectra) and
is ignored here. Our maps are generally not large enough to cover the
bow shocks at the tip of the outflow, ruling out option (v).

For these reasons, the far-infrared line emission seen here most
likely originates in the {\it currently} shocked and UV-heated gas
along the cavity walls in the protostellar environment (options (iii)
and (iv)). This warm and hot gas should be distinguished from the
cooler entrained outflow gas with rotational temperatures of $\sim$100
K traced by $^{12}$CO line wings up to $J_\mathrm{u}\approx$10
\citep[][option (ii)]{Yi12}.  \cite{Vi11} developed a physical model
that includes a passively heated envelope, UV-heated outflow cavity
walls and heating by small-scale C-shocks.  The model successfully
reproduces emission from the central spaxel in three sources (emission
from other spaxels was not included in the models). The cavity walls,
carved by the outflows, are illuminated by the UV from the
protostellar accretion shocks at the star-disk boundary although UV
produced by shocks inside the cavity itself can also contribute
\citep{Sp95,vK09}. The small-scale C-type shocks are produced when the
protostellar wind hits the outflow cavity walls and locally increases
the gas temperature to more than 1000 K, driving much of the oxygen
into water \citep{KN96}. The UV-heated gas can provide $^{12}$CO
emission up to $J_{u}\sim20$ depending on source parameters. Evidence
for UV-heated gas also comes from strong extended $^{13}$CO mid$-J$
emission along the cavity walls \citep{vK09,Yi12}. In contrast, shocks
are predicted to produce the bulk of high-$J$ CO ($J\geq20$) and all
of the H$_2$O in the PACS range \citep{Vi11}. The importance of the
UV-heating was suggested to increase with evolutionary stage in the
three sources studied by \cite{Vi11}, whereas the shock emission
weakens for the more evolved sources. Qualitative analysis of the
[\ion{O}{i}] and OH distribution in one case, HH46, argued for a
separate dissociative $J$-type shock centered on the source to explain
the detected emission \citep{vK10}.

In the following subsections, we seek to test these model predictions
qualitatively against the new observations and for a larger sample. As
we will see, the UV-heated component contributes less to the observed
PACS lines than thought before. Also, evolutionary trends will be
explored and compared to previous ISO observations \citep{Gi01,Ni02}.

%===========================
\subsection{Spatial extent of line emission and correlations}
%===========================

The molecular and atomic far-IR emission is commonly observed to be
extended along the outflow direction which is consistent with the
above picture of the emission being associated with the cavity walls.
The physical scales of the extent are of the order of $\sim10^4$ and
$\sim10^3$ AU for the \textit{extended} and \textit{compact}
groups, respectively. 

Most of the extended sources are well-known Class 0 objects primarily
known for their large-scale outflows. The extended emission seen in
high$-J$ CO and H$_2$O transitions is therefore likely associated
directly with shocks that are currently taking place in the
outflows. A few Class 0 sources famous for their large-scale outflows,
e.g., BHR71 and NGC1333-IRAS2A, are found in the compact group. Those
outflows extend well beyond the PACS field of view mapped here and
their active outflow hot spots are not covered in our data.  The
excitation conditions for high$-J$ CO and H$_2$O emission in the
outflow shocks have been determined recently in several studies based
on PACS and HIFI observations \citep[e.g., ][Tafalla et
  al. subm.]{Sa12,Va12,Be12,Bj12,Le12,Co12}. All studies conclude that
only a combination of high density and temperature can account for the
observed intensities. In that respect, the dense, extended envelopes
surrounding the Class 0 objects provide the ideal excitation
conditions for extended sources \citep[][]{Kr12}.

The compact group contains most Class I objects in our sample
  and objects with outflows seen nearly face-on
  (e.g. Ced110-IRS4). The envelopes surrounding Class I objects are
  less dense than around Class 0 objects \citep{Jo02,Kr12}, and thus
  less conducive to high$-J$ CO and H$_2$O excitation except close to
  the source itself. Their outflows are generally compact.

The fact that both the CO 14-13 and 24-23
fluxes as well as the H$_2$O 2$_{12}$-1$_{01}$ and CO
14-13 fluxes correlate very strongly with each other, both in the
5$\times$5 and central spaxels (Figure \ref{new}), and also
  show a similar spatial distribution, strengthens the conclusion
that the H$_2$O and mid/high-$J$ CO emission originate in the same
physical component. Such strong H$_2$O emission can only originate in
non-dissociative shocks \citep{ND89}. Thus, the observed correlation
suggests that most of the H$_2$O and CO emission arises in shocks and
that only a minor fraction of the CO 14-13 emission originates in the
UV-heated gas. Velocity-resolved line profiles of the CO 14-13 (or
neighboring CO lines such as CO 16-15) are needed to quantify the
relative contributions of UV heating and shocks to the mid-$J$ CO line
profiles. For the single case for which such HIFI data are available
in the WISH program, Ser SMM1, the line profile decomposition
indicates that $\sim$25$\pm$4\% of the flux comes from UV heated gas,
but this number may vary from source to source (Kristensen et al. in
prep.).

The fact that H$_2$O and CO emission is sometimes extended in a different
manner (L1527, L483 and NGC1333-IRAS4A) (see \S 3.2)
could be due to a variety of reasons, including asymmetric densities
and UV fields across the core.  Different densities on the two sides
could be caused by irregularities or gradients of the surrounding
cloud material distribution \citep{Liu12,To11} and can lead to different
excitation of the molecules.  The density also affects the penetration
of the UV field, which in turn results in abundance variations.  For
example, the UV field from the star-disk boundary can readily
photodissociate the H$_2$O molecule but is not hard enough to
dissociate CO. Higher extinction towards the red outflow lobe, as
found for NGC1333-IRAS4B \citep{He12}, can also result in different
fluxes towards the two outflow lobes, but this should affect all
species similarly as long as lines at similar wavelengths are
compared.

The [\ion{O}{i}] emission is usually extended along the outflow
direction and resembles the pattern of the OH extent.  In the two
exceptions, L1527 and HH46, the [\ion{O}{i}] is extended but the OH
emission is seen only at the source position (see Table
\ref{pattern}). For the {\it compact} sources, both [\ion{O}{i}] and
OH consistently peak together at the central position.
Together with a weaker correlation of OH with H$_2$O and no
correlation of [\ion{O}{i}] with H$_2$O (Figure \ref{new}) as well as
a strong correlation between [\ion{O}{i}] and OH \citep{Wa12}, this
 confirms the early conclusion that [\ion{O}{i}] and
at least part of the OH have a common origin. The [\ion{O}{i}]
  and OH emission is likely produced by dissociative shocks, in
  contrast to the non-dissociative shocks which explain the CO and
  H$_2$O emission (see Section 5.3).

%%%%===========================
\subsection{Excitation}
%%%%===========================
%===========================
\begin{figure}[tb]
\begin{center}
 \includegraphics[angle=90,height=6cm]{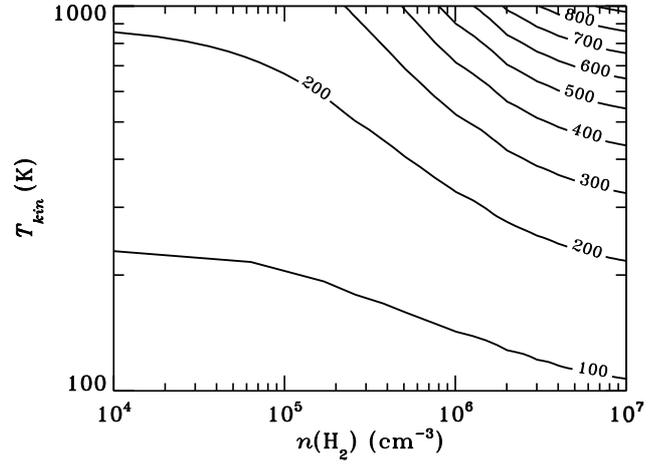}
\caption{\label{radex}CO rotational temperature $T_\mathrm{rot}$ from $J$=15--14 to 25--24 
as a function of density $n_\mathrm{H}$ and temperature of the emitting
  gas derived from non-LTE excitation calculations.}
\end{center}
\end{figure}
%===========================
A remarkable observational finding is the rather constant CO rotational
temperature of $\sim$300 K for the warm component, as well as the
common presence of a hotter component (see also Manoj et al. 2012, Green et al. subm.). 
In order to assess the physical conditions that could reproduce the
observed $T_{\rm rot}$ of the warm component, non-LTE molecular
excitation models were run for hydrogen densities of
$n$(H$_2$)=$10^4-10^7$ cm$^{-3}$ and kinetic temperatures of
$T_\mathrm{kin}=10^2-10^3$ K using RADEX \citep{RADEX} (see Figure
\ref{radex}). The ratio of column density and the line width
\textit{N}(CO)/$\Delta V$, which enters the escape probability
calculation, is chosen such that the emission is optically thin. CO
transitions from $J=15-14$ to $J=25-24$ are modelled using the
high$-T$ collisional rate coefficients from \cite{Ya10}, extended by
\cite{Ne12}. The transitions are plotted as rotational
diagrams and fitted with one rotational temperature, corresponding to
the warm component.

The inferred physical conditions cover a broad range of
$T_\mathrm{kin}$ and $n$(H$_2$) (Figure \ref{radex}). The range of
possible scenarios can be described by two limiting solutions: (i) CO
is sub-thermally excited in hot ($T_\mathrm{kin}\geq10^3$ K),
low-density ($n$(H$_2$)$\leq10^5$ cm$^{-3}$) gas (Neufeld 2012, Manoj
et al.\ 2012); or (ii) CO is close to LTE in warm ($T_\mathrm{kin}$
$\sim$ $T_\mathrm{rot}$) and dense ($n$(H$_2$)$> n_\mathrm{crit} \sim
10^{6}$ cm$^{-3}$) gas. We favor the latter interpretation based on
the strong association with H$_2$O emission and the high densities
needed to excite H$_2$O (see \S 5.1 and \citealt{He12}).
%=================
\begin{figure}[tb]
\begin{center}
%\hspace{+6ex}
\includegraphics[angle=90,height=6cm]{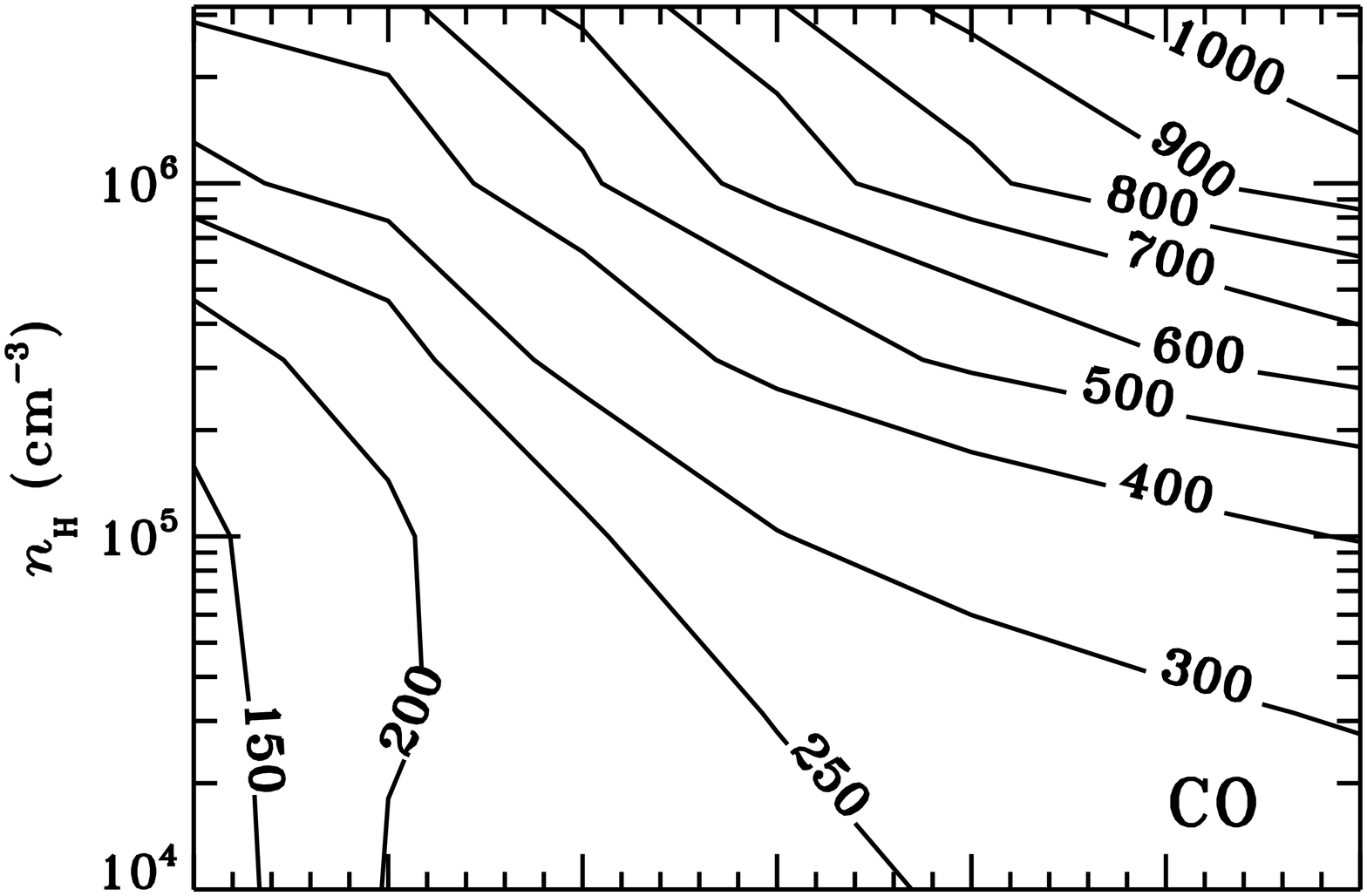}% co_trot_post.eps
\vspace{+1ex}
\includegraphics[angle=90,height=6cm]{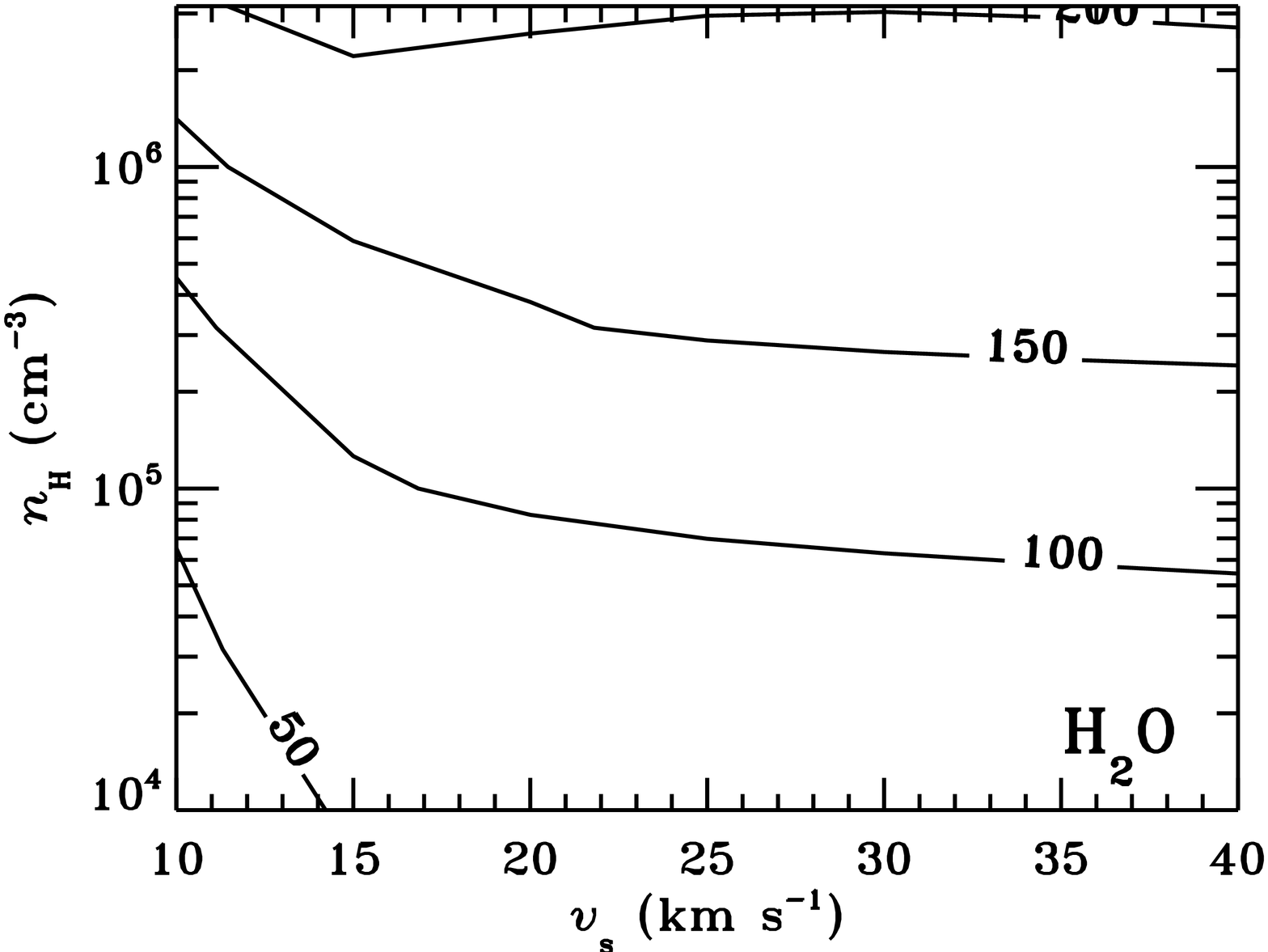} % h2o_trot_nohigh.eps
\caption{\label{kn} Rotational temperature of CO (top) and H$_2$O (bottom) as a 
function of different 
  shock velocities and pre-shock densities from the shock model results of 
  \cite{KN96}. The rotational temperature is
  calculated from J=15-14 to 25-24 for CO and for the eight
    H$_2$O lines commonly observed in this sample.}
\end{center}
\end{figure}

If H$_2$O and CO indeed co-exist, the low rotational
temperatures of H$_2$O ($T_\mathrm{rot}\sim 100-200$ K) as compared to
those of CO ($T_\mathrm{rot}\sim300-750$ K) indicate sub-thermal
excitation of H$_2$O, because the critical density of H$_2$O is higher
than that of high$-J$ CO by 1--2 orders of magnitude (depending on $J$
level; e.g. $n_\mathrm{crit}\sim10^{6}$ for CO 14-13 and $\sim10^{8}$ cm$^{-3}$
 for H$_2$O 2$_{12}$-1$_{01}$).
 RADEX simulations of optically thick H$_2$O (see Herczeg et al.\ 2012 and Figure B.1 in
 Kristensen et al.\ 2011) indicate densities $n_\mathrm{H}>10^{6}$
 cm$^{-3}$ to reproduce our observed H$_2$O rotational
 temperatures. This clearly favors the high-density, moderately
 high-temperature solution (ii) for the CO excitation.  The actual gas
 density likely lies between the CO and H$_2$O critical densities if
 CO is assumed to be close to thermally excited, i.e.,
 between a few $10^{6}$ cm$^{-3}$ to $10^{8}-10^{9}$
 cm$^{-3}$. Such high densities are only found within a few
 hundred AU of the protostar or at shock positions where the gas is
 strongly compressed \citep[][Tafalla et al., subm.]{Sa12,Va12}.
 The hot CO component is not modeled in this work because it is less
 well defined than in other studies, which have larger samples of
 sources with complete 55-200 $\mu$m PACS spectra (see Manoj et
 al. 2012). However, as is clear from Fig.~\ref{radex},
 $T_\mathrm{rot}\sim 450-800$~K requires even warmer and higher
 density gas in our scenario (ii). 

In the alternative solution (i) in which the entire CO far-infrared
ladder comes from low-density, very high-temperature, high column
density material (Neufeld 2012), the question arises where this gas
is located in the protostellar environment.
The only region close to the protostar (within the PACS central
spaxel) where gas densities are lower than $\sim10^{5}$ cm$^{-3}$ is
{\it inside} the outflow cavities. If the CO and H$_2$O emission were
to originate from the jet itself, velocity shifts of the high$-J$ CO
and H$_2$O lines would be expected in the short-wavelength parts of
PACS spectra, where the spectral resolution is the highest. As discussed in
\S 3.3,
we do not see clear evidence for velocity shifts of 100 km s$^{-1}$ or
more in short-wavelength H$_2$O and CO lines. Low velocity emission
from inside the cavity can originate from dusty disk winds
\citep{Pa12}, but more modeling is required to determine whether the
line fluxes and profiles can be reproduced in this scenario.
Moreover, this explanation would raise the question why there is no
contribution from shocks along the higher density cavity walls, which
surely must be present as well.

%%%%===========================
\subsection{Comparison with shock models}
%%%%===========================

The association of the CO and H$_2$O emission with warm dense shocked
gas can be further strengthened by comparison with shock models.
H$_2$O and CO are efficiently excited in non-dissociative
\mbox{C-type} shocks \citep{KN96,Be98,FP10}.
%, for example in shocks
The models by \citet{Vi11} used the output of the \cite{KN96} models
to compute line fluxes along the cavity walls. To test our full data set
against these models and further explore parameter space,
rotational temperatures were computed from the grid of \cite{KN96}
model fluxes and presented in Figure \ref{kn}. The results show that
CO excitation temperatures $T_\mathrm{rot}\approx 300$~K for
\mbox{$J$=15--25} are readily found for pre-shock densities of
$\sim3\cdot10^{4}-10^6$ cm$^{-3}$ and a wide range of shock
velocities. Similar results are obtained for the H$_2$O
  excitation temperatures: the observed values of 100--200 K correspond
  to the pre-shock densities of $\sim10^{5}-10^6$ cm$^{-3}$ and do not
  constrain the shock velocities.
 
 The density probed by the CO and H$_2$O lines is not the pre-shock
 density; by the time CO and H$_2$O cooling becomes important, the gas
 has been compressed by the shock front and a more relevant density is
 the post-shock density \citep{FP10}.  The compression factor, $n_{\rm
   post}$ / $n_{\rm pre}$, is $\sqrt{2} \times M_{\rm A}$, where
 $M_{\rm A}$ is the Alfven Mach number, $\varv_{\rm shock}$ /
 $\varv_{\rm A}$, and $\varv_{\rm A}$ is the Alfven velocity
 \citep{Dr93}. The Alfven velocity is $\varv_{\rm A}$ = $B$ / $\sqrt{4
   \pi \rho}$ = 2.18 km\,s$^{-1}$ $b$ / $\sqrt{\mu_{\rm H}}$ where
 $\rho$ is the pre-shock density, $\rho$ = $\mu_{\rm H}$ $m_{\rm H}$
 $n_{\rm pre}$; the latter relation comes from the assumption that the
 magnetic field is frozen into the pre-shock gas and is $B$ = $b$
 $\sqrt{n_{\rm pre}\ [{\rm cm}^{-3}]}$ $\mu$Gauss.  For a mean atomic
 weight, $\mu_{\rm H}$, of 1.4, the compression factor is
\begin{align}
\frac{n_{\textrm post}}{n_{\textrm pre}} &= \sqrt{2} M_{\rm A} = \sqrt{2} \frac{\varv_{\rm shock}}{\varv_{\rm A}} \\
&= \sqrt{2} \frac{\varv_{\rm shock}}{2.18\, b / \sqrt{\mu_{\rm H}}} = 0.78 \times \varv_{\rm shock} / b\ .
\end{align}
For a standard value of $b$ = 1, the compression factor is thus $\gtrsim$ 10 for all the shock velocities
 considered here, implying that the relevant post-shock densities are all $\gtrsim$ 10$^6$ cm$^{-3}$ as
  also shown by the RADEX results (Figure 13). At high velocities, $\sim$ 50 km\,s$^{-1}$, the post-shock
   densities are greater than 10$^7$ cm$^{-3}$ and the emission is in LTE.
  
Density is clearly the critical parameter: higher $T_{\rm rot}$
require higher pre-shock densities which in turn lead to higher
post-shock densities. Thus, this shock model analysis is consistent
with that presented in \S 5.2 and with the typical pre-shock densities
found in the inner envelope.
%As noted above, the strong correlation of CO 14-13 with 24-23 argues
%against UV radiation being the dominant heating agent of the CO
%emitting gas. However, 
The presence of UV radiation in the outflow cavity may affect the
shock structure, however, and new irradiated models are required to
fully test this scenario (Kaufman et al., in prep.).
%by changing the ion abundance and thus the ion-neutral
%coupling length. Work is in progress to develop a new set of shock
%models that apply to this situation (Kaufman et al. in prep.). 
With the addition of UV, some OH emission is likely also produced by
photodissociation of H$_2$O.

A large fraction of the [\ion{O}{i}] and some OH emission must
originate in a different physical component. Because the emission from
these species does not correlate with that of CO and H$_2$O,
especially in the central spaxel, their origin is likely in
dissociative shocks rather than non-dissociative C- or J-type
shocks. In particular, the absolute flux of [\ion{O}{i}] is generally
too high to be produced in non-dissociative, C-type shocks
\citep{FP10}. Dissociative shocks are typically found at the terminal
bow shock where the $>$100 km s$^{-1}$ jet rams into the surrounding
cloud, but they can also be located closer to the protostar where the
protostellar wind impacts directly on the dense inner envelope near
the base of the wind \citep{vK10}.  In dissociative shocks,
[\ion{O}{i}] and OH emission greatly dominate over that of H$_2$O as
molecules are gradually reformed in the dense post-shock gas
\citep{ND89}. Some emission may also arise in the jet itself, as
evidenced by the velocity shifts of $\sim$100 km s$^{-1}$ of the
[\ion{O}{i}] line in a few sources (Figure \ref{oshift}), but this is
usually a minor component according to our data.

%%%%===========================
\subsection{Evolution from Class 0 to Class I}
%%%%===========================
Class 0 and Class I objects are relatively short phases of
protostellar evolution \citep[0.16 and 0.54 Myr, respectively,
see][]{Ev09}, yet our far-IR observations reveal a significant change
in their spectral line characteristics. In the following, evolutionary
trends of line fluxes, spatial extent, excitation and far-infrared gas
cooling are discussed. 

%=======
\begin{figure}[!tb]
\begin{center}
 \includegraphics[angle=90,height=6cm]{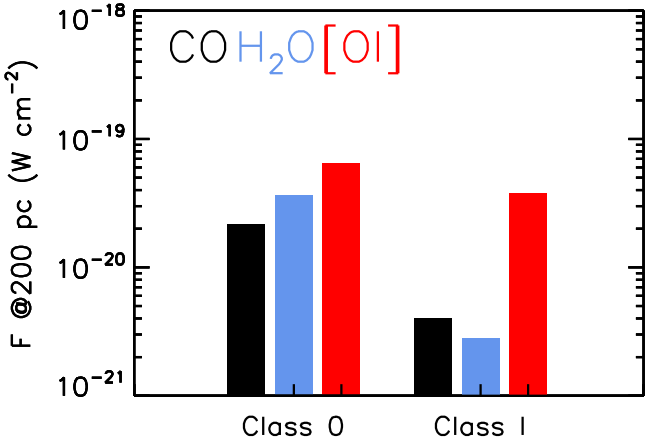}
 \caption{\label{hist} Median line flux of the CO
   14-13, H$_2$O \mbox{2$_{12}$-1$_{01}$} and [\ion{O}{i}]
   \mbox{$^3P_{1}-^{3}P_{2}$} lines for Class 0 and I objects from our
   sample are shown from left to right for each class in black, blue and red, respectively.}
\end{center}
\end{figure}
%=======
\begin{figure}[!tb]
\begin{center}
 \includegraphics[angle=0,height=10cm]{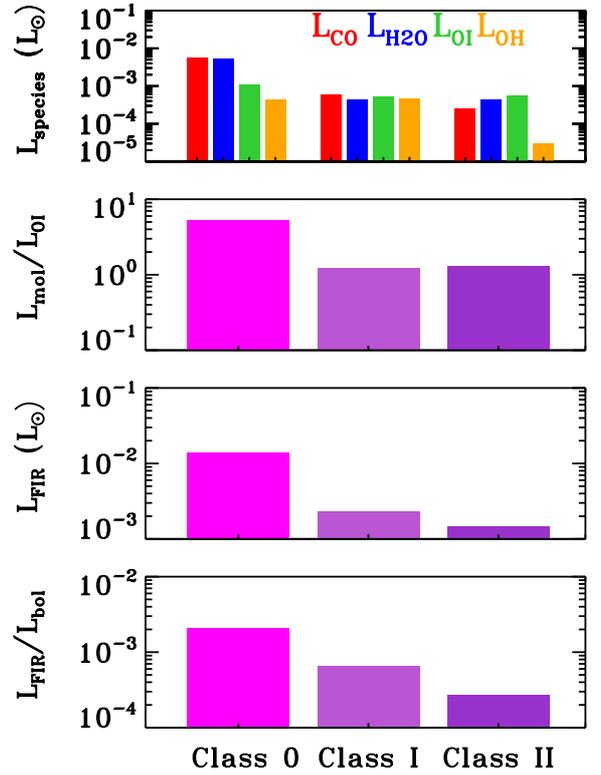}
 \caption{\label{hist2}{\it Panel 1:} Median cooling in CO, H$_2$O, OH and
   [\ion{O}{i}] lines for Class 0, I and II objects are shown from left to right for 
   each class in red, blue, orange and green colors, respectively. (Class II data from Podio et al. in press.).
  {\it Panel 2:} Evolution of the ratio of molecular and atomic cooling. {\it Panel 3:} Evolution of 
  the median total far-IR gas cooling. {\it Panel 4:} Evolution of the 
  median total far-IR gas cooling over bolometric luminosity.}
\end{center}
\end{figure}
%==========
\begin{table}
\begin{minipage}[t]{\columnwidth}
\caption{Median molecular and atomic luminosities for Class 0, I and II sources.}
\label{coolnew}
\centering
\renewcommand{\footnoterule}{}  % to avoid a line before footnotes
\begin{tabular}{llllllll}
\hline \hline
Species & Class 0 & Class I & Class II \\
\hline
$L_\mathrm{CO}$ & 5.7 & 0.6 & 0.25 \\
$L_\mathrm{H2O}$& 5.3 & 0.5 & 0.45 \\
$L_\mathrm{OH}$ & 0.4 & 0.5 & 0.03  \\
$L_\mathrm{OI}$ & 1.1 & 0.5  & 0.55 \\
\hline
$L_\mathrm{mol}$ & 11.0 & 1.6 & 0.73 \\
$L_\mathrm{FIR}$& 14.0 & 2.3 & 1.45 \\
$L_\mathrm{mol}$/L$_\mathrm{OI}$ & 5.2 & 1.2 & 1.3 \\
$L_\mathrm{FIR}$/L$_\mathrm{bol}$ & 2.1 & 0.6 & 0.27 \\
\hline
\end{tabular}
\end{minipage}
\tablefoot{
Luminosities are expressed in $10^{-3}$ L$_{\odot}$. 
Percent of the total FIR cooling is given for each species
 and total molecular cooling in brackets. 
Class II data are taken from Podio et al. (in press).}
\end{table}
 
 Figure \ref{hist} shows the median fluxes of the CO 14-13, H$_2$O
2$_{12}$-1$_{01}$ and [\ion{O}{i}] $^3P_{1}-^{3}P_{2}$ lines for the
12 Class 0 and 6 Class I sources. Although the Class I sample is
small, a factor of 5 decrease is observed in the CO and H$_2$O
fluxes for the more evolved sources. These trends indicate that
the molecular emission decreases quickly with evolution, whereas atomic emission
remains similar. One explanation could be that molecular abundances are
lower due to enhanced photodissociation as the envelope becomes more
dilute. When the average density of the envelope drops
(see Kristensen et al. 2012 for our sources), UV photons penetrate more readily
and over larger spatial scales, thus dissociating more H$_2$O and OH
to form O. This scenario was originally put forward by \cite{Ni02}. 

%===========================
% \subsubsection{Excitation}
%===========================
An alternative explanation could be that the molecules are less
excited because of the lower densities. The strong correlation of the
line fluxes with $M_{\rm env}$ would be consistent with this
option. Also, comparison of CO rotational temperatures for Class 0 and
I sources shows that the former are characterized by higher
temperatures of the hot component, possibly due to the higher
densities. The warm component rotational temperatures of CO are
remarkably constant for all sources, however. The rotational
temperature of H$_2$O tends to increase for the Class I sources and is
mainly due to the hot water detections in the Taurus objects. Since
there are only 6 Class I sources in this sample it is difficult to
examine how robust the high H$_2$O rotational temperatures are for the
Class I sources, although these lines are now also seen in some Class
II sources \citep{RM12}. Dust extinction could hide the highly-excited
H$_2$O lines at shorter wavelengths in the more embedded Class 0
sources \citep{He12}. Disks may also contribute to the far-IR emission
seen from some Class I sources but are likely entirely obscured at
short wavelengths by the more massive envelopes around Class 0
sources.

% \subsubsection{Cooling budget}
%===========================
The evolution of the far-infrared gas cooling in CO, H$_2$O, OH and
[\ion{O}{i}] is presented in Figure \ref{hist2}. Our calculations for 
Class 0 and I sources are supplemented by results from Podio et al.\ (2012)
for the Class II sources (average cooling of DG Tau A and RW Aur) that drive prominent
jets. Similar to the selected line flux trends, a factor of 10 decrease in the CO and H$_2$O
cooling is detected for the Class I sources as compared to the Class 0
objects. Further decrease, although much smaller, is seen for the Class II 
objects. The absolute values of the OH cooling are similar in the Class 0 and I objects, 
but decrease by a factor of 10 for the Class II objects. Cooling in [\ion{O}{i}] 
does not change that dramatically in the three classes; it is a factor of 
2 lower for Class I and II sources with respect to the Class 0 sources.

As a result, the relative contribution to the total cooling by
different species changes significantly in the course of the embedded
phase of evolution. In the earliest phase, CO and H$_2$O dominate the
total gas cooling with only a minor contribution from OH and
[\ion{O}{i}]. In the more evolved Class I phase, all species
considered here contribute almost equally to the total gas cooling.
The evolution of the total molecular cooling, $L_\mathrm{mol}$, where
$L_\mathrm{mol}=L_\mathrm{CO}+L_\mathrm{H_{2}O}+L_\mathrm{OH}$ with
respect to atomic cooling, $L_\mathrm{OI}$, shows a significant
decrease of a factor of $\sim5$ between Class 0 and I sources.  In the
Class II sources, [\ion{O}{i}] is the largest coolant among the
considered species, followed by CO and H$_2$O. These conclusions are
robust within the uncertainties of our calculations of the total
luminosities. As discussed in \S 4.2, this discussion excludes any 
potential contribution from H$_2$ cooling.

Similarly, the total far-infrared gas cooling and the total cooling
over the bolometric luminosity decrease with evolutionary stage, most
drastically between the Class 0 and Class I phases.  These results are
in agreement with the \textit{ISO} observations of 17 Class 0 and 11
Class I sources \citep{Gi01,Ni02}. The decrease in molecular emission
was attributed to the less powerful jet impacting the lower-density
envelope of the more evolved sources \citep{Ni02}. Our results show
that density may be more important than shock velocity in controlling
the emission (see Figure \ref{kn}).

%===========================
\section{Conclusions}
%===========================
We have characterized the \textit{Herschel}/PACS spectra of 18
deeply embedded low-mass protostars. The spatially resolved information
allows us to link the emission in different species with the physical
components of a young stellar object and study the spatial scales over
which the object interacts with its surroundings. The conclusions
are as follows:
\begin{enumerate}

\item Emission from the H$_2$O $2_{12}-1_{01}$ line at 179.5
  $\mu$m ($E_\mathrm{u}/k_\mathrm{B}=114$ K) is detected towards all Class 0 and I YSOs from our
  sample. The highly excited H$_2$O 8$_{18}$-7$_{07}$ line ($E_\mathrm{u}/k_\mathrm{B}=1071$
  K) at 63.3 $\mu$m is detected in 7 out of 18 sources, in both Class
  0 and I objects. CO lines are detected from $J=14-13$ up to $J=48-47$.

\item Emission in the H$_2$O, CO, OH and [\ion{O}{i}] lines is
  extended along the outflow directions mapped in lower-$J$ CO
  emission.  The extent of emission covers the PACS array in the
  outflow direction for half of our sources (\textit{extended}
  sample), which accounts for a region of $\sim10^4$ AU. The rest of
  the sources show compact emission limited mostly to the central
  spaxel (a region of $\sim10^3$ AU) (\textit{compact} sample). The
  {\it extended} sample is dominated by Class 0 sources with evidence
  for active shocks (`hot spots' ) currently taking place along the
  outflow.

\item Fluxes of the H$_2$O $2_{12}-1_{01}$ line and the CO
  $14-13$ and $24-23$ lines are strongly correlated with each other as
  well as with the physical source parameters $L_\mathrm{bol}$ and
  $M_\mathrm{env}$ suggesting they arise in the
  same physical component. In contrast, H$_2$O and CO fluxes 
  correlate less strongly with the OH, in particular on small scales, 
  and not at all with the [\ion{O}{i}] line fluxes.  

\item Rotational diagrams of CO show two distinct components -- the
  \textit{warm component} with $T_\mathrm{rot1}\sim$350 K and the
  \textit{hot component} with $T_\mathrm{rot2}\sim700$ K. 
  The hot component is weaker for Class I
  sources. Rotational diagrams of H$_2$O show scatter due to 
  sub-thermal excitation and optical depth effects and are
  characterized by $T_\mathrm{rot1}\sim150$ K for all sources,
  with a tendency for higher temperatures for Class I sources.

\item CO and H$_2$O are argued to arise in non-dissociative shocks
  along the outflow walls with a range of pre-shock densities. UV
  heating appears to play a minor role in the excitation of these PACS
  lines but affects the shock structure and chemistry by dissociating
  H$_2$O to OH. These shocks are likely responsible for the
  warm CO component. The origin of the hot CO component requires 
  further data and analysis.
  
\item {}[\ion{O}{i}] and part of the OH originate largely in
  dissociative shocks at the point of direct impact of the wind on the
  dense envelope. Only a few sources show high-velocity [\ion{O}{i}]
  emission tracing a hidden atomic jet.

\item H$_2$O is a major coolant among the observed far-IR lines and
  accounts for 25 to 50\% of the total far-IR gas cooling,
  $L_\mathrm{FIR}$. CO is the other important coolant with a
  contribution of 5 to 50 \% to $L_\mathrm{FIR}$. The [\ion{O}{i}] 
  contribution to the total cooling accounts for
  5-30\% and increases for the more evolved sources, whereas 1-15\%
  is radiated in the OH lines.

\item  Weaker emission in the molecular lines for the more evolved sources
 can result either from lower envelope densities resulting in less
 excitation or from lower abundances of the molecules due to enhanced
 photodissociation. Both aspects likely play a role, but density is plausibly 
 the critical factor in controlling the line emission.

\end{enumerate}

\begin{acknowledgements}
The authors thank Linda Podio for making her results concerning cooling in Class II 
sources available for us prior to the publication and the referee for numerous suggestions which improved the paper.
AK acknowledges support from the Christiane N\"{u}sslein-Volhard-Foundation, the L$^\prime$Or\'{e}al  
Deutschland and the German Commission for UNESCO via the `Women in Science' prize. JRG is 
supported by a Ram\'on y Cajal research contract and 
thanks the Spanish MICINN for funding support through grants AYA2009-07304 and
CSD2009-00038. DJ acknowledges support from an NSERC Discovery Grant.
Herschel is an ESA space observatory with science instruments provided
by European-led Principal Investigator consortia and with important
participation from NASA. Astrochemistry in Leiden is supported by the Netherlands Research
School for Astronomy (NOVA), by a Spinoza grant and grant 614.001.008
from the Netherlands Organisation for Scientific Research (NWO), and
by the European Community's Seventh Framework Programme FP7/2007-2013
under grant agreement 238258 (LASSIE).
\end{acknowledgements}   
   
\bibliographystyle{aa}
\bibliography{biblio2009}

\clearpage
\clearpage
\appendix
%================================
\section{Targeted lines and measurements}

Table \ref{linestable} lists the species and transitions observed in
the PACS range with the line spectroscopy mode in the decreasing
wavelength order. Information about the upper level energies,
$E_\mathrm{u}/k_\mathrm{B}$, the Einstein coefficients, $A$, the weights,
$g_\mathrm{up}$, and frequencies, $\nu$, are obtained from the
Cologne Database for Molecular Spectroscopy \citep{CDMS2,CDMS}, the
Leiden Atomic and Molecular Database \citep{LAMDA} and the JPL Catalog
\citep{JPL}.

Tables \ref{obsn1} and \ref{obsn2} list line fluxes for all our sources in units of 10$^{-20}$ W cm$^{-2}$.
The uncertainties are 1$\sigma$ measured in the continuum on both sides of each line; 
calibration uncertainty of 30\% of the flux should be included for comparisons with other 
modes of observations or instruments.

\begin{table}
\begin{minipage}[t]{\columnwidth}
\caption{\label{linestable}Lines observed in the line spectroscopy mode.}             
\centering     
\renewcommand{\footnoterule}{}  % to avoid a line before footnotes
{\tiny
\begin{tabular}{l l l l l l l l l l}     % 7 columns 
\hline\hline       
Species & $\lambda_\mathrm{lab}$ ($\mu$m) & $E_\mathrm{u}$/$k_\mathrm{B}$ (K) & $A$ (s$^{-1}$) &
$g_\mathrm{up}$ & $\nu$ (GHz)\\
\hline           
CO 14-13 & 185.999 & 580.49 & 2.739(-4) & 29 & 1611.7935 \\
o-H$_2$O 2$_{21}$-2$_{12}$ & 180.488 & 194.093 & 3.065(-2) & 15 & 1661.0076 \\
o-H$_2$O 2$_{12}$-1$_{01}$ & 179.527 & 114.377 & 5.599(-2) & 15 & 1669.9048 \\
o-H$_2$O 3$_{03}$-2$_{12}$ & 174.626 & 196.769 & 5.059(-2) & 21 & 1716.7697 \\
OH $\frac{3}{2}$,$\frac{1}{2}$-$\frac{1}{2}$,$\frac{1}{2}$ & 163.398 & 269.761 & 2.121(-2) & 4 &
1834.7355 \\
OH $\frac{3}{2}$,$\frac{1}{2}$-$\frac{1}{2}$,$\frac{1}{2}$ & 163.131 & 270.134 & 2.133(-2) & 4 &
1837.7466 \\
CO 16-15 & 162.812 & 751.72 & 4.050(-4) & 33 & 1841.3455 \\
{[\ion{O}{i}]} $^{3}P_{0}-^{3}P_{1}$ &  145.525 &   326.630  &  1.750(-5)  &  1 & 2060.0691 \\
CO 18-17 & 144.784 & 944.97 & 5.695(-4) & 37 & 2070.6160 \\
p-H$_2$O 8$_{44}$-7$_{53}$ & 138.641 & 1628.371 & 1.188(-2) & 17 & 2162.3701 \\
p-H$_2$O 3$_{13}$-2$_{02}$ & 138.528 & 204.707 & 1.253(-1) & 7 & 2164.1321 \\
p-H$_2$O 4$_{04}$-3$_{13}$ & 125.354 & 319.484 & 1.730(-1) & 9 & 2391.5728 \\
OH $\frac{5}{2}$,$\frac{3}{2}$-$\frac{3}{2}$,$\frac{3}{2}$ & 119.442 & 120.460 & 1.361(-2) & 6 &
2509.9355 \\
OH $\frac{5}{2}$,$\frac{3}{2}$-$\frac{3}{2}$,$\frac{3}{2}$ & 119.235 & 120.750 & 1.368(-2) & 6 &
2514.2988 \\
CO 22-21 & 118.581 & 1397.39 & 1.006(-3) & 45 & 2528.1721 \\
p-H$_2$O 5$_{33}$-5$_{24}$ & 113.948 & 725.097 & 1.644(-1) & 11 & 2630.9595 \\
o-H$_2$O 4$_{14}$-3$_{03}$ & 113.537 & 323.492 & 2.468(-1) & 27 & 2640.4736 \\
CO 23-22 & 113.458 & 1524.20 & 1.139(-3) & 47 & 2642.3303 \\
CO 24-23 & 108.763 & 1656.48 & 1.281(-3) & 49 & 2756.3875 \\
o-H$_2$O 2$_{21}$-1$_{10}$ & 108.073 & 194.093 & 2.574(-1) & 15 & 2773.9766 \\
CO 29-28 & 90.163 & 2399.84 & 2.127(-3) & 59 & 3325.0054 \\
p-H$_2$O 7$_{44}$-7$_{35}$ & 90.050 & 1334.815 & 3.549(-1) & 15 & 3329.1853 \\
p-H$_2$O 3$_{22}$-2$_{11}$ & 89.988 & 296.821 & 3.539(-1) & 7 & 3331.4585 \\
CO 30-29 & 87.190 & 2564.85 & 2.322(-3) & 61 & 3438.3645 \\
o-H$_2$O 7$_{16}$-7$_{07}$ & 84.767 & 1013.206 & 2.131(-1) & 45 & 3536.6667 \\
OH $\frac{7}{2}$,$\frac{3}{2}$-$\frac{5}{2}$,$\frac{3}{2}$ & 84.596 & 290.536 & 4.883(-1) & 8 &
3543.8008 \\
OH $\frac{7}{2}$,$\frac{3}{2}$-$\frac{5}{2}$,$\frac{3}{2}$ & 84.420 & 291.181 & 2.457(-2) & 8 &
3551.1860 \\
CO 31-30 & 84.411 & 2735.30 & 2.525(-3) & 63 & 3551.5923 \\
o-H$_2$O 6$_{16}$-5$_{05}$ & 82.032 & 643.496 & 7.517(-1) & 39 & 3654.6033 \\
CO 32-31 & 81.806 & 2911.18 & 2.735(-3) & 65 & 3664.6843 \\
CO 33-32 & 79.360 & 3092.47 & 2.952(-3) & 67 & 3777.6357 \\
OH $\frac{1}{2}$,$\frac{1}{2}$-$\frac{3}{2}$,$\frac{3}{2}$ & 79.182 & 181.708 & 2.933(-2) & 4 &
3786.1318 \\
OH $\frac{1}{2}$,$\frac{1}{2}$-$\frac{3}{2}$,$\frac{3}{2}$ & 79.116 & 181.936 & 5.818(-3) & 4 &
3789.2703 \\
p-H$_2$O 6$_{15}$-5$_{24}$ & 78.928 & 781.120 & 4.555(-1) & 13 & 3798.2817 \\
o-H$_2$O 4$_{23}$-3$_{12}$ & 78.7423 & 432.154 & 4.865e-01 & 27 & 3807.2585 \\
CO 36-35 & 72.843 & 3668.82 & 3.639(-3) & 73 & 4115.6055 \\
p-H$_2$O 8$_{17}$-8$_{08}$ & 72.032 & 1270.28 & 3.050(1) & 17 & 4161.9189 \\
o-H$_2$O 7$_{07}$-6$_{16}$ & 71.947 & 843.47 & 1.161(0) & 45 & 4166.8511 \\
CO 38-37 & 69.074 & 4080.03 & 4.120(-3) & 77 & 4340.1382 \\
o-H$_2$O 8$_{18}$-7$_{07}$ & 63.324 & 1070.683 & 1.759 & 51 & 4734.2959 \\
{[\ion{O}{i}]} $^{3}P_{1}-^{3}P_{2}$ & 63.184 & 227.713 & 8.914(-5) & 3 & 4744.7773 \\
p-H$_2$O 4$_{31}$-4$_{04}$ & 61.809 & 552.263 & 2.383(2) & 9 & 4850.3345 \\
p-H$_2$O 7$_{26}$-6$_{15}$ & 59.987 & 1020.967 & 1.350 & 15 & 4997.6133 \\
CO 44-43 & 59.843 & 5442.39 & 5.606(-3) & 89 & 5009.6079 \\
CO 46-45 & 57.308 & 5939.20 & 6.090(-3) & 93 & 5231.2744 \\
p-H$_2$O 4$_{31}$-3$_{22}$ & 56.325 & 552.263 & 1.463 & 9 & 5322.5459 \\
o-H$_2$O 5$_{32}$-5$_{05}$ & 54.507 & 732.066 & 3.700(-2) & 33 & 5500.1006 \\
CO 49-48 & 53.898 & 6724.17 & 6.777(-3) & 99 & 5562.2583 \\
\hline                  
\end{tabular}}
\end{minipage}
\end{table}

\begin{landscape}
\begin{table}
\caption{\label{obsn1} Line fluxes of Class 0 and I sources in 10$^{-20}$ W cm$^{-2}$.}             
%\centering     
\renewcommand{\footnoterule}{}  % to avoid a line before footnotes
\resizebox{1.3\textheight}{!}
{\begin{tabular}{llccccccccc}     % 7 columns 
\hline\hline       
Species & Trans. & $\lambda_\mathrm{lab}$ ($\mu$m) & \multicolumn{8}{c}{Full array flux (10$^{-20}$ W cm$^{-2}$)}\\     
~ & ~ & ~ &  IRAS2A & IRAS4A  & L1527 & CedIRS4 & BHR71 &  IRAS15398 & L483 & SerSMM4 \\
\hline
CO & 14-13 & 185.999 &  1.65$\pm$0.28 & 9.42$\pm$0.08 & 0.52$\pm$0.14 & 0.14$\pm$0.01 & 7.79$\pm$0.15 & 4.19$\pm$0.10 & 1.31$\pm$0.02 & 5.75$\pm$0.10\\
CO & 16-15 & 162.812 &  2.11$\pm$0.38 & 8.00$\pm$0.09 & 0.57$\pm$0.07 & 0.16$\pm$0.01 &     \ldots      & 4.24$\pm$0.12 & 1.10$\pm$0.02 &     \ldots     \\
CO & 18-17 & 144.784\tablefootmark{a} &  2.76$\pm$0.29 & 6.74$\pm$0.09 & 0.51$\pm$0.08 & 0.24$\pm$0.02 & 8.54$\pm$0.16 & 3.11$\pm$0.09 & 1.16$\pm$0.05 & 3.68$\pm$0.12\\
CO & 22-21 & 118.581 &      \ldots      & 3.12$\pm$0.27 & 0.31$\pm$0.11 &     $<$0.20      &     \ldots      & 2.21$\pm$0.20 & 0.61$\pm$0.02 &     \ldots     \\
CO & 23-22 & 113.458\tablefootmark{b}  &  3.16$\pm$0.37 &     \ldots      & 0.95$\pm$0.12 & 0.59$\pm$0.02 &     \ldots      & 3.56$\pm$0.14 & 2.55$\pm$0.07 &     \ldots     \\
CO & 24-23 & 108.763 &      \ldots      & 3.92$\pm$0.28 &     $<$0.26    & 0.20$\pm$0.03 & 5.65$\pm$0.33 & 1.13$\pm$0.11 & 0.97$\pm$0.04 & 1.17$\pm$0.18\\
CO & 29-28 & 90.163\tablefootmark{c} &     $<$0.30    &     \ldots      & 0.10$\pm$0.02 &     \ldots      &     \ldots      & 0.42$\pm$0.04 & 0.59$\pm$0.04 &     \ldots     \\
CO & 30-29 & 87.190 &      $<$0.67      & 0.97$\pm$0.12 &     $<$0.33      &     $<$0.35      &     \ldots      & 0.36$\pm$0.03 & 0.41$\pm$0.05 &     \ldots     \\
%CO & 31-30 &  84.411 &      n.d.      &     n.d.      & 0.39$\pm$0.05 & 0.61$\pm$0.04 &     n.d.      &     n.d.      & 1.25$\pm$0.08 &     n.d.     \\
CO & 32-31 &  81.806 &      $<$0.37      & 0.45$\pm$0.05 &     $<$0.21      & 0.16$\pm$0.04 &     \ldots      & 0.26$\pm$0.04 & 0.43$\pm$0.05 &     \ldots     \\
CO & 33-32 &  79.360 &      $<$0.54      & 0.47$\pm$0.09 & 0.13$\pm$0.03 & 0.14$\pm$0.05 &     \ldots      & 0.40$\pm$0.05 & 0.60$\pm$0.07 &     \ldots     \\
CO & 36-35 &  72.843\tablefootmark{d} &      $<$0.42      &    $<$0.09      &     $<$0.36      &    $<$0.08      & 0.89$\pm$0.08 &     $<$0.16      & 0.19$\pm$0.04 & 0.40$\pm$0.09\\
\hline
o-H$_2$O & 2$_{21}$--2$_{12}$ & 180.488\tablefootmark{d} &  2.72$\pm$0.32 &     \ldots      & 0.11$\pm$0.04 &     \ldots      &     \ldots      &     \ldots      & 0.50$\pm$0.02 &     \ldots     \\
o-H$_2$O & 2$_{12}$--1$_{01}$ & 179.527 &  3.13$\pm$0.34 & 13.03$\pm$0.08 & 0.77$\pm$0.06 &     $<$0.13      &     \ldots      & 2.18$\pm$0.06 & 1.45$\pm$0.02 &     \ldots     \\
o-H$_2$O & 3$_{03}$--2$_{12}$ & 174.626 &  4.88$\pm$0.47 & 10.54$\pm$0.11 & 0.61$\pm$0.07 & 0.24$\pm$0.01 &     \ldots      & 1.93$\pm$0.09 & 1.30$\pm$0.02 &     \ldots     \\
p-H$_2$O & 3$_{13}$--2$_{02}$ & 138.528 &  2.60$\pm$0.33 & 6.91$\pm$0.08 & 0.39$\pm$0.07 & 0.12$\pm$0.02 &     \ldots      & 1.10$\pm$0.08 & 1.00$\pm$0.02 &     \ldots     \\
p-H$_2$O & 4$_{04}$--3$_{13}$ & 125.354\tablefootmark{d} &     $<$0.31      &     \ldots      & 0.17$\pm$0.06 &     \ldots      &     \ldots      &     \ldots      & 0.59$\pm$0.03 &     \ldots     \\
o-H$_2$O & 2$_{21}$--1$_{10}$ & 108.073 &  3.84$\pm$0.51 & 10.23$\pm$0.21 & 0.78$\pm$0.17 & 0.26$\pm$0.04 & 4.88$\pm$0.22 & 1.82$\pm$0.19 & 1.33$\pm$0.05 & 2.16$\pm$0.17\\
p-H$_2$O & 3$_{22}$--2$_{11}$ &  89.988 &  $<$0.30       & 2.37$\pm$0.12 & 0.10$\pm$0.03 & 0.13$\pm$0.04 &     \ldots      & 0.30$\pm$0.02 & 0.61$\pm$0.04 &     \ldots     \\
o-H$_2$O & 7$_{16}$--7$_{07}$ &  84.767 &  $<$0.44      &     $<$0.14      &     $<$0.36      &     $<$0.28      &     \ldots      &    $<$0.10      & 0.14$\pm$0.05 &     \ldots     \\
p-H$_2$O & 6$_{15}$--5$_{24}$ &  78.928 &  $<$0.65      &     $<$0.30      &     $<$0.41      &     $<$0.48     &     \ldots      &     $<$0.13      &     $<$0.54     &     \ldots     \\
o-H$_2$O & 4$_{23}$--3$_{12}$ &  78.742\tablefootmark{c} &    $<$0.65     & 2.03$\pm$0.19 & 0.33$\pm$0.03 & 0.24$\pm$0.06 &     \ldots      &     \ldots      & 1.34$\pm$0.08 &     \ldots     \\
o-H$_2$O & 7$_{07}$--6$_{16}$ &  71.947\tablefootmark{e} &    $<$0.62      & 0.78$\pm$0.08 &     $<$0.26      &     \ldots      & 0.32$\pm$0.04 &     \ldots      & 0.60$\pm$0.05 &     $<$0.26     \\
o-H$_2$O & 8$_{18}$--7$_{07}$ &  63.324 &      $<$0.77      &     $<$0.16      &     $<$0.47      &     $<$0.57     & 0.50$\pm$0.11 &     $<$0.19      & 0.41$\pm$0.05 &     $<$0.44     \\
\hline
\ion{O}{i} & $^{3}P_{0}-^{3}P_{1}$ & 145.525 &      $<$0.2      & 0.42$\pm$0.10 & 1.07$\pm$0.10 & 0.87$\pm$0.10 & 1.33$\pm$0.12 & 1.13$\pm$0.11 & 0.78$\pm$0.10 & 2.36$\pm$0.06\\
\ion{O}{i} & $^{3}P_{1}-^{3}P_{2}$ &  63.184 &  9.01$\pm$1.83 & 2.33$\pm$0.28 & 11.29$\pm$0.30 & 8.92$\pm$0.46 & 20.91$\pm$0.53 &  19.07$\pm$0.47  & 8.16$\pm$0.62 & 23.21$\pm$0.41\\
\hline
\end{tabular}}
\tablefoot{
NGC1333-IRAS4B and Serpens SMM1 fluxes are published separately in Herczeg et al. (2012) and 
Goicoechea et al. (2012), respectively. CO 31-30 and OH 84.6 $\mu$m fluxes are presented in \citet{Wa12}. 
NGC1333-IRAS2A and  NGC1333-IRAS4A full PACS range fluxes will appear in Karska et al. (in prep.),
 but IRAS2A central spaxel only fluxes were listed in Visser et al. (2012). Non-observed lines
  are marked with ellipsis dots (\ldots). The uncertainties are 1$\sigma$ measured in the
  continuum on both sides of each line; 
calibration uncertainty of 30\% of the flux should be included for comparisons with other 
modes of observations or instruments. 1$\sigma$ upper limits calculated using 
wavelength dependent values of full-width high maximum for a point source observed with PACS
are listed for non-detections.
\tablefoottext{a}{The baseline from one side is affected by p-H$_2$O 4$_{13}$-3$_{22}$ line at 144.518 $\mu$m, 
which falls in the edge of the scan.}
\tablefoottext{b}{A blend with the o-H$_2$O 4$_{14}$-3$_{03}$ line at 113.537 $\mu$m.}
\tablefoottext{c}{Lines at the edges of the scans, for many cases not possible to measure. Selected objects 
were observed in dedicated scan for the CO 29-28 line.}
\tablefoottext{d}{Lines observed in dedicated scans for selected objects only.}
\tablefoottext{e}{The line falls in the leakage region of PACS and therefore the flux is less reliable.}}
\end{table}
\end{landscape}

\begin{landscape}
\begin{table}
\caption{\label{obsn2} Line fluxes of Class 0 and I sources in 10$^{-20}$ W cm$^{-2}$.}             
%\centering     
\renewcommand{\footnoterule}{}  % to avoid a line before footnotes
\resizebox{1.3\textheight}{!}
{\begin{tabular}{llccccccccc}     % 7 columns 
\hline\hline       
Species & Trans. & $\lambda_\mathrm{lab}$ ($\mu$m) & \multicolumn{8}{c}{Full array flux (10$^{-20}$ W cm$^{-2}$)}\\     
%~ & ~ & ~ &  IRAS2A & IRAS4A  & L1527 & CedIRS4 & BHR71 &  IRAS15398 & L483 & SerSMM4 \\
~ & ~ & ~ &  SerSMM3 & L723  & L1489 & TMR1 & TMC1A &  TMC1 & HH46 & RNO91 \\
\hline
CO & 14--13 & 185.999 &  7.06$\pm$0.11 & 0.64$\pm$0.02 & 0.83$\pm$0.03 & 1.22$\pm$0.11 & 0.54$\pm$0.08 & 0.78$\pm$0.10 & 1.06$\pm$0.08 & 0.24$\pm$0.02\\
CO & 16--15 & 162.812 &  6.86$\pm$0.09 & 0.61$\pm$0.01 & 0.98$\pm$0.02 & 1.58$\pm$0.06 & 0.59$\pm$0.02 & 0.92$\pm$0.05 & 0.86$\pm$0.06 & 0.27$\pm$0.01\\
CO & 18--17 & 144.784\tablefootmark{a} &  4.98$\pm$0.18 & 0.62$\pm$0.03 & 1.05$\pm$0.01 & 1.50$\pm$0.07 & 0.44$\pm$0.08 & 0.73$\pm$0.06 & 0.65$\pm$0.08 & 0.25$\pm$0.01\\
CO & 22--21 & 118.581 &  3.42$\pm$0.13 & 0.35$\pm$0.03 & 0.82$\pm$0.02 & 0.81$\pm$0.12 & 0.60$\pm$0.11 & 0.41$\pm$0.08 & 0.72$\pm$0.14 & 0.08$\pm$0.02\\
CO & 23--22 & 113.458\tablefootmark{b} &  6.75$\pm$0.19 & 0.67$\pm$0.03 & 1.95$\pm$0.02 & 2.30$\pm$0.15 & 0.69$\pm$0.09 & 1.24$\pm$0.09 & 0.68$\pm$0.11 & 0.33$\pm$0.03\\
CO & 24--23 & 108.763 &  2.28$\pm$0.20 & 0.46$\pm$0.04 & 0.83$\pm$0.03 & 1.35$\pm$0.10 & 0.39$\pm$0.01 & 0.60$\pm$0.11 & 0.34$\pm$0.04 &      $<$0.18   \\
CO & 29--28 & 90.163\tablefootmark{c} &  1.96$\pm$0.10 & 0.14$\pm$0.03 & 0.57$\pm$0.04 & 0.37$\pm$0.07 &      $<$0.12     & 0.25$\pm$0.05 & 0.17$\pm$0.04 & $<$0.08    \\
CO & 30--29 & 87.190 &  2.00$\pm$0.21 & 0.32$\pm$0.06 & 0.49$\pm$0.05 & 0.42$\pm$0.08 &      $<$0.11      & 0.24$\pm$0.04 & 0.28$\pm$0.04 & 0.07$\pm$0.03\\
%CO & 31--30 & 84.411 &  2.29$\pm$0.24 & 0.70$\pm$0.05 & 1.03$\pm$0.06 & 0.89$\pm$0.09 &      n.d.     & 0.56$\pm$0.06 &      n.d.     &      n.d.    \\
CO & 32--31 & 81.806 &  1.06$\pm$0.32 &      $<$0.27     &     $<$0.32    &  $<$0.20     & $<$0.12     & 0.14$\pm$0.03 &      $<$0.20     &      $<$0.04    \\
CO & 33--32 & 79.360 &  1.30$\pm$0.28 &      $<$0.28     & 0.17$\pm$0.04 & 0.14$\pm$0.04 &      $<$0.11     & 0.16$\pm$0.03 &      $<$0.26     &      $<$0.07    \\
CO & 36--35 & 72.843 &  0.77$\pm$0.10 &      $<$0.10     &     $<$0.10     & 0.26$\pm$0.04 &      $<$0.03      & 0.07$\pm$0.02 &     $<$0.07     &      $<$0.02   \\
\hline
o-H$_2$O & 2$_{21}$--2$_{12}$ & 180.488\tablefootmark{d} &  1.23$\pm$0.11 & 0.14$\pm$0.02 &      \ldots     &      \ldots     &      \ldots     &      \ldots     &      \ldots     &      \ldots    \\
o-H$_2$O & 2$_{12}$--1$_{01}$ & 179.527 &  4.90$\pm$0.09 & 0.14$\pm$0.01 & 0.73$\pm$0.02 & 0.69$\pm$0.12 &      $<$0.13     & 0.29$\pm$0.04 & 1.16$\pm$0.10 &      $<$0.15    \\
o-H$_2$O & 3$_{03}$--2$_{12}$ & 174.626 &  4.59$\pm$0.08 & 0.30$\pm$0.02 & 0.83$\pm$0.03 & 0.50$\pm$0.09 &      $<$0.12     & 0.24$\pm$0.07 & 0.57$\pm$0.09 & 0.20$\pm$0.02\\
p-H$_2$O & 3$_{13}$--2$_{02}$ & 138.528 &  2.59$\pm$0.07 & 0.36$\pm$0.02 & 0.70$\pm$0.02 & 0.53$\pm$0.07 & 0.11$\pm$0.06 & 0.33$\pm$0.06 & 0.38$\pm$0.07 & 0.11$\pm$0.01\\
p-H$_2$O & 4$_{04}$--3$_{13}$ & 125.354\tablefootmark{d} &  1.38$\pm$0.08 & 0.10$\pm$0.02 &      \ldots     &      \ldots     &      \ldots     &      \ldots     &      \ldots     &      \ldots    \\
o-H$_2$O & 2$_{21}$--1$_{10}$ & 108.073 &  3.84$\pm$0.21 & 0.23$\pm$0.04 & 0.92$\pm$0.04 & 0.79$\pm$0.07 &      $<$0.31     & 0.41$\pm$0.14 & 0.21$\pm$0.04 & 0.09$\pm$0.02\\
p-H$_2$O & 3$_{22}$--2$_{11}$ & 89.988 &  1.23$\pm$0.11 & 0.13$\pm$0.04 & 0.49$\pm$0.04 & 0.46$\pm$0.07 &      $<$0.09     & 0.15$\pm$0.04 & 0.16$\pm$0.03 &      $<$0.05    \\
o-H$_2$O & 7$_{16}$--7$_{07}$ & 84.767 &       $<$0.27     &      $<$0.21 & $<$0.29 &      $<$0.15     &      $<$0.09     &      $<$0.10     &      $<$0.17     &     $<$0.04    \\
p-H$_2$O & 6$_{15}$--5$_{24}$ & 78.928 &       $<$0.42     &      $<$0.44 & 0.24$\pm$0.04 &    $<$0.19    &      $<$0.14     &      $<$0.12     &      $<$0.18     &     $<$0.06    \\
o-H$_2$O & 4$_{23}$--3$_{12}$ & 78.742\tablefootmark{c} &  2.93$\pm$0.25 & 0.24$\pm$0.07 & 0.93$\pm$0.10 & 0.55$\pm$0.04 &      \ldots    & 0.30$\pm$0.05 &      \ldots     & 0.19$\pm$0.06\\
o-H$_2$O & 7$_{07}$--6$_{16}$ & 71.947\tablefootmark{e} &  0.59$\pm$0.10 & 0.14$\pm$0.04 &      \ldots     &      \ldots     &      \ldots     &      \ldots     &      \ldots     &      \ldots    \\
o-H$_2$O & 8$_{18}$--7$_{07}$ & 63.324 &  0.75$\pm$0.50 &      $<$0.35     & 0.61$\pm$0.04 & 0.59$\pm$0.07 &      $<$0.13     & 0.17$\pm$0.05 &      $<$0.23     &      $<$0.05    \\
%p-H$_2$O & 4$_{31}$--4$_{04}$ & 61.809 &       n.d.     &      n.d.     &      n.d.     & 0.19$\pm$0.05 &      n.d.     &      n.d.     &      n.d.     &      n.d.    \\
\hline
\ion{O}{i} & $^{3}P_{0}-^{3}P_{1}$ & 145.525 &  2.26$\pm$0.08 & 0.50$\pm$0.11 & 0.24$\pm$0.08 & 0.67$\pm$0.13 & 0.64$\pm$0.09 & 0.97$\pm$0.13 & 2.35$\pm$0.13 & 0.47$\pm$0.08\\
\ion{O}{i} & $^{3}P_{1}-^{3}P_{2}$ & 63.184 & 21.08$\pm$0.56 & 3.19$\pm$0.45 & 6.34$\pm$0.47 & 7.00$\pm$0.39 & 8.53$\pm$0.56 & 14.12$\pm$0.41 & 33.76$\pm$0.63 & 9.38$\pm$0.47\\
\hline
\end{tabular}}
\tablefoot{
NGC1333-IRAS4B and Serpens SMM1 fluxes are published separately in Herczeg et al. (2012) and 
Goicoechea et al. (2012), respectively. CO 31-30 and OH 84.6 $\mu$m fluxes are presented in \citet{Wa12}. 
NGC1333-IRAS2A and  NGC1333-IRAS4A full PACS range fluxes will appear in Karska et al. (in prep.),
 but IRAS2A central spaxel only fluxes were listed in Visser et al. (2012). Non-observed lines
  are marked with ellipsis dots (\ldots). The uncertainties are 1$\sigma$ measured in the continuum on both sides of each line; 
calibration uncertainty of 30\% of the flux should be included for comparisons with other 
modes of observations or instruments. 1$\sigma$ upper limits calculated using 
wavelength dependent values of full-width high maximum for a point source observed with PACS
are listed for non-detections.
\tablefoottext{a}{The baseline from one side is affected by p-H$_2$O 4$_{13}$-3$_{22}$ line at 144.518 $\mu$m, 
which falls in the edge of the scan.}
\tablefoottext{b}{A blend with the o-H$_2$O 4$_{14}$-3$_{03}$ line at 113.537 $\mu$m.}
\tablefoottext{c}{Lines at the edges of the scans, for many cases not possible to measure. Selected objects 
were observed in dedicated scan for the CO 29-28 line.}
\tablefoottext{d}{Lines observed in dedicated scans for selected objects only.}
\tablefoottext{e}{The line falls in the leakage region of PACS and therefore the flux is less reliable.}}
\end{table}
\end{landscape}

\section{Extended source correction method}
To account for the combination of real spatial extent in the emission and the wavelength-dependent
point spread function, we developed an `extended source correction' method.
%For point sources, Point Spread Function (PSF) correction factors are supplied by the
%Herschel Science Center to account for these effects. 
We first inspected the $5\times5$ spectral (or contour) maps. Contributions from NGC1333-IRAS4A, 
Ser SMM6, and an unlabelled object were subtracted from the observations of NGC1333-IRAS4B, Ser SMM3, and SMM4
\citep[e.g.] [Dionatos et al. subm.]{Yi12}.
In the next step, we used two long-wavelength lines of CO and H$_{2}$O (CO 14-13 at 
185.999 $\mu$m and H$_{2}$O $2_{12}-1_{01}$ at 179.527 $\mu$m) to visualise the spatial extent of
the line emission attributed to each object and summed all the spaxels that contained emission.
 Since all of our lines, except the [\ion{O}{i}] line at 63.2 $\mu$m, are spectrally unresolved by PACS,
  single or double (for OH doublets, closeby and blended lines) 
Gaussian fits to the resulting spectra are used to calculate the line fluxes of the detected lines. 

\begin{figure}[tb]
 %\hspace{+10ex}
\begin{center}
 \includegraphics[angle=0,height=10cm]{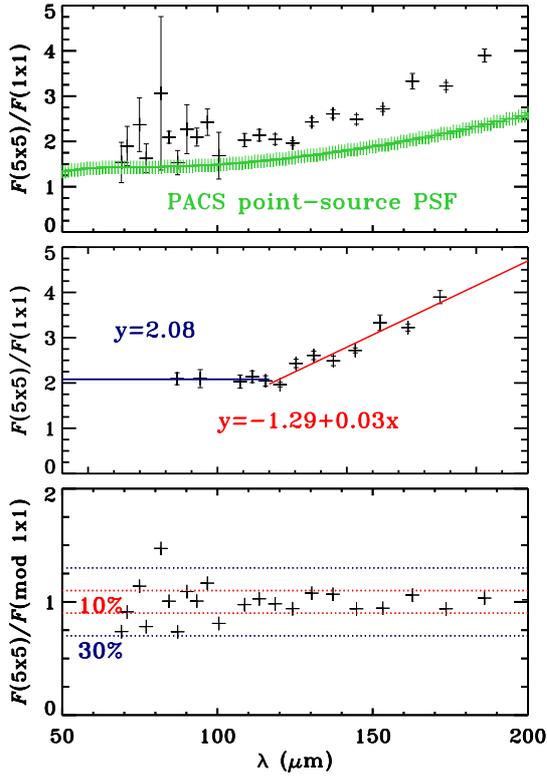}
 %\vspace{+5ex
\vspace{+6ex}
\caption{\label{nicecurve} Illustration of the correction curve method. 
\textit{Top:} CO fluxes of Serpens SMM1 measured over 5$\times$5 array 
divided by the central spael fluxes are plotted versus wavelengths. PACS PSF 
for a point source is shown in green. CO emission is clearly extended for this sources. 
\textit{Middle:} Best signal-to-noise measurements are used to make a fit to the data 
and derive the wavelength-dependent correction factors. \textit{Bottom:} CO fluxes 
measured over 5$\times$5 array are divided by central spaxel measurements corrected for 
the extended emission using calculated correction factors are plotted versus wavelength. 
Accuracy longward $\sim$100 $\mu$m is better than 10\%, whereas for short-wavelength lines 
($\leq$100 $\mu$m) the accuracy is $\sim$30\%.}
\end{center}
\end{figure}

Summing the spectra from all 25 spaxels increases the noise and often
prevents the detection of weak lines. Using only those spaxels which
contain most of the emission results in a much higher signal-to-noise
and ultimately a higher detection rate for lines but fails to include
emission that leaks out of those spaxels because of real spatial
extent in the line and the instrumental point spread function.
Therefore, our `extended source correction' method provides a
wavelength-dependent correction factor to account for the missing
flux.  The main idea of the method is to use the brightest spaxels,
which contain most of the emission, to measure the line fluxes and
then correct the value for the missing flux, contained in the omitted
spaxels. This method assumes that weak lines are similarly distributed
to the strong ones, with observed differences in spatial distributions
caused only by the wavelength dependence in the PSF.

The correction factors are derived using the strongest lines,
i.e. those that can be measured in the brightest spaxels as well
as in all spaxels that contain emission from the object (usually 25 of
them). The ratio of flux in the small, bright extraction region and
the large extraction region yields a wavelength-dependent
\textit{correction curve} (see Figure \ref{nicecurve}). A 0th-order
fit (horizontal line) is used for the short-$\lambda$ part of the
spectrum, whereas a 1st- or 2nd-order polynomial is used for the
long-$\lambda$ part of the spectrum ($\geq$100-110 $\mu$m).  All line
fluxes are then measured in only the brightest spaxels and 
multiplied by this correction factor. This method was used primarily for the
full spectral scans.

\section{Spectral energy distributions}

\begin{figure}[tb]
 %\hspace{+10ex}
\begin{center}
 \includegraphics[angle=0,height=10cm]{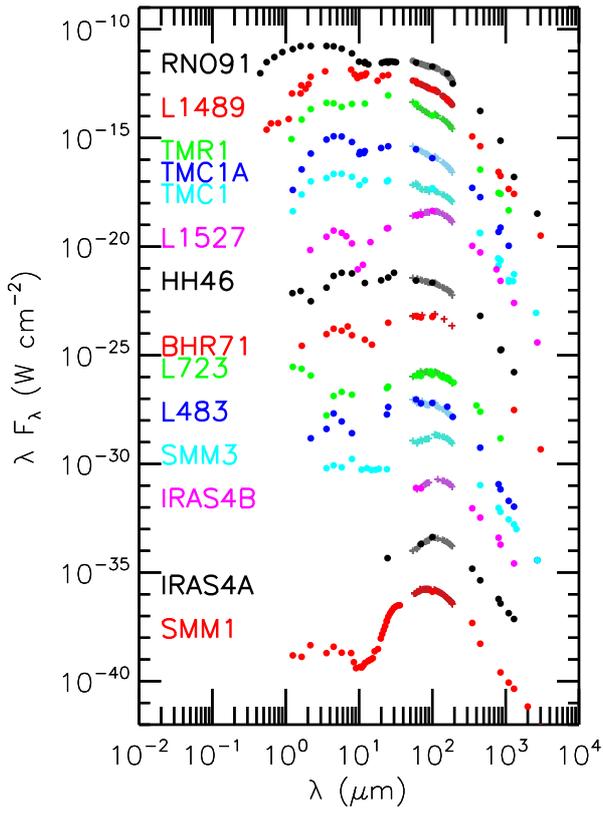}
 %\vspace{+5ex
\vspace{+6ex}

\caption{\label{seds}Spectral energy distribution shapes for most of our sources. Literature 
observations are shown as filled circles, whereas our PACS observations are marked with crosses
and drawn in a different hue (see Table \ref{tab:cont}). The maxima of the SEDs lie between $10^{-16}$ and $10^{-19}$ W 
cm$^{-2}$ for all objects and thus, for better shape vizualisation, the SEDs are offset by several orders of magnitude.
Objects are shown in the sequence of decreasing evolutionary parameter $L_\mathrm{bol}^{0.6}/M_\mathrm{env}$.}
\end{center}
\end{figure}

Figure \ref{seds} shows the spectral energy distributions for all of
our sources obtained from our PACS spectroscopy and literature
measurements from {\it Spitzer}-IRAC and MIPS \citep{Ev09}, 2MASS \citep{Sk06}, 
SCUBA \citep{Sh00,Fr08}, as well as APEX/LABOCA, Bolocam, SEST, ISO and IRAS telescopes.
The PACS measurements cover the peak of dust emission and are in good agreement with 
the previous observations \citep[for more details, see][]{Kr12}.

For PACS, the overlap regions of different orders cause the regions
of 70-73 $\mu$m, 98-105 $\mu$m and 190-220 $\mu$m to be less
reliable in terms of continuum shapes and flux densities. These regions were
 thus excluded from our spectral energy distribution analysis. New values of
$L_\mathrm{bol}$ and $T_\mathrm{bol}$ are calculated following the
standard definitions of the two physical quantities
\citep[e.g.,][]{MD10}. Several methods of interpolation were tested
for consistency of the results including linear interpolation,
midpoint, prismodial method and trapezoidal summation. Among them the
trapezoidal summation offered the most stable values and is used in
this study.

\begin{landscape}
\begin{table}
\caption{\label{tab:cont} Continuum measurements for Class 0 and I sources.}             
%\centering     
\renewcommand{\footnoterule}{}  % to avoid a line before footnotes
%\resizebox{1.3\textheight}{!}
{\begin{tabular}{lcccccccccccccccccccccc}     % 7 columns 
\hline\hline       
$\lambda$ ($\mu$m) & \multicolumn{18}{c}{Continuum (Jy)}\\     
 ~ &  I2A & I4A  & I4B & L1527 & CedIRS4 & BHR71 &  I15398 & L483 & SMM1 & SMM4 & SMM3 & L723  & L1489 & TMR1 & TMC1A &  TMC1 & HH46 & RNO91 \\
\hline
  62.050 & 289 & 16 &  9 & 17 & 15 & 79 & 12 & 85 &  153 & \ldots & 11 &  9 & 43 & 37 & 38 &  8 & 34 & 37 \\
  62.700 & 235 & \ldots & \ldots & \ldots & \ldots & \ldots & \ldots & 66 & 159 & \ldots & \ldots & \ldots & \ldots & \ldots & \ldots & \ldots & \ldots & \ldots \\
  63.184 & 243 & 20 & \ldots & 18 & 14 & 85 & 12 & 85 & 166 & \ldots & 14 &  9 & 42 & 37 & 41 & 10 & 35 & 36 \\
  69.300 & 303 & 34 & 13 & 31 & 17 & \ldots & 21 & 76 & 224 & \ldots & 21 & 15 & 49 & 38 & 43 &  7 & 39 & 42\\
  72.843 & 349 & 38 & 15 & 27 & 18 & 125 & \ldots & 113 & 245 &  2 & 20 & 14 & 48 & 36 & 41 &  8 & 39 & 43\\
  79.160 & 394 & 54 & 24 & 40 & 21 & \ldots & 24 & 93 & 325 & \ldots & 29 & 17 & 51 & 35 & 42 &  9 & 42 & 47\\
  81.806 & 412 & 61 & 28 & 41 & 21 & \ldots & 26 & 98 & 346 & \ldots & 31 & 20 & 52 & 37 & 43 & 10 & 44 & 47\\
  84.600 & 436 & 68 & 32 & 46 & 23 & \ldots & 27 & 102 & 373 & \ldots & 33 & 22 & 52 & 36 & 43 & 10 & 47 & 49\\
  87.190 & 457 & 78 & 34 & 50 & 24 & \ldots & 28 & 110 & 399 & \ldots & 37 & 21 & 53 & \ldots & 45 & 10 & 48 & 52\\
  89.990 & 460 & 83 & 37 & 51 & 24 & \ldots & 27 & 113 & 417 & \ldots & 40 & 21 & 51 & 35 & 42 &  9 & 48 & 50\\
 108.070 & 427 & 157 & \ldots & 78 & 45 & 302 & 54 & 206 & 551 & 22 & 88 & 34 & 63 & 49 & 56 & 15 & 67 & 69\\
 108.760 & 426 & 156 & \ldots & 79 & 45 & 304 & 54 & 206 & 552 & 22 & 88 & 34 & 62 & 48 & 54 & 15 & 67 & 69\\
 113.458 & 417 & \ldots & \ldots & 93 & 47 & \ldots & 56 & 162 & 558 & \ldots & 92 & 35 & 61 & 48 & 53 & 15 & 67 & 70\\
 118.581 & 410 & 178 & 93 & 97 & 50 & \ldots & 59 & 165 & 561 & \ldots & 98 & 36 & 61 & 49 & 53 & 15 & 68 & 71\\
 125.354 & 416 & \ldots & \ldots & 100 & \ldots & \ldots & \ldots & 165 &  571 & \ldots & 105 & 37 & \ldots &  \ldots & \ldots & \ldots &  68 &  \ldots\\
 138.528 & 402 & 206 & 110 & 106 & 57 & \ldots & 66 & 168 & 570 & \ldots & 117 & 38 & 55 & 48 & 50 & 16 & 70 & 71\\
 145.525 & 412 & 211 & 115 & 98 & 58 & 326 & 68 &  201 & 587 & 45 & 122 & 39 & 54 & 47 & 48 & 17 & 71 & 70\\
 157.700 & 397 & 220 & 121 & 107 & 60 & \ldots & 71 & 163 & 558 & \ldots & 128 & 40 & 52 & 46 & 46 & 17 & 74 & 70\\
 162.812 & 387 & 216 & 121 & 103 & 59 & \ldots & 69 & 157 & 536 & \ldots & 124 & 40 & 50 & 43 & 43 & 16 & 72 & 68\\
 169.100 & 364 & 201 & 113 & 95 & 53 & \ldots & 65 & 145 & 510 & \ldots & 117 & 36 & 45 & 39 & 38 & 15 & 68 & 62\\
 174.626 & 357 & 206 & 119 & 98 & 57 & \ldots & 69 & 149 & 468 & \ldots & 121 & 37 & 46 & 39 & 40 & 16 & 72 & 63\\
 179.527 & 325 & 198 & 116 & 88 & 51 & \ldots & 64 & 139 & 441 & \ldots & 115 & 36 & 42 & 35 & 35 & 15 & 67 & 58\\
 185.999 & 287 & 191 & 103 & 79 & 46 & 263 & 60 & 156 & 401 & 56 &  103 & 31 & 38 & 31 & 31 & 14 & \ldots & 51\\
\hline
\end{tabular}}
\tablefoot{
Non-observed spectral regions are marked with ellipsis dots (\ldots). The 
calibration uncertainty of 30\% of the flux should be included for comparisons with other 
modes of observations or instruments.}
\end{table}
\end{landscape}

%=========================================
\section{Spatial extent of line emission}
\begin{figure*}[!tb]
  \begin{center}  
   \includegraphics[angle=0,height=21cm]{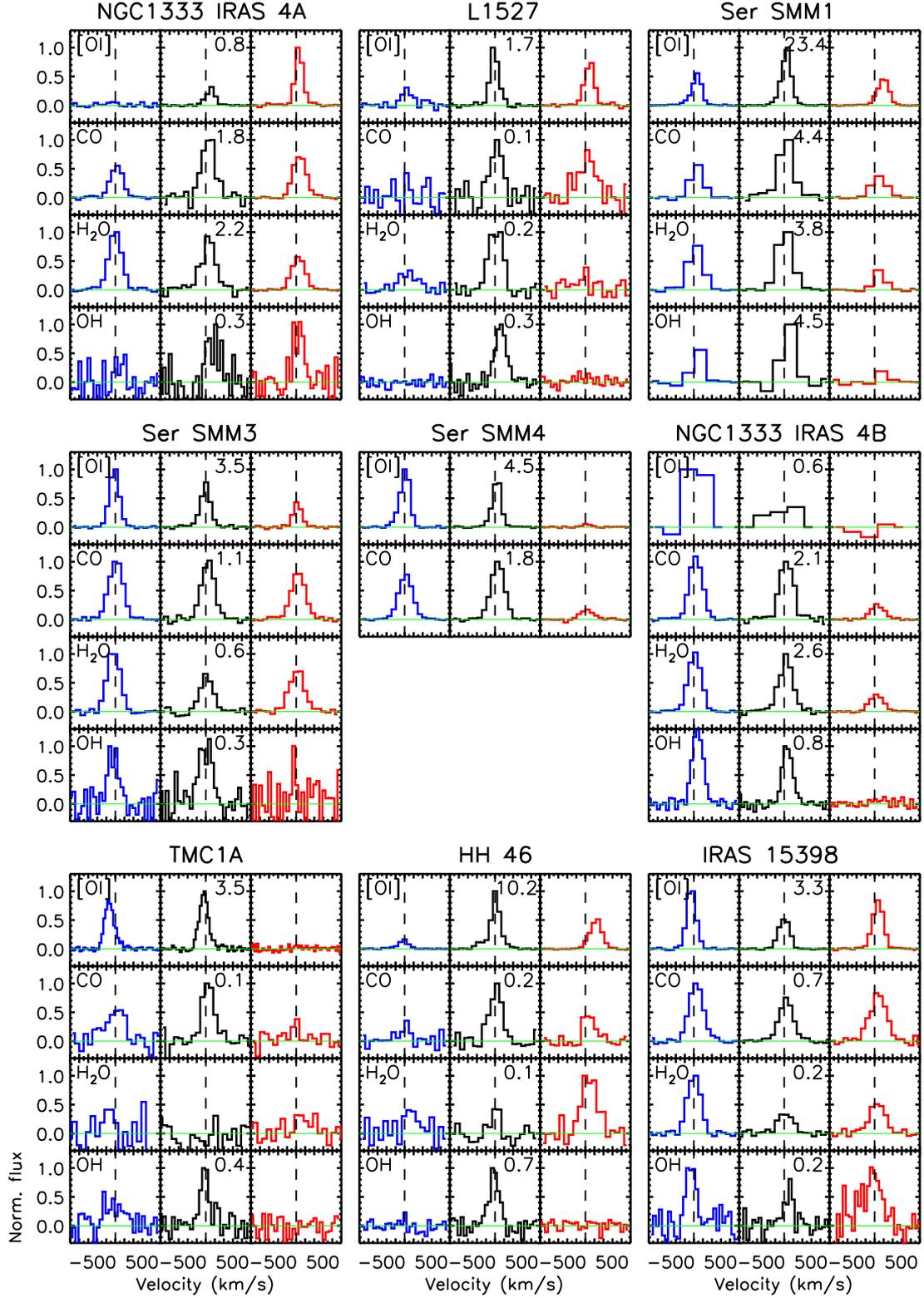}
   \end{center}
    \caption{\label{liness2}Spectra of the [\ion{O}{i}] 63.2 $\mu$m, CO $14-13$ 186.0 $\mu$m, H$_2$O
2$_{12}$-1$_{01}$ 179.5 $\mu$m
and OH 84.6  $\mu$m lines in the selected blue outflow, on-source and red outflow positions
(marked with blue, green and red frames around the spaxels e.g. in Figure \ref{specmap}) for the 
\textit{extended} sources (see \S 3.2). The figure shows relative emission 
at different positions for each species separately. Measured line fluxes at central spaxel
position in units of 10$^{-20}$ W cm$^{-2}$ are written next to the corresponding spectra.}
\end{figure*}

\begin{figure*}[!tb]
  \begin{center}  
    \includegraphics[angle=0,height=21cm]{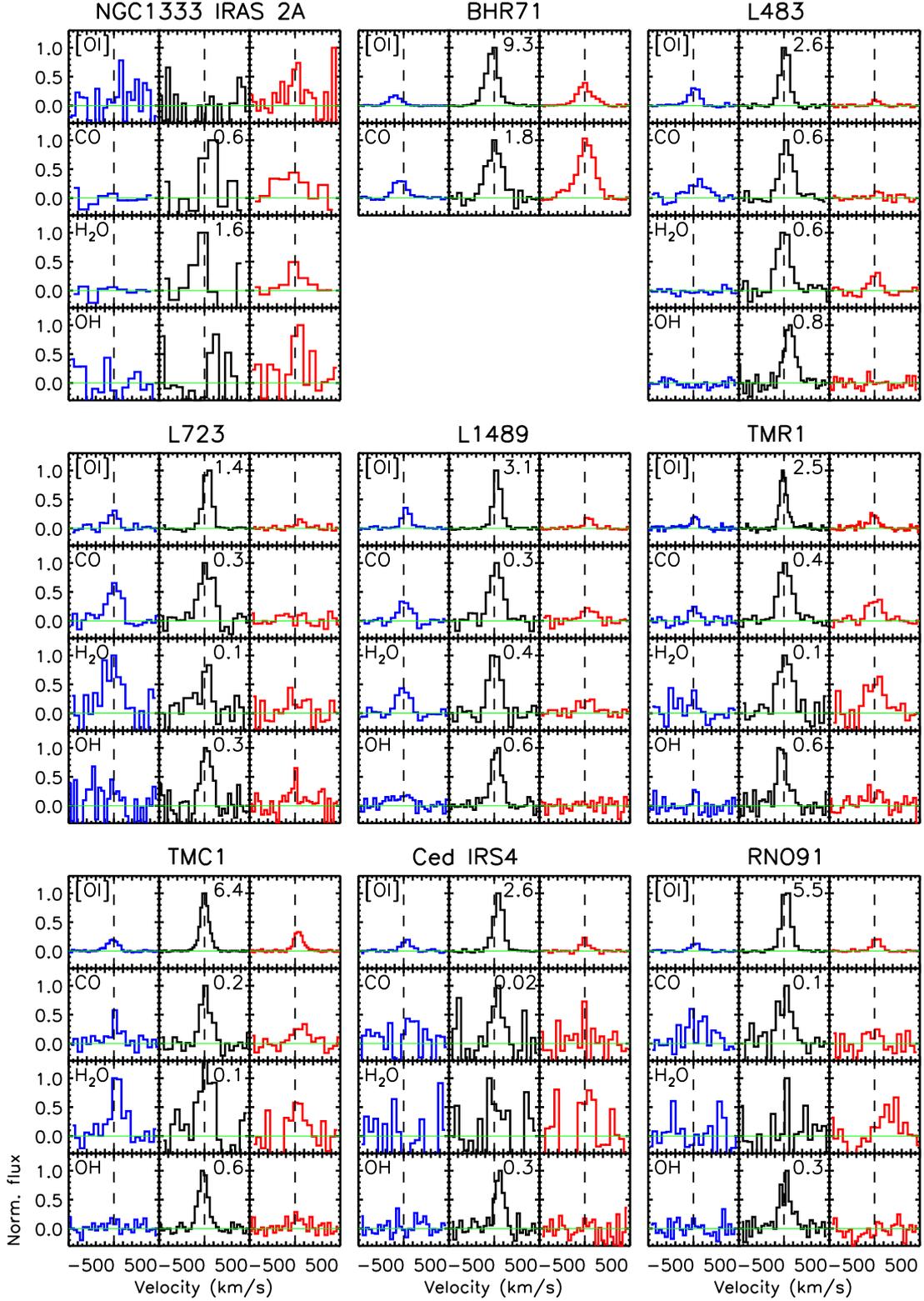}%  
    \end{center}
    \caption{\label{liness1}The same as Figure \ref{liness2} but for the 
    \textit{compact} sources.}
\end{figure*}

Figures \ref{liness2} and \ref{liness1} show the spectra in the on-source and outflow positions
for objects with \textit{extended} emission and objects with \textit{compact}
 emission, respectively (see \S 3.2). 

\begin{figure}[tb]
	\begin{center}
	\includegraphics[angle=90,height=7.5cm]{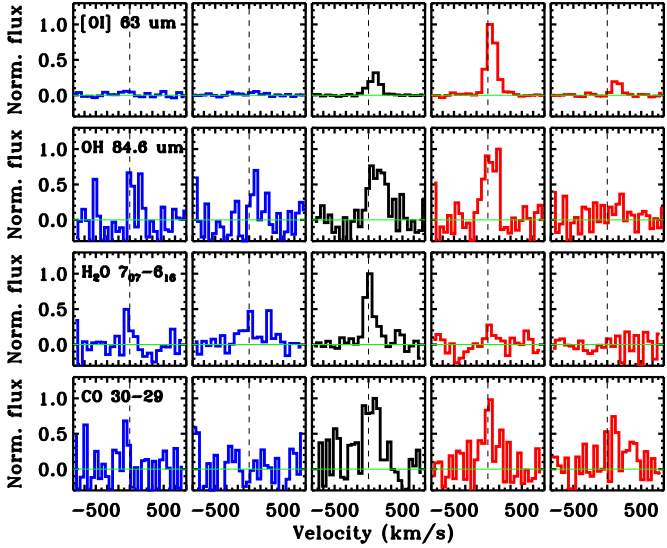}
	      \vspace{-4ex} 

	\end{center}
	\caption{\label{5spectra} NGC1333-IRAS4A spectra in the [\ion{O}{i}] 63.2 $\mu$m line, 
	the OH 84.6 $\mu$m line, the H$_2$O 7$_{07}$-6$_{16}$ line and the CO $30-29$ line. 
	Two blue outflow, on-source and two red outflow positions are shown, corresponding to the 
	the colored spaxels in Figure \ref{specmap}.}
\end{figure}

Figure \ref{5spectra} presents the spectra of the [\ion{O}{i}] 63.2 $\mu$m line, 
	the OH 84.6 $\mu$m line, the H$_2$O 7$_{16}$-6$_{07}$ line and the CO $30-29$ line
for the central target, two spaxels in the red-shifted outflow, and two spaxels in the
blue-shifted outflow of NGC1333-IRAS4A. 
%-----------
%-----------
%===================

\section{Comparing PACS and ISO far-IR spectra}
Higher sensitivity of Herschel/PACS compared to ISO-LWS allows us to improve 
the detection rate of H$_2$O (15 out of 16 Class 0/I sources) and higher$-J$ CO transitions
 (14 out of 16 sources detected in CO 24-23). In particular, water detections in
Class I sources are now possible for the majority of the sources. Detection of the 
more highly excited CO transitions allows to distinguish the \textit{hot component}
on the rotational diagram, which was not possible with ISO \citep{Ni10}.

The chopping capabilities and the spatial resolution of \textit{Herschel} at the
distances of our objects allow us to distinguish the YSO--related
atomic emission from the emission from the nearby objects or the surrounding cloud. 
Herschel observations show that only two objects (Ser SMM1 and TMC1) show [\ion{C}{ii}] emission
associated with the YSO. The [\ion{O}{i}] emission, on the other hand, is clearly extended
in the outflow direction and most probably traces the hidden jet. For some of the 
outflow-dominated emission sources, the Herschel beam does not cover the full extent of 
the [\ion{O}{i}] emission, whereas the ISO beam can suffer from the cloud or nearby sources
 emission. Table \ref{compISO} shows the comparison between the [\ion{O}{i}] emission for the sources 
 observed with both instruments.
 
 \begin{table}
\begin{minipage}[t]{\columnwidth}
\caption{\label{compISO} Comparison between ISO and Herschel line emission in $10^{-20}$ W cm$^{-2}$.}
\centering
\renewcommand{\footnoterule}{}  % to avoid a line before footnotes
\begin{tabular}{lccccccccc}
\hline \hline
Object & \multicolumn{2}{c}{[\ion{O}{i}] 63 um}  & \multicolumn{2}{c}{[\ion{O}{i}] 145 um}   \\
 & LWS & PACS & LWS & PACS \\
\hline \hline
IRAS2  	& $29.0\pm2.9$ & 9.01$\pm$1.83 & $<4.5$ &  $<0.2$  \\
IRAS4 (A+B) & $24.3\pm1.6$ & $4.2\pm0.4$  & $<3.6$ &  0.4$\pm$0.1   \\
L1527 	& $13.4\pm2.0$ & $11.3\pm0.3$ & $4.8\pm0.7$ &  $1.1\pm0.1$   \\
L483  	& $18.8\pm2.0$ & $8.2\pm0.6$   & $3.7\pm1.0$ & $0.8\pm0.1$   \\
L723  	& $14.5\pm3.6$ & $3.19\pm0.5$ & $3.1\pm0.4$ & $0.5\pm0.1$    \\
\hline
\end{tabular}
\end{minipage}
\end{table}

%-----------
\section{Rotational diagrams}
\begin{figure*}[!tb]
  \begin{minipage}[t]{.5\textwidth}
  \begin{center}  
      \includegraphics[angle=90,height=6cm]{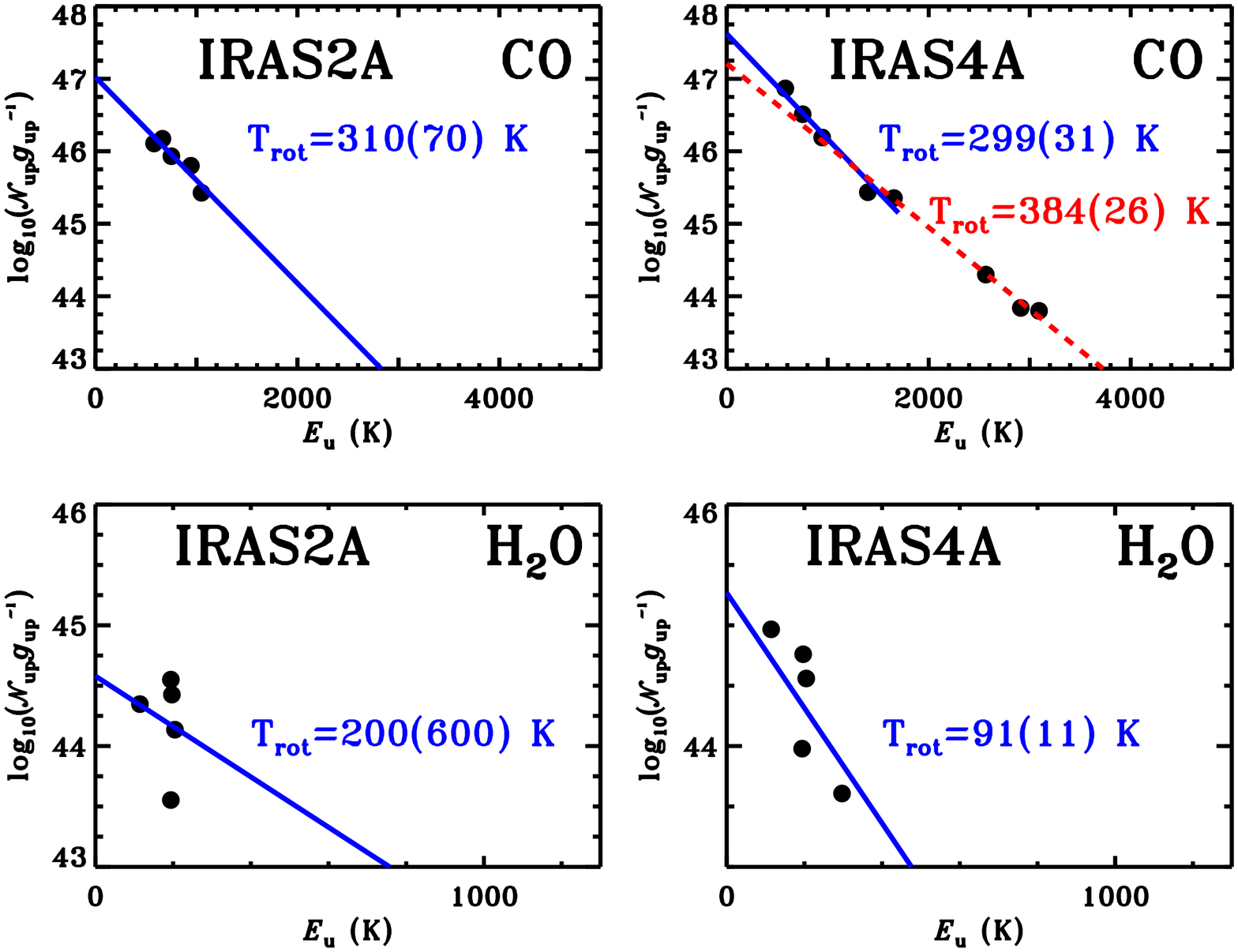}
                     \vspace{+3ex}
    
    \includegraphics[angle=90,height=6cm]{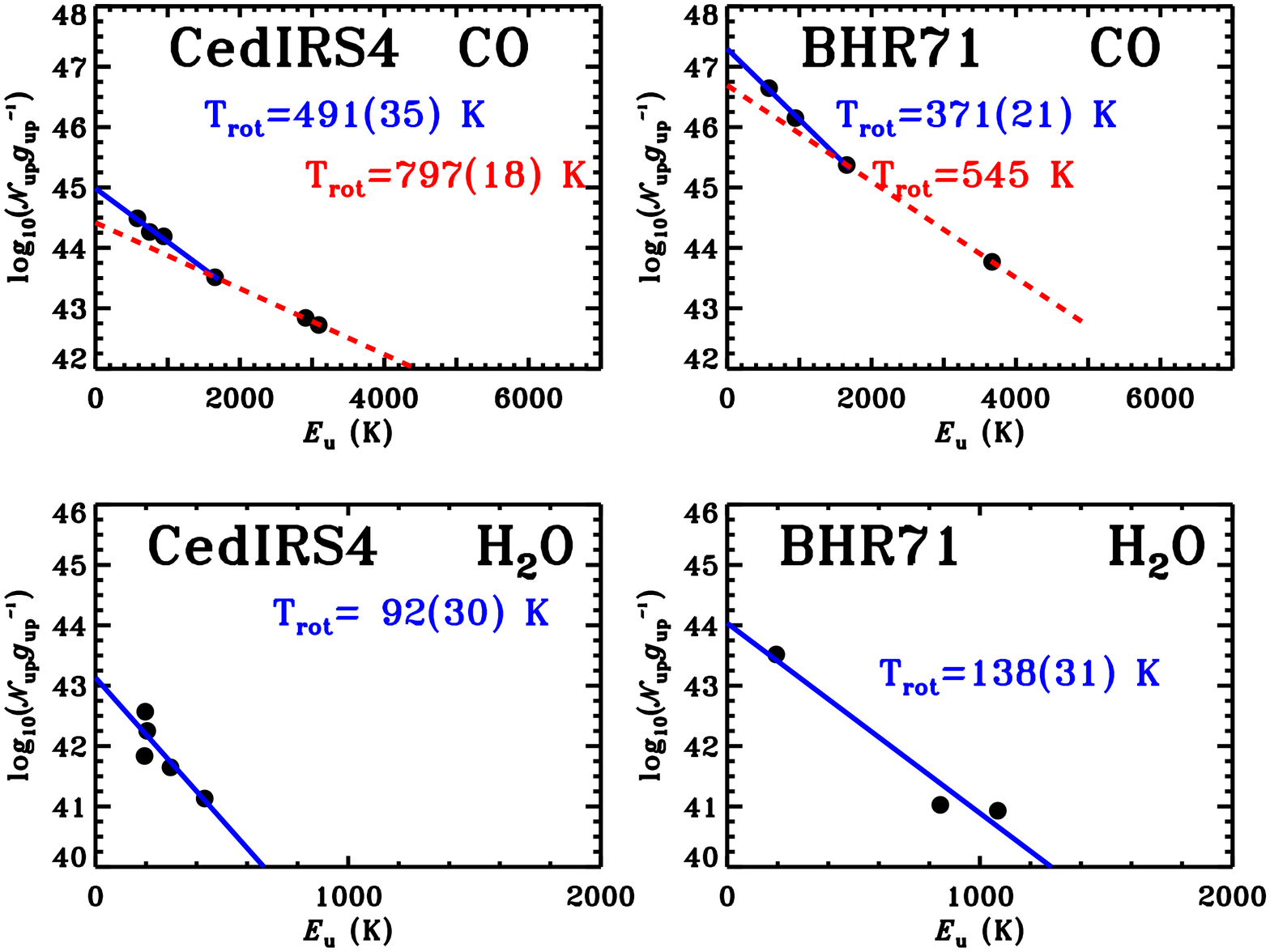}
                     \vspace{+3ex}
       
    \includegraphics[angle=90,height=6cm]{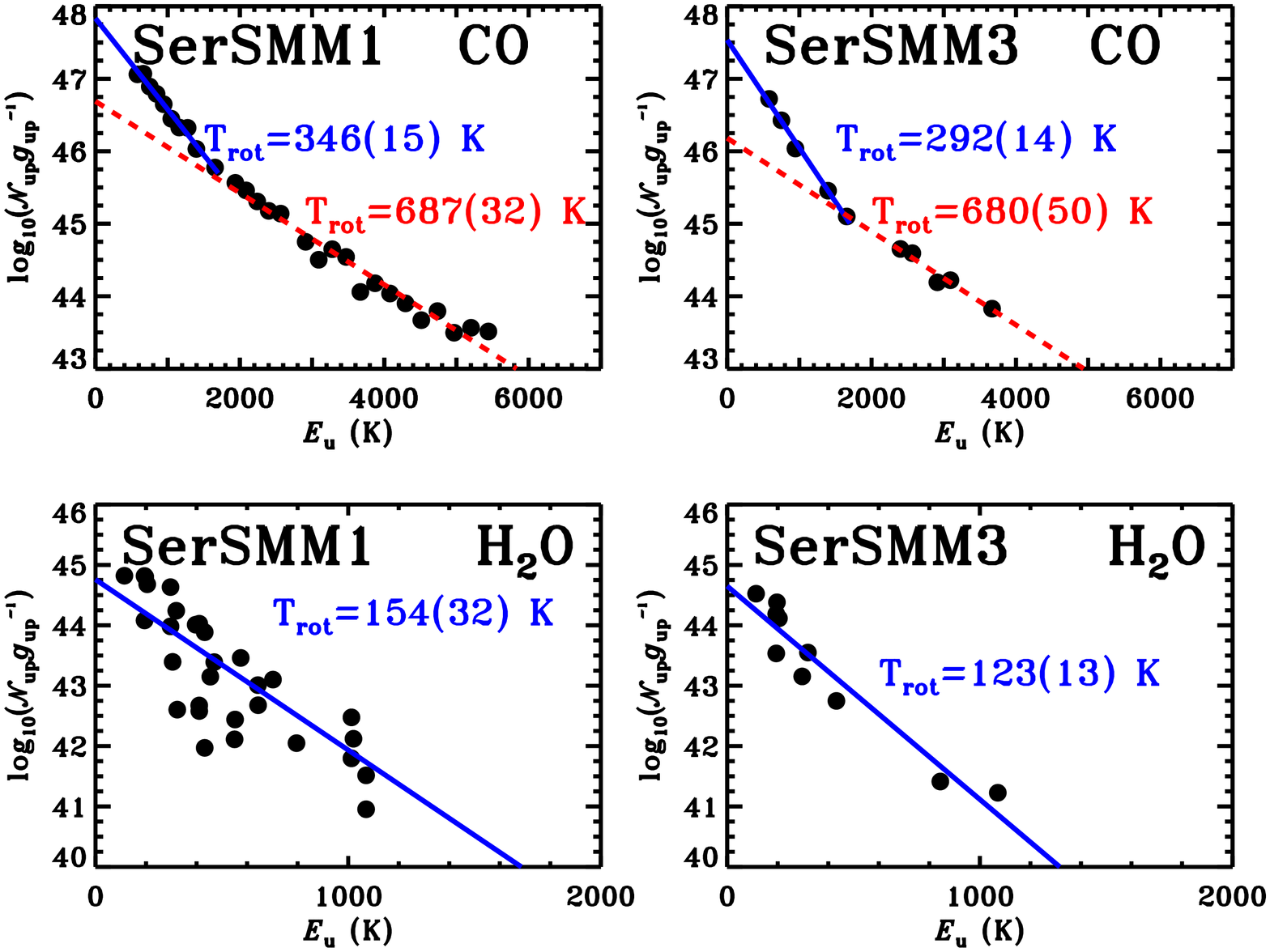}
    \end{center}
  \end{minipage}
  \hfill
  \begin{minipage}[t]{.5\textwidth}
  \begin{center}         
    \includegraphics[angle=90,height=6cm]{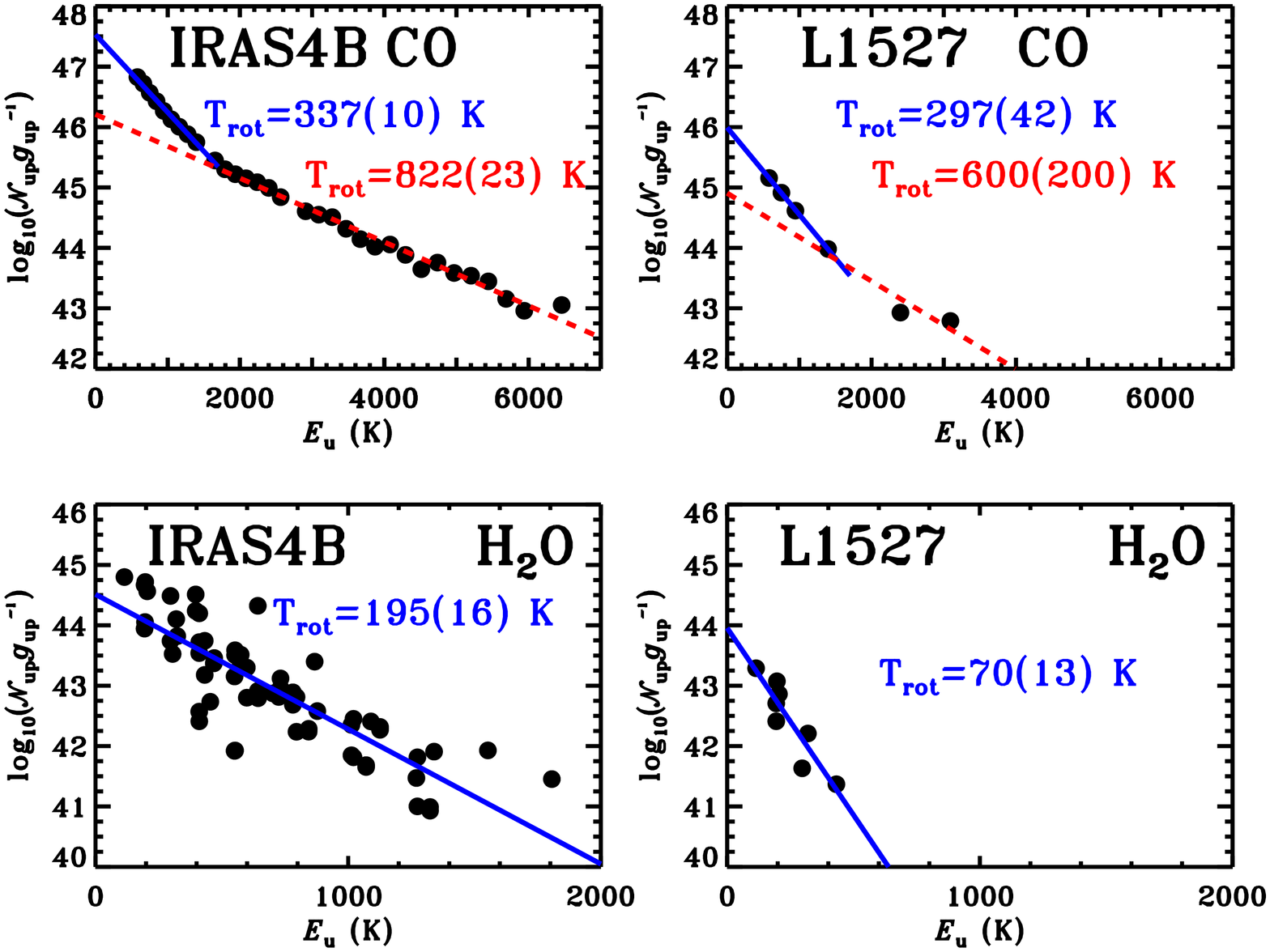} 
                   \vspace{+3ex}
                
     \includegraphics[angle=90,height=6cm]{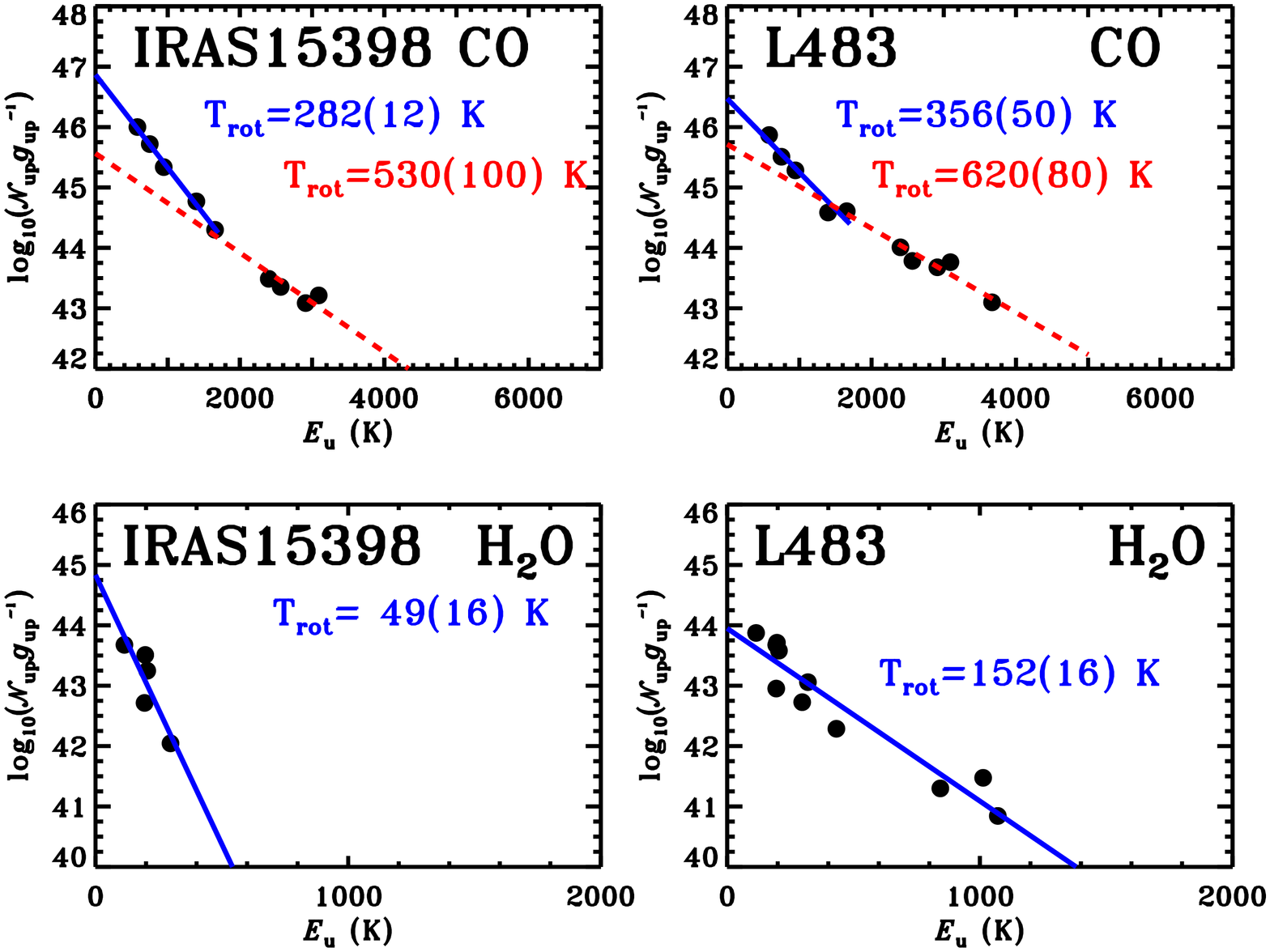} 
                   \vspace{+3ex}
    
    \includegraphics[angle=90,height=6cm]{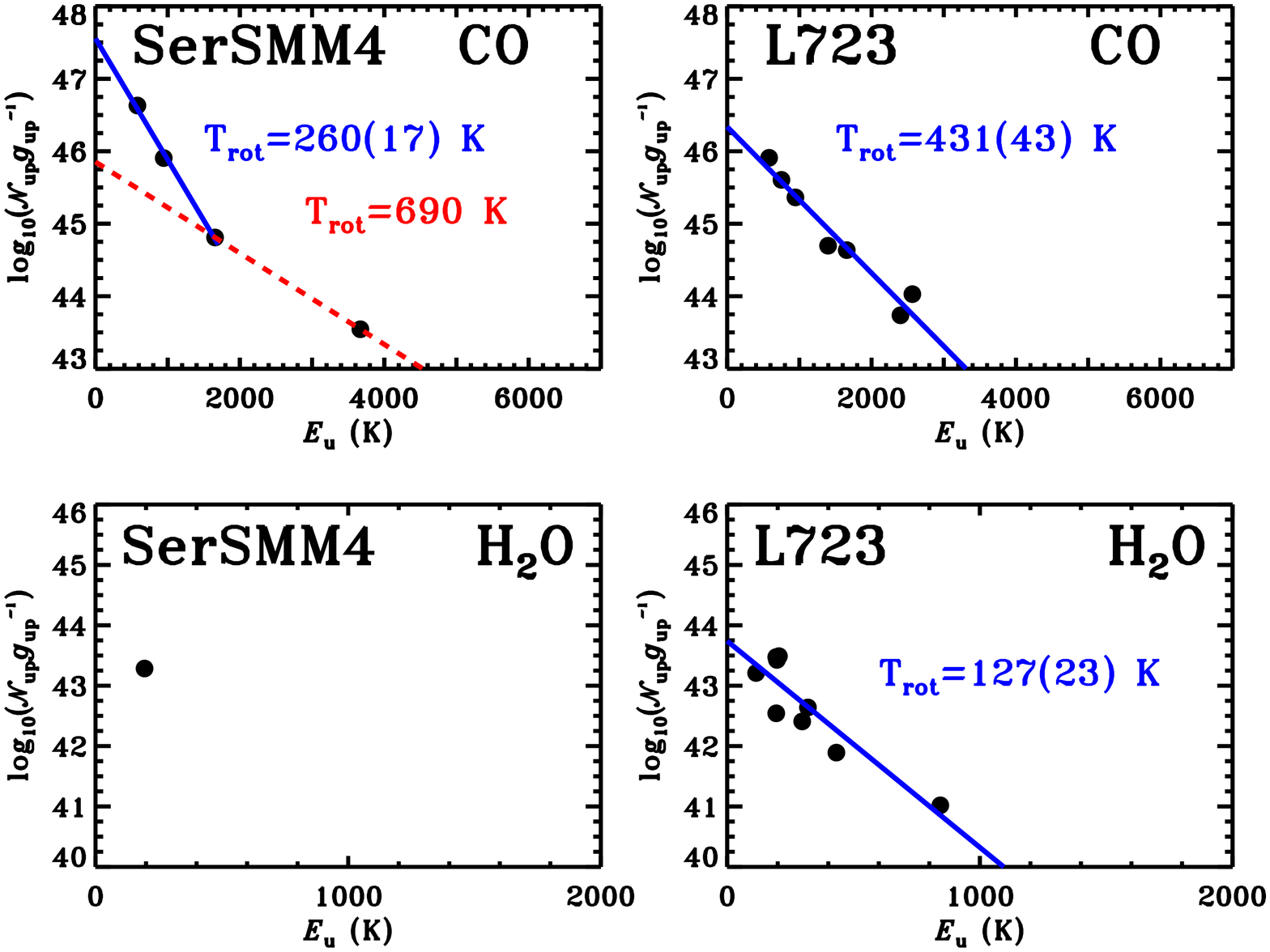} 
    \end{center}
  \end{minipage}
    \hfill
        \caption{\label{codiag} Rotational diagrams of CO and H$_2$O for Class 0 
sources. Blue and red lines show linear fits to warm and hot components, respectively. 
The corresponding rotational temperatures are written in the same colors. Errors 
associated with the fit are shown in the brackets. Warm component only is seen towards 
L723 in our diagram.}
\end{figure*}

\begin{figure*}[!tb]
  \begin{minipage}[t]{.5\textwidth}
  \begin{center}  
    \includegraphics[angle=90,height=6cm]{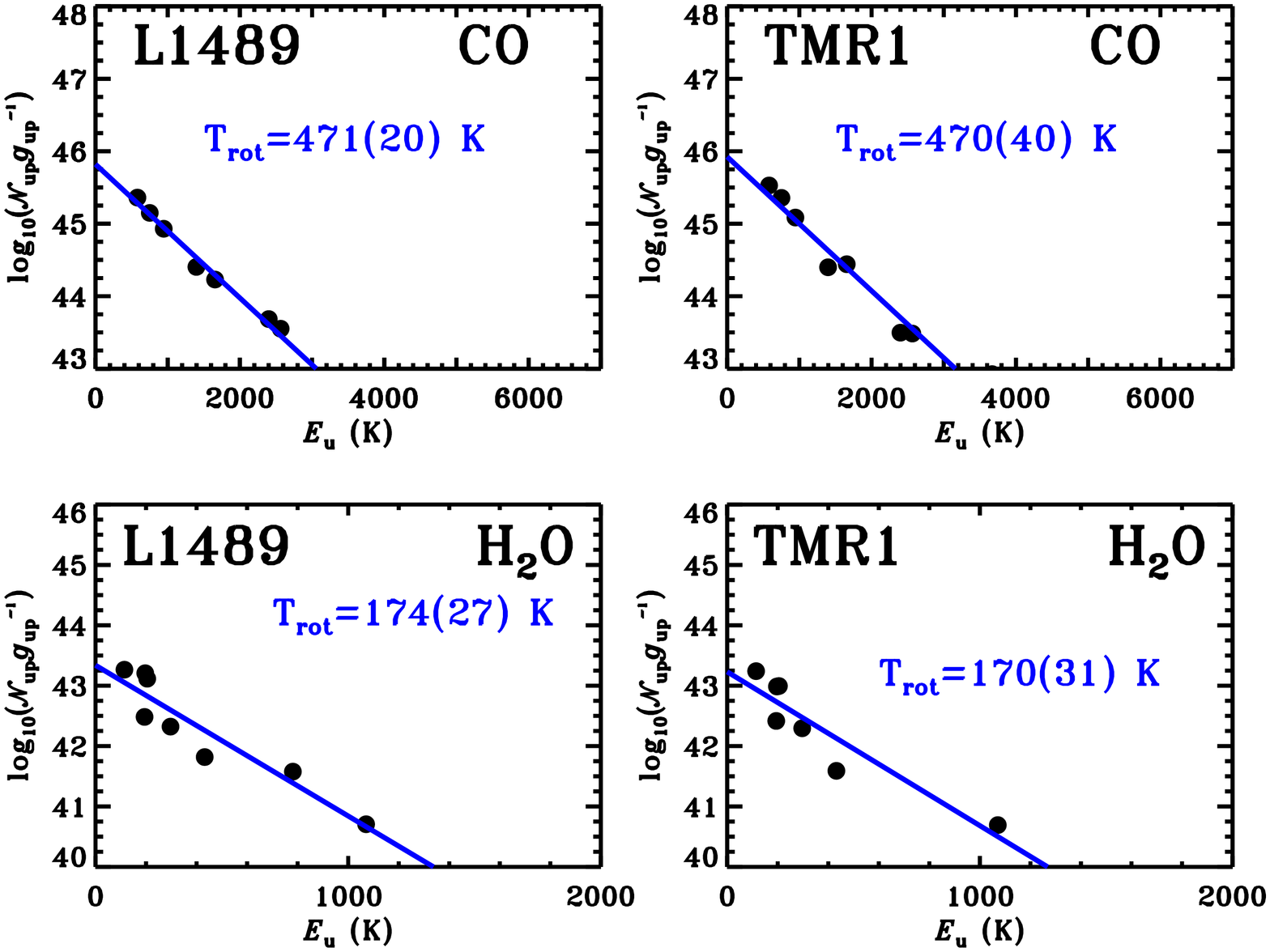}
                     \vspace{+3ex}
       
    \includegraphics[angle=90,height=6cm]{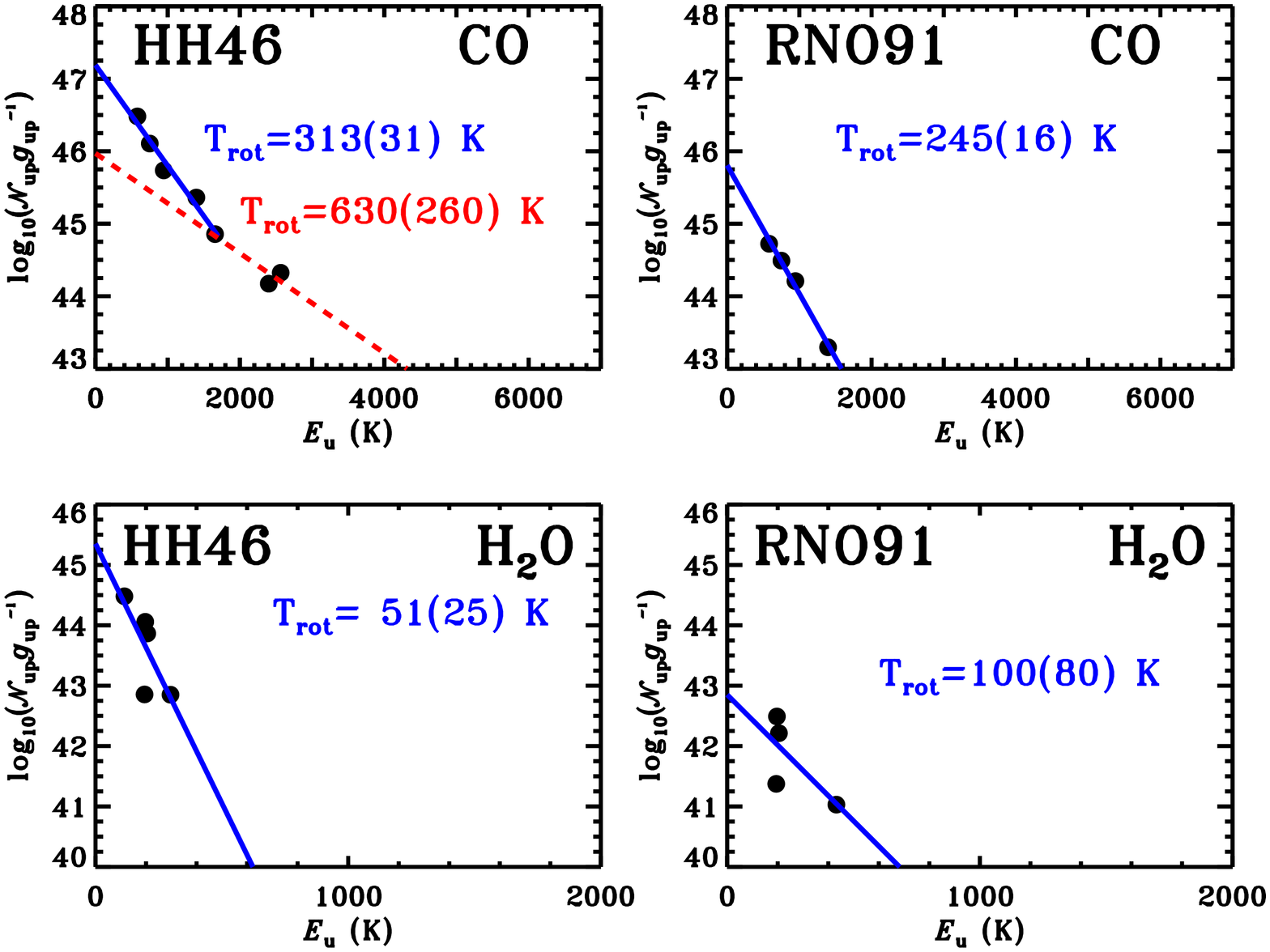}
    \end{center}
  \end{minipage}
  \hfill
  \begin{minipage}[t]{.5\textwidth}
  \begin{center}         
    \includegraphics[angle=90,height=6cm]{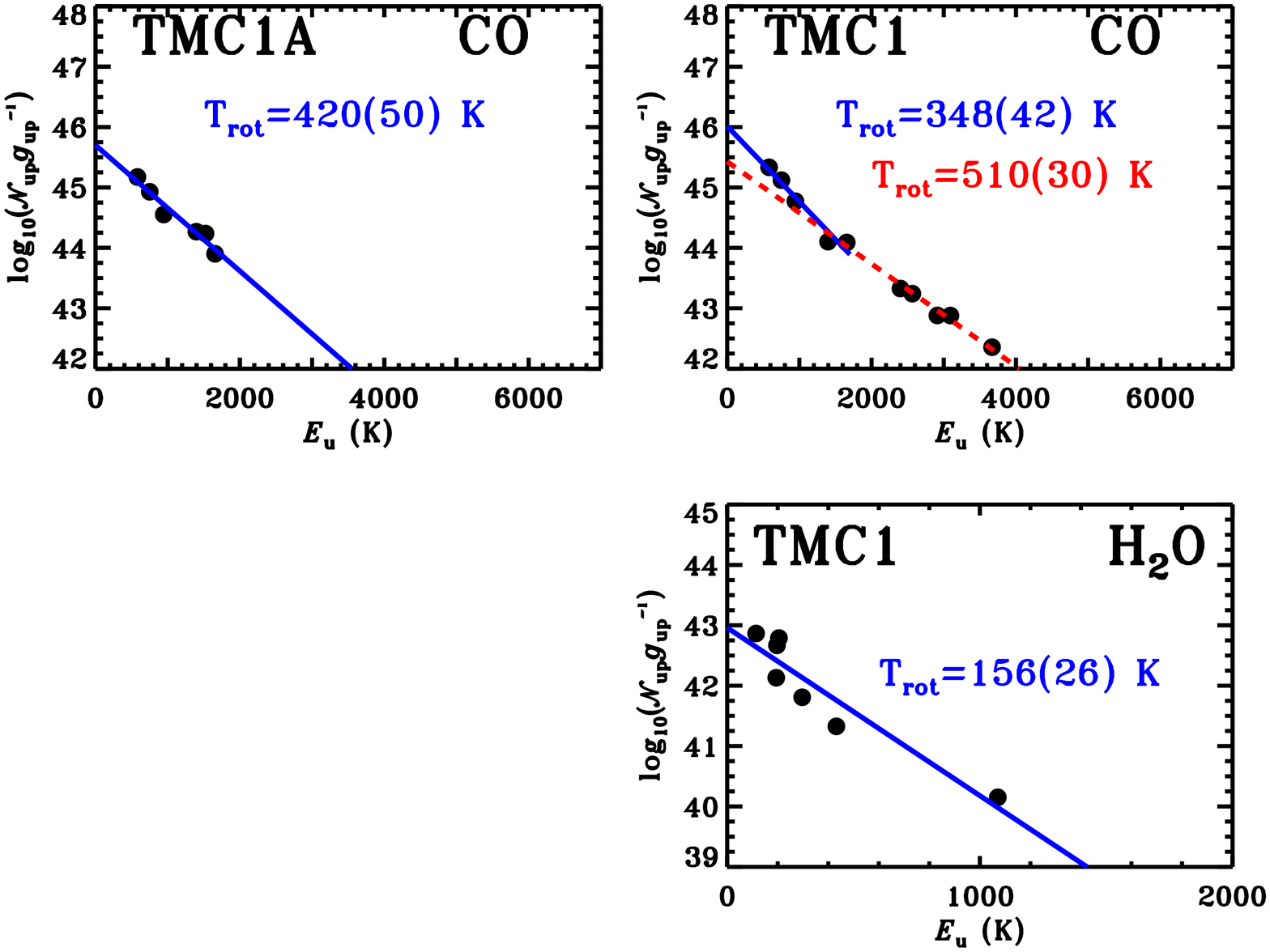} 
    \end{center}
  \end{minipage}
    \hfill
        \caption{\label{wdiag} Rotational diagrams of CO H$_2$O for Class I
sources. Blue lines show linear fits to the data. The corresponding rotational
 temperatures are given. Errors associated with the fit are shown in the brackets. 
  Warm components only are seen towards L1489, TMR1 and TMC1A in our diagrams.}
\end{figure*}

%-----------
Figures \ref{codiag} and \ref{wdiag} show CO and H$_2$O rotational diagrams for all sources
from our sample. Two-component fits are used for the CO diagrams and one-component fits 
for the H$_2$O diagrams. These fits are used to determine the total 
cooling budget in the two molecules for each objects, as discussed in \S 4.
The errors in the temperatures reflect the statistical error of the fit taking the
 uncertainties in individual line fluxes as listed in Table A.2 into account.
They do not include the absolute flux uncertainties since the relative fluxes
between lines within a single spectrum have much lower uncertainties.

\section{Correlations}
\begin{figure}[!tb]
\begin{center}
\vspace{+3ex}

 \includegraphics[angle=0,height=12cm]{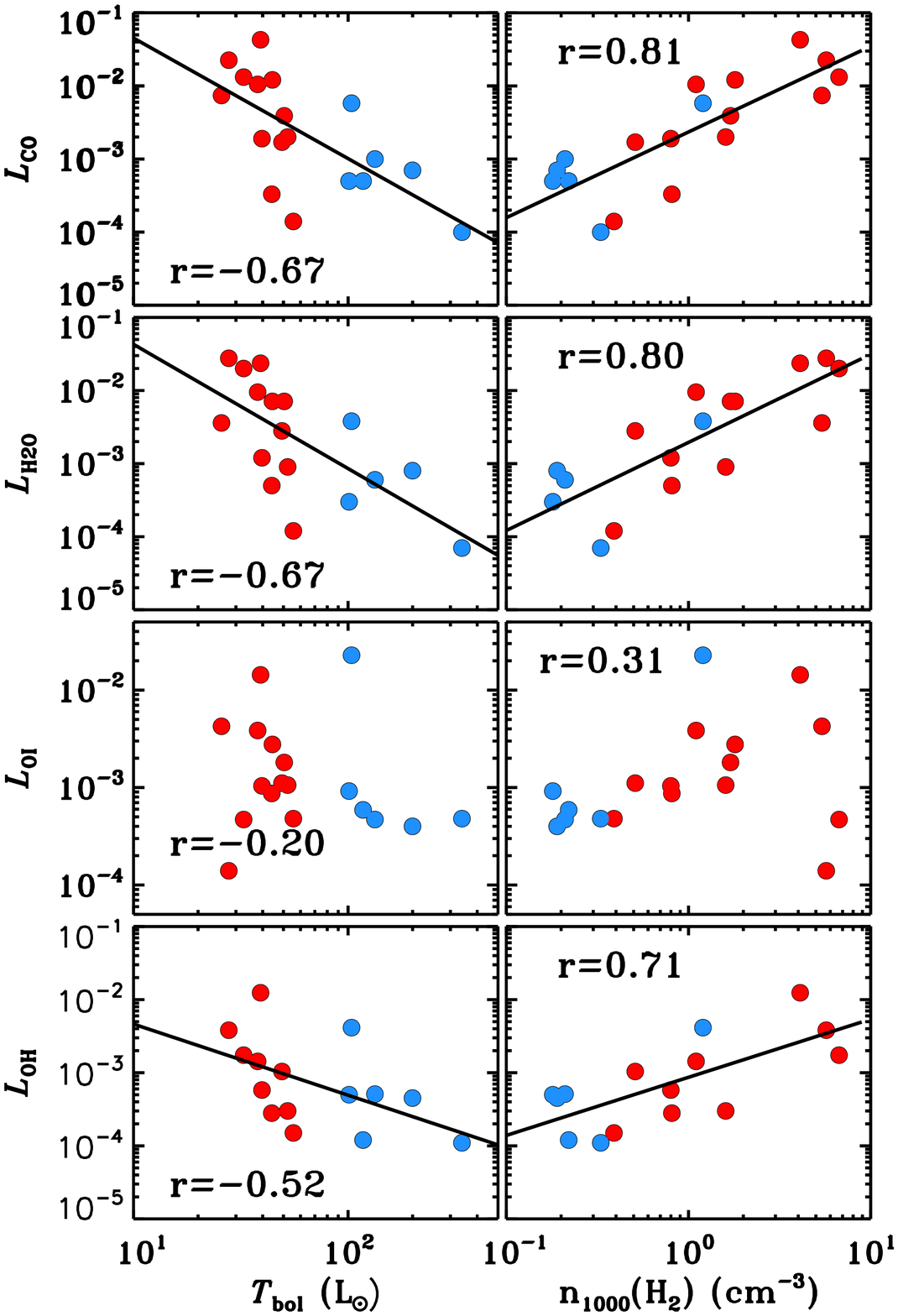}
 \vspace{+9ex}
 
 \caption{\label{flx3} Correlations between bolometric temperature
   (left column) and envelope density at 1000 AU (right column) and (from top to
   bottom): CO 14-13, H$_2$O 2$_{12}$-1$_{01}$, [\ion{O}{i}] at 63.18
   $\mu$m and OH 84.6 $\mu$m line luminosities.}
\end{center}
\end{figure}

Figure \ref{flx3} shows correlations between selected line luminosities 
and bolometric temperature ($T_\mathrm{bol}$) and density at 1000 AU ($n_\mathrm{H2}$). 
Strong correlations are found with the latter quantity.
%===========================
%===========================

\section{Rotational temperature uncertainties}
\label{sec:unc}
\begin{figure*}[!tb]
  \begin{minipage}[t]{.5\textwidth}
  \begin{center}  
      \includegraphics[angle=90,height=6cm]{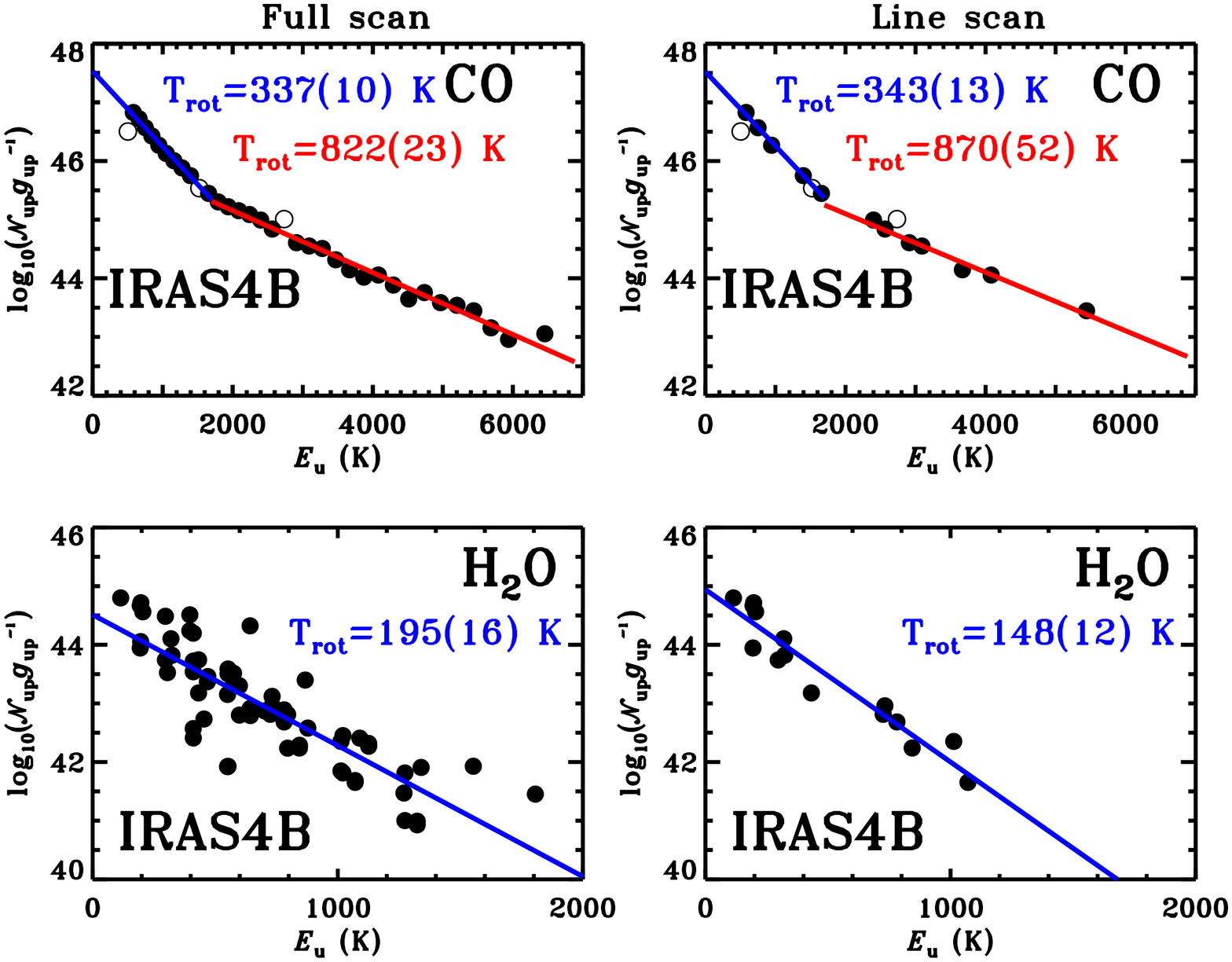}
    \end{center}
  \end{minipage}
  \hfill
  \begin{minipage}[t]{.5\textwidth}
  \begin{center}         
    \includegraphics[angle=90,height=6cm]{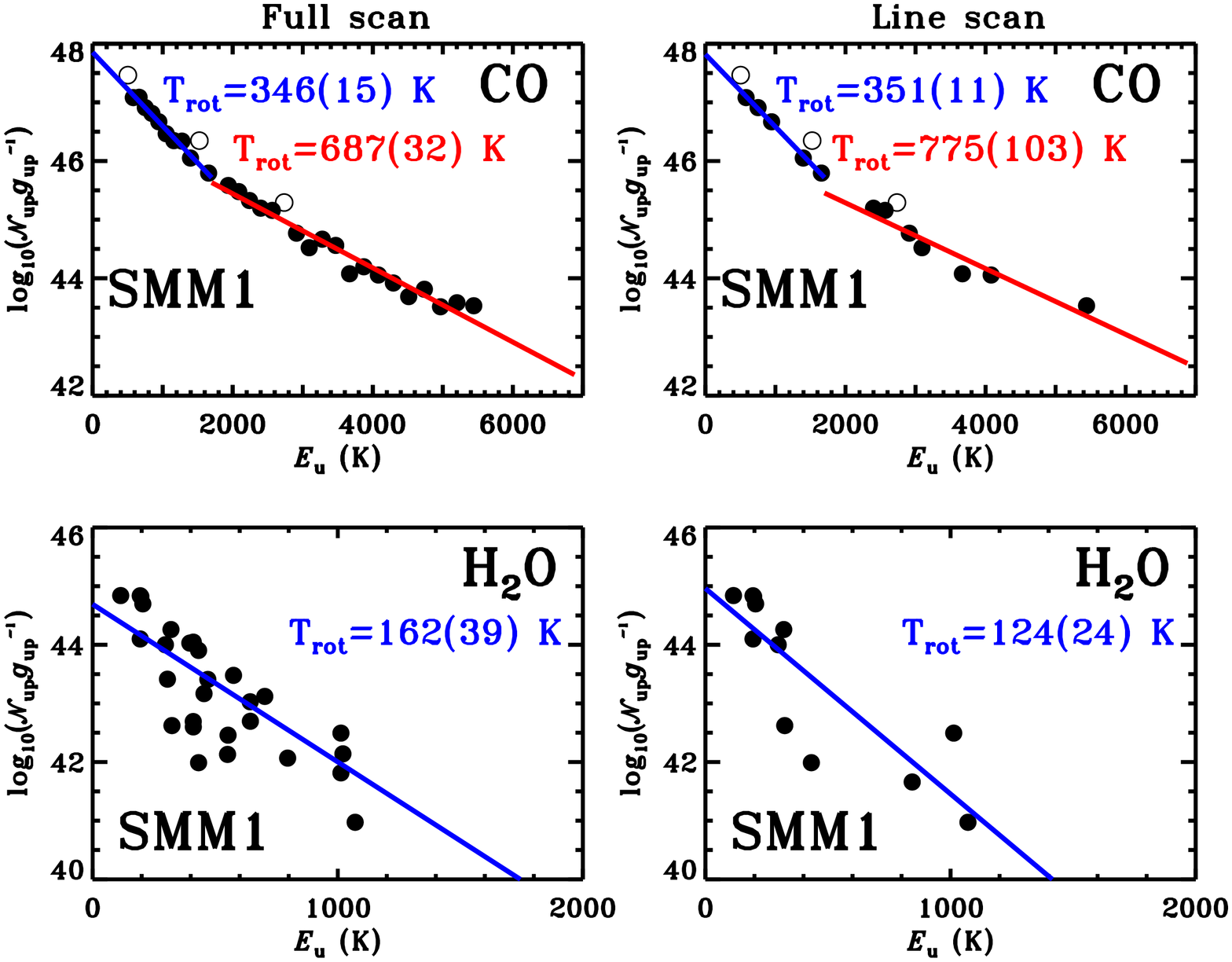} 
    \end{center}
  \end{minipage}
    \hfill
        \caption{\label{unc} The CO and H$_2$O rotational diagrams for NGC1333-IRAS4B and Serpens 
SMM1. Both the full spectroscopy line and the selected lines observed in the line scan mode are 
shown. Two component fit is done to the CO diagram with the break at 1700 K
(i.e. transitions $J\geq24$ correspond to the warm component) and a single component fit 
to the H$_2$O diagram. Errors associated with the fit are shown in the brackets.}
\end{figure*}

Figure \ref{unc} shows the CO and H$_2$O rotational diagrams for the NGC1333-IRAS4B and Serpens 
SMM1, using the data published in Herczeg et al. (2012) and Goicoechea et al. (2012). 
The full spectroscopy data is shown, with the full line coverage in the PACS range, 
as well as the selected lines only, typically observed in our line spectroscopy mode for 16 
sources in our sample.

Rotational temperatures calculated from the rotational diagrams constructed using the 
full and limited line configurations are in good agreement for the CO. For the assumed 
ranges of the two components, the warm component 
$T_\mathrm{rot}$ error of the fit is $\pm15$ K and the hot component $T_\mathrm{rot}$ 
error is $\pm50-100$ K. 

The change of the energy break in a wide range of 
transitions results in $\pm20$ K error for the $T_\mathrm{rot}$(warm) and $\pm40$ K 
for the $T_\mathrm{rot}$(hot) for the full spectroscopy data. Those ranges are not 
well determined by line spectroscopy data only and thus in this work we always use 1700 K 
for this kind of observations. 

H$_2$O rotational temperatures, on the other hand, are less accurately determined 
for the line spectroscopy observations than the formal error of the fit would imply. 
The scatter due to the subthermal excitation and likely large opacities of the water lines
results in significant differences in $T_\mathrm{rot}$ calculation, depending on the choice 
of observed lines. The fit to the water lines chosen in our program underestimates the 
resulting temperature by about 50-80 K for the NGC1333-IRAS4B and Serpens SMM1. 

\section{Cooling budget calculations}
\label{sec:cool}
Since different methods of cooling budget calculation exist in the literature and will appear 
due to the availability of new Herschel observations, we perform here a comparison between 
the methods and estimate the differences between the resulting budgets. 

\subsection{Carbon monoxide}
In order to calculate the CO cooling, \citet{Ni02} calculated LVG models that reproduced 
the detected transitions and used them to determine the fluxes for the first 60 
transitions of CO. Due to the limited \textit{ISO} sensitivity, CO transitions from 
$J=$14-13 to $J=$22-21 (Class I) and $J=$29-28 (Class 0) were available for the brightest 
sources only which did not allow them to distinguish the \textit{hot} component. 
Their method corresponds to a single-component fit to the excitation diagrams. 

We use the entire PACS array line fluxes of NGC1333-IRAS4B and Ser SMM1 
from Herczeg et al. (2012) and Goicoechea et al. (2012) to compare the observed CO 
total luminosities with those calculated using fits to the excitation diagrams (see Table
\ref{accu}). In particular,
we show the results of the two-components fits for PACS data and three-components fits for PACS and 
SPIRE data, available for Ser SMM1. Three wavelength ranges are included: (i) the PACS range 
from 54-60 to 190 $\mu$m; (ii) the Nisini et al. 2002 range, namely 44-2601 $\mu$m; (iii) the PACS + SPIRE 
range, from 60 to 650 $\mu$m (for SMM1 only). Additionally, we include the fitting results 
to the observed lines only, in order to check how good our two/three-component linear fits reproduce the
observations.

The calculations in Table \ref{accu} show that multi-component linear fits to the excitation
diagrams agree well
 with the observed values of the total CO luminosity from the detected lines (rows 1+2 and 5+6). 
 The uncertainties correspond to different choice of the break energy for the two components. 
The fits are then used to extrapolate the fluxes of the lines which are either blends or fall in
the region of $\sim$100 $\mu$m, where the measured line fluxes are less reliable (row 3 for PACS
range and row 7 for PACS+SPIRE range). The resultant total CO luminosities are a good measure 
of the far-IR CO cooling (for the PACS range) and total CO cooling (for PACS+SPIRE range). 

The example of SMM1 shows that the additional CO emission from the entrained outflow gas
increases the CO luminosity by a factor of 1.3 with respect to the extrapolated values 
from the warm CO component ($5.23\cdot10^{-2}$ L$_{\odot}$ versus $6.93\cdot10^{-2}$
 L$_{\odot}$).

 The relative CO luminosity in different
 spectral regions (A: $J$=4-3 to $J$=13-12 B: $J$=14-13 to $J$=24-23 and C: $J$=25-24 to $J$=44-43)
 for SMM1 is $\sim$2:2:1 (A:B:C) and for IRAS4B is $\sim$2:1 (B:C), when the additional 
 cold CO component is included. Thus, approximately 80\% 
 of the CO luminosity comes from transitions lower than CO 24-23, roughly equally in both SPIRE
 and PACS ranges. The poorer determination of the hot component rotational temperature 
 does not affect significantly the total cooling determination. 

\begin{table}
\begin{minipage}[t]{\columnwidth}
\caption{\label{accu}CO total luminosities for IRAS4B and SMM1 in $10^{-4}$ L$_{\odot}$.}
\centering 
\renewcommand{\footnoterule}{}  % to avoid a line before footnotes
\begin{tabular}{lrrrrrrr}
\hline \hline
Method & Range ($\mu$m)&								   IRAS4B    	&   SMM1 \\
\hline 
Obs. line fluxes\tablefootmark{a} & 54-190       	 &  212 		&	396		\\
2-comp. fit for obs. lines only & 54-190 			 & 	210$\pm$2   &	394$\pm$5\\
\textbf{2-comp. fit, PACS range\tablefootmark{b}} & \textbf{54-190} &	\textbf{225$\pm$3}	& \textbf{428$\pm$8}\\
2-comp. fit, Nisini+2002 range\tablefootmark{c} & 44-2601   &	271$\pm$6	&	523$\pm$4\\
\hline
Obs.PACS+SPIRE line fluxes & 60-650	 			     & 		--		&  650 \\
3-comp. fit for obs. lines only 	& 60-650				 &		--		&  648 \\
3-comp. fit, PACS range\tablefootmark{d}  & 60-650	 &		--		&  690 \\
3-comp. fit, Nisini+2002 range   & 44-2601  				 &	  	--		&  693 \\
\hline
\end{tabular}
\end{minipage}
\tablefoot{Errors correspond 
to standard deviation of the total CO cooling calculated using the break-point upper energies from
1000 to 2200 K.
\tablefoottext{a}{From full PACS range, excluding CO 23-22 and CO 31-30 which are blended with the H$_2$O $4_{14}-3_{03}$
 and OH $^{2}\Pi_{\nicefrac{3}{2}}$ $J=\nicefrac{7}{2}-\nicefrac{5}{2}$, respectively.}
\tablefoottext{b}{Includes CO transitions from $J=$14-13 to $J=$48-47 (IRAS4B) or $J=$44-43 (SMM1). This is the method 
used to determine cooling budget in Table 4.}
\tablefoottext{c}{Includes CO transitions from $J=$1-0 to $J=$60-59, used in Nisini et al. (2002).}
\tablefoottext{d}{Includes CO transitions from $J=$4-3 to $J=$44-43.}
}
\end{table}

\subsection{Water}
Nisini et al. (2002) have calculated H$_2$O luminosities based on the assumption that H$_2$O
emission arises from the same gas as CO, for which large velocity gradient (LVG) models 
determined the gas temperature and density. The model prediction were used to extrapolate
 H$_2$O line fluxes for rotational transitions with $J<10$ and $E_\mathrm{u}/k_\mathrm{B}<2031$ K. 
 
Total H$_2$O cooling in this work (see \S 4.2) is calculated based on the full spectroscopy data for 
the NGC1333-IRAS4A and Serpens SMM1 (Herczeg et al. 2012 and Goicoechea et al. 2012).
The average scaling factor than transfers the luminosity observed in the selected lines 
in the line spectroscopy mode to the total water luminosities (as observed in the range 
spectroscopy) is 2.4$\pm0.3$.

An alternative method considered for the water cooling calculation is the extrapolation of the 
non-observed line fluxes based on the H$_2$O rotational temperature. Table \ref{accu:h2o} compares
 the results
of this method with the values obtained when the scaling factor was used. The extrapolation
is done for (i) the 328 lowest rotational transitions of water ($J<10$, $E_\mathrm{u}/k_\mathrm{B}<2031$ K, 
so called 'Nisini et al. range'); (ii) the same transitions but for PACS range only. 
Molecular information is obtained from the JPL and CDMS catalogs \citep{JPL,CDMS2,CDMS}. 

Calculations for Serpens SMM3 show that the extrapolation of the fluxes based on the 
fitted rotational temperature results in a factor of $\sim 2$ larger total water luminosities than the value
calculated using the scaling factor of 2.4. Even higher values are obtained when we extend 
the range used in Nisini et al. (2002). Additionally, 
the rotational temperature of H$_2$O derived from the line scan data is likely underestimated by 
a factor of $\sim1.3-1.6$ (see Appendix \ref{sec:unc}). This uncertainty has an effect 
on the derived, extrapolated, H$_2$O total cooling. 
 
In summary, from the comparisons it is concluded that the total luminosity of both CO
and H$_2$O in the PACS range is accurate to 30\% or better.

\begin{table}
\begin{minipage}[t]{\columnwidth}
\caption{\label{accu:h2o} Different methods of H$_2$O luminosities calculation 
(luminosities in $10^{-3}$ L$_\mathrm{\odot}$).}
\centering
\renewcommand{\footnoterule}{}  % to avoid a line before footnotes
\begin{tabular}{lccc}
\hline \hline
Method & 											 Ser SMM3 \\
\hline \hline
Observed line fluxes (line spec mode)    	 &  	   4.0		\\
Scaling to the total PACS range scan flux (factor: 2.4)		 & 	  	   9.6		\\
\hline
$T_\mathrm{rot}$=125 K & \\
Extrapolation for PACS range 		     	 &		   18.3		\\
Extrapolation for Nisini et al. range 		 &		   23.5		\\
\hline
\end{tabular}
\end{minipage}
\end{table}

\Online
\section{PACS maps for all sources}
\begin{figure*}[!tb]
  \begin{minipage}[t]{.33\textwidth}
  \begin{center}  
      \includegraphics[angle=90,height=7cm]{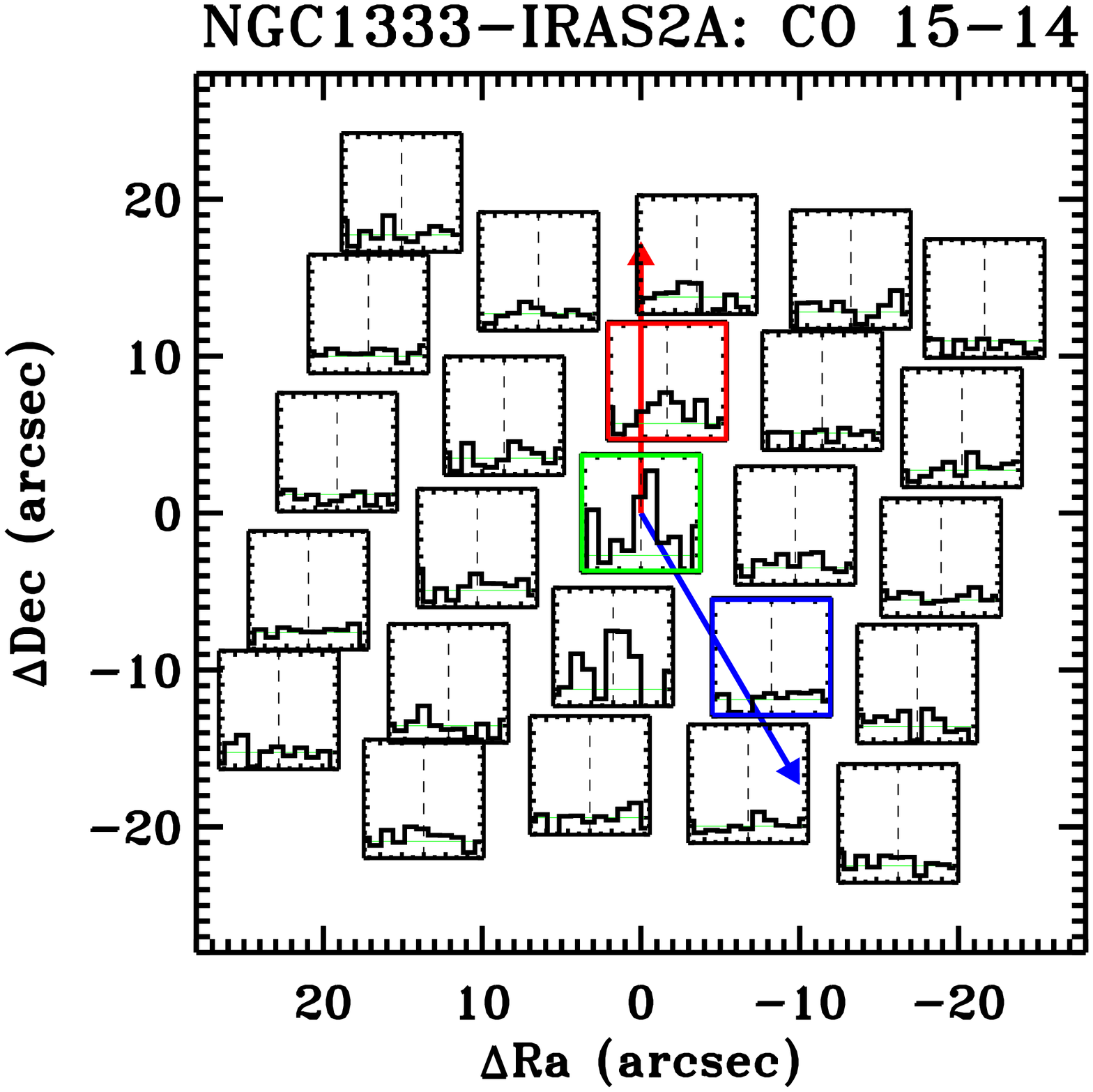}
    \end{center}
  \end{minipage}
  \hfill
  \begin{minipage}[t]{.33\textwidth}
  \begin{center}         
    \includegraphics[angle=90,height=7cm]{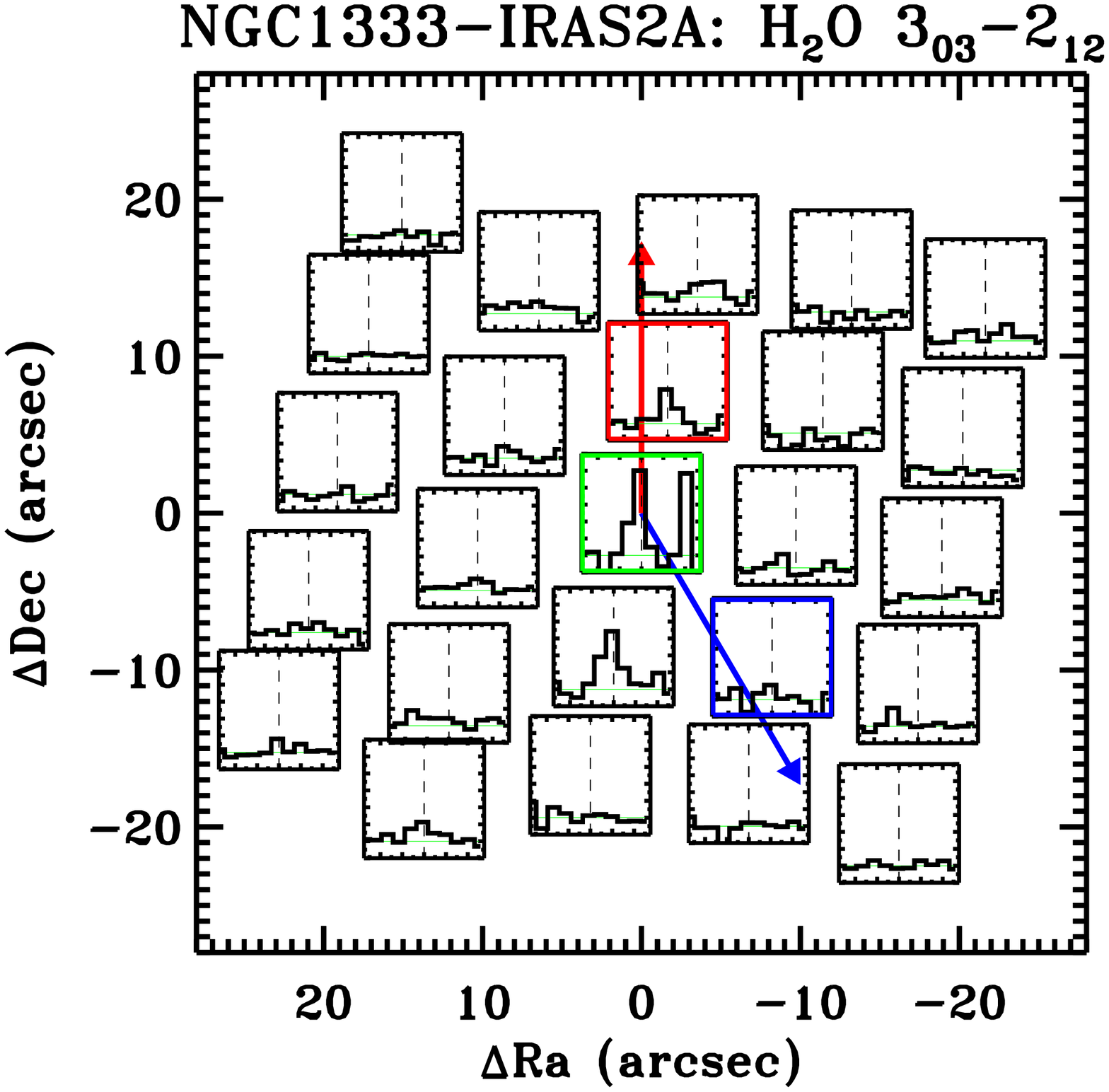} 
    \end{center}
  \end{minipage}
    \hfill
  \begin{minipage}[t]{.33\textwidth}
  \begin{center}  
     \includegraphics[angle=90,height=7cm]{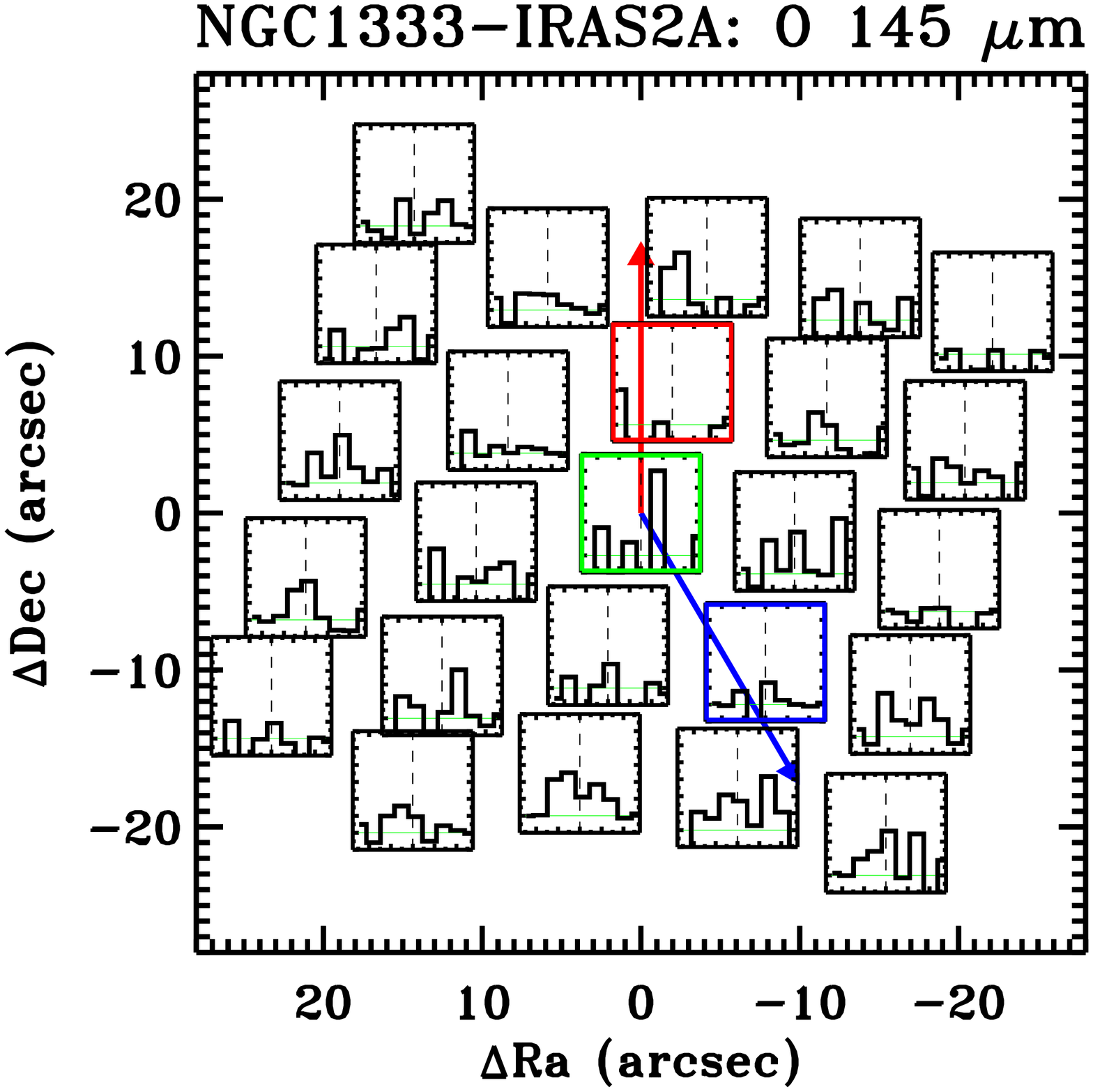}
     \end{center}
  \end{minipage}
 %  \vspace{+3ex}
        \caption{\label{i2amap} NGC1333-IRAS2A maps in the CO 15--14 line at 173.6 $\mu$m, 
        the H$_2$O 3$_{03}$-2$_{12}$ line at 174.6 $\mu$m and 
        the [\ion{O}{i}] $^3P_{0}-^{3}P_{1}$ line at 145.5 $\mu$m.}
\end{figure*}

\begin{figure*}[!tb]
  \begin{minipage}[t]{.5\textwidth}
  \begin{center}  
      \includegraphics[angle=90,height=9cm]{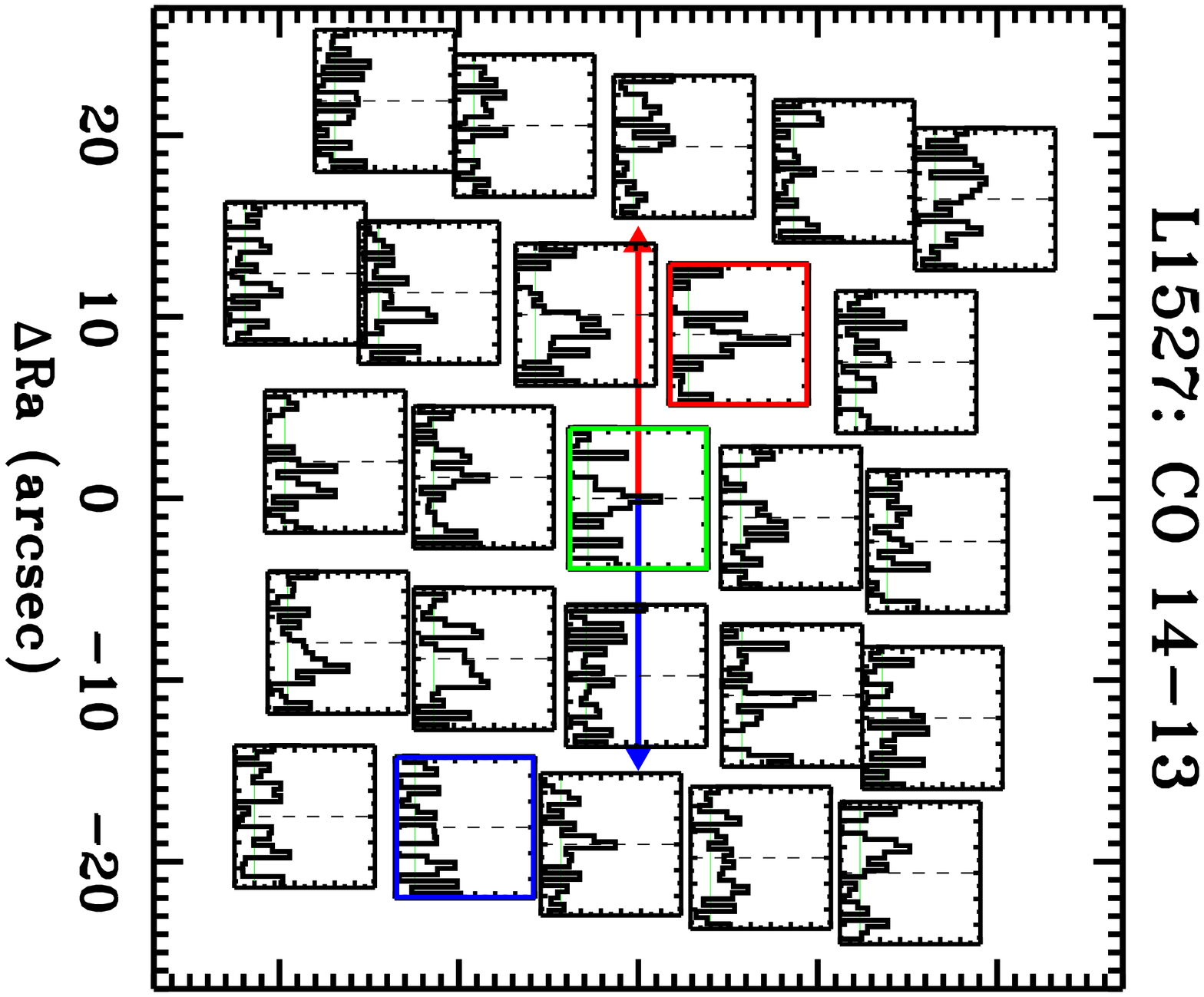}
               \vspace{+3ex}
       
      \includegraphics[angle=90,height=9cm]{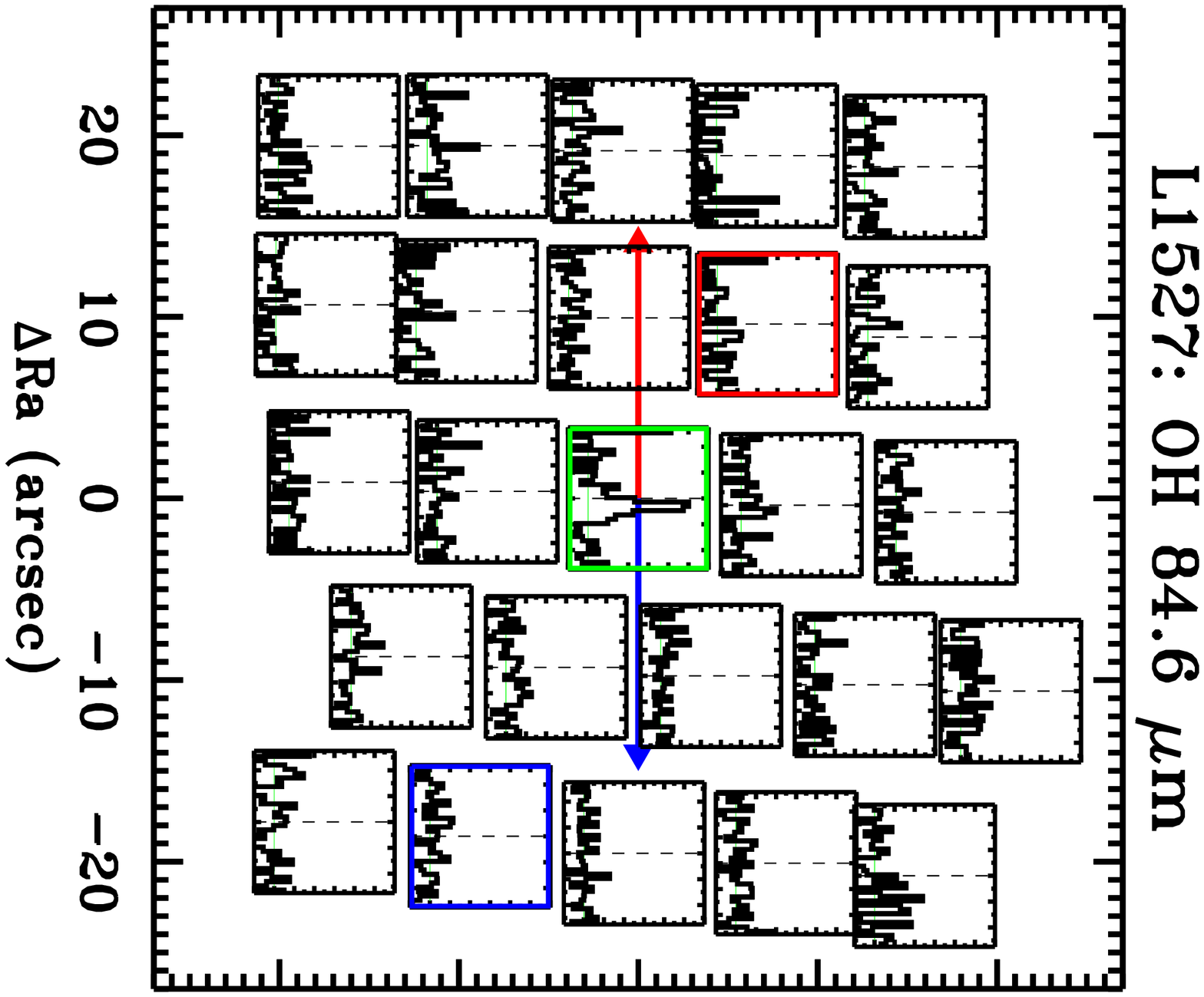}
    \end{center}
  \end{minipage}
  \hfill
  \begin{minipage}[t]{.5\textwidth}
  \begin{center}         
      \includegraphics[angle=90,height=9cm]{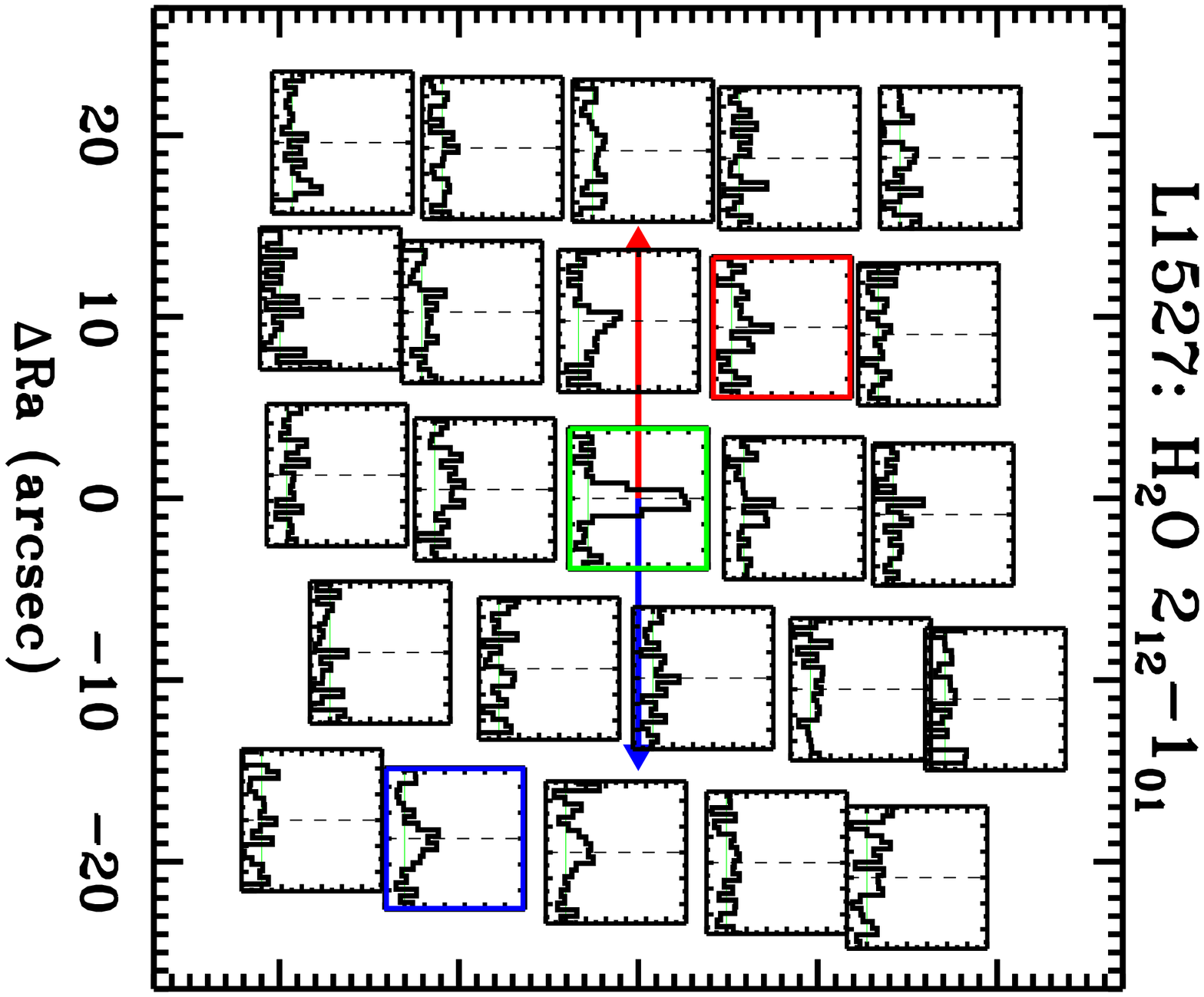}
               \vspace{+3ex}
       
    \includegraphics[angle=90,height=9cm]{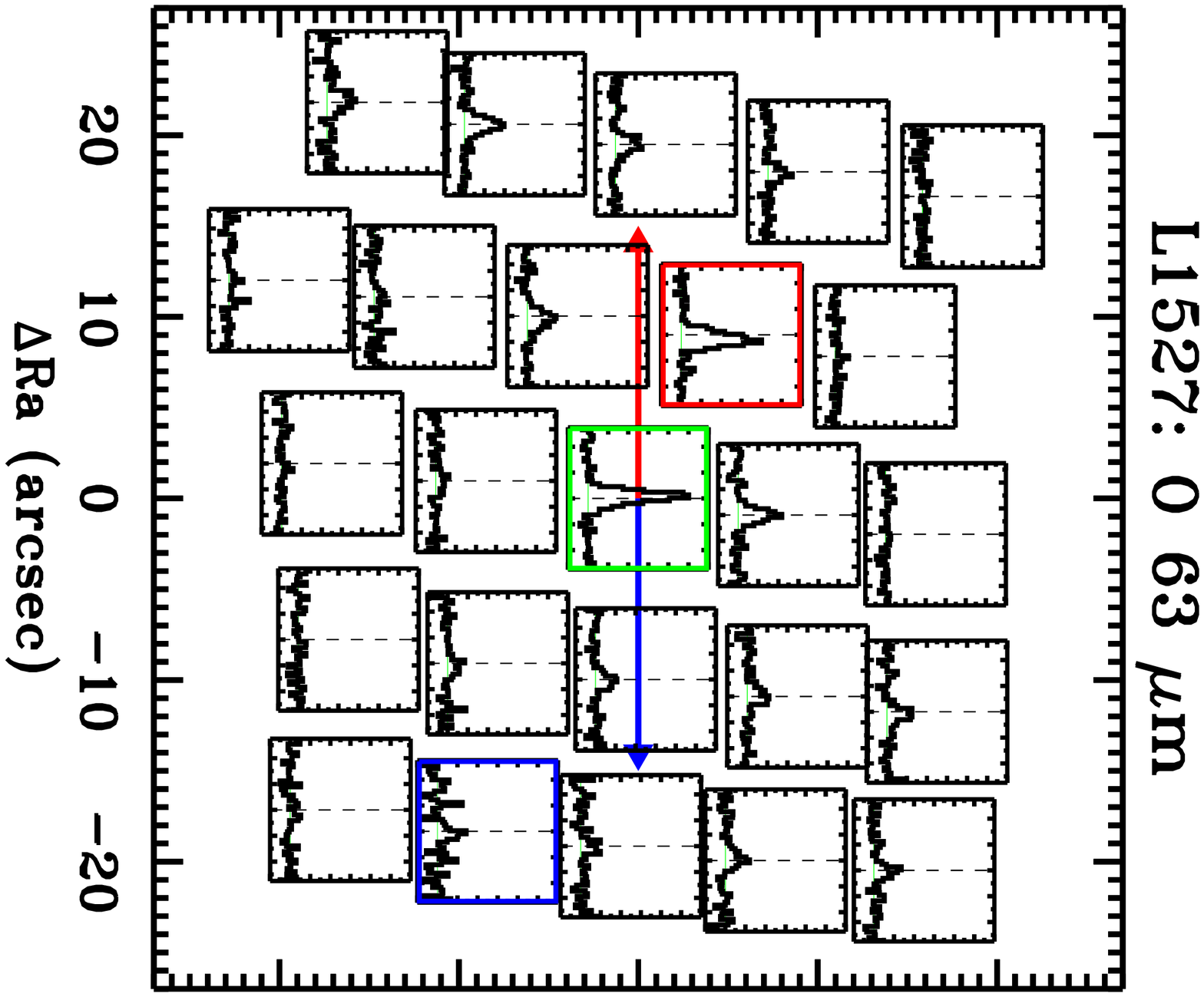}
    \end{center}
  \end{minipage}
 %  \vspace{+3ex}
        \caption{\label{l1527map} L1527 maps in the [\ion{O}{i}] $^3P_{1}-^{3}P_{2}$ line
        at 63.2 $\mu$m, the H$_2$O 2$_{12}$-1$_{01}$ line at 179.5 $\mu$m, the 
        CO 14-13 at 186.0 $\mu$m and the OH $^{2}\Pi_{\nicefrac{3}{2}}$
        $J=\nicefrac{7}{2}-\nicefrac{5}{2}$ line at 84.6 $\mu$m.}
\end{figure*}

\begin{figure*}[!tb]
  \begin{minipage}[t]{.5\textwidth}
  \begin{center}  
      \includegraphics[angle=90,height=9cm]{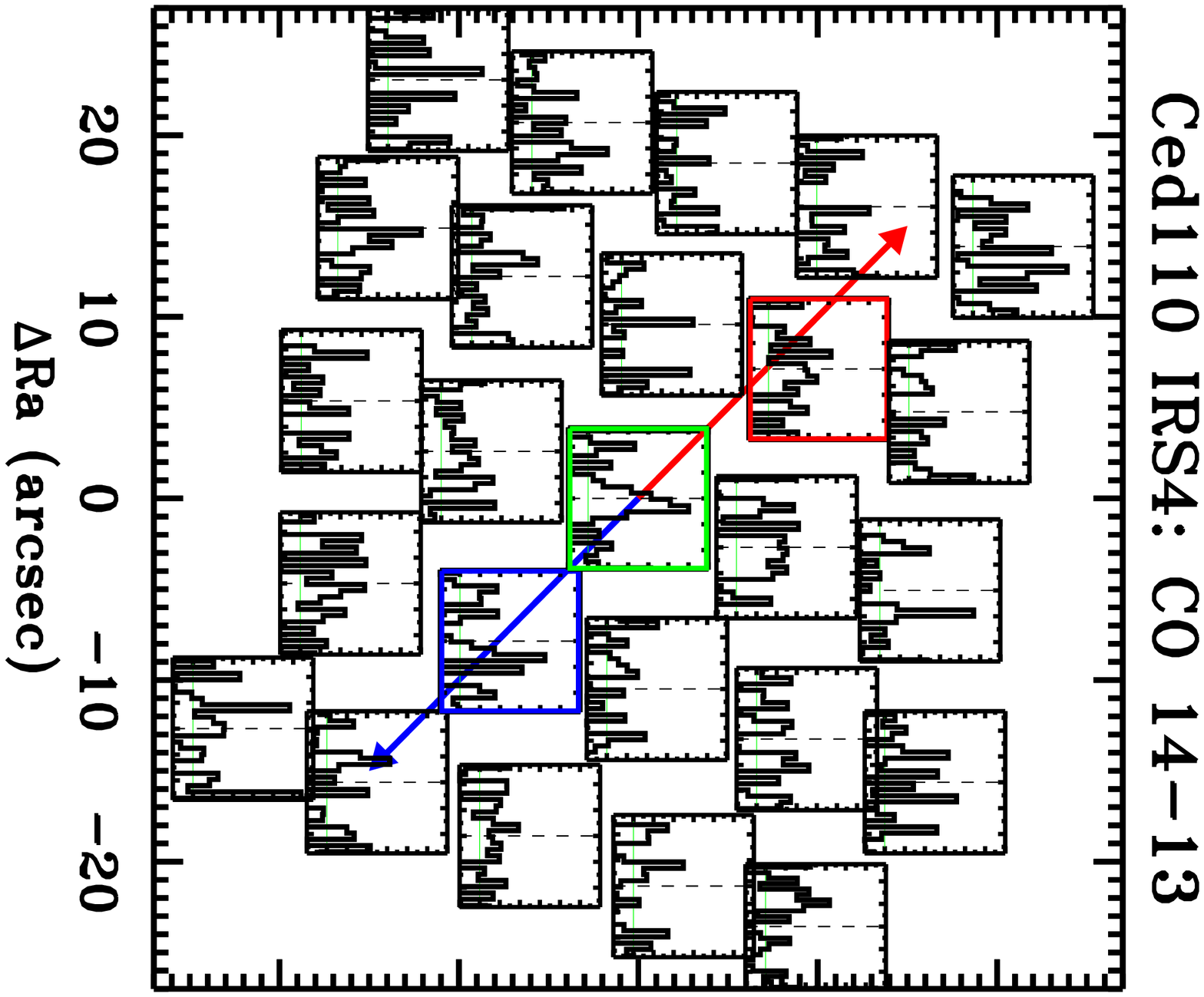}
               \vspace{+3ex}
       
      \includegraphics[angle=90,height=9cm]{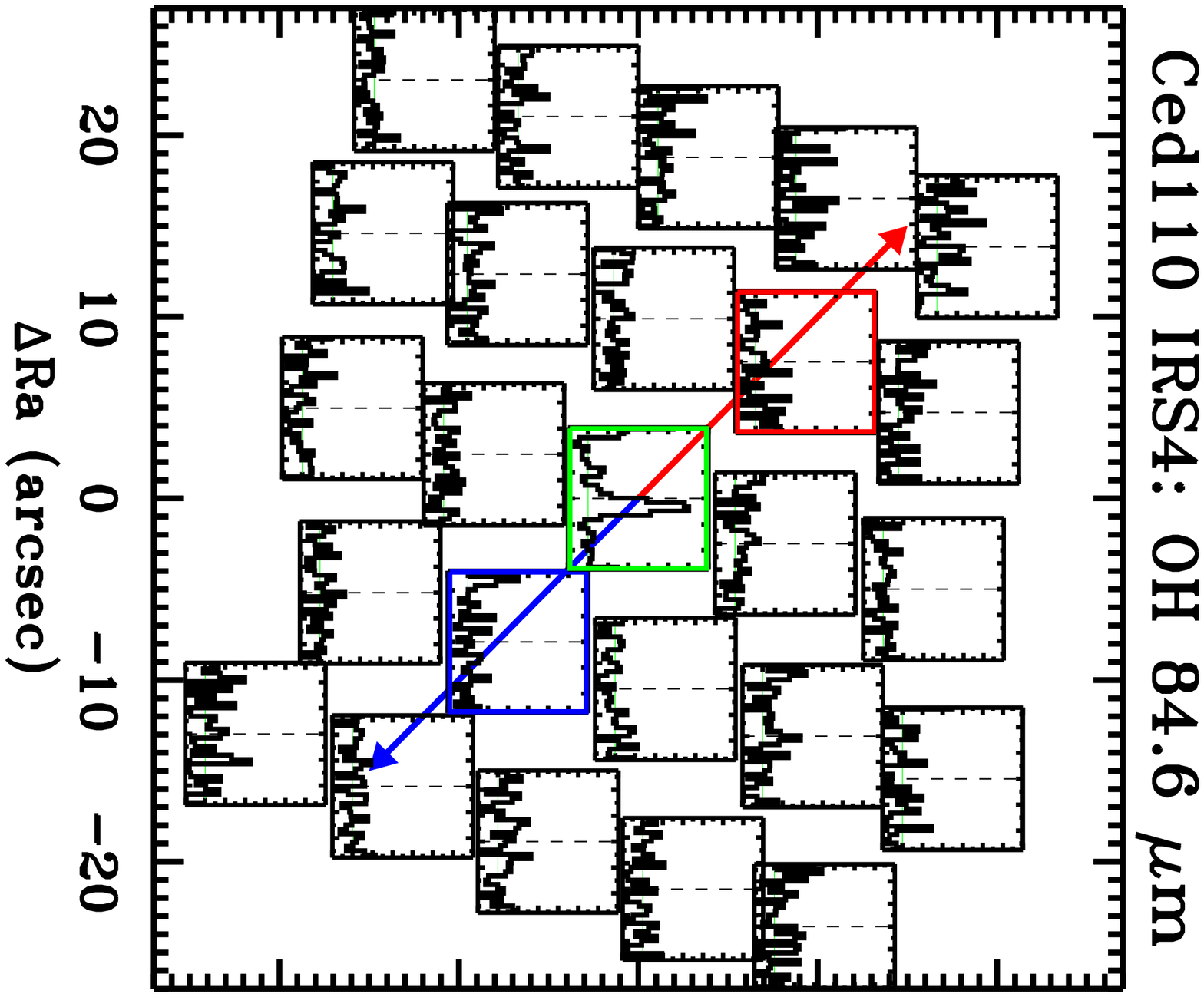}
    \end{center}
  \end{minipage}
  \hfill
  \begin{minipage}[t]{.5\textwidth}
  \begin{center}         
      \includegraphics[angle=90,height=9cm]{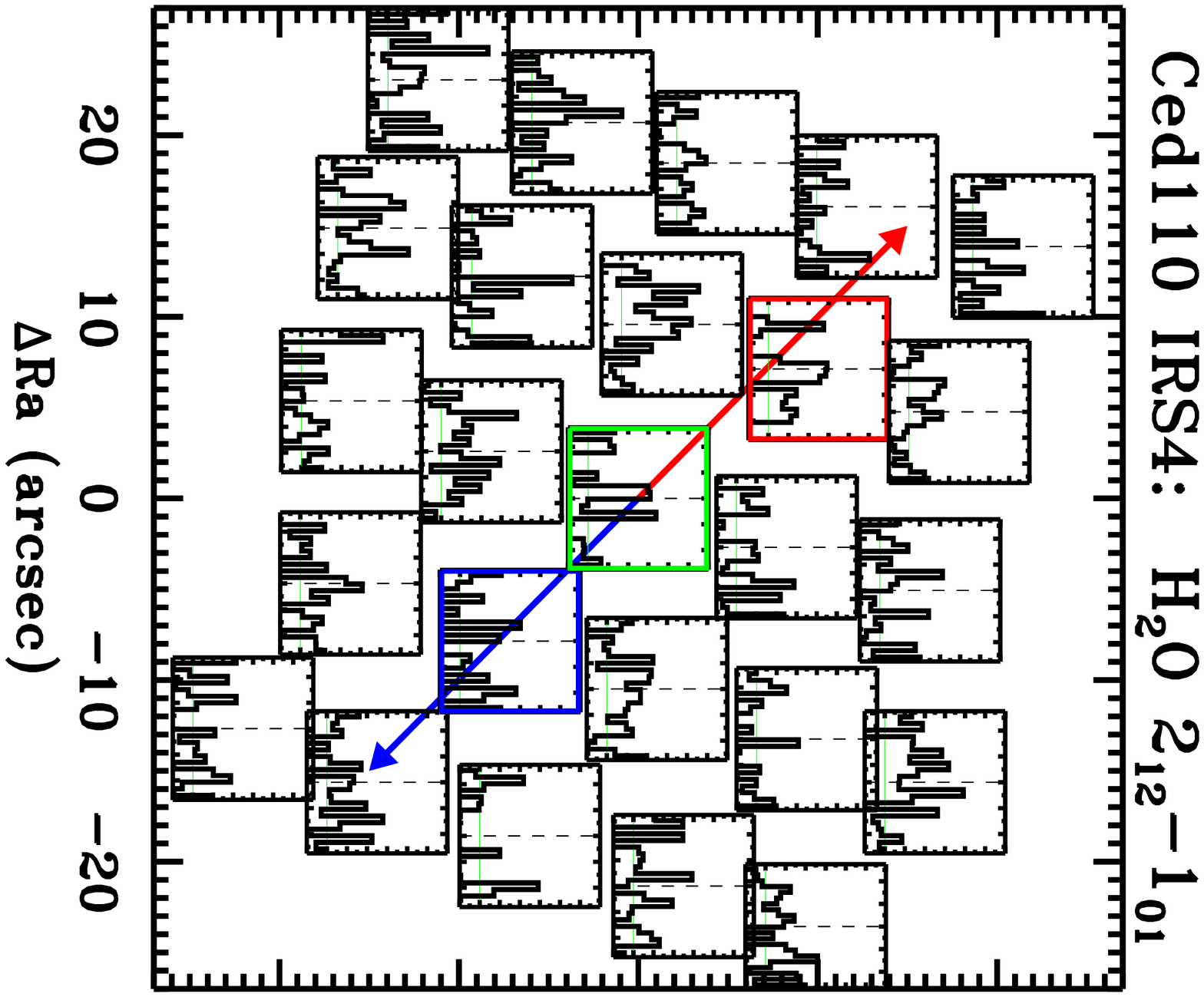}
               \vspace{+3ex}
       
    \includegraphics[angle=90,height=9cm]{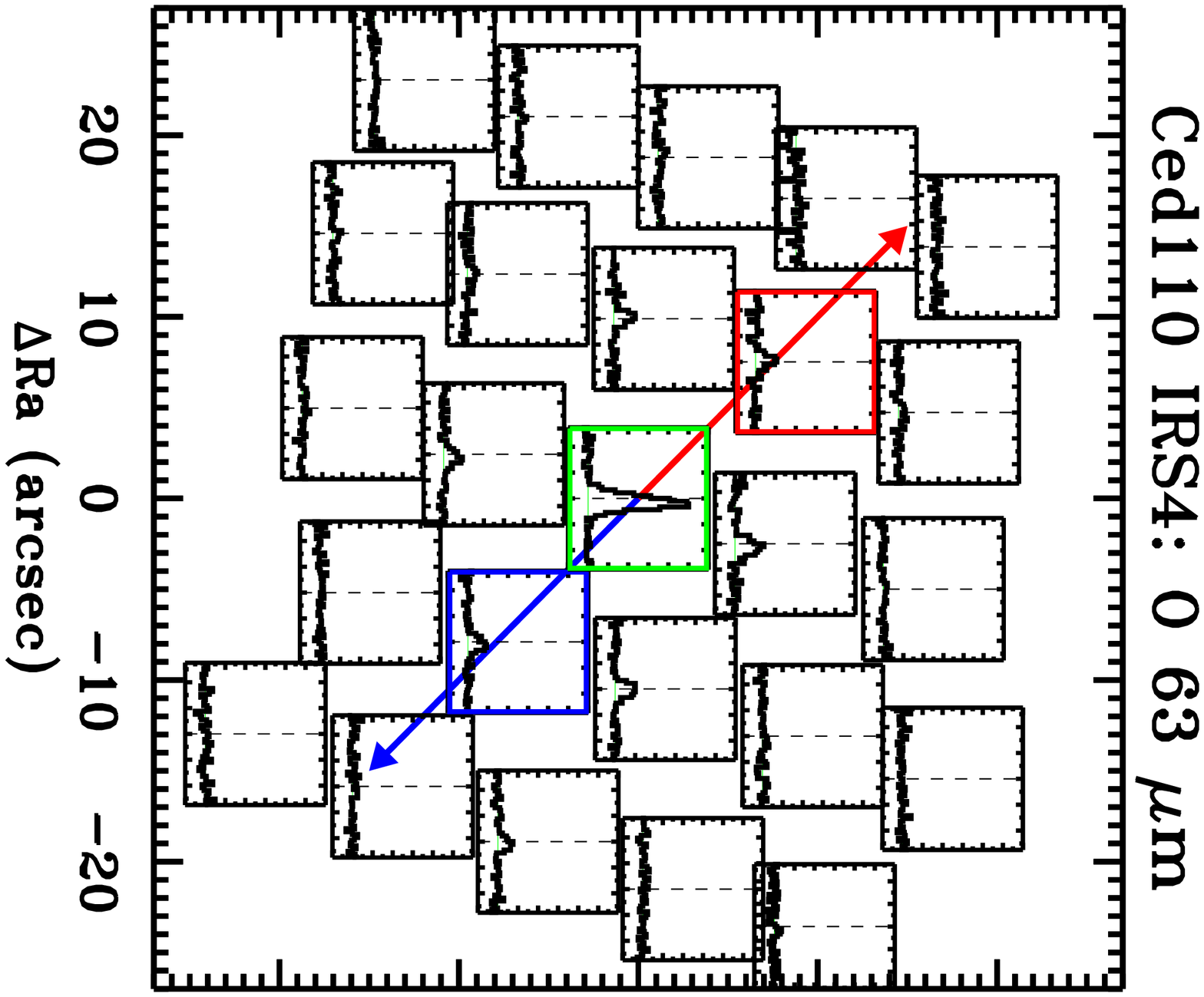}
    \end{center}
  \end{minipage}
 %  \vspace{+3ex}
        \caption{\label{ced_map} Ced110-IRS6 maps in the [\ion{O}{i}] $^3P_{1}-^{3}P_{2}$ line
        at 63.2 $\mu$m, the H$_2$O 2$_{12}$-1$_{01}$ line at 179.5 $\mu$m, the 
        CO 14-13 at 186.0 $\mu$m and the OH $^{2}\Pi_{\nicefrac{3}{2}}$
        $J=\nicefrac{7}{2}-\nicefrac{5}{2}$ line at 84.6 $\mu$m.}
\end{figure*}

\begin{figure*}[!tb]
  \begin{minipage}[t]{.5\textwidth}
  \begin{center}  
      \includegraphics[angle=90,height=9cm]{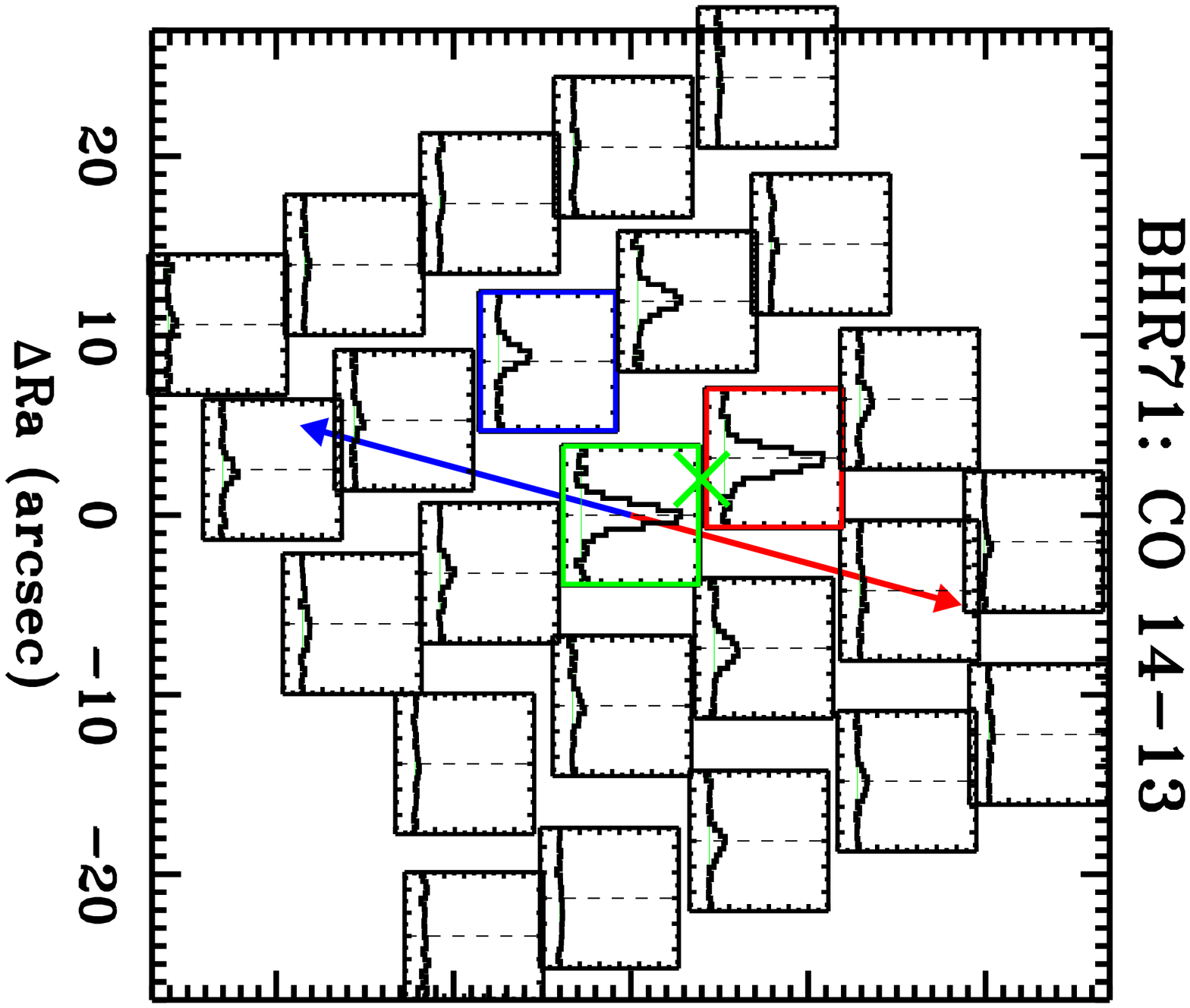}
               \vspace{+3ex}
       
     \includegraphics[angle=90,height=9cm]{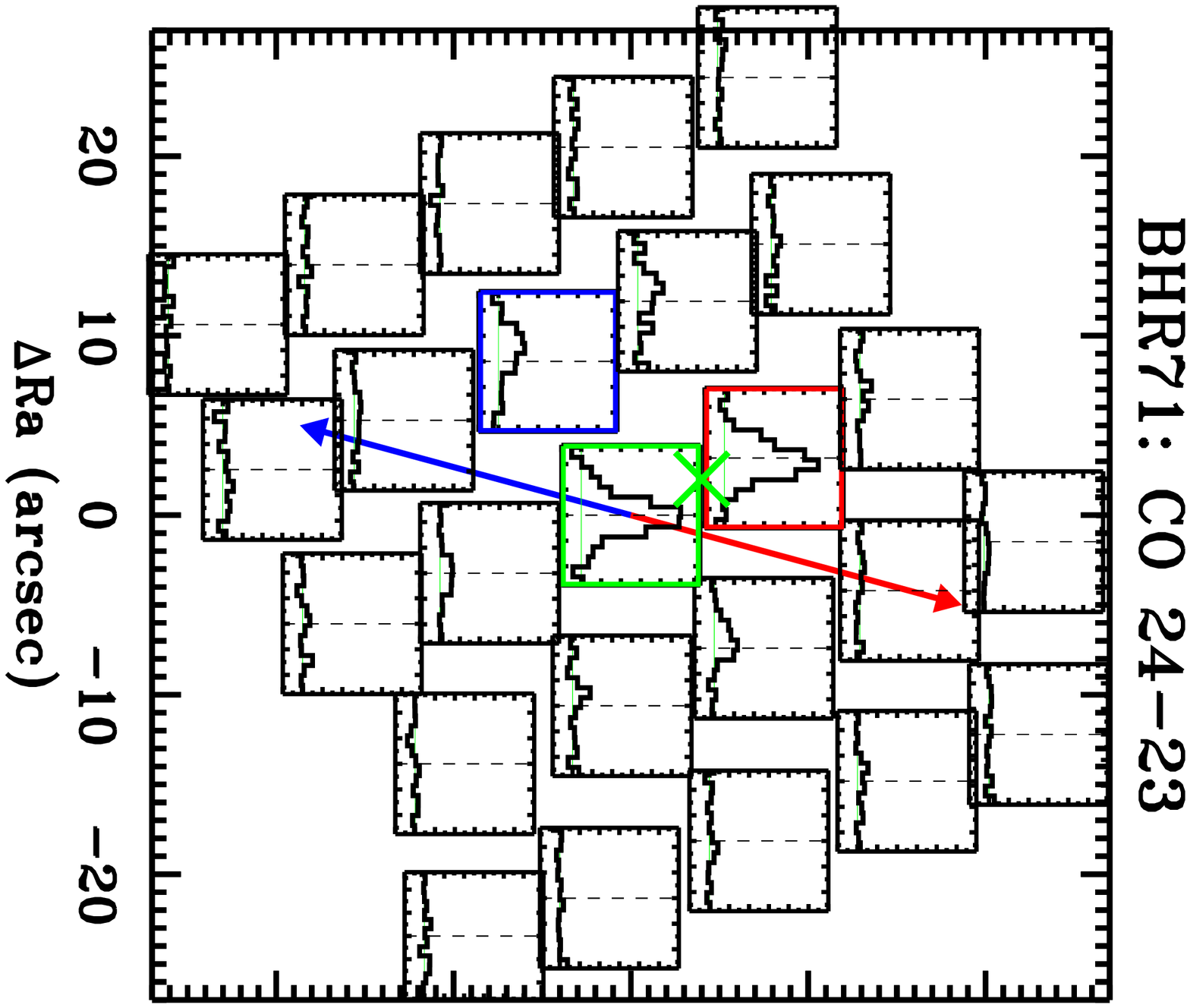}
    \end{center}
  \end{minipage}
  \hfill
  \begin{minipage}[t]{.5\textwidth}
  \begin{center}         
    \includegraphics[angle=90,height=9cm]{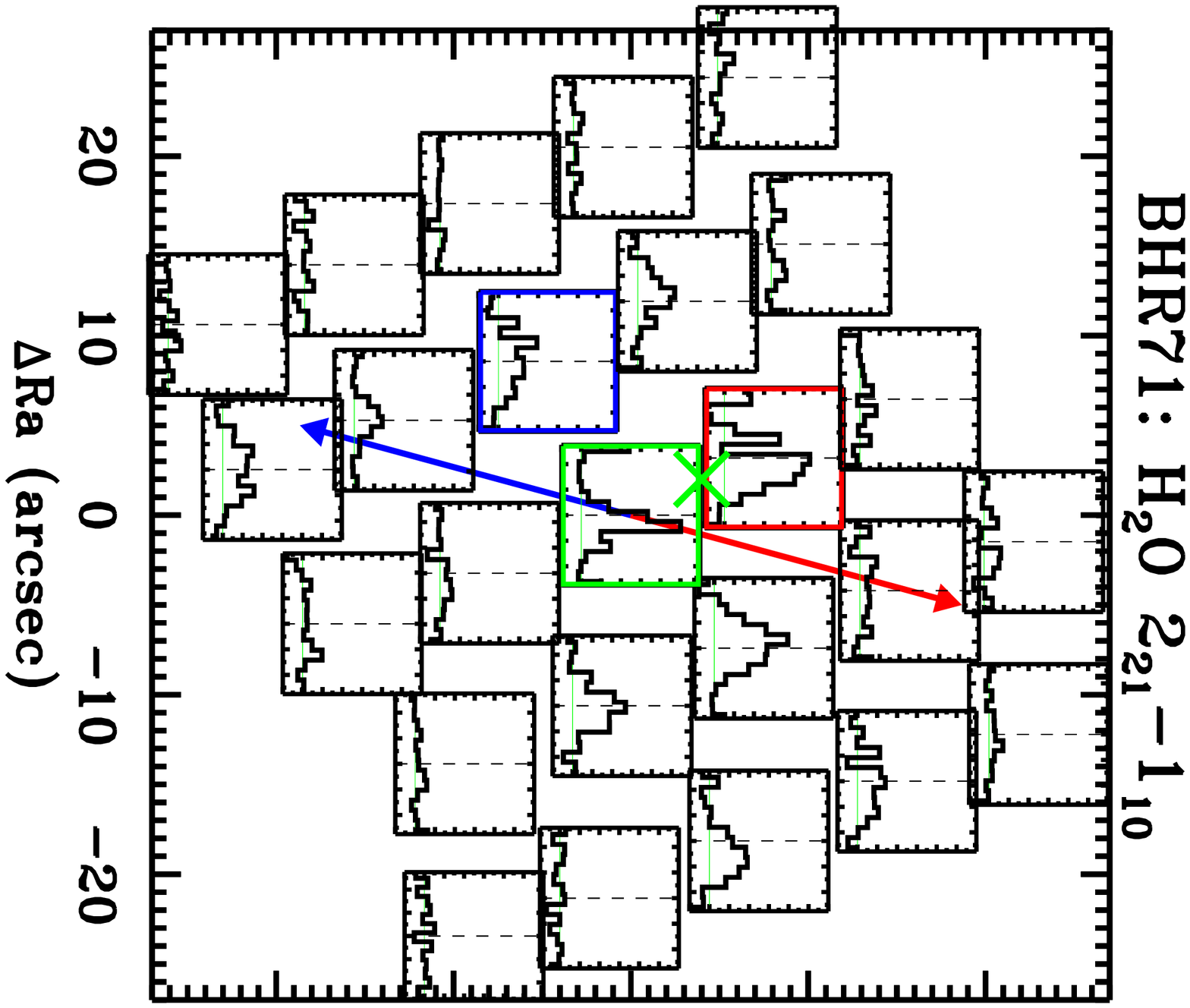}
             \vspace{+3ex}
       
    \includegraphics[angle=90,height=9cm]{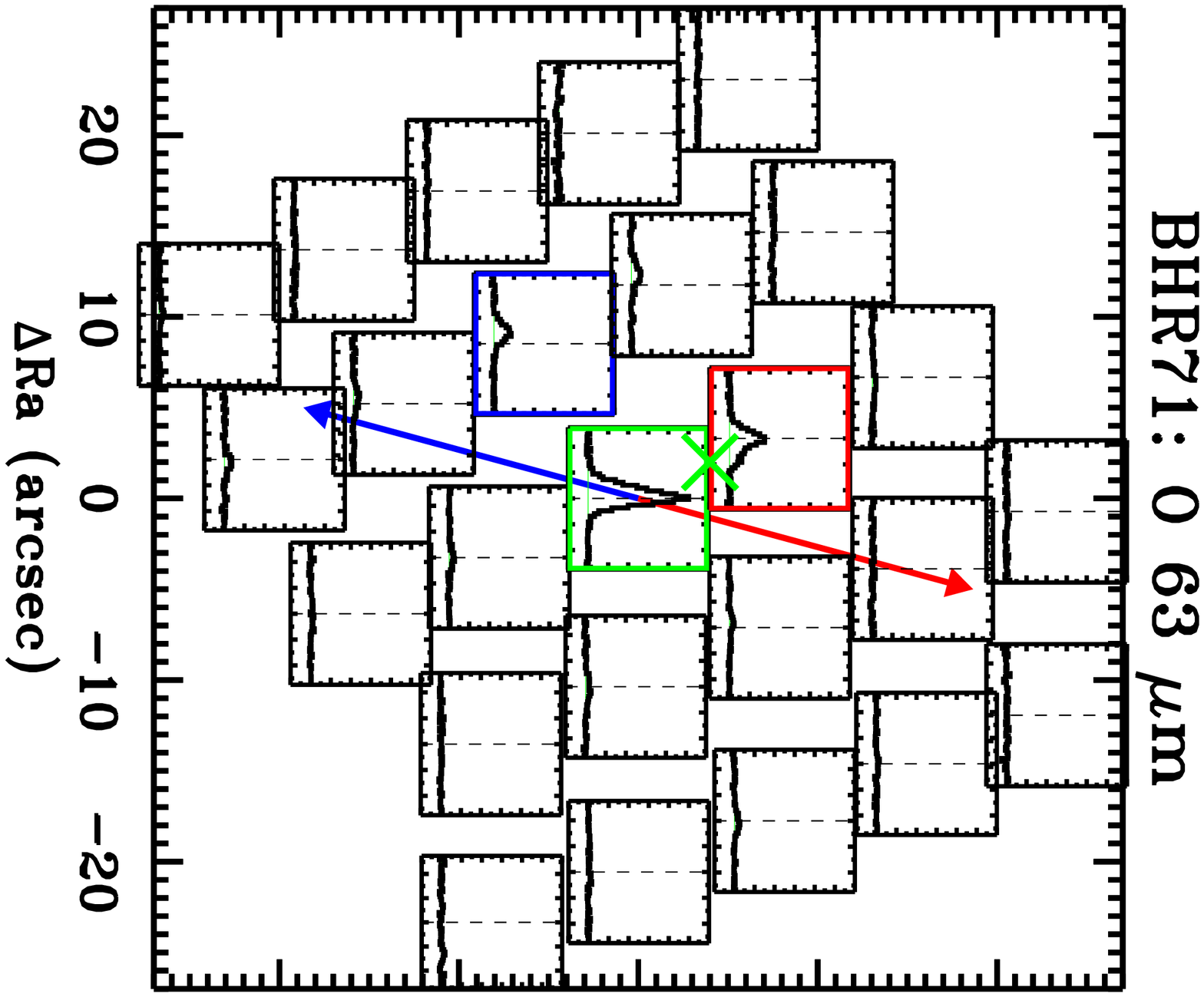}
    \end{center}
  \end{minipage}
        \caption{\label{bhr71map} BHR71 maps in the [\ion{O}{i}] $^3P_{1}-^{3}P_{2}$ line
        at 63.2 $\mu$m, the H$_2$O 2$_{21}$-1$_{10}$ at 108.1 $\mu$m, the CO 14-13 at 
        186.0 $\mu$m and the CO 24-23 at 108.7 $\mu$m.}
\end{figure*}

\begin{figure*}[tb]
  \begin{minipage}[t]{.5\textwidth}
  \begin{center}  
    \includegraphics[angle=90,height=7cm]{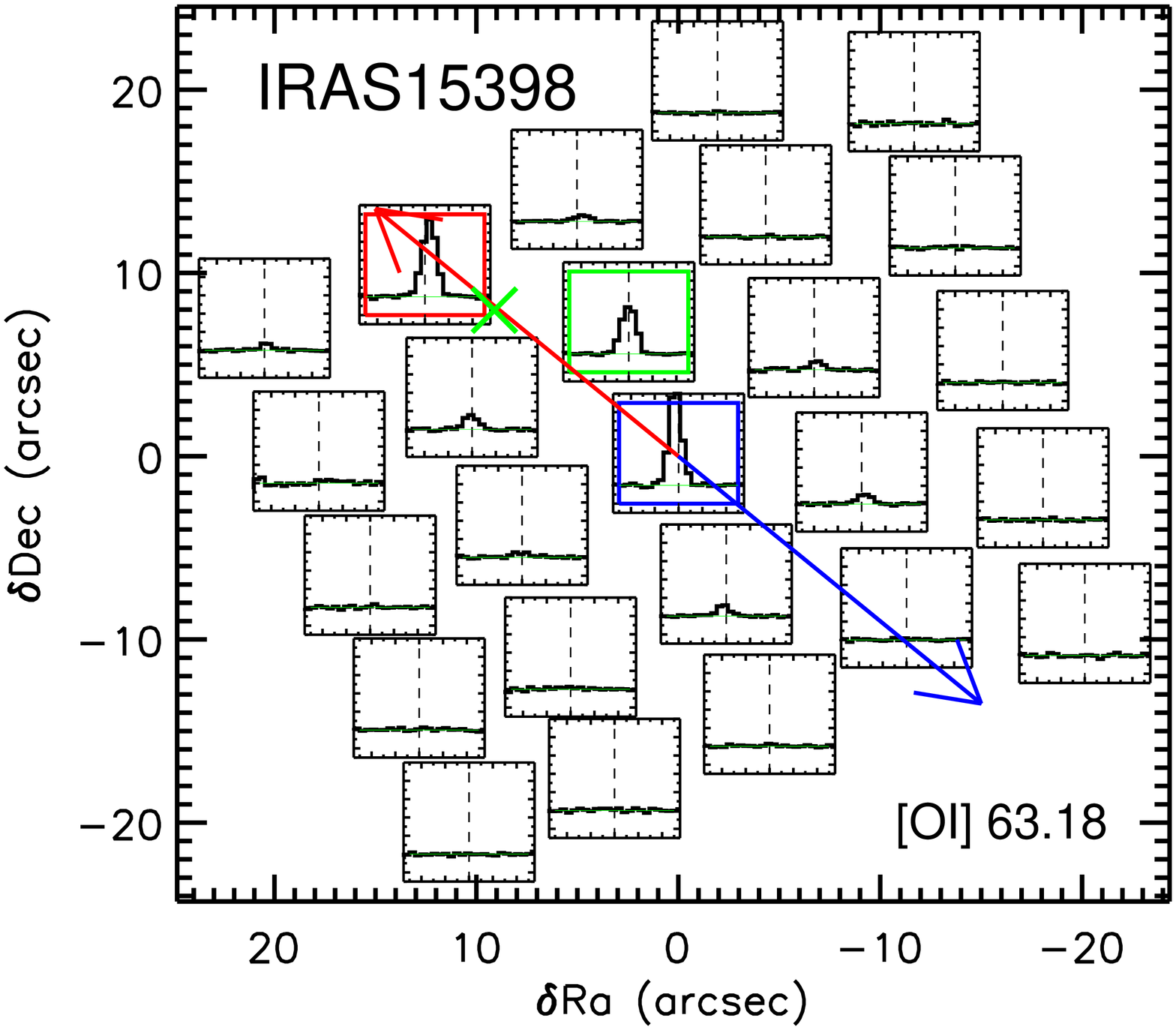}
        % \hspace{+5ex}
                 \vspace{+5ex}
     
     \includegraphics[angle=90,height=7cm]{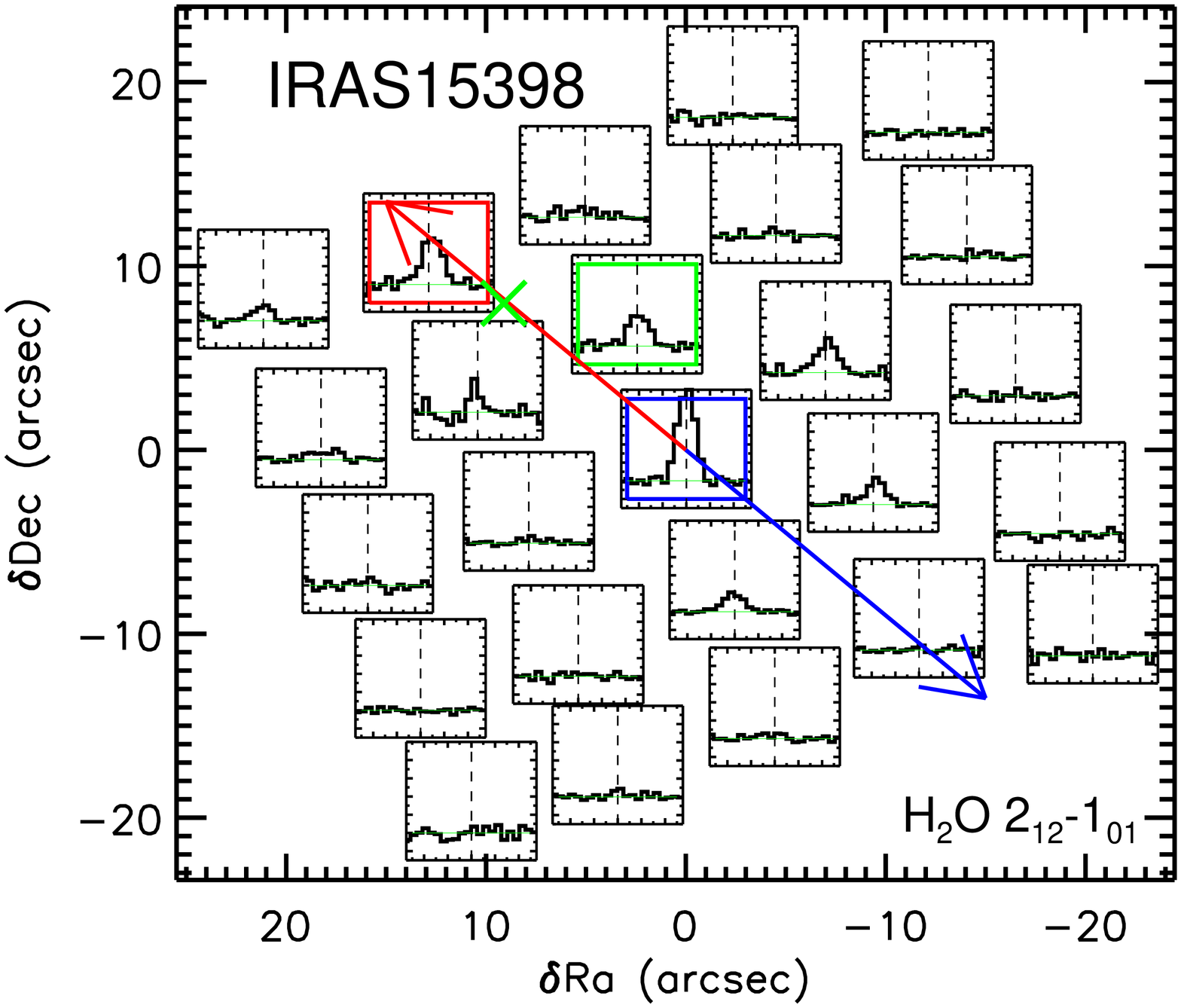}
    \end{center}
  \end{minipage}
  \hfill
  \begin{minipage}[t]{.5\textwidth}
  \begin{center}  
    \includegraphics[angle=90,height=7cm]{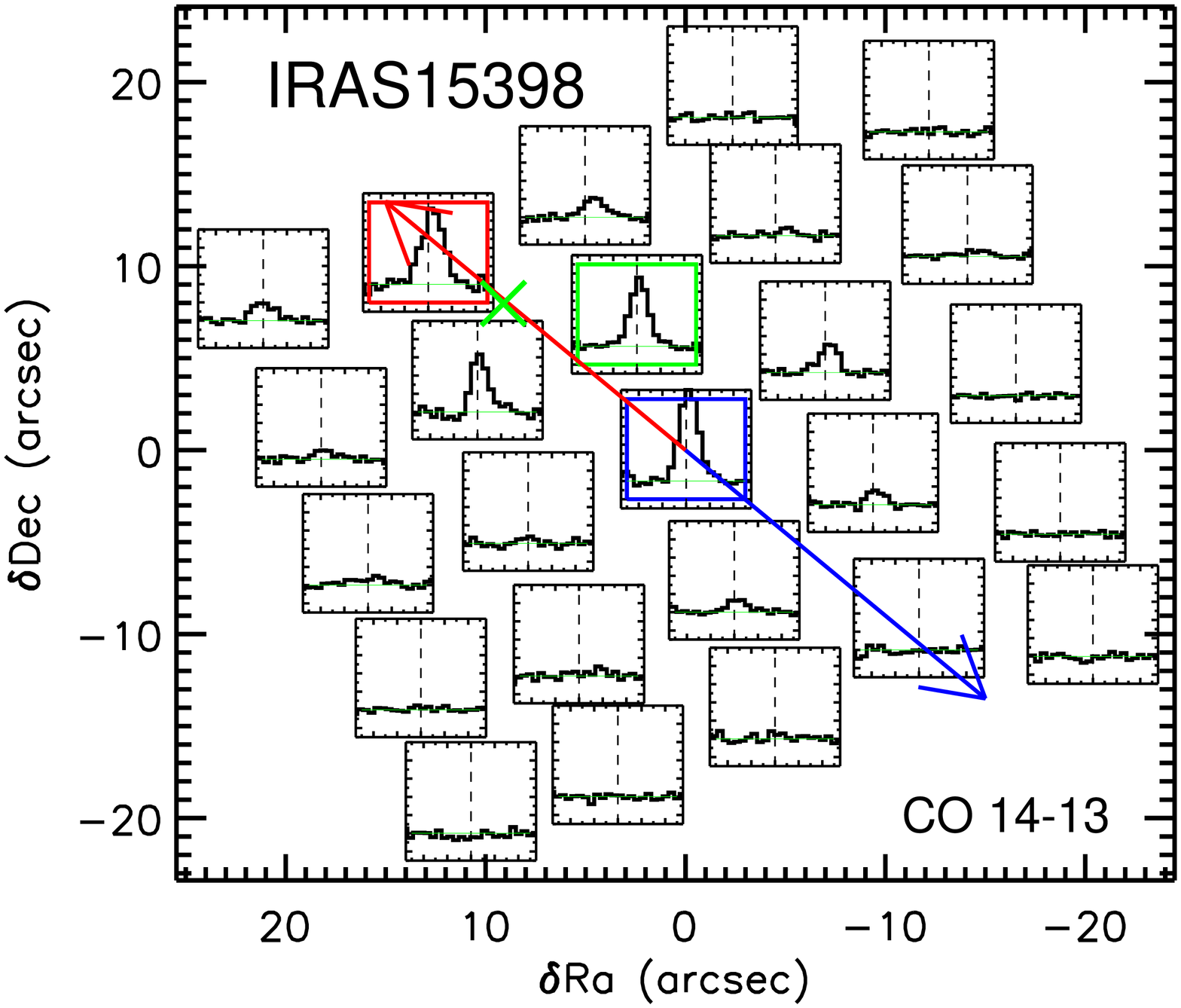}
   %  \hspace{+5ex}
            \vspace{+5ex}
     
     \includegraphics[angle=90,height=7cm]{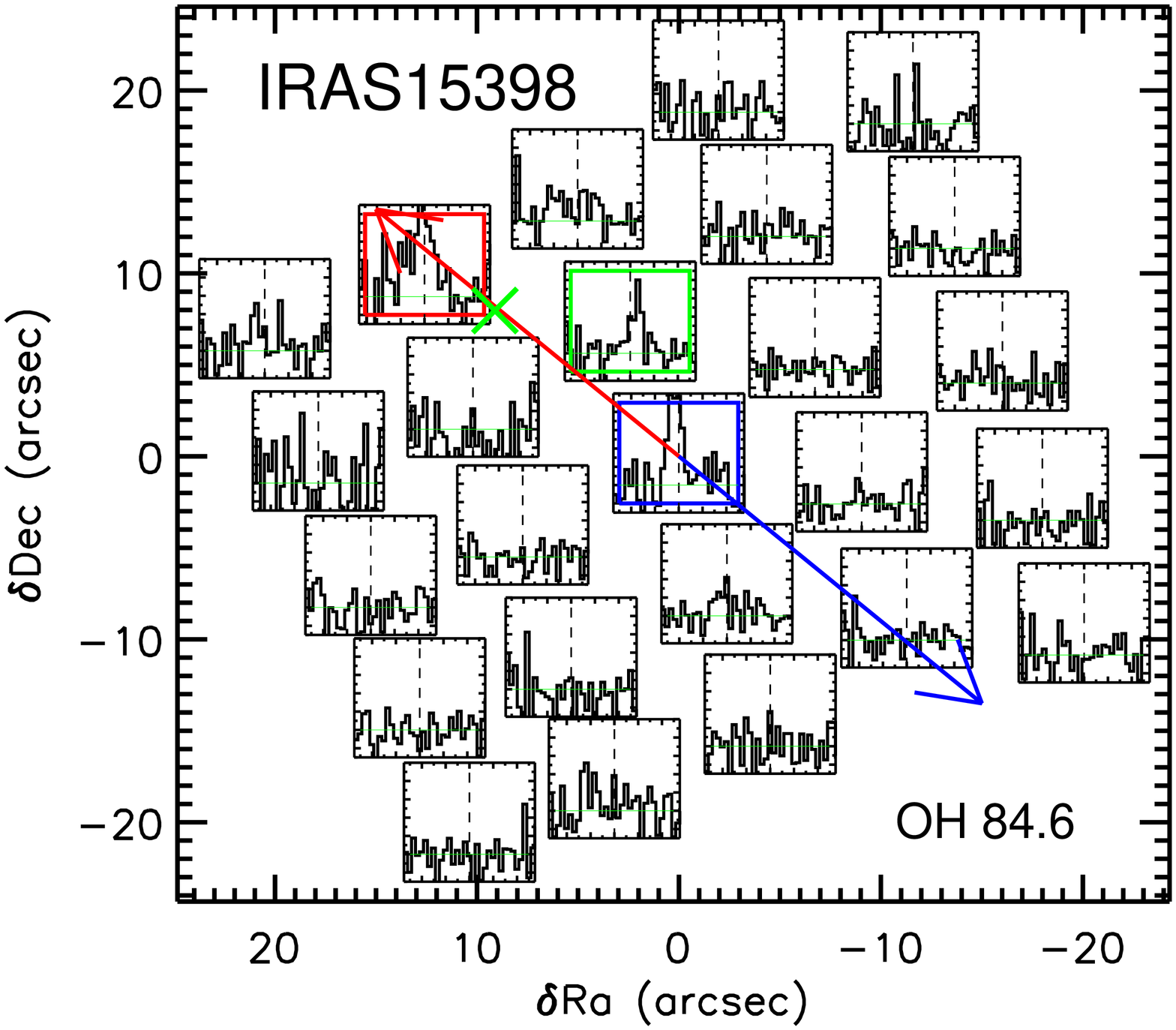}
    \end{center}
  \end{minipage}
 %  \vspace{+3ex}
    \caption{\label{iras15map}IRAS15398 maps in the [\ion{O}{i}] $^3P_{1}-^{3}P_{2}$ line
        at 63.2 $\mu$m, the H$_2$O 2$_{12}$-1$_{01}$ line at 179.5 $\mu$m, the 
        CO 14-13 at 186.0 $\mu$m and the OH $^{2}\Pi_{\nicefrac{3}{2}}$
        $J=\nicefrac{7}{2}-\nicefrac{5}{2}$ line at 84.6 $\mu$m.}
\end{figure*}

\begin{figure*}[!tb]
  \begin{minipage}[t]{.5\textwidth}
  \begin{center}  
      \includegraphics[angle=90,height=9cm]{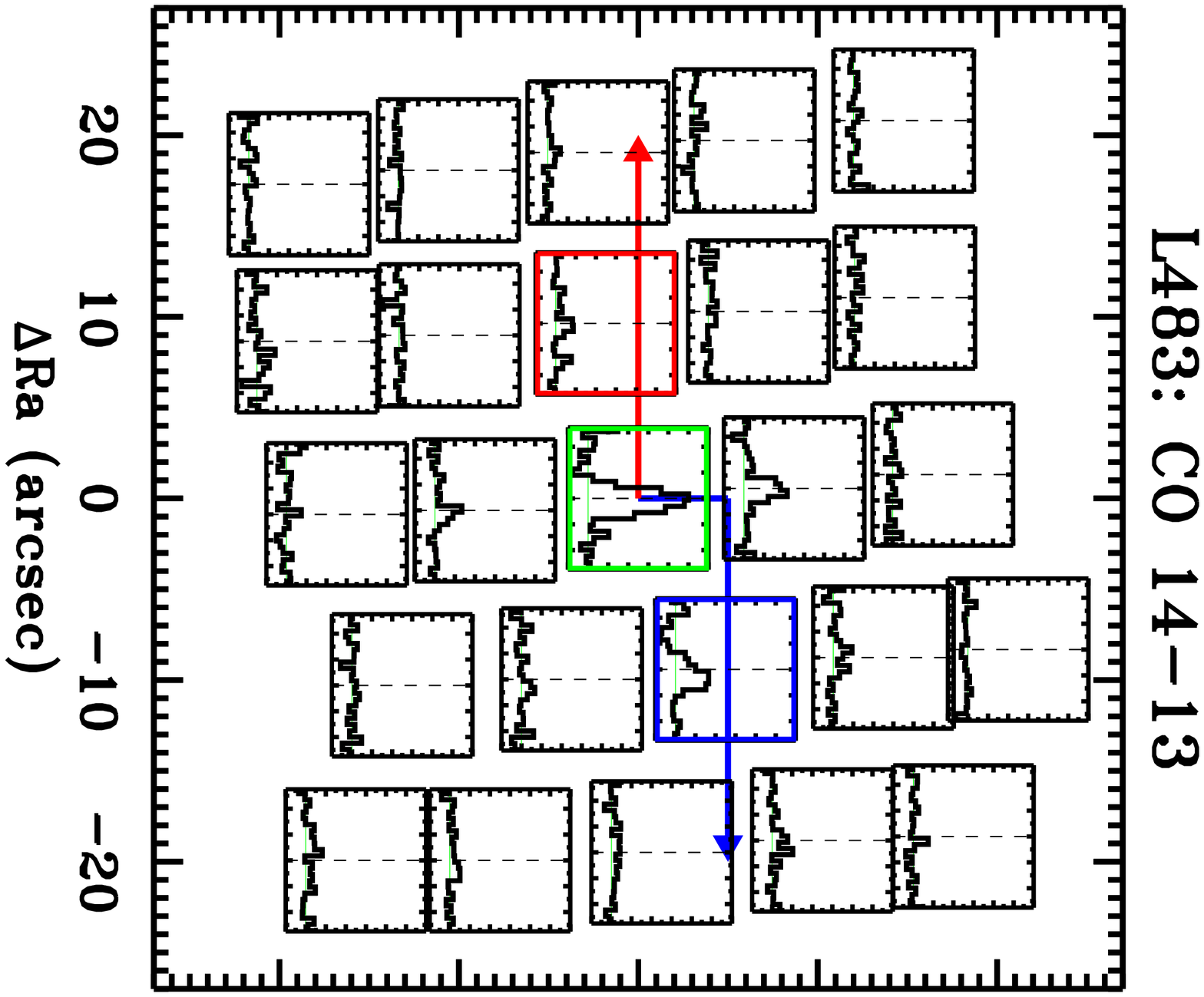}
               \vspace{+3ex}
       
      \includegraphics[angle=90,height=9cm]{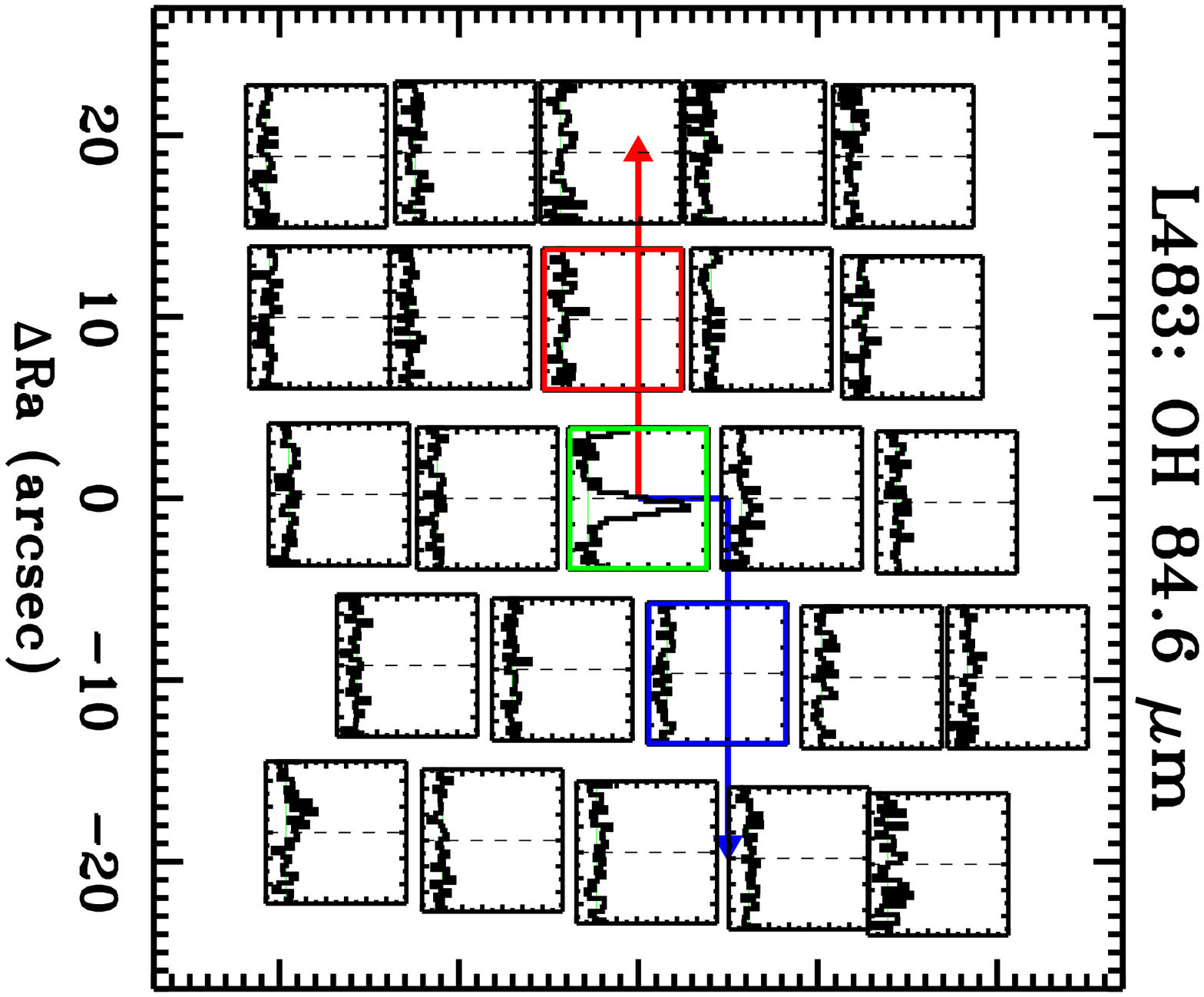}
    \end{center}
  \end{minipage}
  \hfill
  \begin{minipage}[t]{.5\textwidth}
  \begin{center}         
      \includegraphics[angle=90,height=9cm]{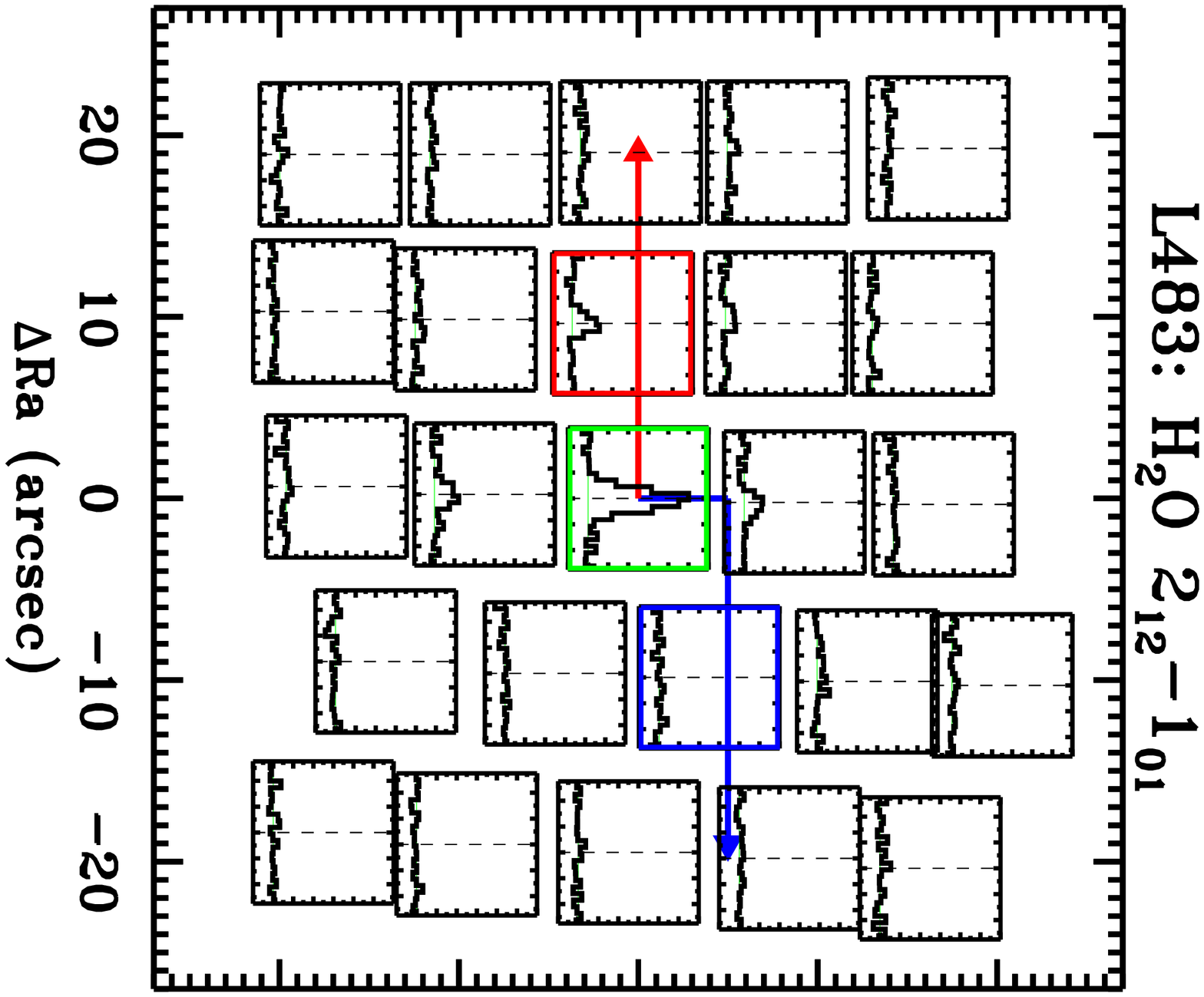}
               \vspace{+3ex}
       
    \includegraphics[angle=90,height=9cm]{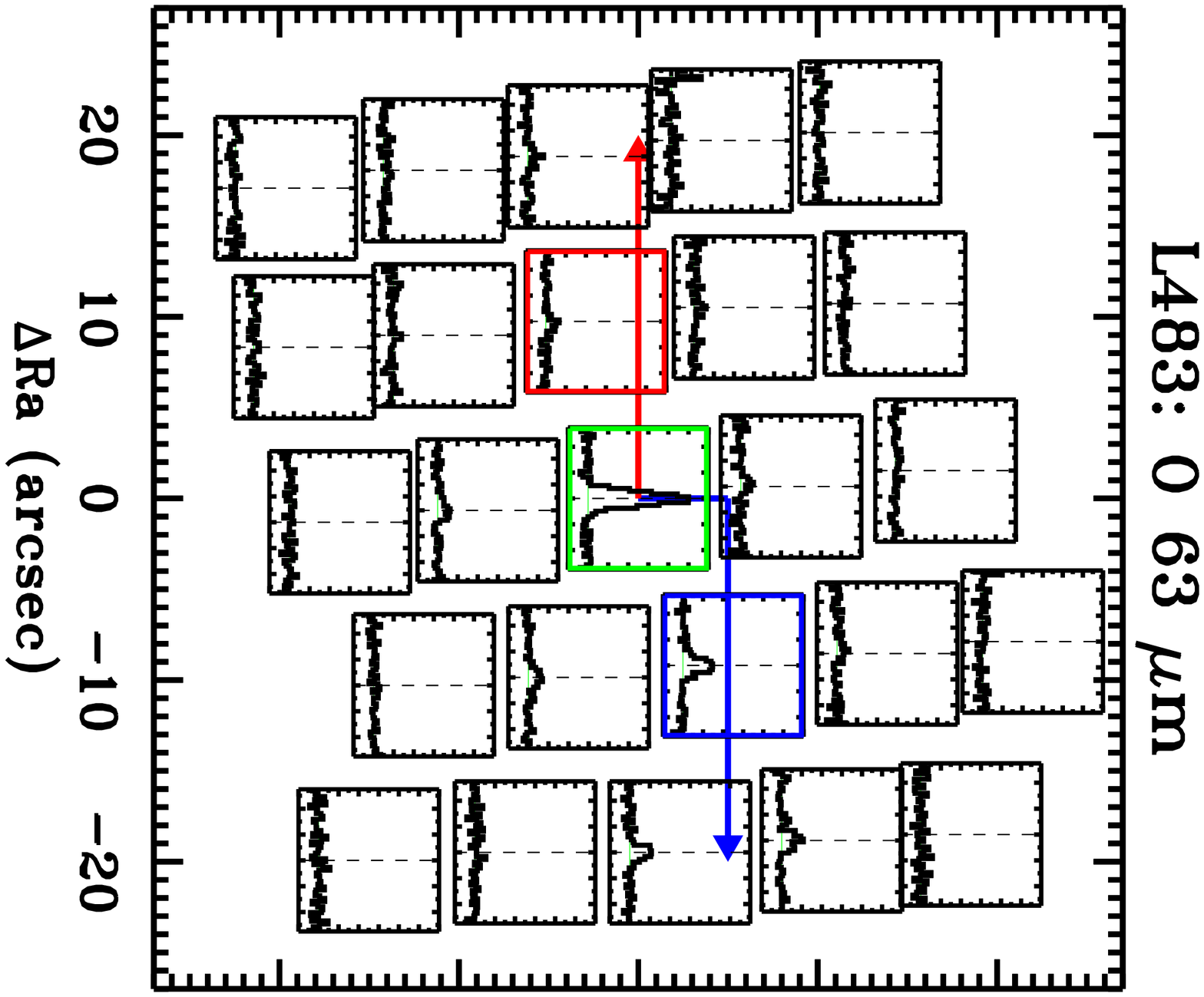}
    \end{center}
  \end{minipage}
 %  \vspace{+3ex}
        \caption{\label{l483map} L483 maps in the [\ion{O}{i}] $^3P_{1}-^{3}P_{2}$ line
        at 63.2 $\mu$m, the H$_2$O 2$_{12}$-1$_{01}$ line at 179.5 $\mu$m, the 
        CO 14-13 at 186.0 $\mu$m and the OH $^{2}\Pi_{\nicefrac{3}{2}}$
        $J=\nicefrac{7}{2}-\nicefrac{5}{2}$ line at 84.6 $\mu$m.}
\end{figure*}

\begin{figure*}[!tb]
  \begin{minipage}[t]{.5\textwidth}
  \begin{center}  
      \includegraphics[angle=90,height=9cm]{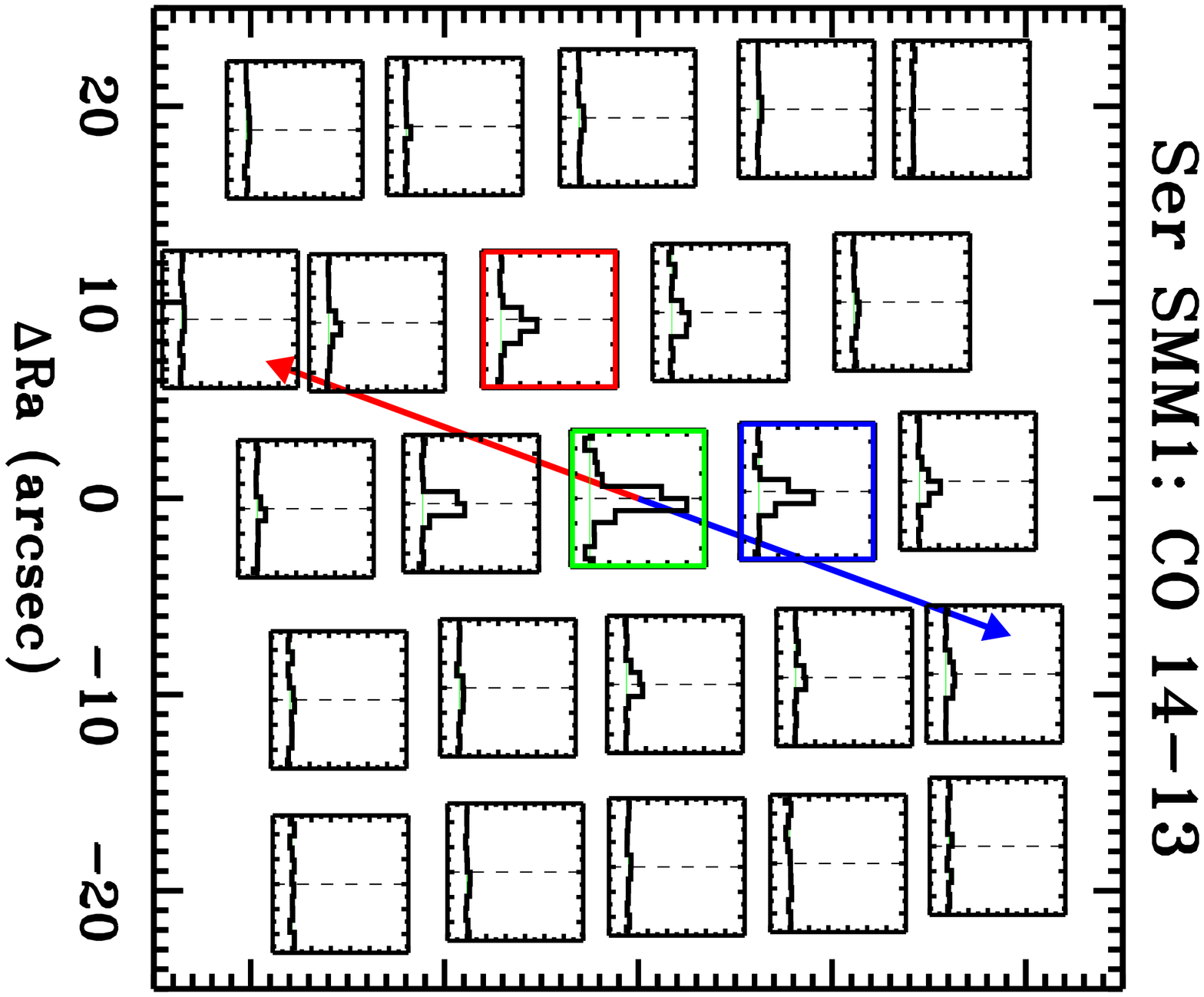}
               \vspace{+3ex}
       
      \includegraphics[angle=90,height=9cm]{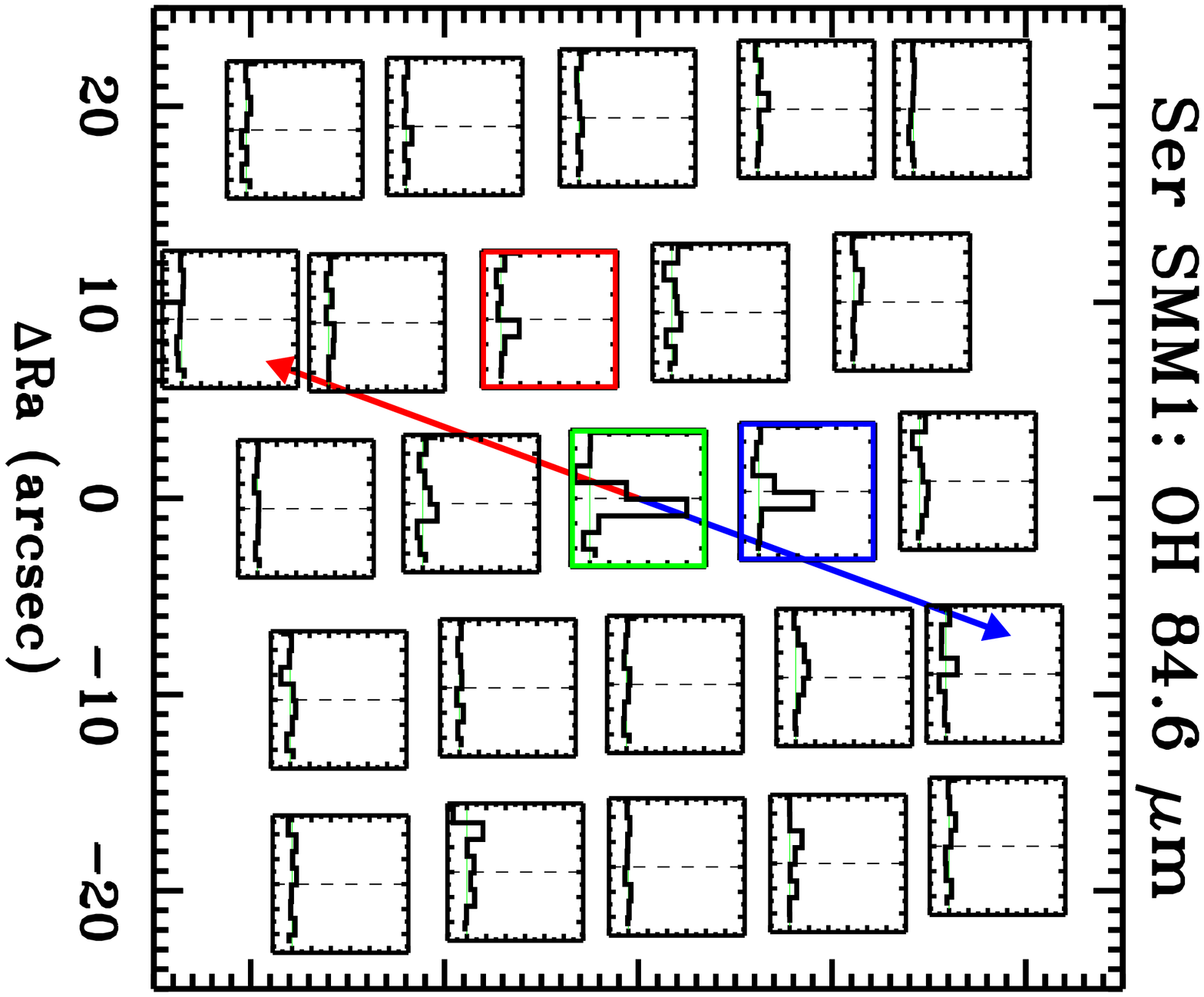}
    \end{center}
  \end{minipage}
  \hfill
  \begin{minipage}[t]{.5\textwidth}
  \begin{center}         
      \includegraphics[angle=90,height=9cm]{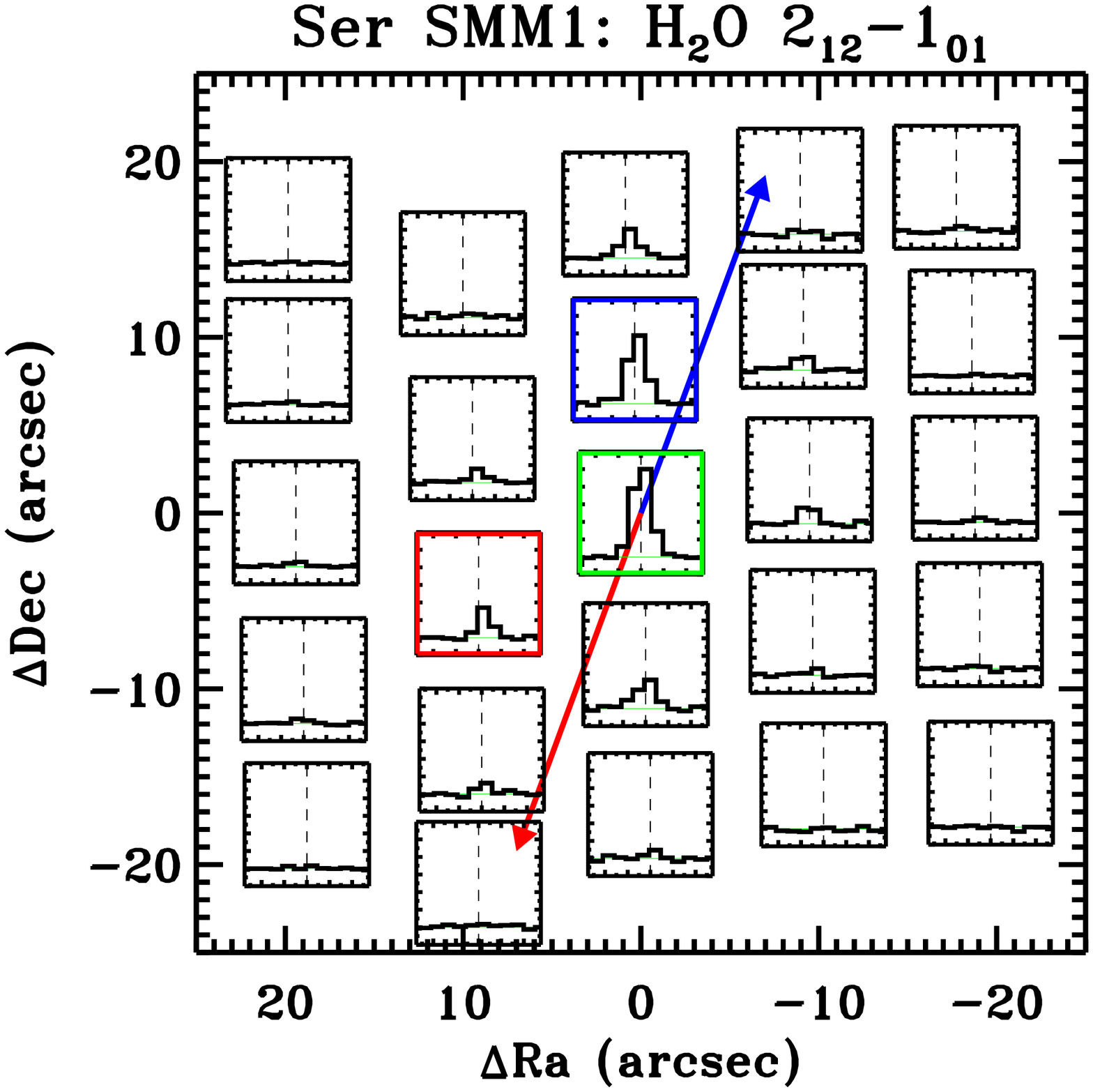}
               \vspace{+3ex}
       
    \includegraphics[angle=90,height=9cm]{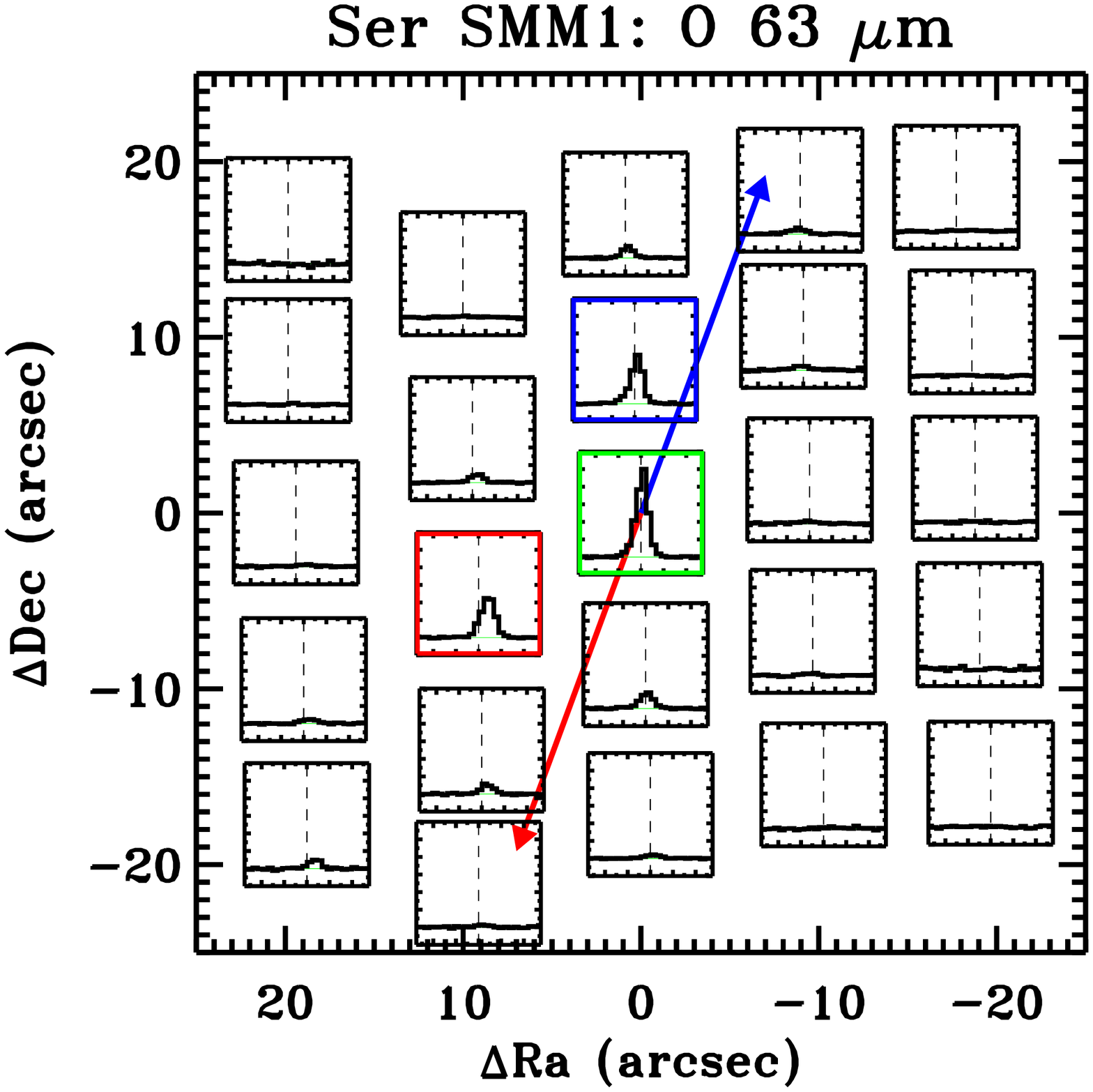}
    \end{center}
  \end{minipage}
        \caption{\label{smm1map} Ser SMM1 maps in the [\ion{O}{i}] $^3P_{1}-^{3}P_{2}$ line
        at 63.2 $\mu$m, the H$_2$O 2$_{12}$-1$_{01}$ line at 179.5 $\mu$m, the 
        CO 14-13 at 186.0 $\mu$m and the OH $^{2}\Pi_{\nicefrac{3}{2}}$
        $J=\nicefrac{7}{2}-\nicefrac{5}{2}$ line at 84.6 $\mu$m.}
\end{figure*}

\begin{figure*}[!tb]
  \begin{minipage}[t]{.5\textwidth}
  \begin{center}  
      \includegraphics[angle=90,height=9cm]{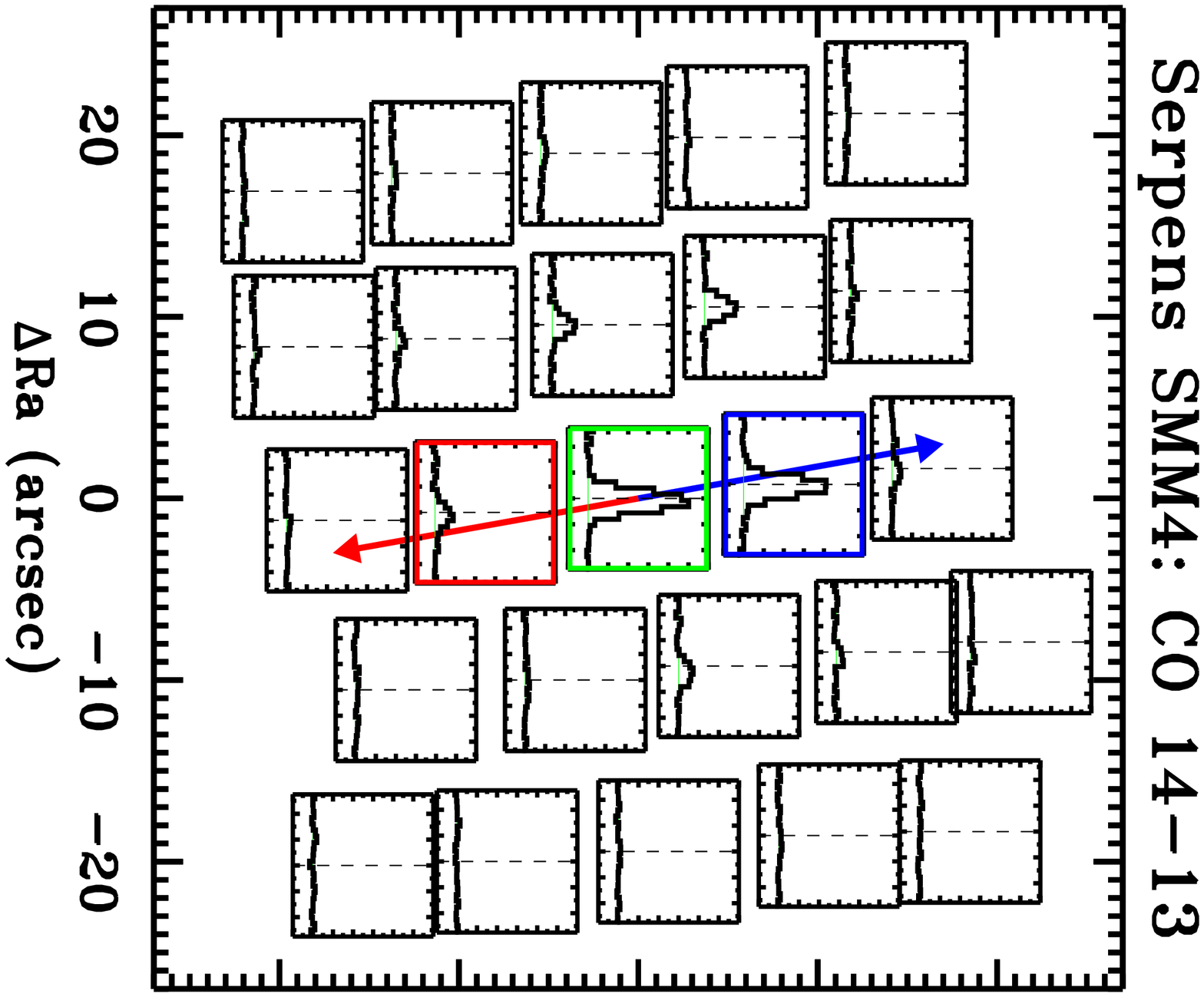}
               \vspace{+3ex}
       
     \includegraphics[angle=90,height=9cm]{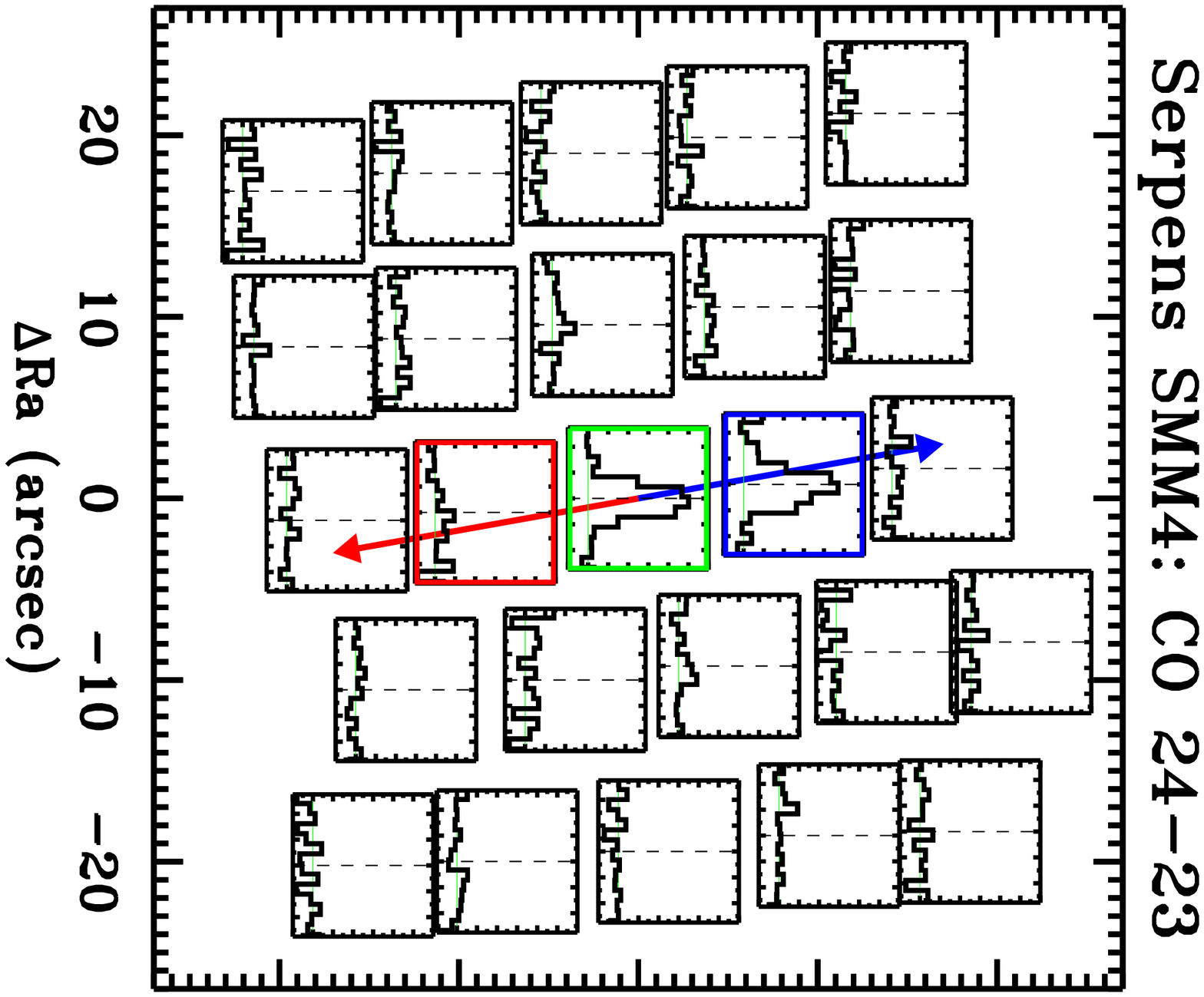}
    \end{center}
  \end{minipage}
  \hfill
  \begin{minipage}[t]{.5\textwidth}
  \begin{center}         
    \includegraphics[angle=90,height=9cm]{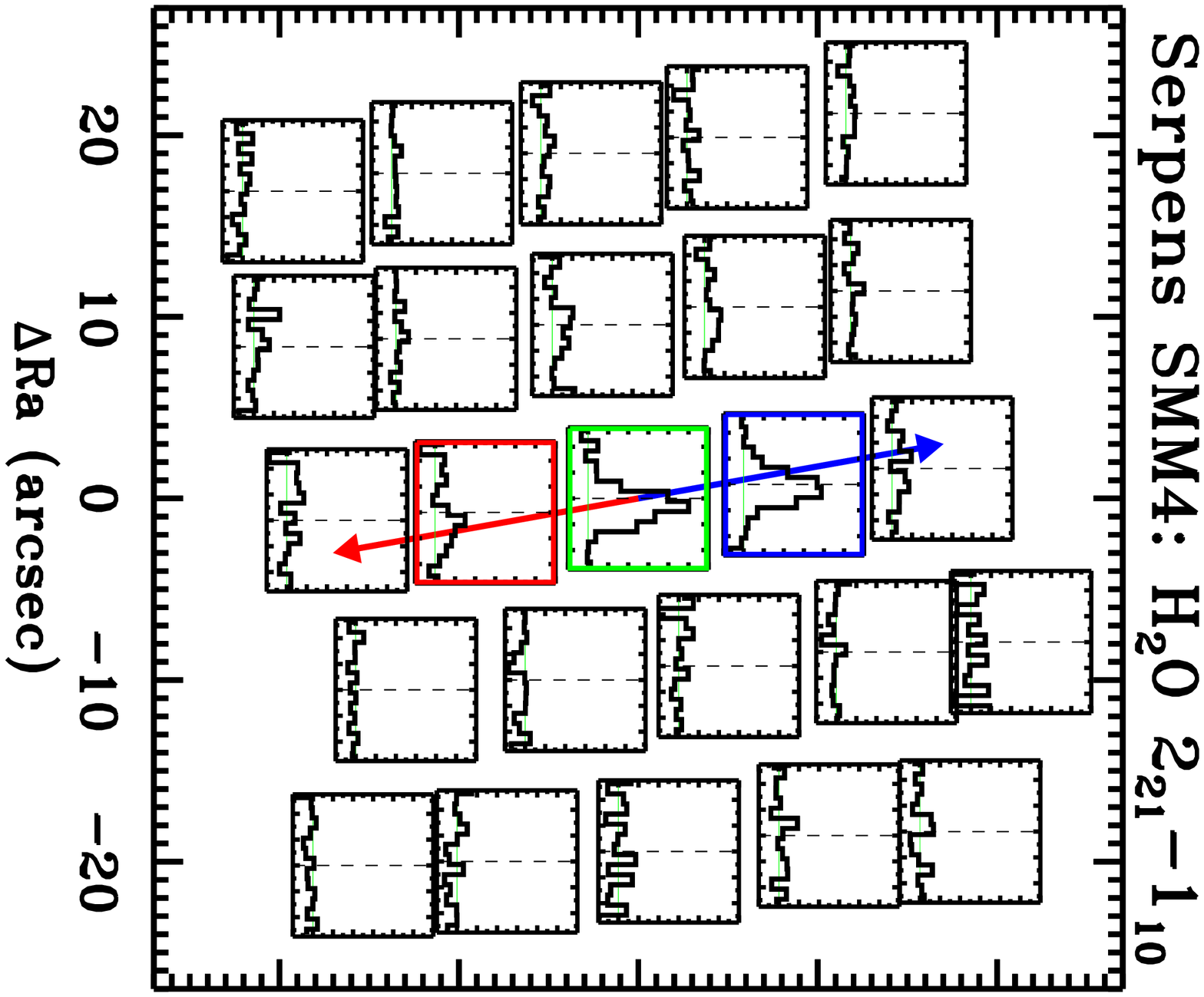}
             \vspace{+3ex}
       
    \includegraphics[angle=90,height=9cm]{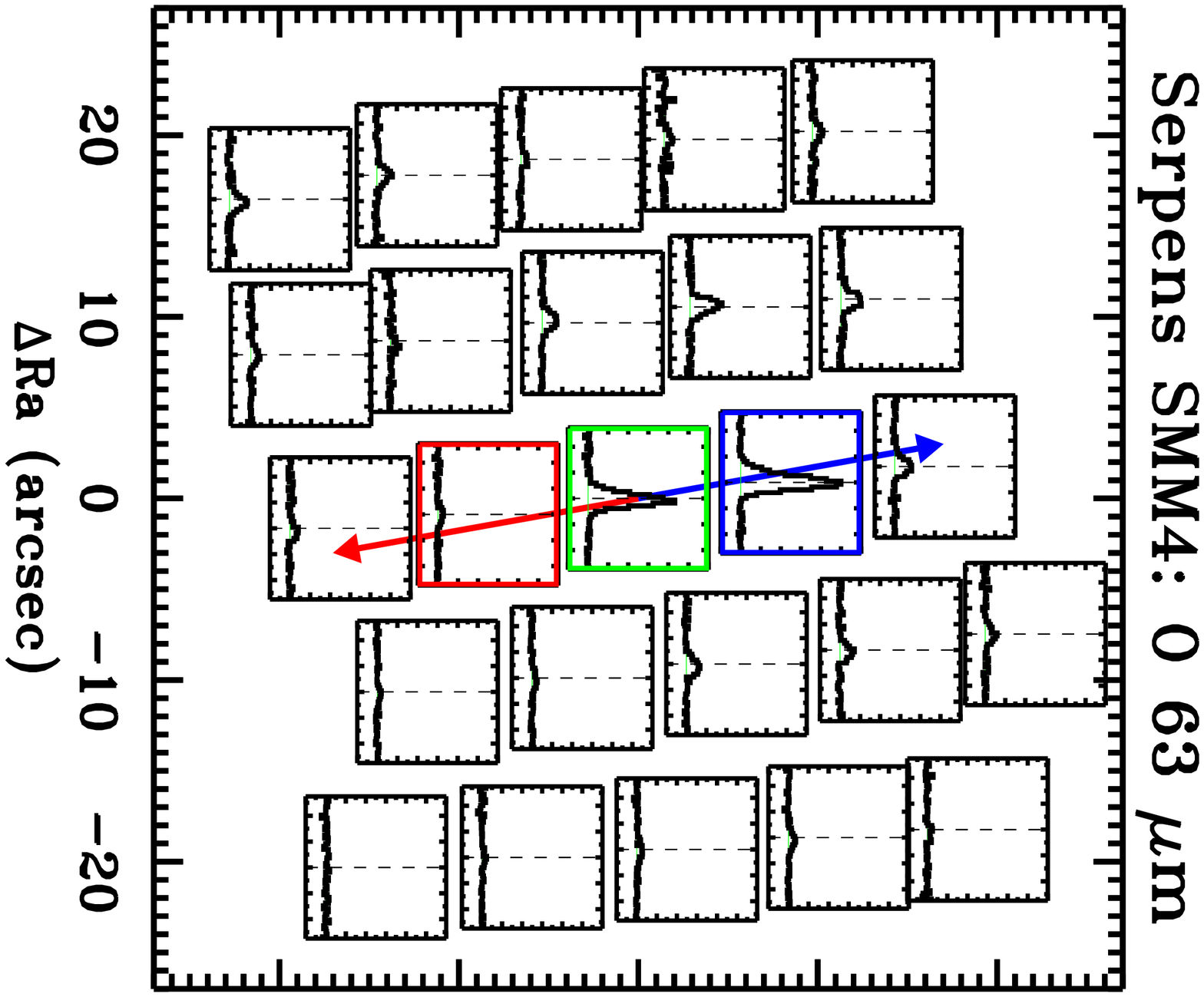}
    \end{center}
  \end{minipage}
 %  \vspace{+3ex}
        \caption{\label{smm4map} Ser SMM4 maps in the [\ion{O}{i}] $^3P_{1}-^{3}P_{2}$ line
        at 63.2 $\mu$m, the H$_2$O 2$_{21}$-1$_{10}$ at 108.1 $\mu$m, the CO 14-13 at 
        186.0 $\mu$m and the CO 24-23 at 108.7 $\mu$m.}
\end{figure*}

\begin{figure*}[!tb]
  \begin{minipage}[t]{.5\textwidth}
  \begin{center}  
      \includegraphics[angle=90,height=9cm]{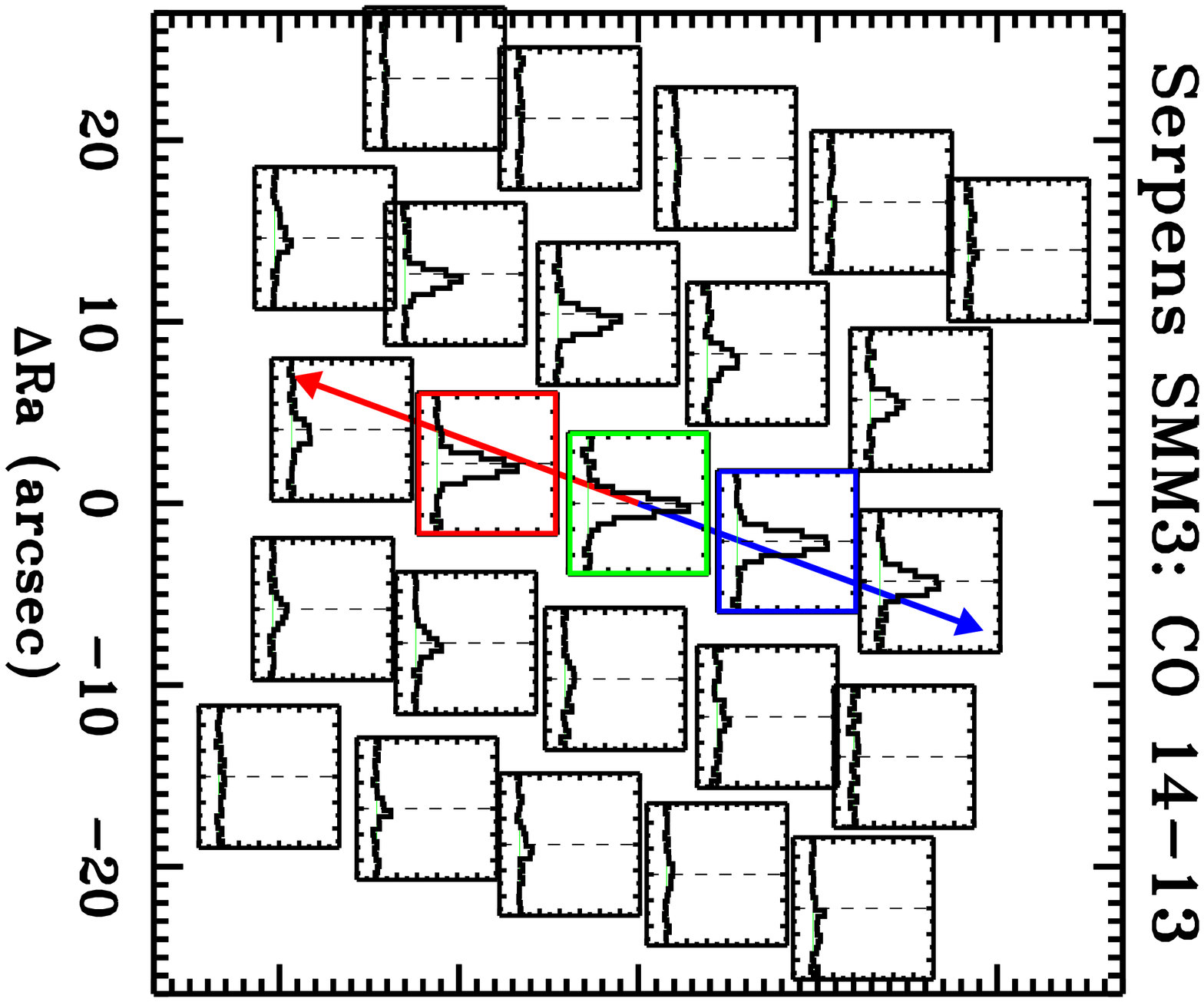}
               \vspace{+3ex}
       
      \includegraphics[angle=90,height=9cm]{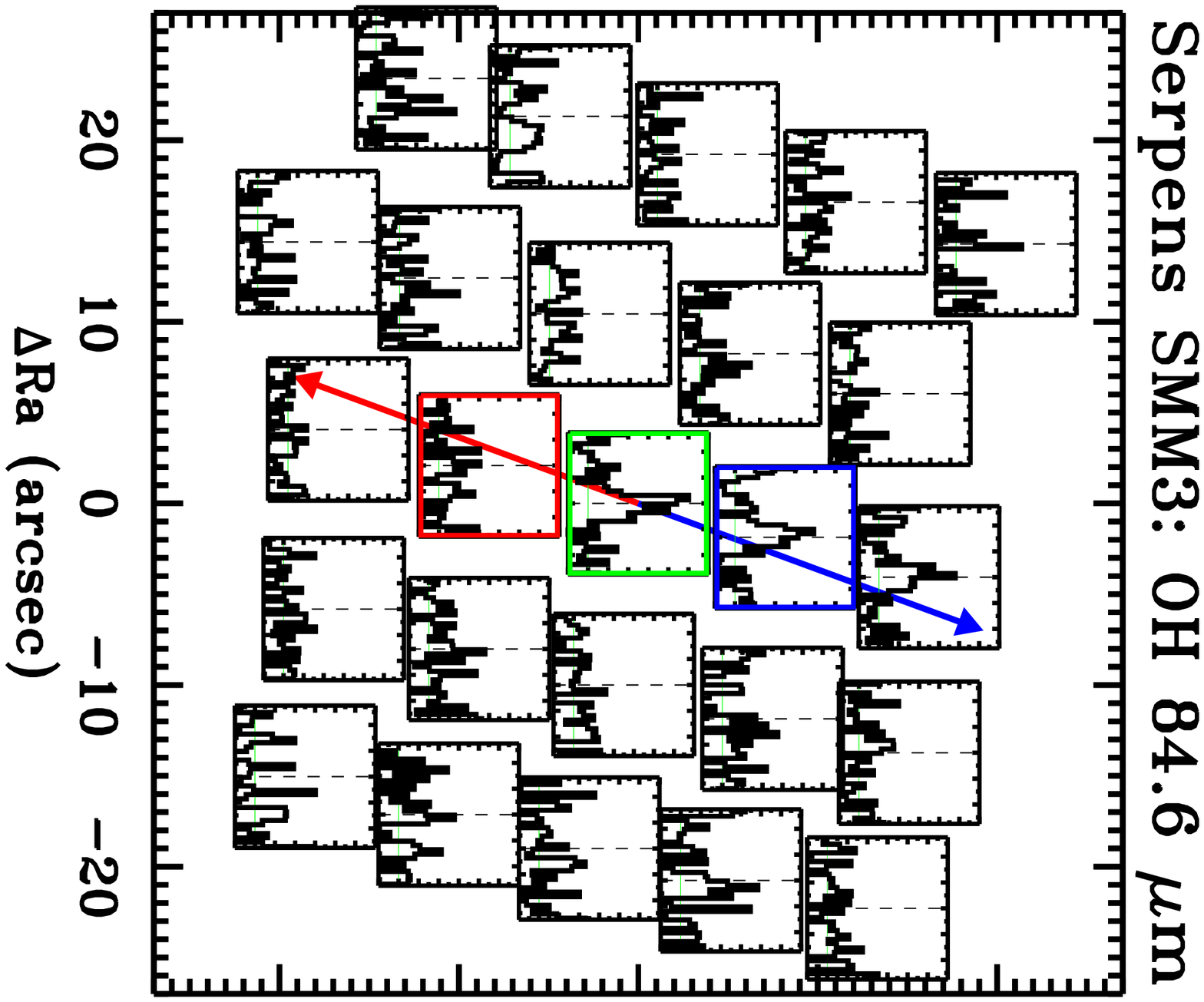}
    \end{center}
  \end{minipage}
  \hfill
  \begin{minipage}[t]{.5\textwidth}
  \begin{center}         
      \includegraphics[angle=90,height=9cm]{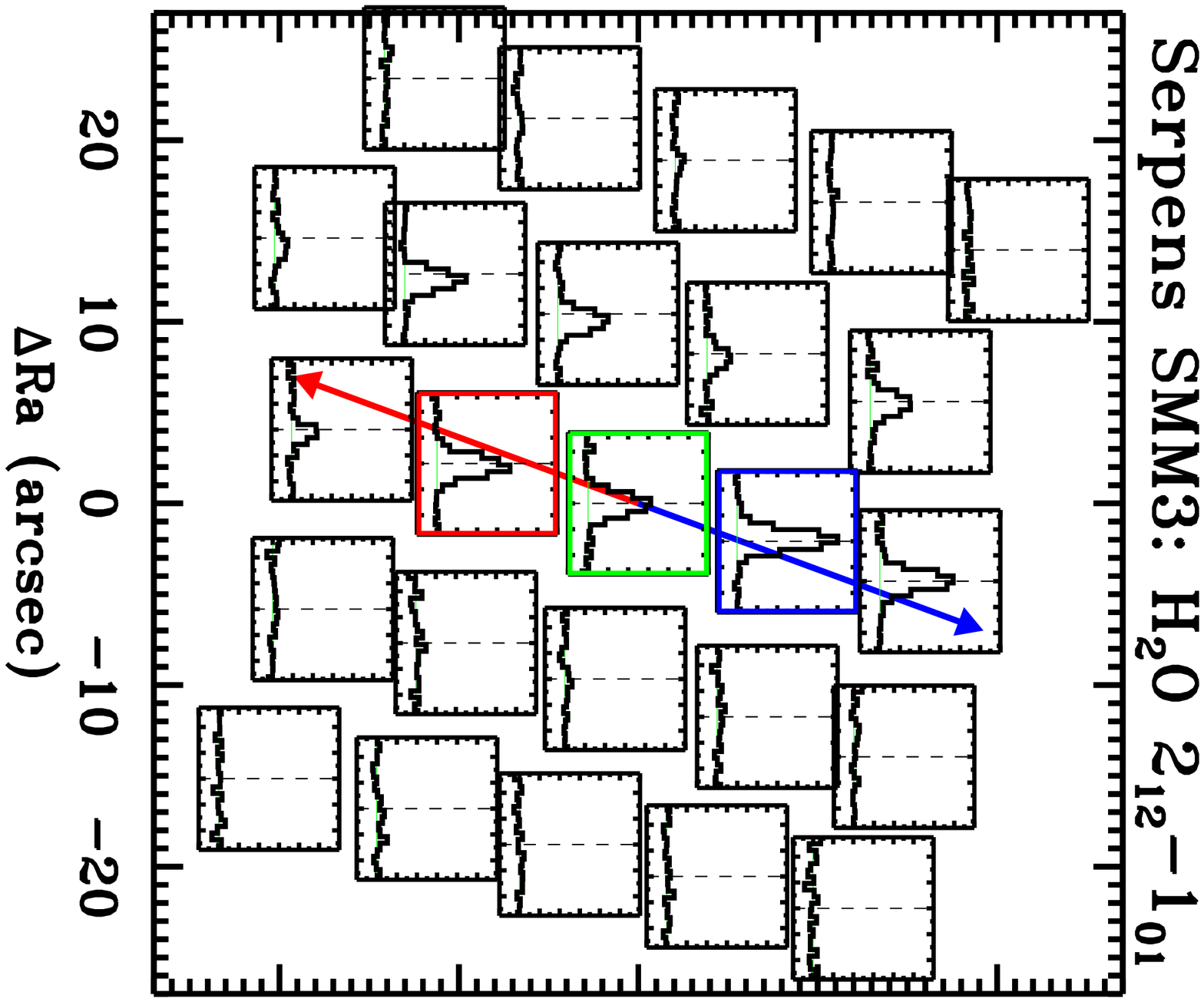}
               \vspace{+3ex}
      
    \includegraphics[angle=90,height=9cm]{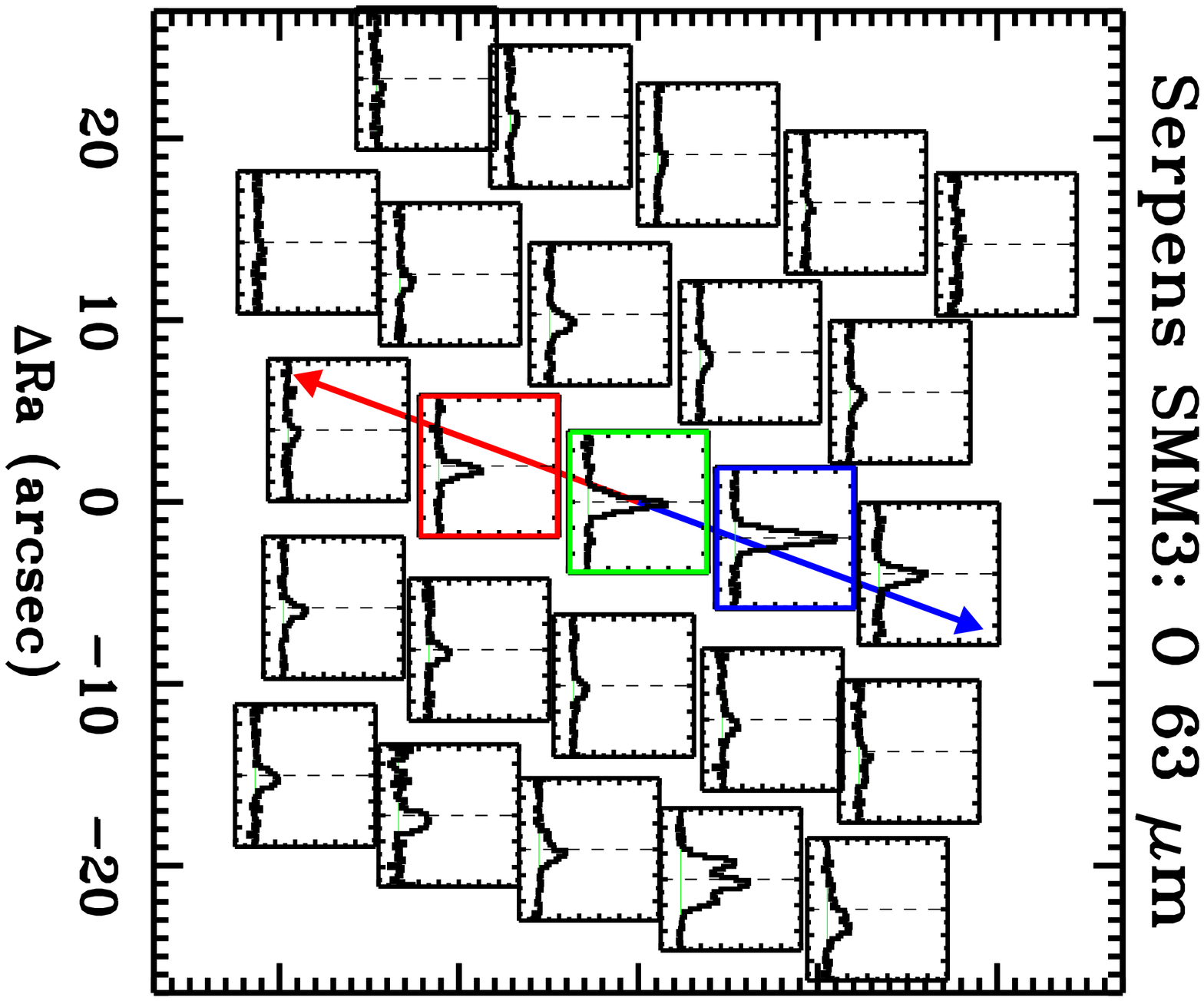}
    \end{center}
  \end{minipage}
  %  \vspace{+3ex}
        \caption{\label{smm3map} Ser SMM3 maps in the [\ion{O}{i}] $^3P_{1}-^{3}P_{2}$ line
        at 63.2 $\mu$m, the H$_2$O 2$_{12}$-1$_{01}$ line at 179.5 $\mu$m, the 
        CO 14-13 at 186.0 $\mu$m and the OH $^{2}\Pi_{\nicefrac{3}{2}}$
        $J=\nicefrac{7}{2}-\nicefrac{5}{2}$ line at 84.6 $\mu$m.}
\end{figure*}

\begin{figure*}[tb]
  \begin{minipage}[t]{.5\textwidth}
  \begin{center}  
    \includegraphics[angle=90,height=7cm]{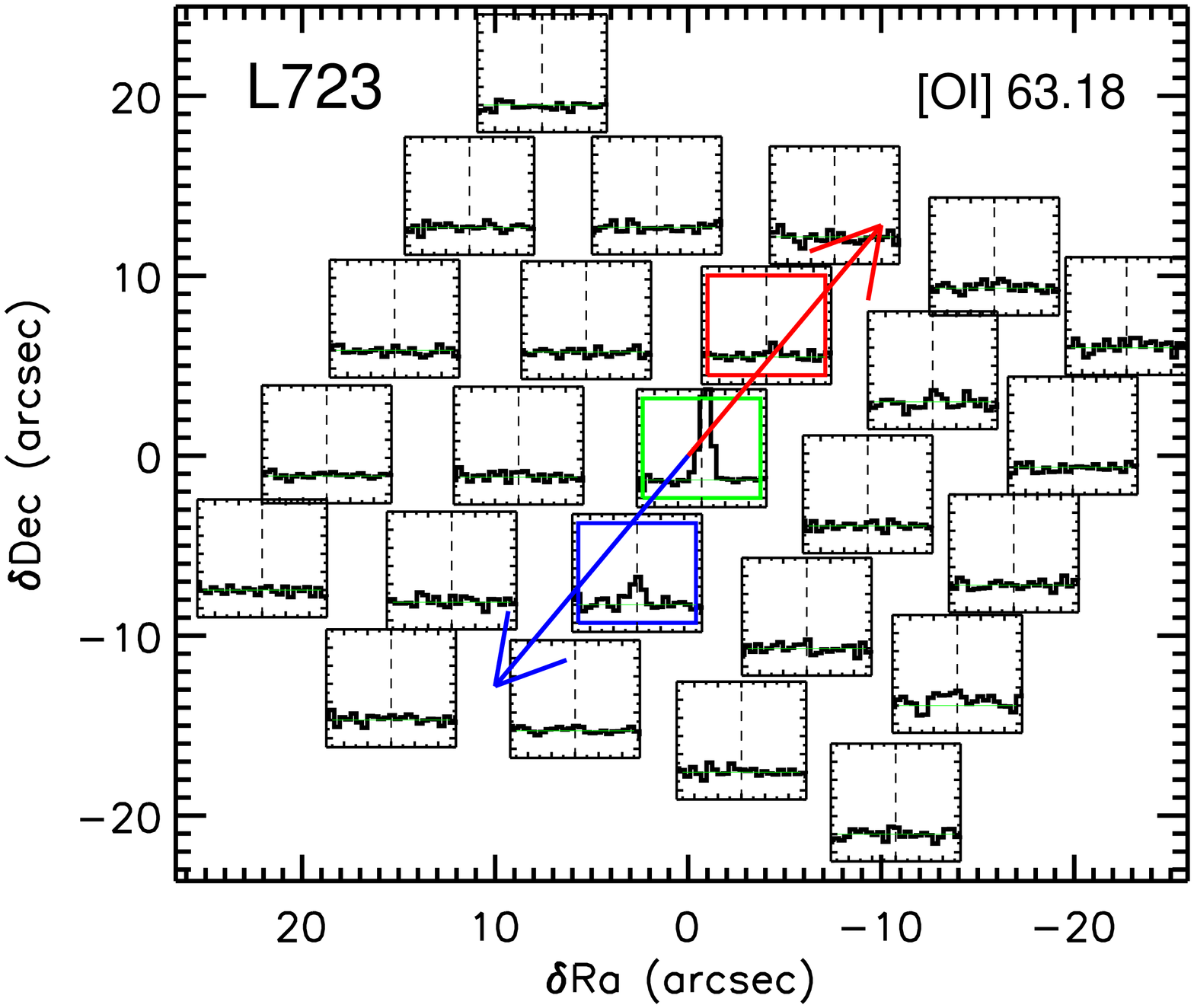}
         \vspace{+5ex}
     
     \includegraphics[angle=90,height=7cm]{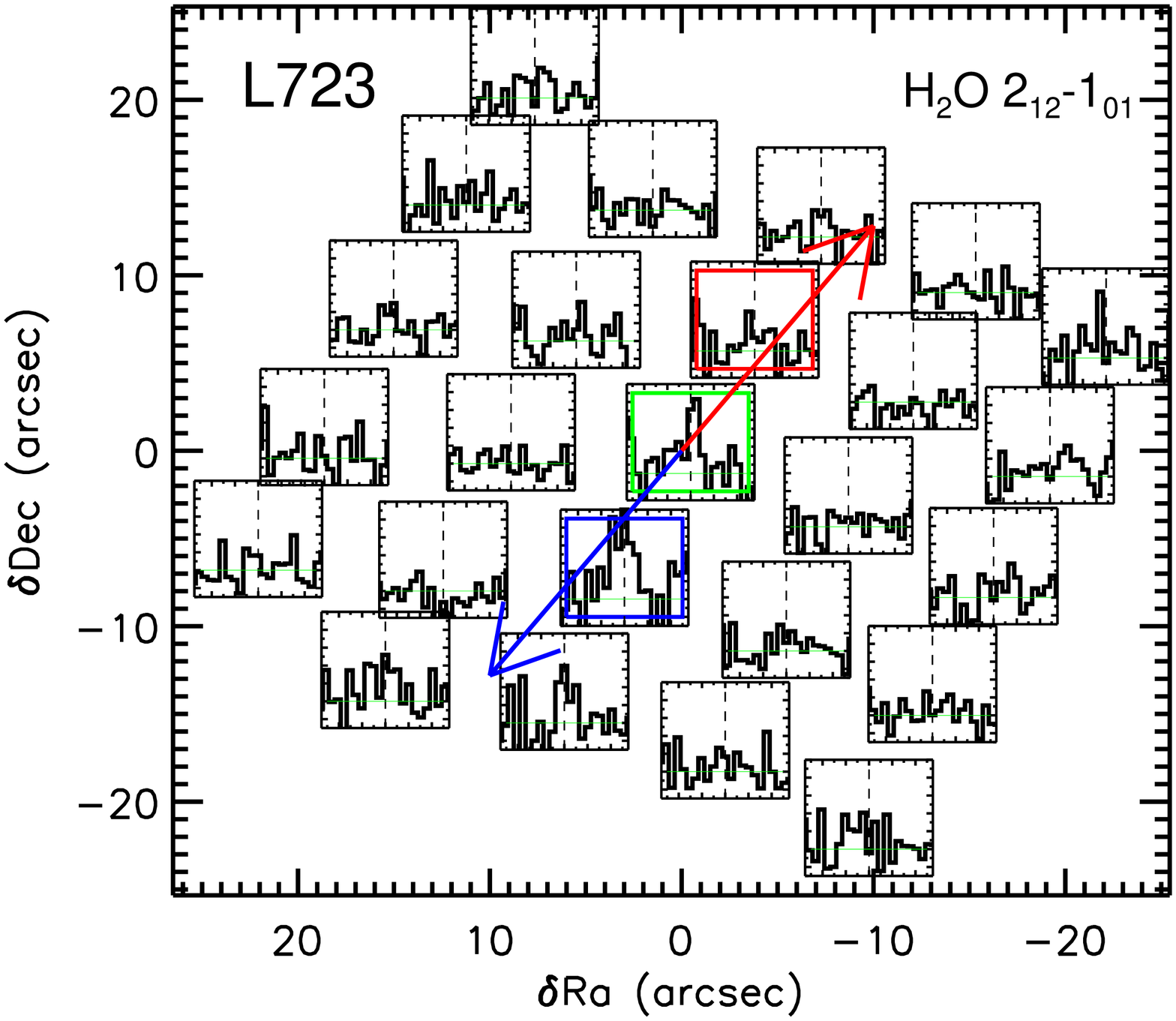}
    \end{center}
  \end{minipage}
  \hfill
  \begin{minipage}[t]{.5\textwidth}
  \begin{center}  
    \includegraphics[angle=90,height=7cm]{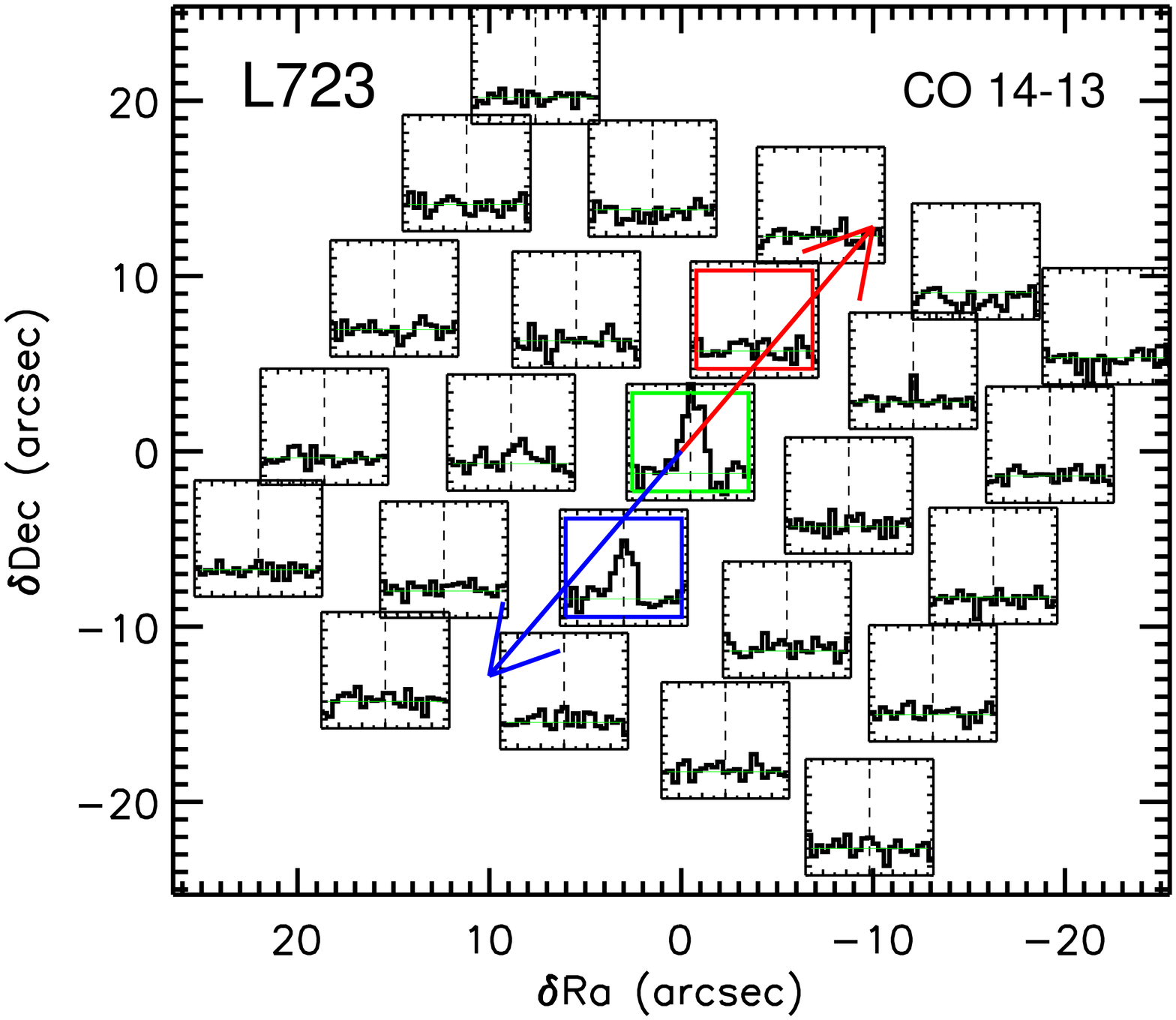}
         \vspace{+5ex}
     
     \includegraphics[angle=90,height=7cm]{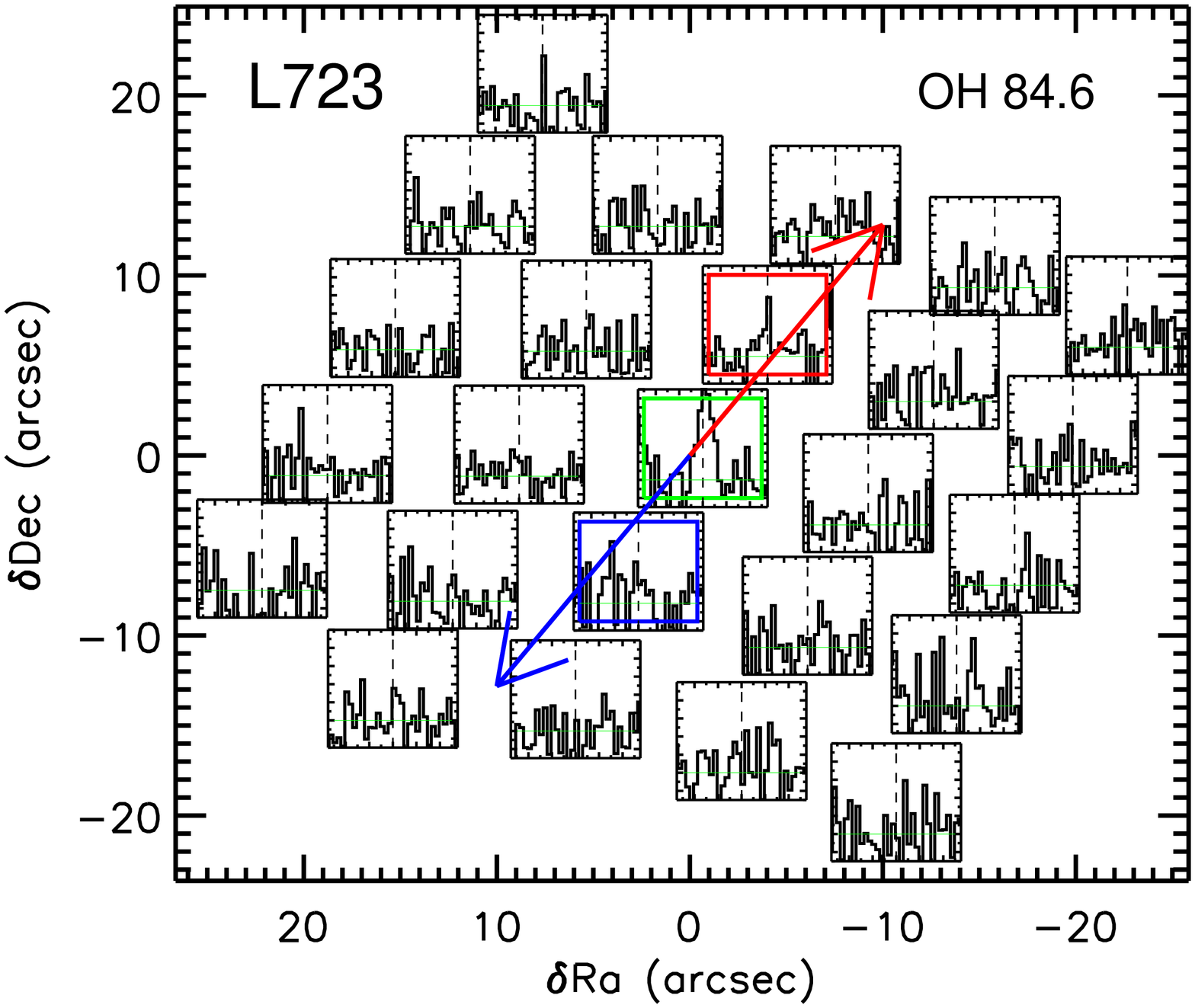}
    \end{center}
  \end{minipage}
 %  \vspace{+3ex}
    \caption{\label{l723map} L723 maps in the [\ion{O}{i}] $^3P_{1}-^{3}P_{2}$ line
        at 63.2 $\mu$m, the H$_2$O 2$_{12}$-1$_{01}$ line at 179.5 $\mu$m, the 
        CO 14-13 at 186.0 $\mu$m and the OH $^{2}\Pi_{\nicefrac{3}{2}}$
        $J=\nicefrac{7}{2}-\nicefrac{5}{2}$ line at 84.6 $\mu$m.}
\end{figure*}

\begin{figure*}[!tb]
  \begin{minipage}[t]{.5\textwidth}
  \begin{center}  
      \includegraphics[angle=90,height=9cm]{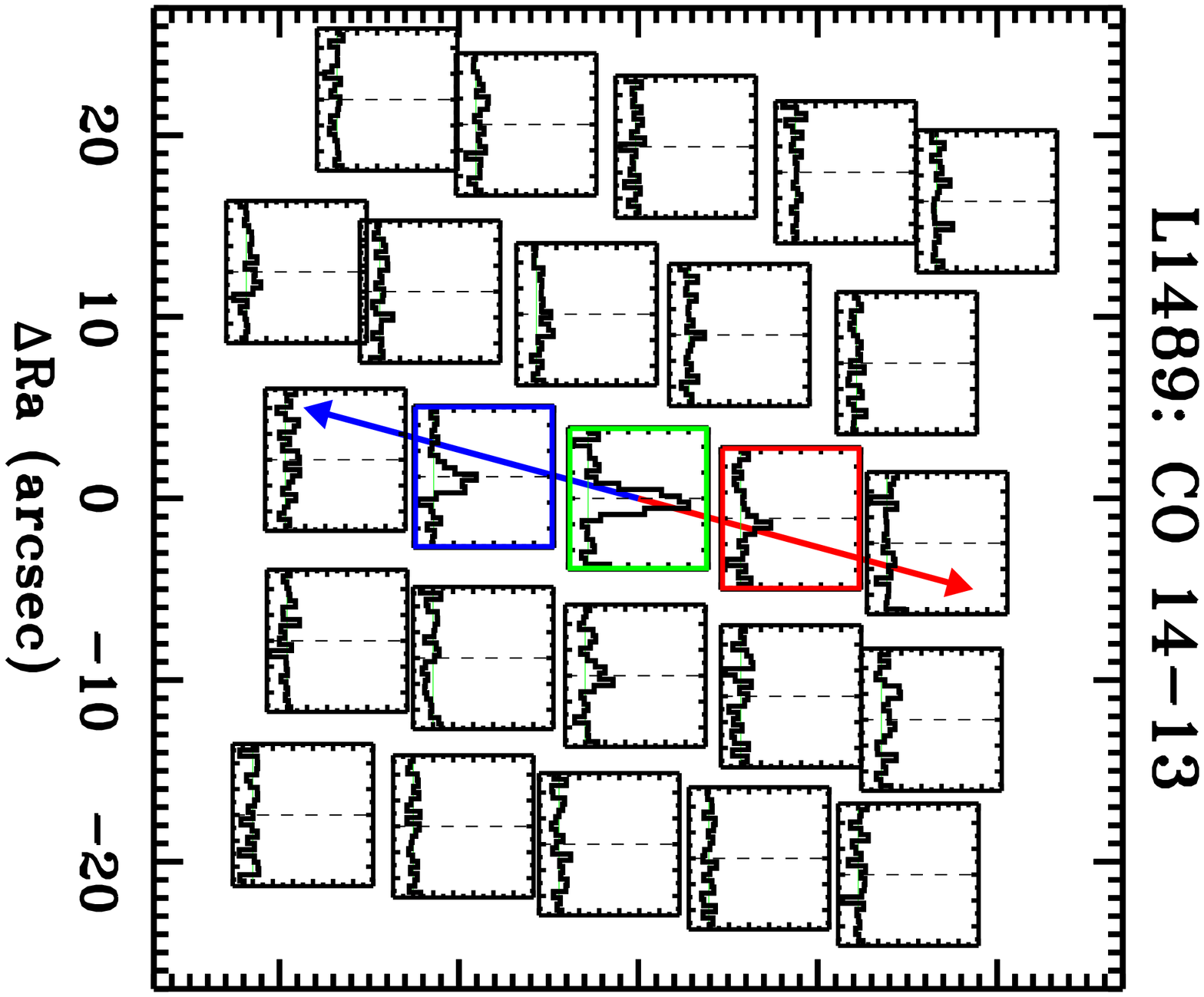}
               \vspace{+3ex}
       
      \includegraphics[angle=90,height=9cm]{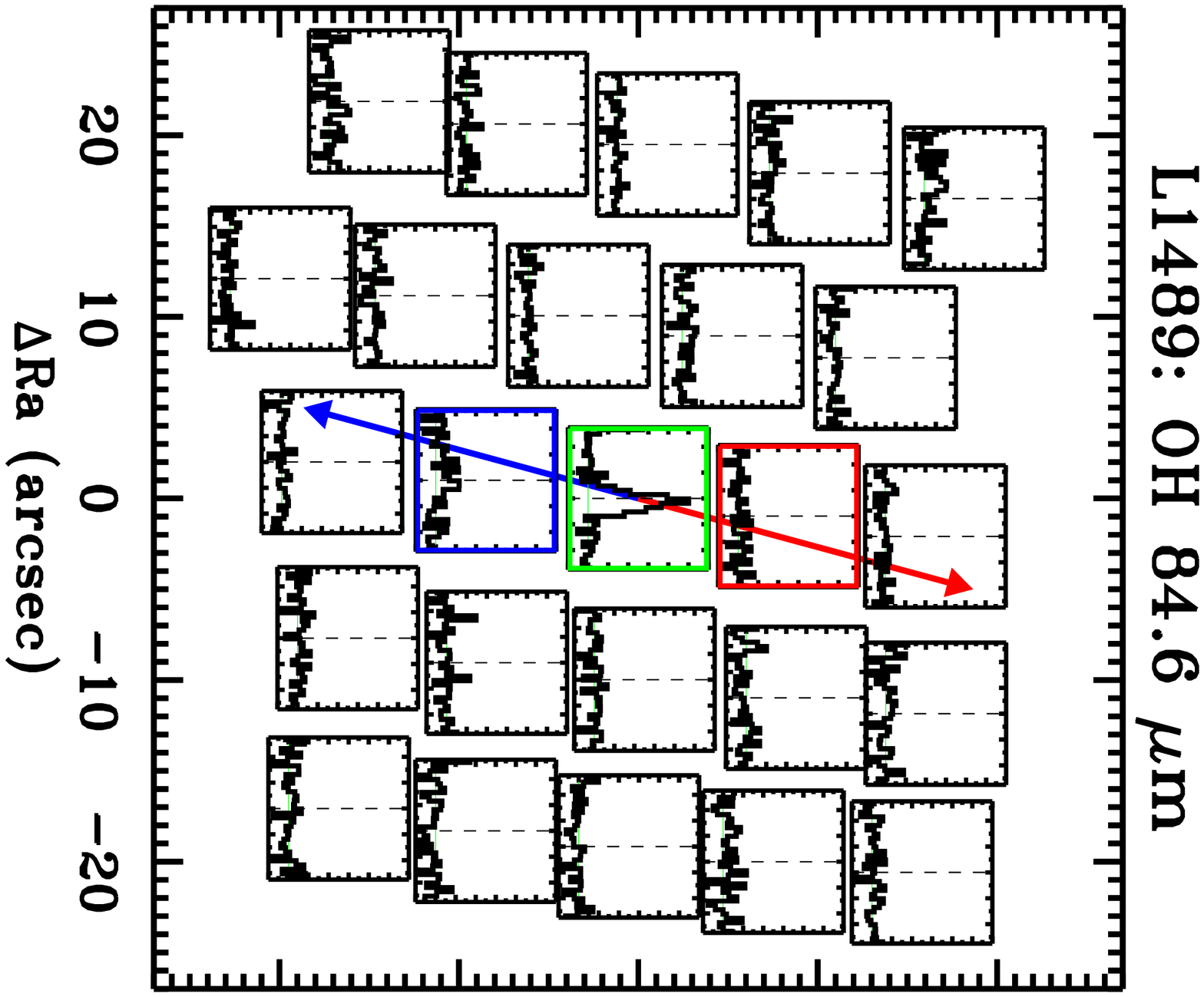}
    \end{center}
  \end{minipage}
  \hfill
  \begin{minipage}[t]{.5\textwidth}
  \begin{center}         
      \includegraphics[angle=90,height=9cm]{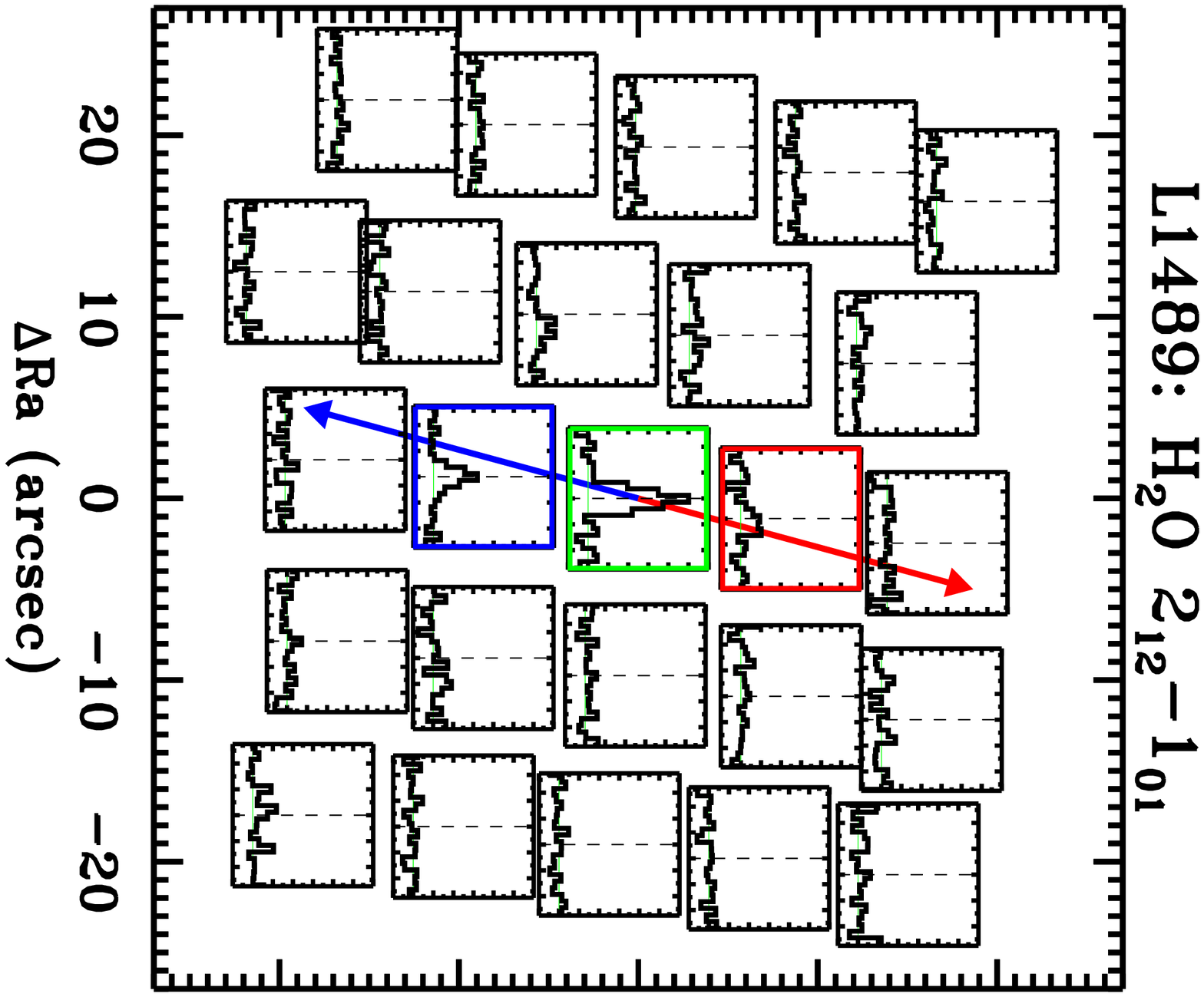}
               \vspace{+3ex}
               
    \includegraphics[angle=90,height=9cm]{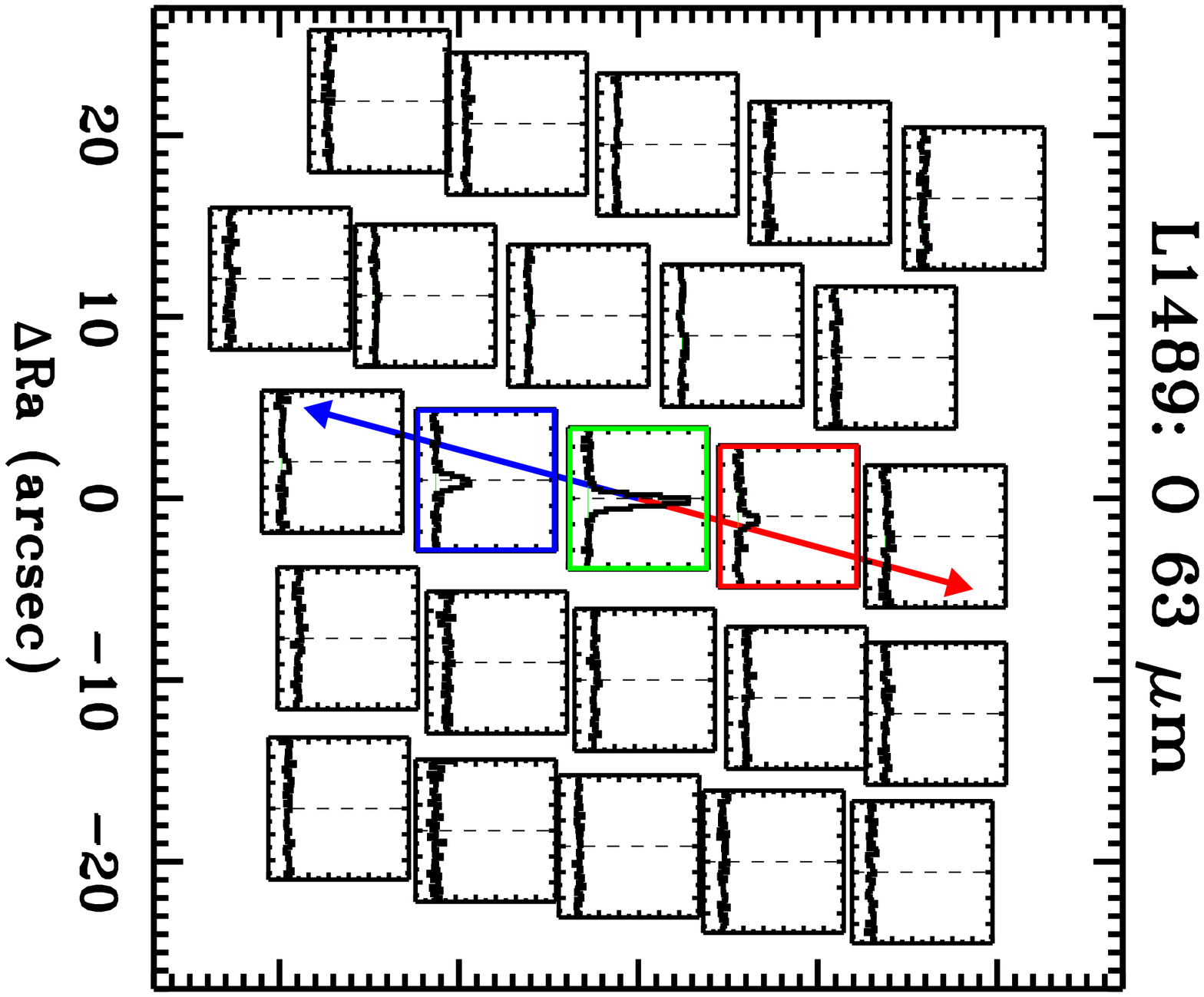}
    \end{center}
  \end{minipage}
        \caption{\label{l1489map} l1489 maps in the [\ion{O}{i}] $^3P_{1}-^{3}P_{2}$ line
        at 63.2 $\mu$m, the H$_2$O 2$_{12}$-1$_{01}$ line at 179.5 $\mu$m, the 
        CO 14-13 at 186.0 $\mu$m and the OH $^{2}\Pi_{\nicefrac{3}{2}}$
        $J=\nicefrac{7}{2}-\nicefrac{5}{2}$ line at 84.6 $\mu$m.}
\end{figure*}

\begin{figure*}[!tb]
  \begin{minipage}[t]{.5\textwidth}
  \begin{center}  
      \includegraphics[angle=90,height=9cm]{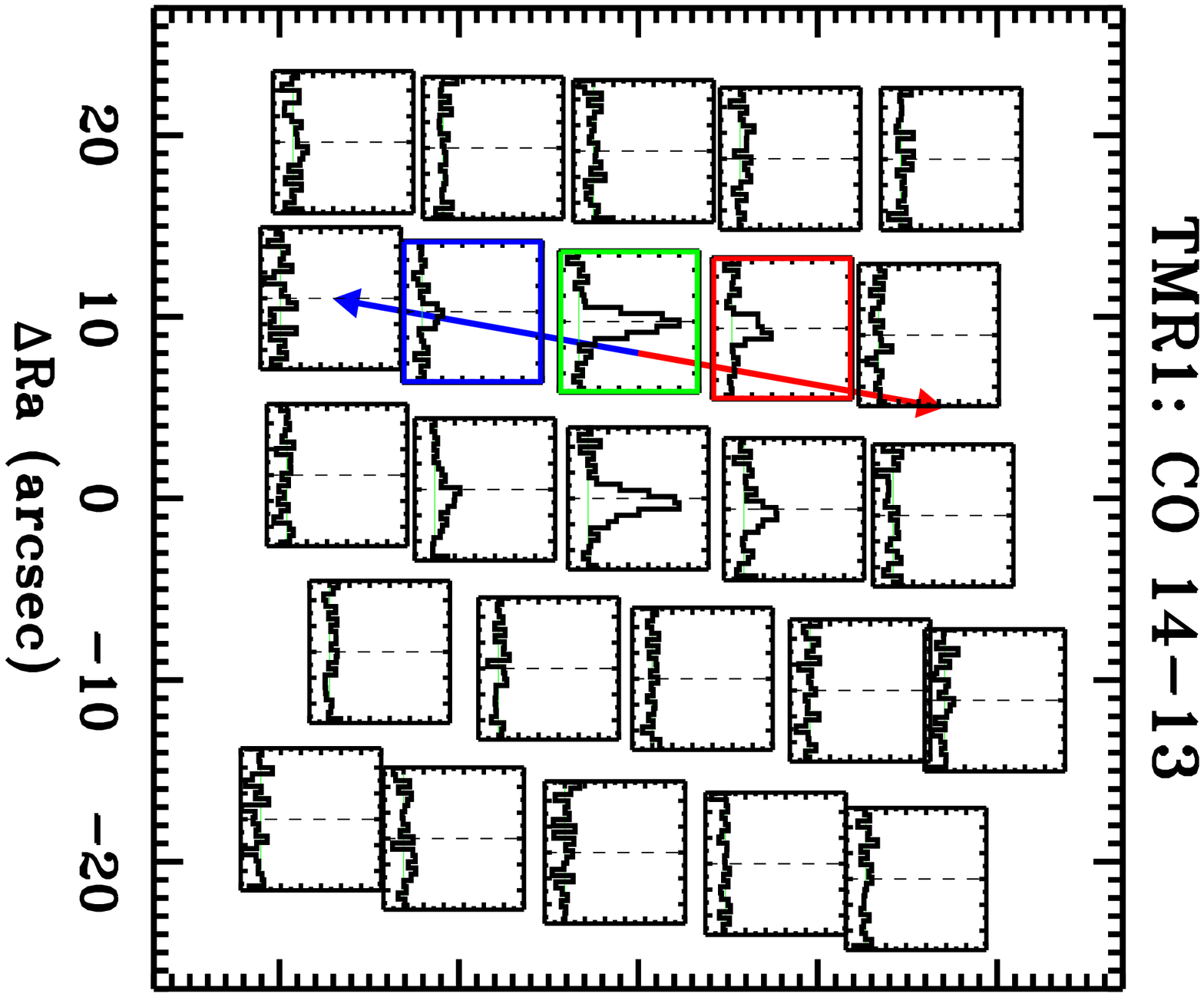}
               \vspace{+3ex}

      \includegraphics[angle=90,height=9cm]{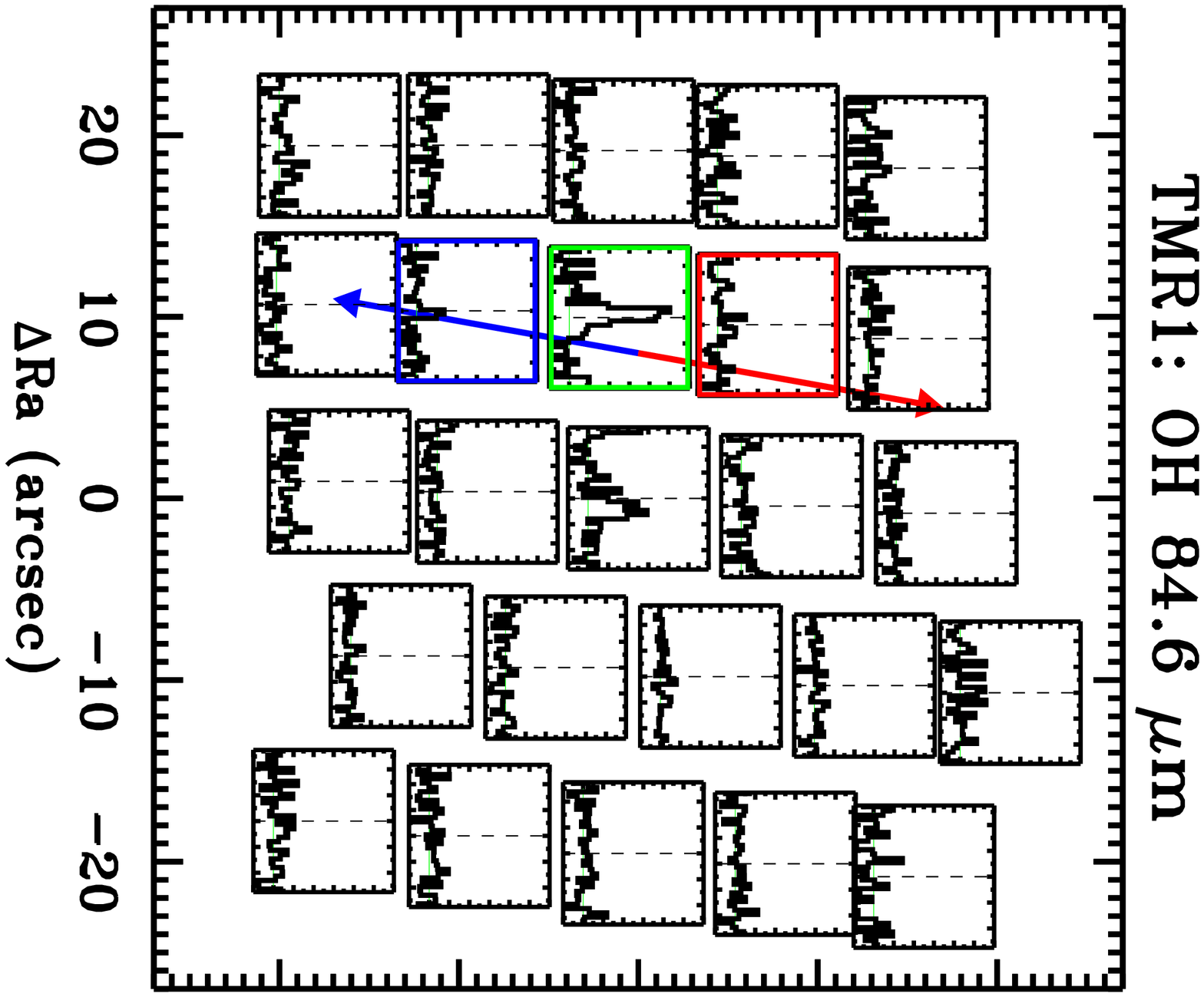}
    \end{center}
  \end{minipage}
  \hfill
  \begin{minipage}[t]{.5\textwidth}
  \begin{center}         
        \includegraphics[angle=90,height=9cm]{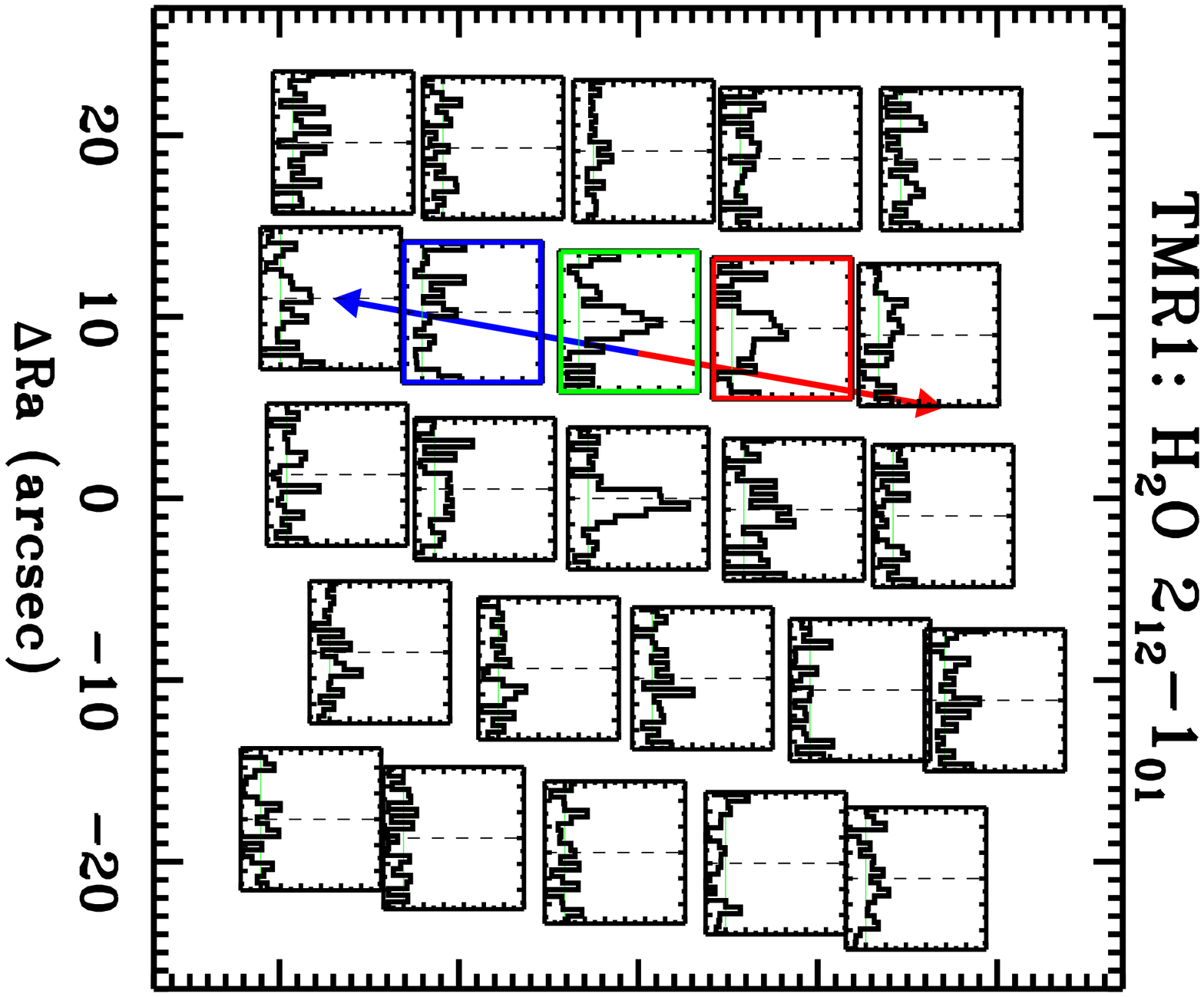}
               \vspace{+3ex}
       
    \includegraphics[angle=90,height=9cm]{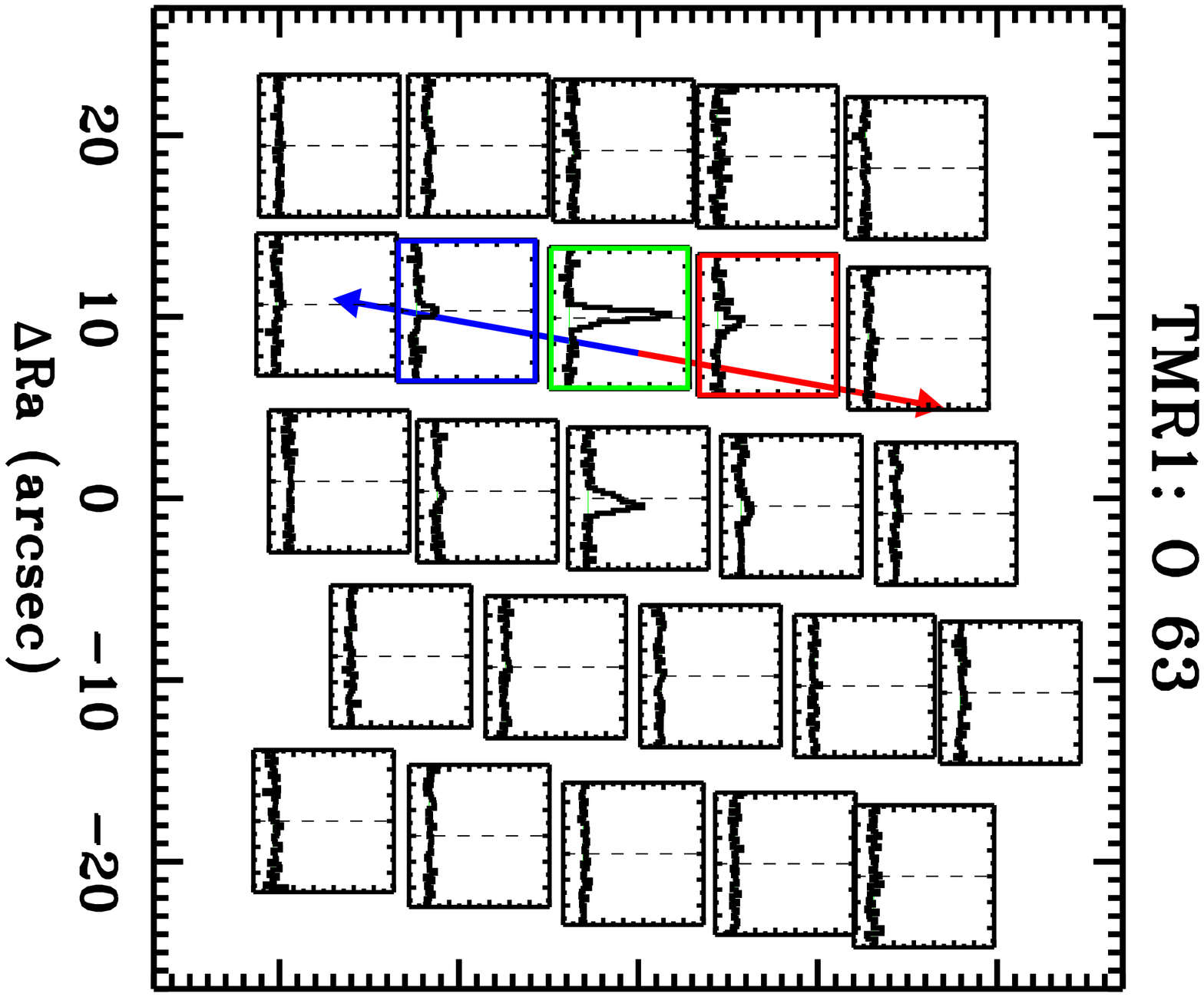}
    \end{center}
  \end{minipage}
         \caption{\label{tmr1map} TMR1 maps in the [\ion{O}{i}] $^3P_{1}-^{3}P_{2}$ line
        at 63.2 $\mu$m, the H$_2$O 2$_{12}$-1$_{01}$ line at 179.5 $\mu$m, the 
        CO 14-13 at 186.0 $\mu$m and the OH $^{2}\Pi_{\nicefrac{3}{2}}$
        $J=\nicefrac{7}{2}-\nicefrac{5}{2}$ line at 84.6 $\mu$m.}
\end{figure*}

Figures \ref{i2amap}-\ref{rno91map} show PACS $5\times5$ maps for all the sources, 
except NGC1333-IRAS4A and L1489 presented in the main text, in the [\ion{O}{i}] $^3P_{1}-^{3}P_{2}$,
H$_2$O 2$_{12}$-1$_{01}$, CO 14-13 and OH $^{2}\Pi_{\nicefrac{3}{2}}$
$J=\nicefrac{7}{2}-\nicefrac{5}{2}$ lines (unless stated otherwise in the captions). In each
 map the CO 6-5 blue and red outflow direction are overplotted for comparison
(Y{\i}ld{\i}z et al.\ in prep.). Color frames show the blue and red outflow positions used 
to create Figures \ref{liness2} and \ref{liness1}. 

%------IRAS 2A -------

%------IRAS 4A--------

%------L1527--------

% =====Ced 

%------BHR71--------

% -=====IRAS 15

%------L483--------

%=====SMM1

%=====SMM4

%=====SMM3

%==== l723

%=====l1489

%======TMR1

%=====TMC1A

\begin{figure*}[tb]
  \begin{minipage}[t]{.5\textwidth}
  \begin{center}  
    \includegraphics[angle=90,height=7cm]{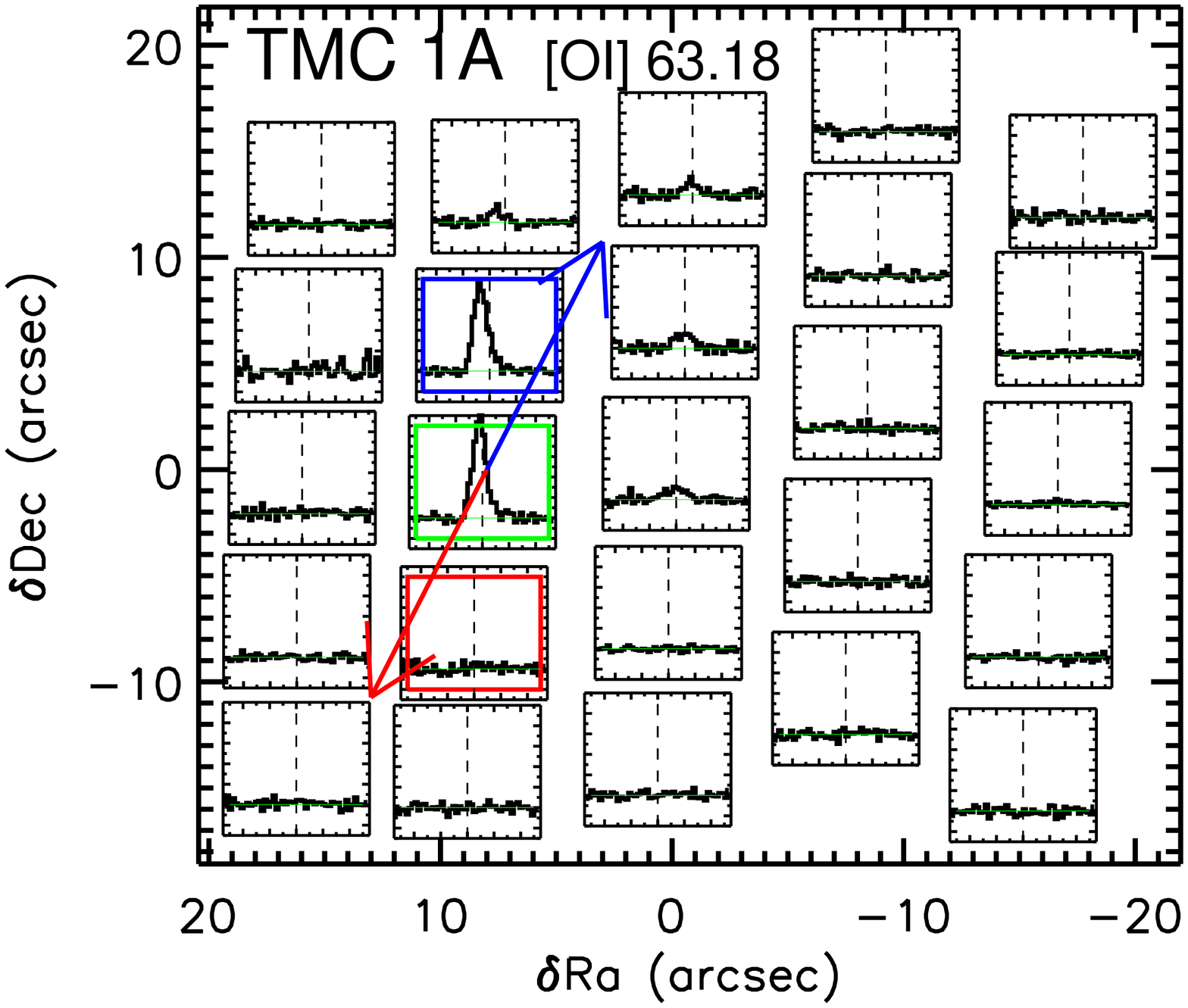}
         \vspace{+5ex}
     
     \includegraphics[angle=90,height=7cm]{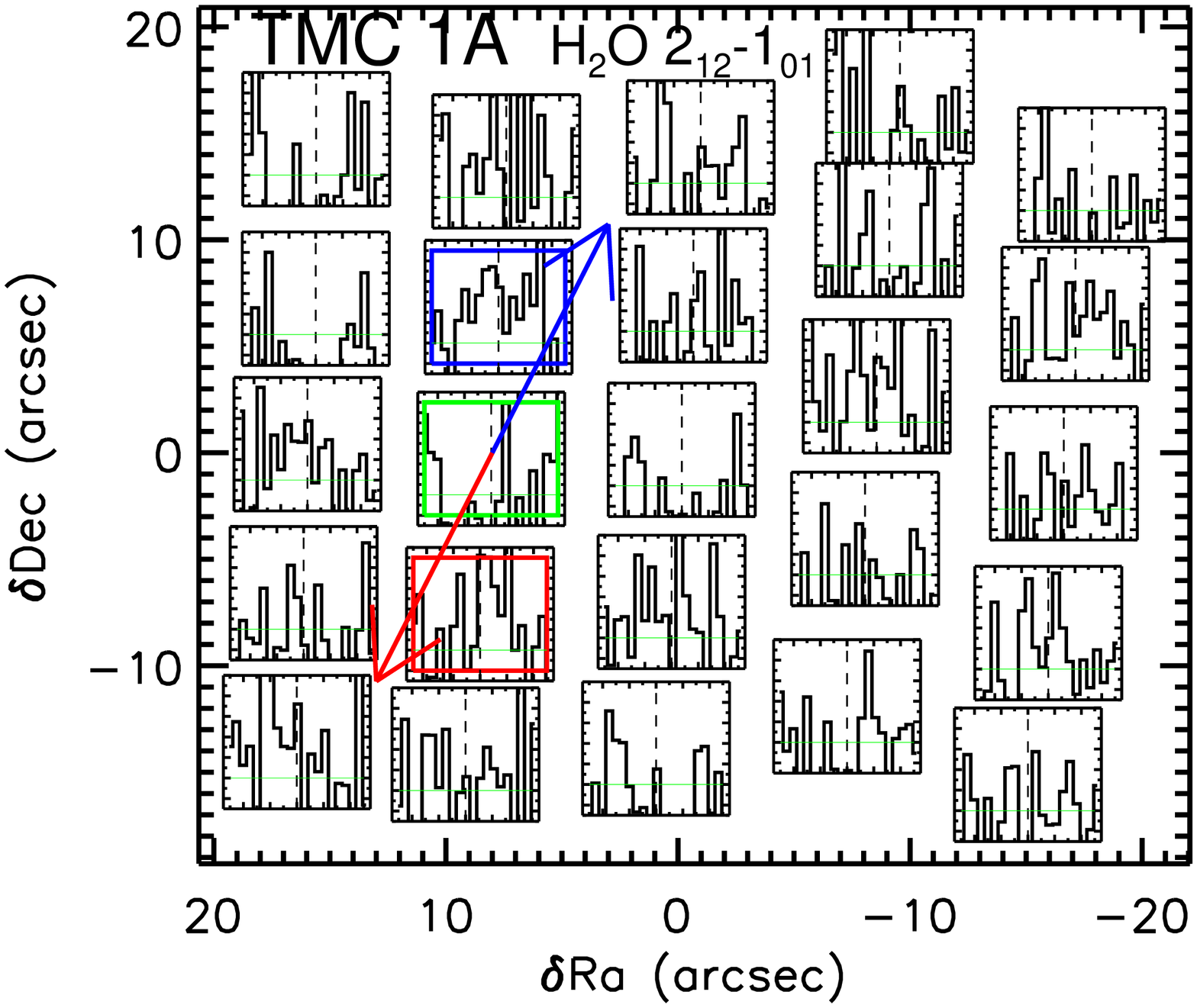}
    \end{center}
  \end{minipage}
  \hfill
  \begin{minipage}[t]{.5\textwidth}
  \begin{center}  
    \includegraphics[angle=90,height=7cm]{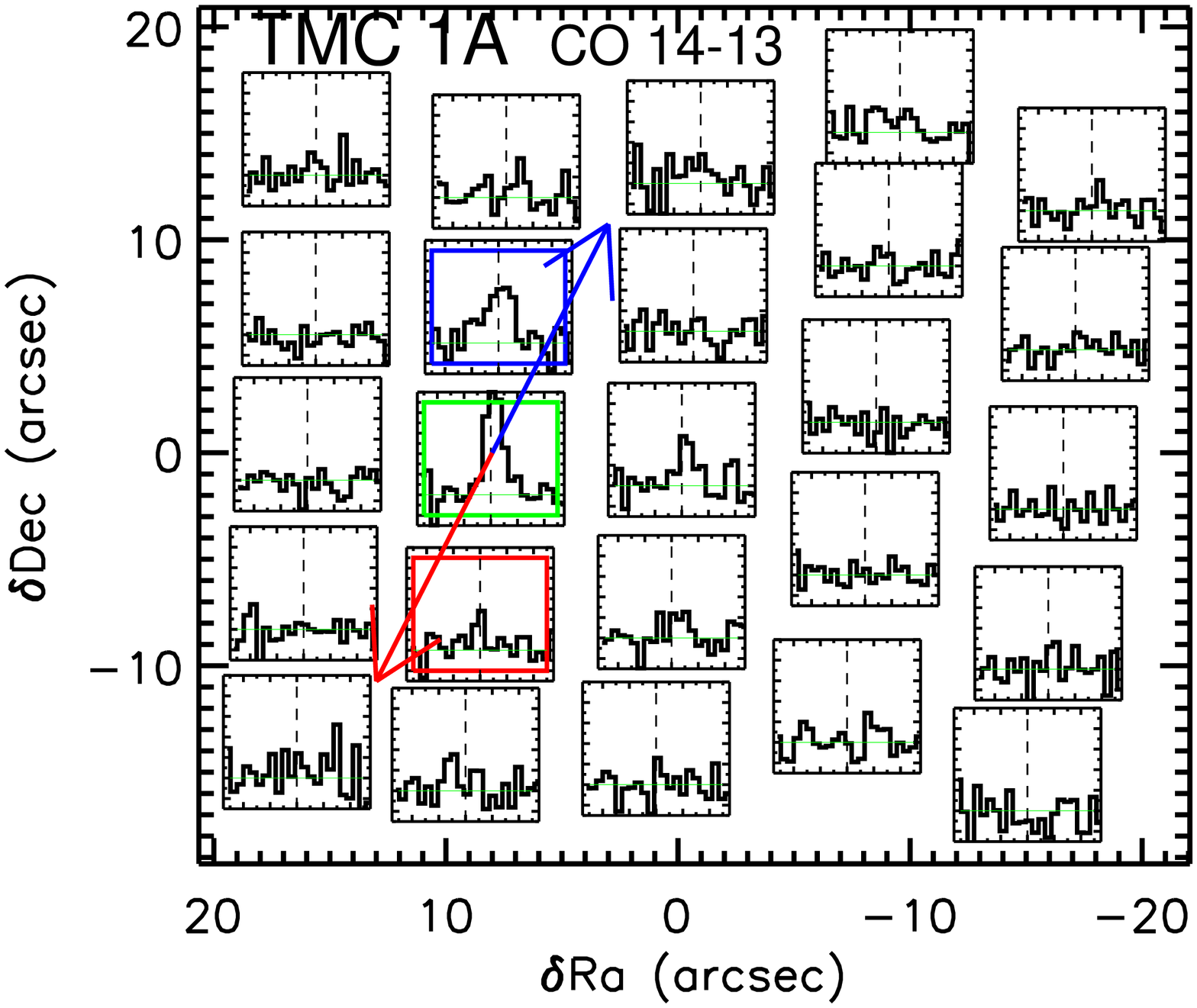}
         \vspace{+5ex}
     
     \includegraphics[angle=90,height=7cm]{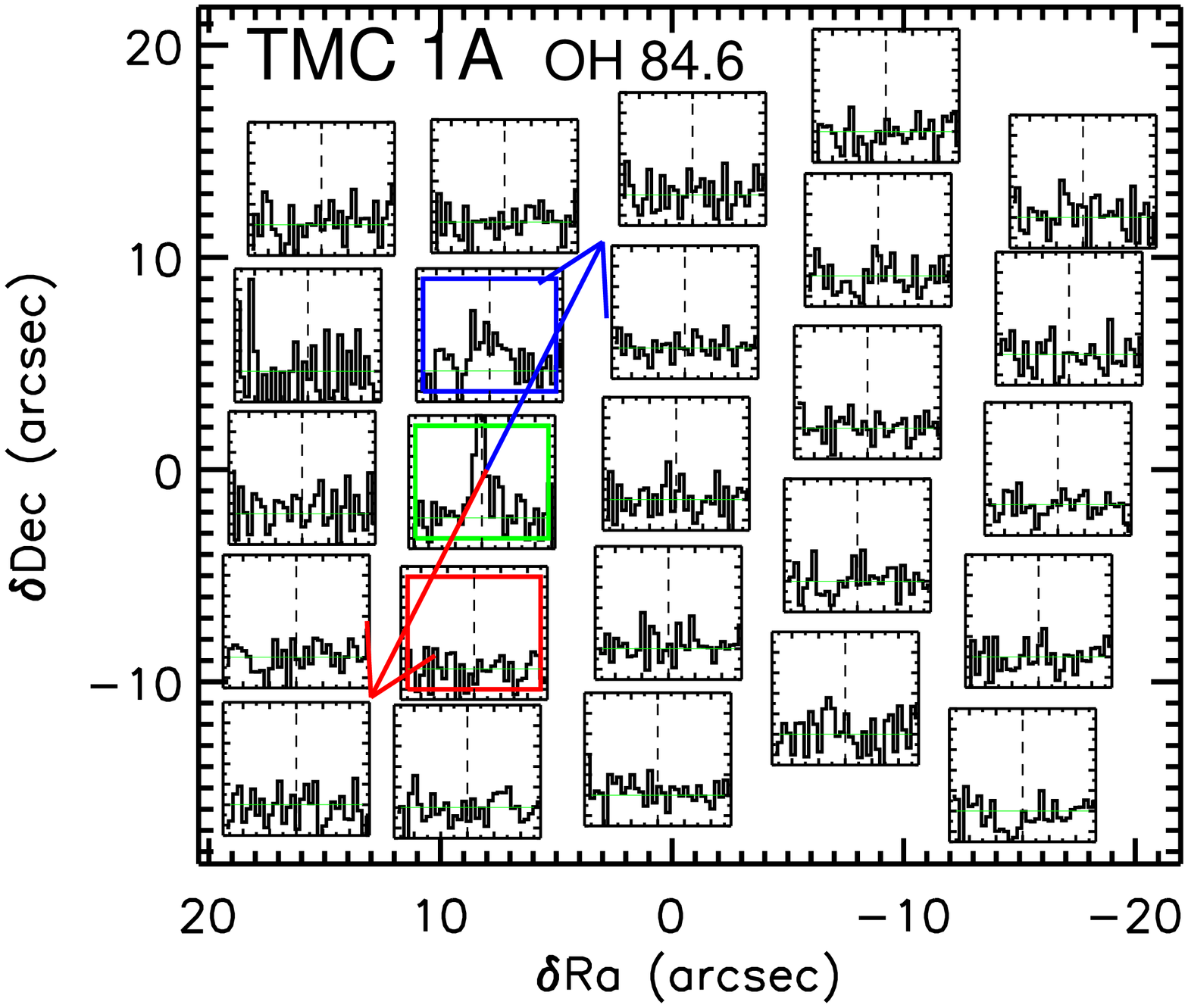}
    \end{center}
  \end{minipage}
 %  \vspace{+3ex}
    \caption{\label{tmc1a}TMC1A maps in the [\ion{O}{i}] $^3P_{1}-^{3}P_{2}$ line
        at 63.2 $\mu$m, the H$_2$O 2$_{12}$-1$_{01}$ line at 179.5 $\mu$m, the 
        CO 14-13 at 186.0 $\mu$m and the OH $^{2}\Pi_{\nicefrac{3}{2}}$
        $J=\nicefrac{7}{2}-\nicefrac{5}{2}$ line at 84.6 $\mu$m.}
\end{figure*}

%=====TMC1
\begin{figure*}[tb]
  \begin{minipage}[t]{.5\textwidth}
  \begin{center}  
    \includegraphics[angle=90,height=7cm]{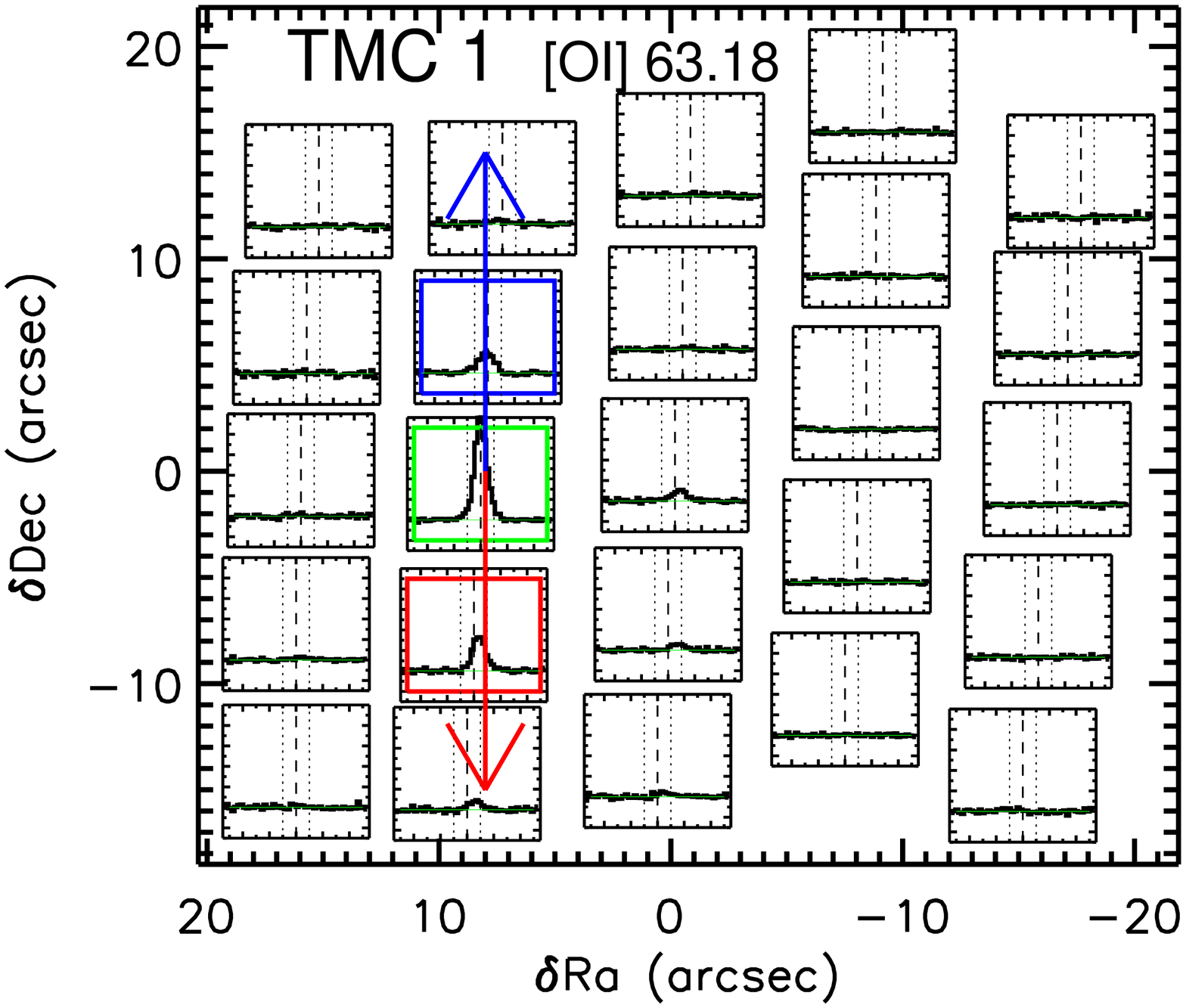}
         \vspace{+5ex}
     
     \includegraphics[angle=90,height=7cm]{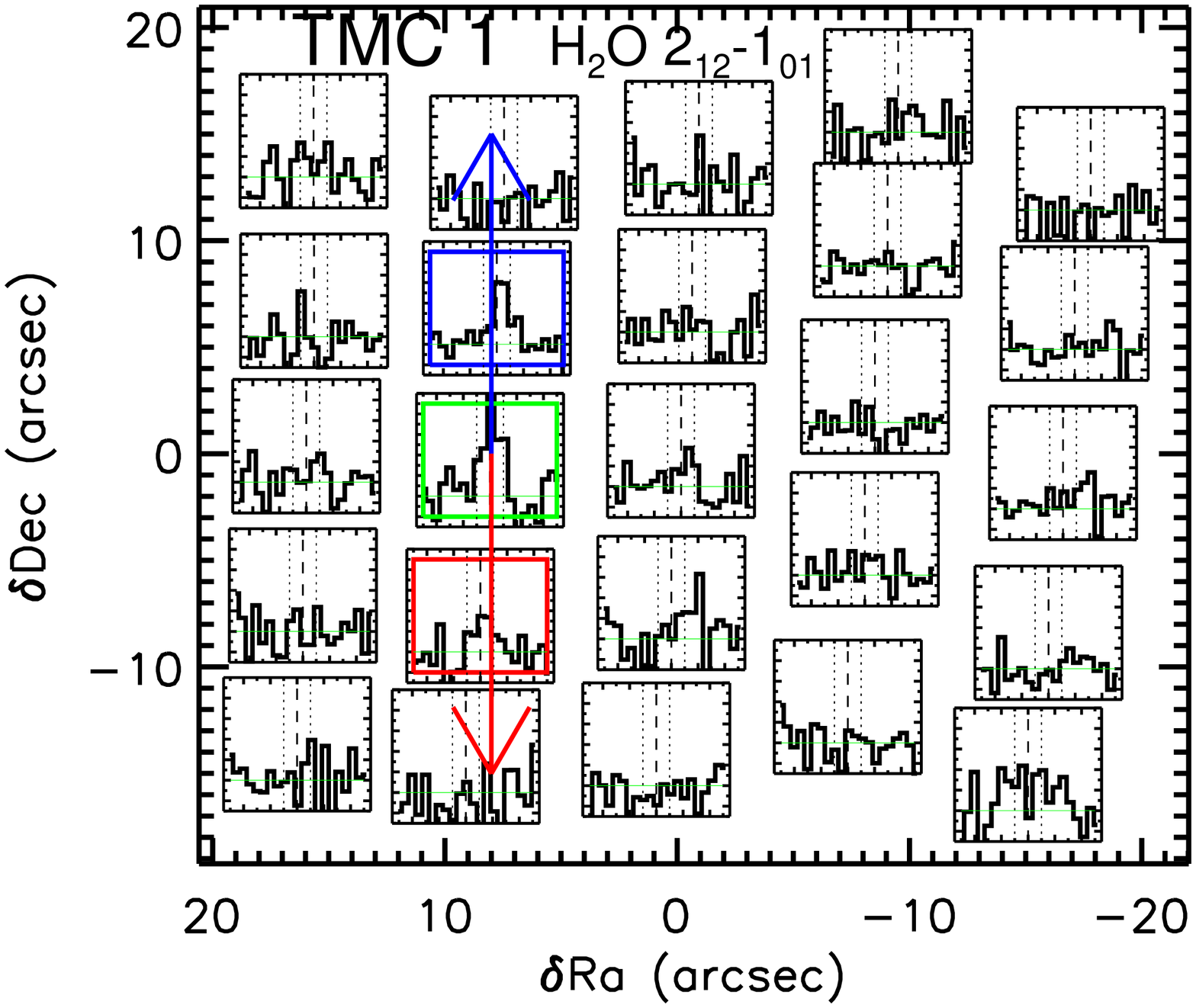}
                   \vspace{+5ex}
     
     \includegraphics[angle=90,height=7cm]{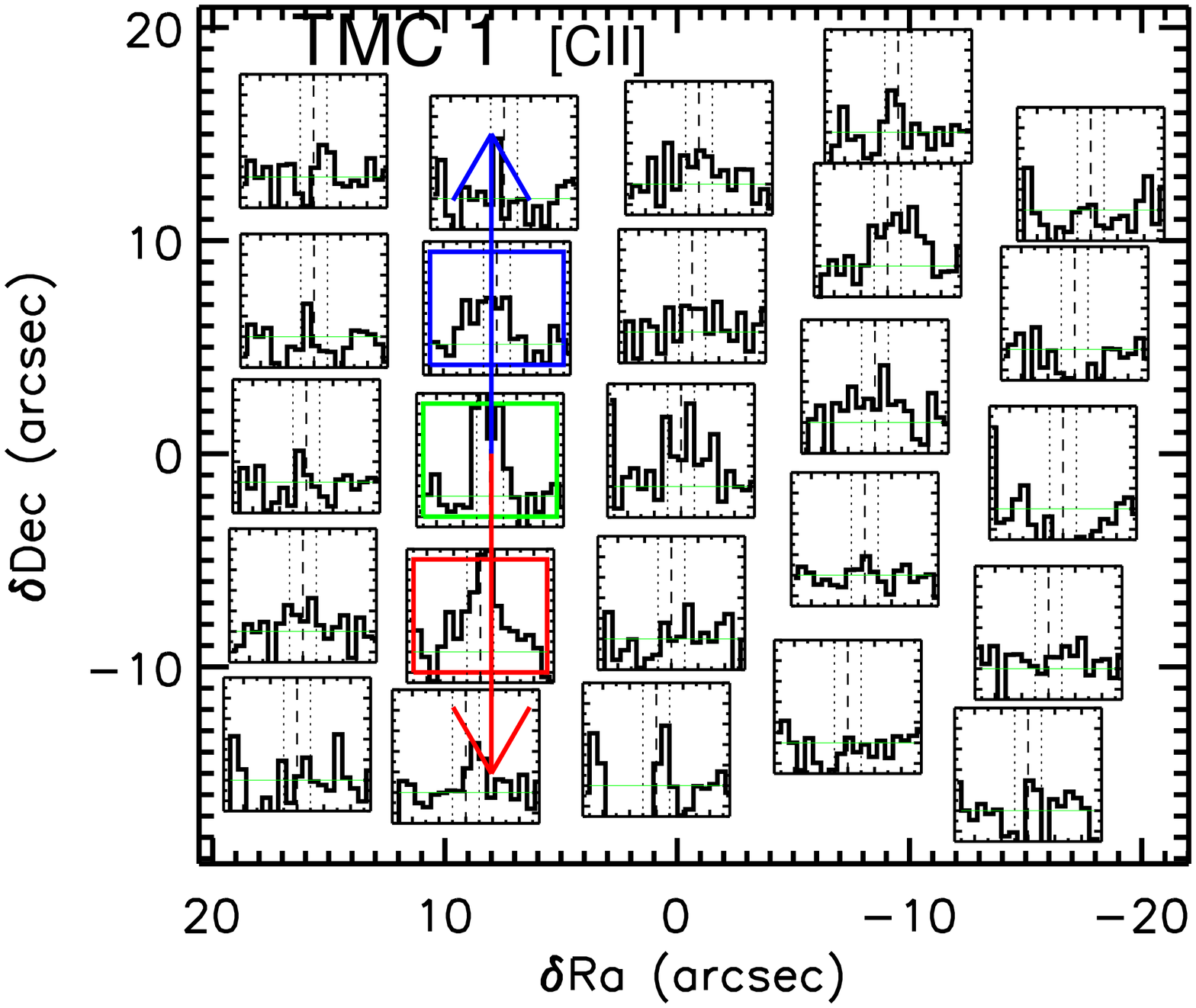}
    \end{center}
  \end{minipage}
  \hfill
  \begin{minipage}[t]{.5\textwidth}
  \begin{center}  
    \includegraphics[angle=90,height=7cm]{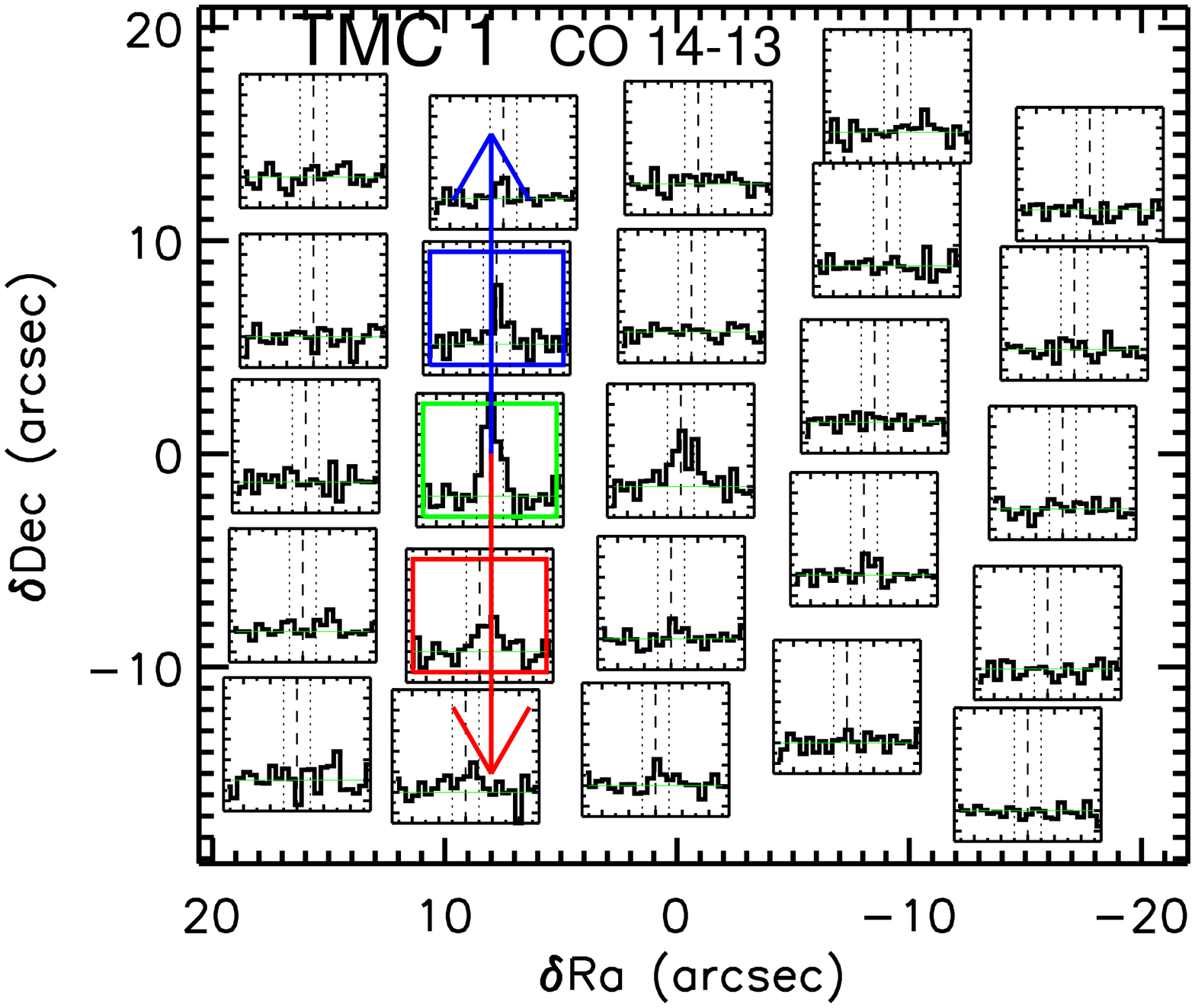}
         \vspace{+5ex}
     
     \includegraphics[angle=90,height=7cm]{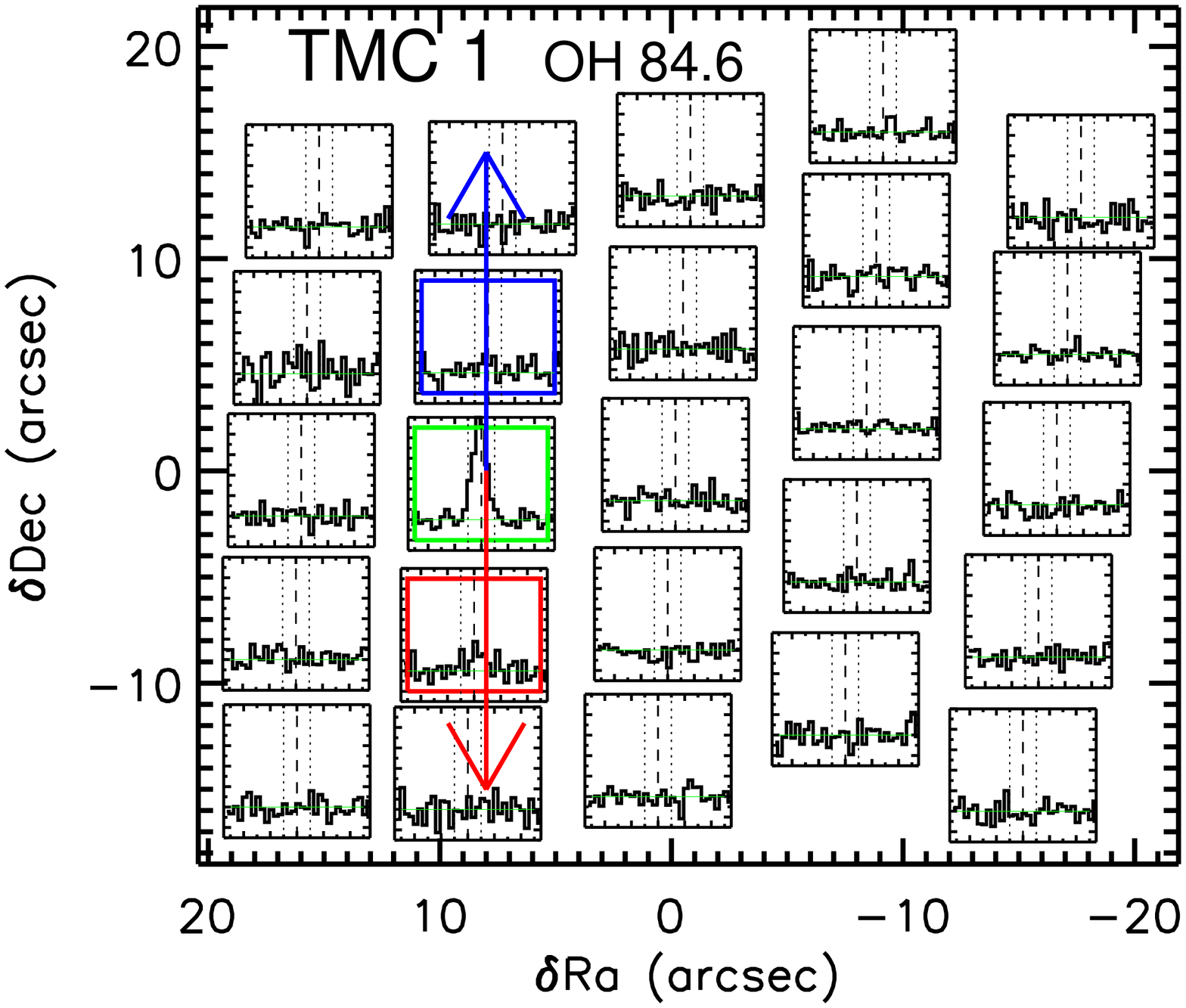}
                   \vspace{+5ex}
     
     \includegraphics[angle=90,height=7cm]{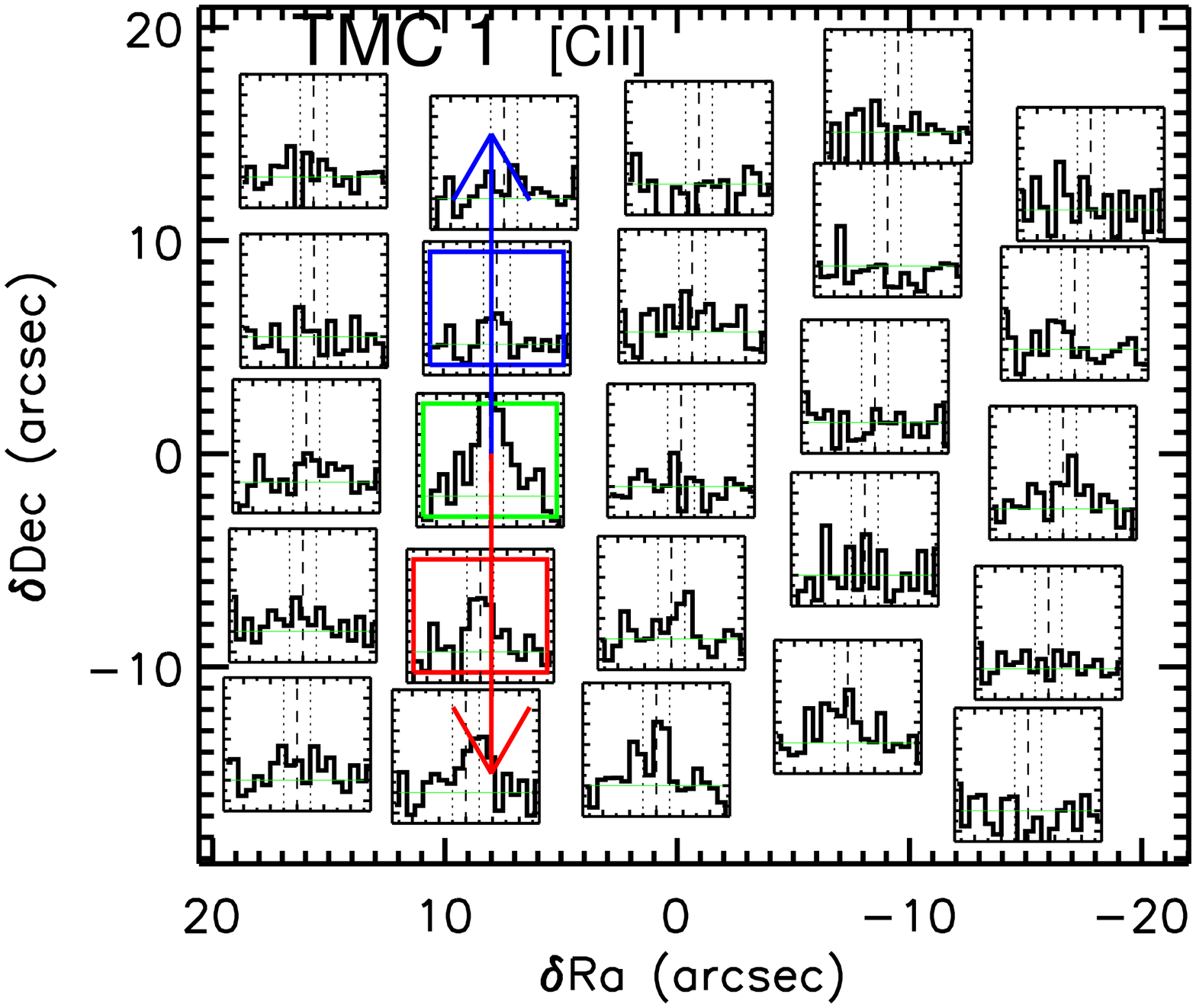}
    \end{center}
  \end{minipage}
 %  \vspace{+3ex}
    \caption{\label{tmc1a}TMC1 maps in the [\ion{O}{i}] $^3P_{1}-^{3}P_{2}$ line
        at 63.2 $\mu$m, the H$_2$O 2$_{12}$-1$_{01}$ line at 179.5 $\mu$m, the 
        CO 14-13 at 186.0 $\mu$m and the OH $^{2}\Pi_{\nicefrac{3}{2}}$
        $J=\nicefrac{7}{2}-\nicefrac{5}{2}$ line at 84.6 $\mu$m.  At the bottom, 
        two nodes of [\ion{C}{ii}] observations are shown.}
\end{figure*}

%=== HH 46

\begin{figure*}[!tb]
  \begin{minipage}[t]{.5\textwidth}
  \begin{center}  
      \includegraphics[angle=90,height=9cm]{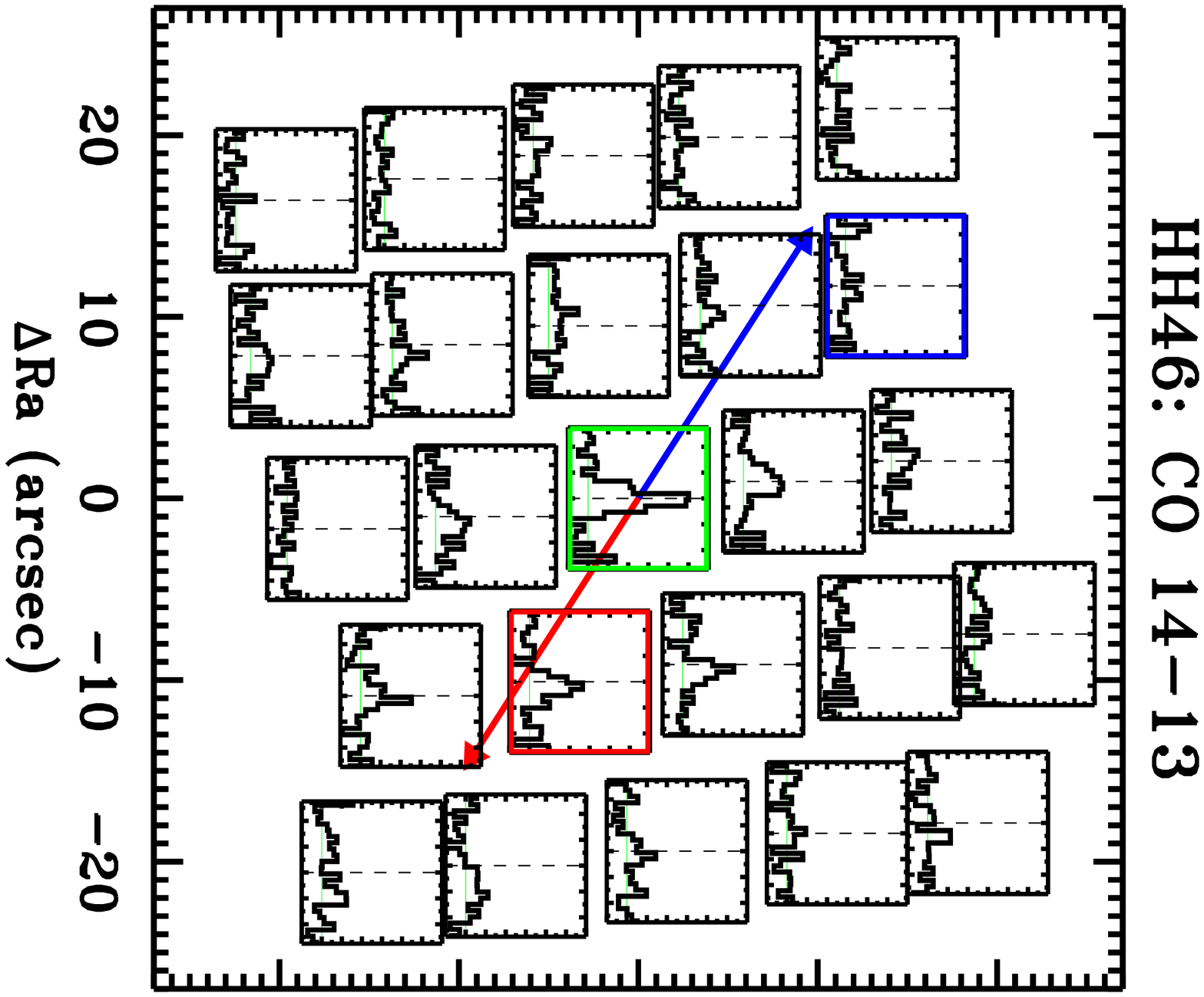}
               \vspace{+3ex}
       
      \includegraphics[angle=90,height=9cm]{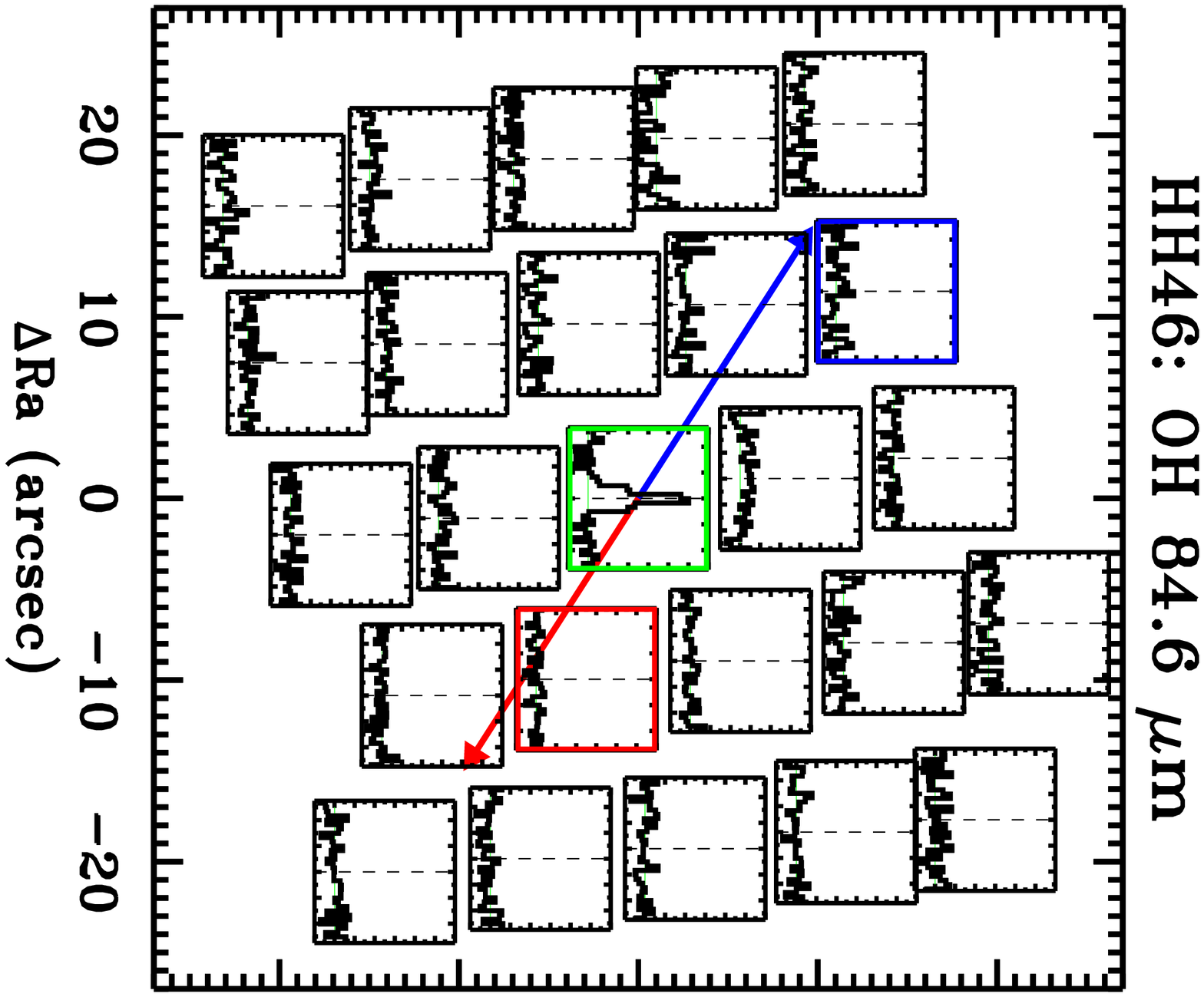}
    \end{center}
  \end{minipage}
  \hfill
  \begin{minipage}[t]{.5\textwidth}
  \begin{center}         
      \includegraphics[angle=90,height=9cm]{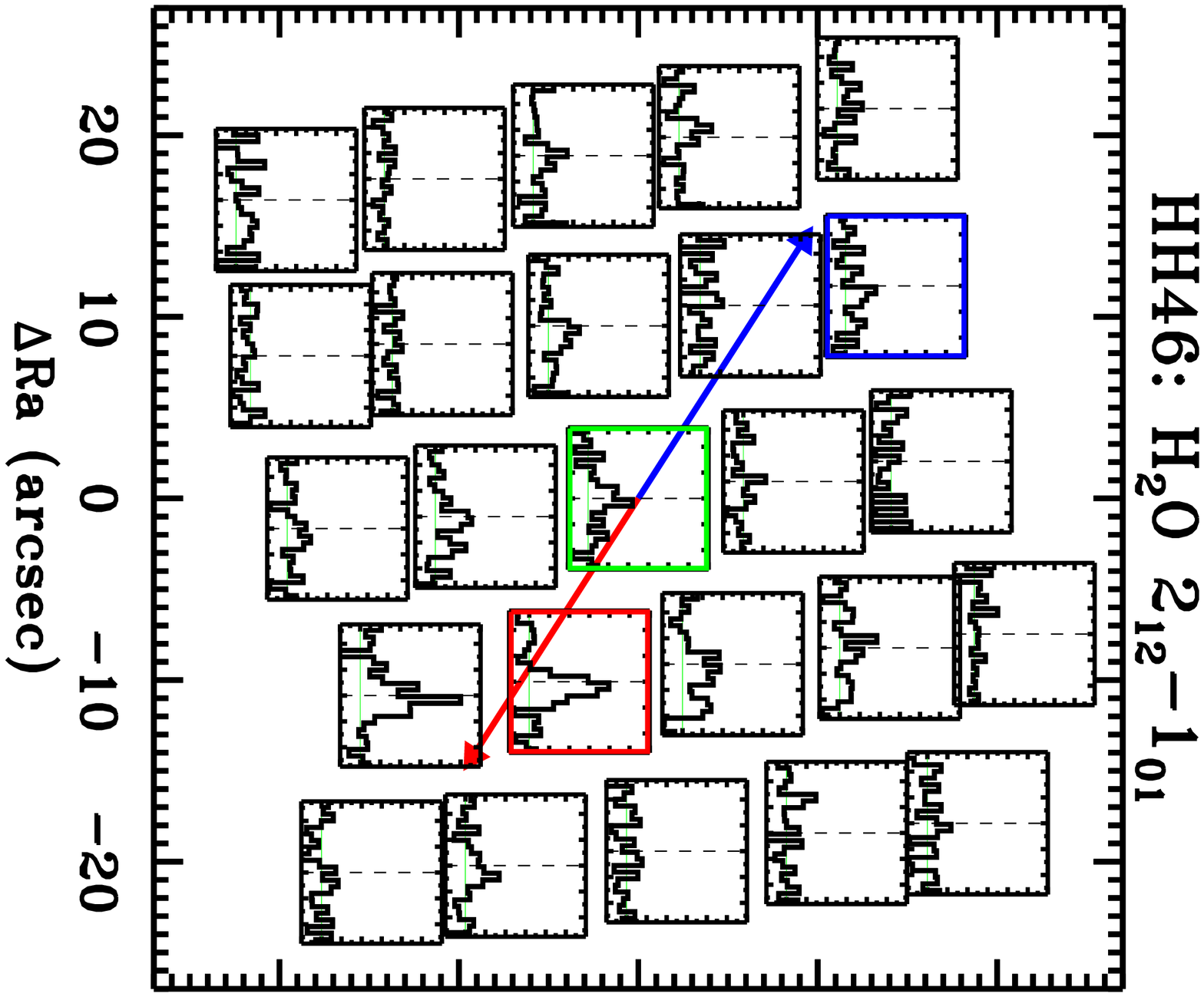}
               \vspace{+3ex}
       
    \includegraphics[angle=90,height=9cm]{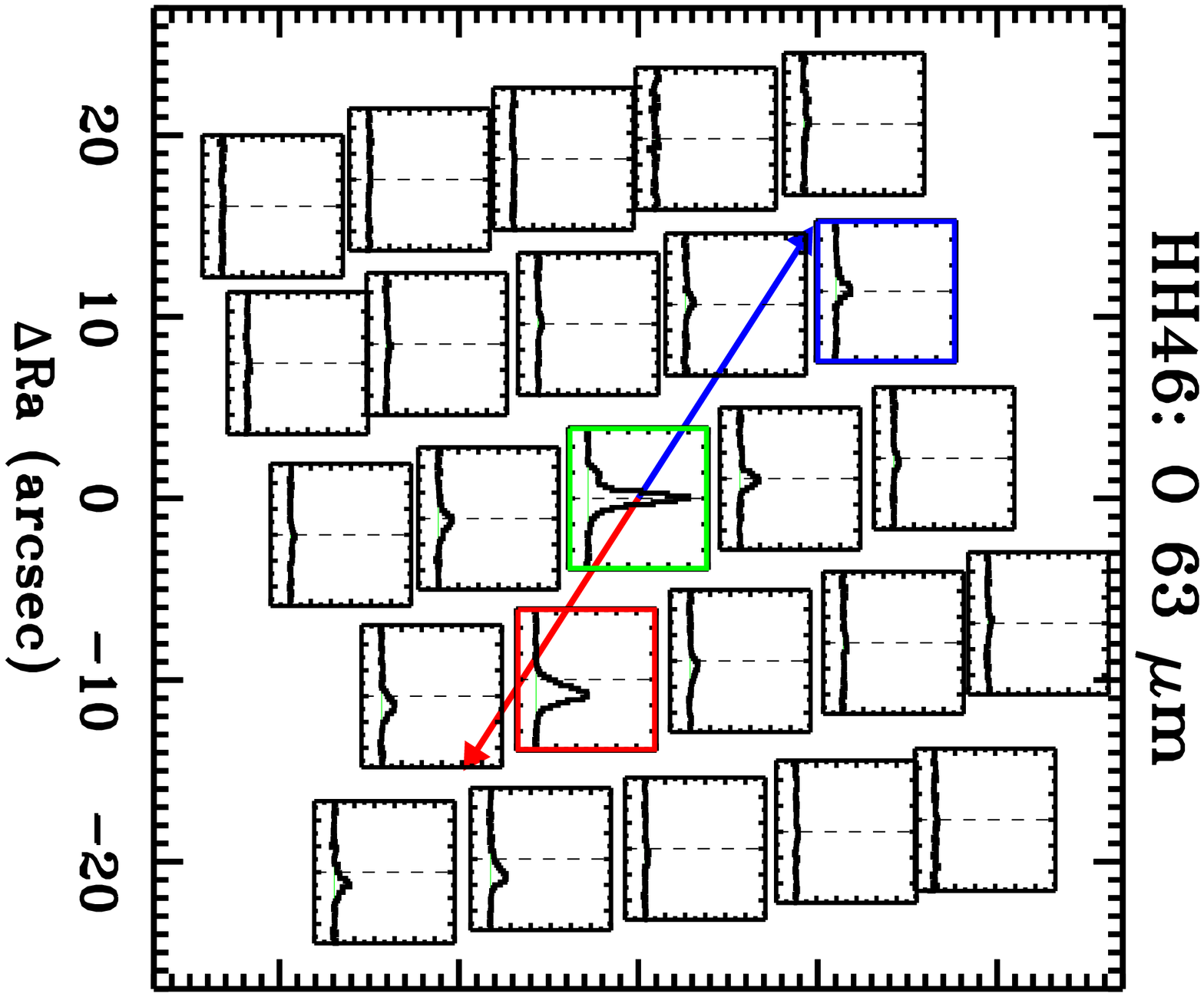}
    \end{center}
  \end{minipage}
        \caption{\label{hh46map} HH46 maps in the [\ion{O}{i}] $^3P_{1}-^{3}P_{2}$ line
        at 63.2 $\mu$m, the H$_2$O 2$_{12}$-1$_{01}$ line at 179.5 $\mu$m, the 
        CO 14-13 at 186.0 $\mu$m and the OH $^{2}\Pi_{\nicefrac{3}{2}}$
        $J=\nicefrac{7}{2}-\nicefrac{5}{2}$ line at 84.6 $\mu$m.}
\end{figure*}

% =====RNO

\begin{figure*}[tb]
  \begin{minipage}[t]{.5\textwidth}
  \begin{center}  
    \includegraphics[angle=90,height=7cm]{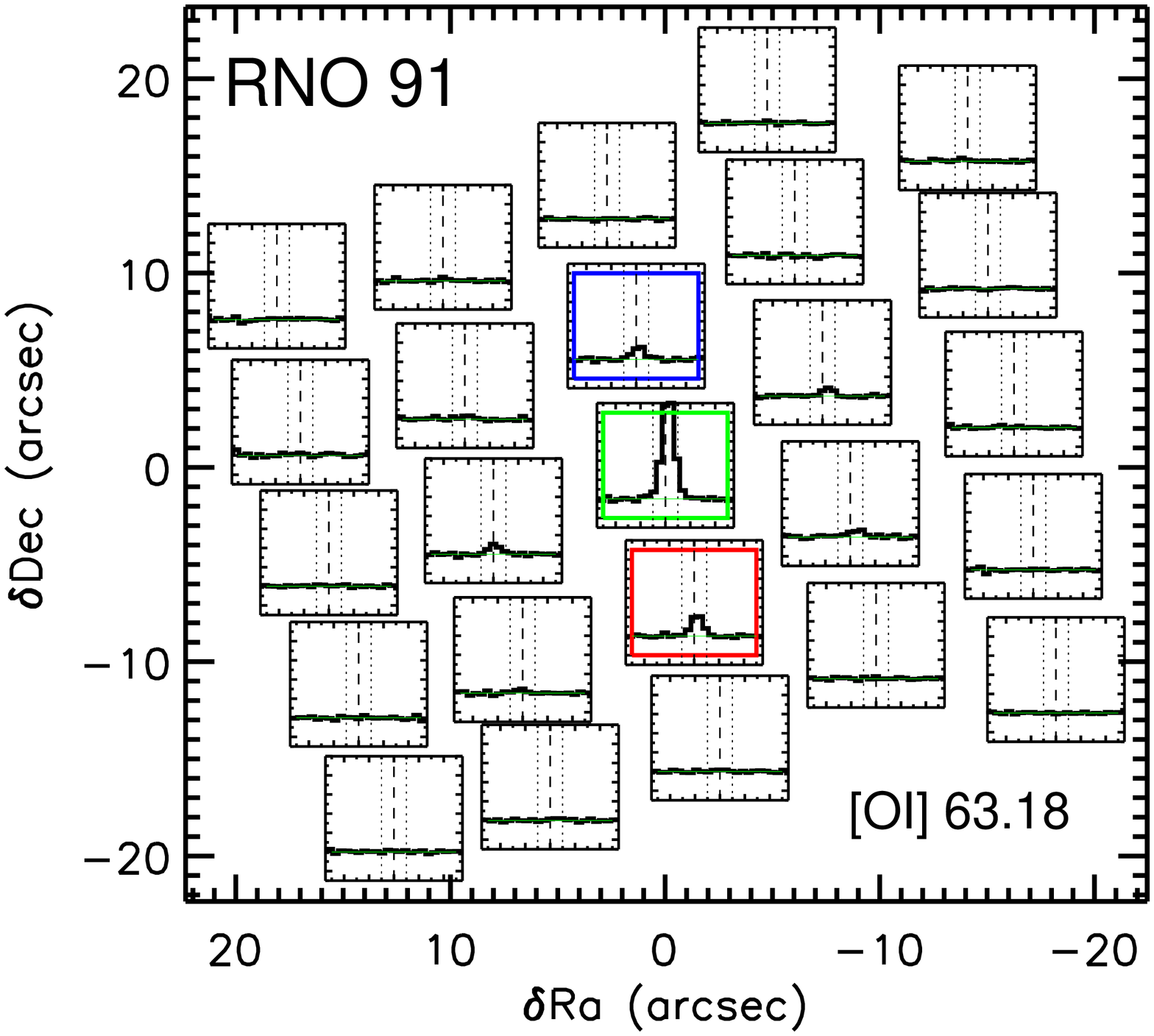}
        % \hspace{+5ex}
                 \vspace{+5ex}
     
     \includegraphics[angle=90,height=7cm]{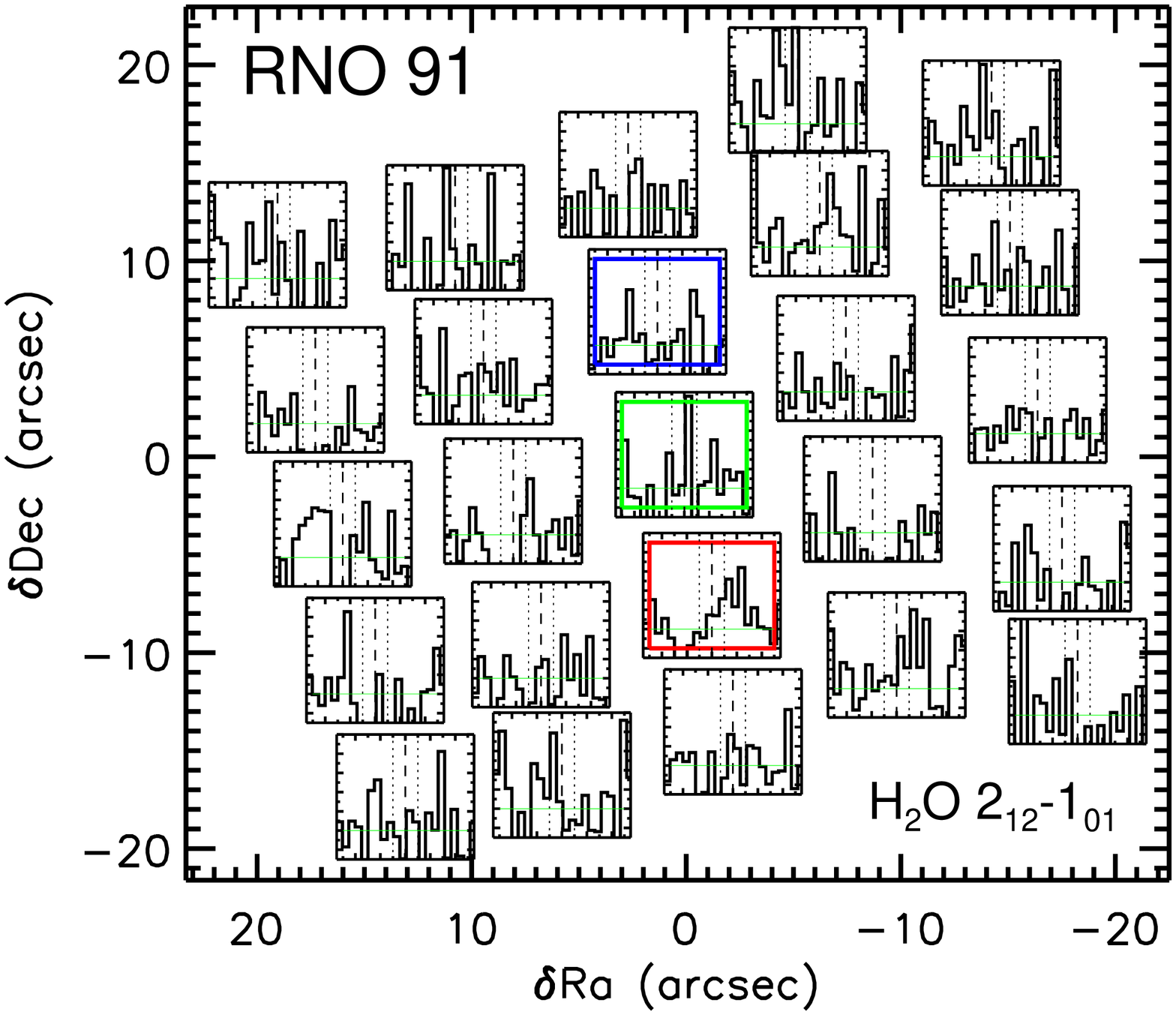}
    \end{center}
  \end{minipage}
  \hfill
  \begin{minipage}[t]{.5\textwidth}
  \begin{center}  
    \includegraphics[angle=90,height=7cm]{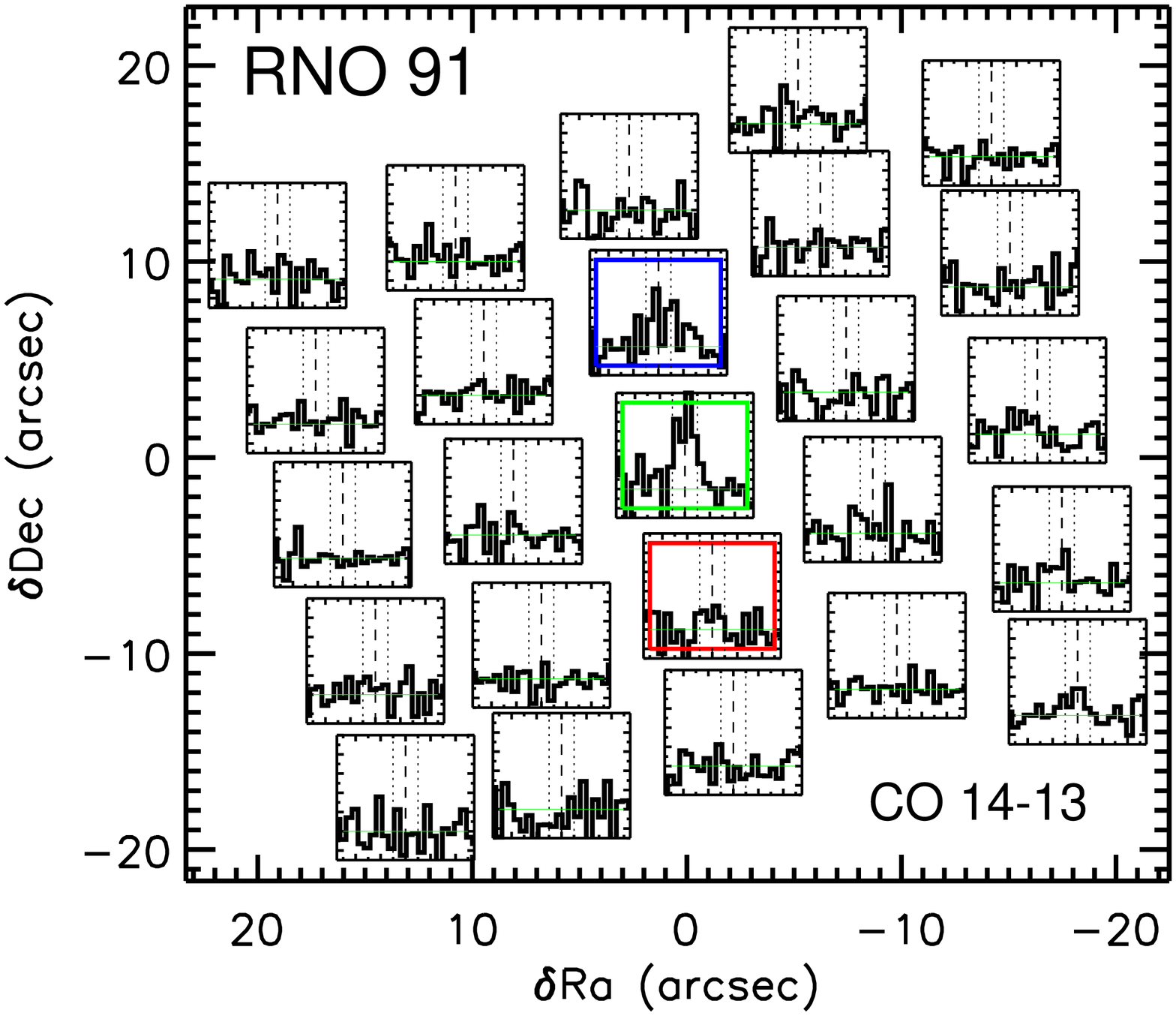}
   %  \hspace{+5ex}
            \vspace{+5ex}
     
     \includegraphics[angle=90,height=7cm]{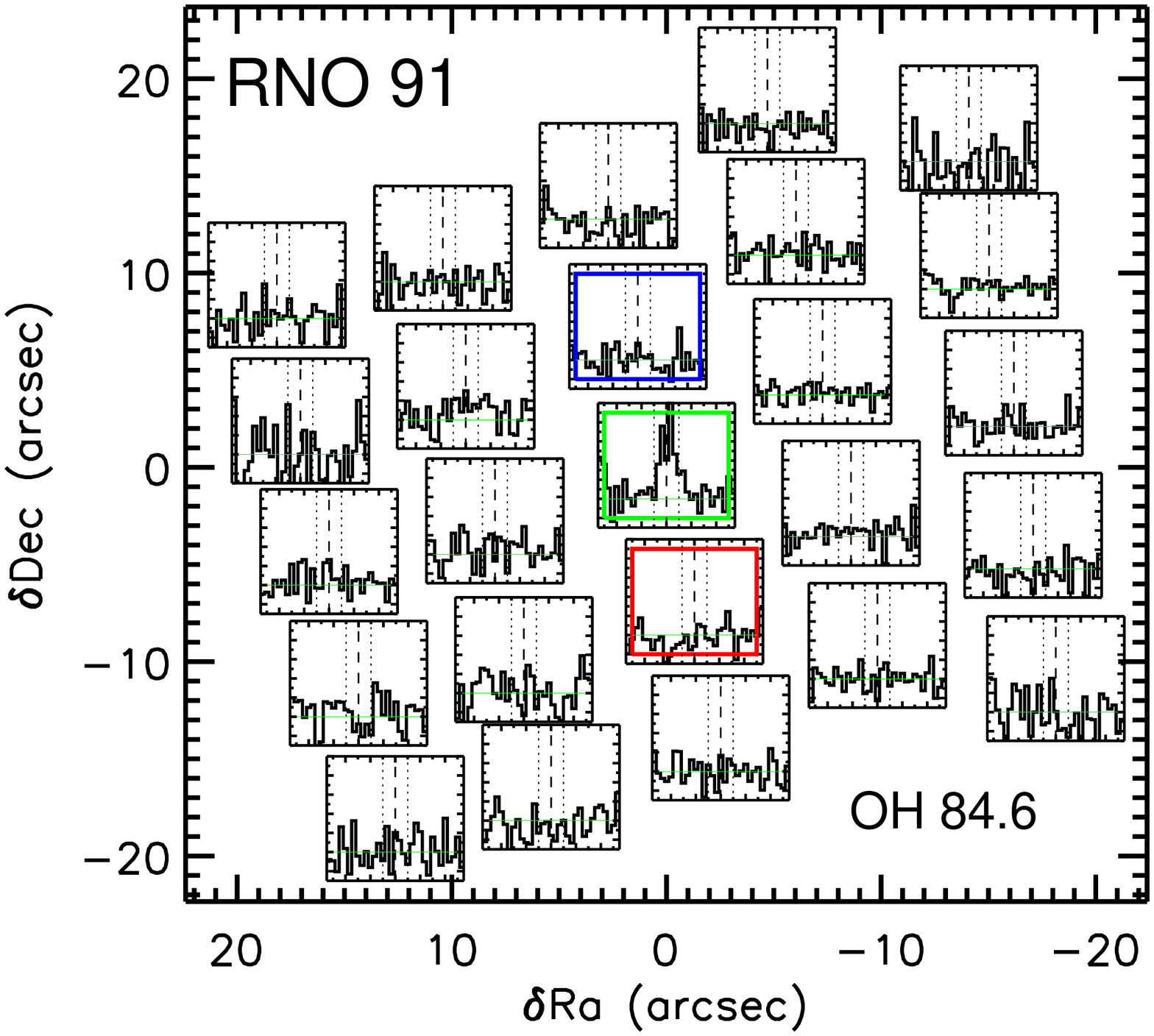}
    \end{center}
  \end{minipage}
 %  \vspace{+3ex}
    \caption{\label{rno91map} RNO91 maps in the [\ion{O}{i}] $^3P_{1}-^{3}P_{2}$ line
        at 63.2 $\mu$m, the H$_2$O 2$_{12}$-1$_{01}$ line at 179.5 $\mu$m, the 
        CO 14-13 at 186.0 $\mu$m and the OH $^{2}\Pi_{\nicefrac{3}{2}}$
        $J=\nicefrac{7}{2}-\nicefrac{5}{2}$ line at 84.6 $\mu$m.}
\end{figure*}

\end{document}